%% file: arXiv_ridesharing.tex
\documentclass[11pt,a4paper]{article}
\pdfoutput=1 
\usepackage[margin=1in]{geometry}

\usepackage{amsfonts}
\usepackage{amsmath}
\usepackage{amssymb}
\usepackage{amsthm}
\usepackage{float}
\usepackage{xspace}
\usepackage[hyphens]{url}

\usepackage[tt=false]{libertine}
\usepackage[libertine,vvarbb]{newtxmath}

\usepackage{bm}

\usepackage{hyperref}
\usepackage[svgnames]{xcolor}
\definecolor{BrightBlue}{RGB}{65, 145, 225}
\definecolor{BlueInFigures}{RGB}{1, 114, 177}
\definecolor{OrangeInFigures}{RGB}{222, 142, 5}
\hypersetup{colorlinks={true},urlcolor={BlueInFigures},linkcolor={BlueInFigures},citecolor=[named]{ForestGreen}}
\usepackage[authoryear,square]{natbib}

\usepackage{microtype}
\usepackage[capitalise,nameinlink,noabbrev]{cleveref}

\usepackage{subcaption}
\usepackage{tikz}
\usetikzlibrary{math}
\usepackage{doi}

\renewcommand{\paragraph}[1]{\medskip\noindent\textbf{#1.\;}}

\renewcommand*{\cite}[1]{\citep{#1}}
\usepackage[utf8]{inputenc}
\usepackage{caption}
\usepackage{graphicx}
\usepackage{booktabs}
\usepackage{algorithm}
\usepackage{thmtools,thm-restate}
\usepackage[noend]{algpseudocode}
\usepackage{enumerate}
\usepackage{enumitem}
\usepackage{cases}
\usepackage{multicol}
\usepackage{multirow}
\usepackage{dsfont}
\usepackage{lscape}
\usepackage{import}
\usepackage{tabularx}

\makeatletter
\newcommand\footnoteref[1]{\protected@xdef\@thefnmark{\ref{#1}}\@footnotemark}
\makeatother

\title{Putting Ridesharing to the Test:\\Efficient and Scalable Solutions and the Power of Dynamic Vehicle Relocation\thanks{A version of this paper has been published in the Artificial Intelligence Review~\citep{Danassis2022Putting}.}}

\author{
\begin{tabular}{c c}
& \\ \textbf{Panayiotis Danassis\thanks{Corresponding author.}} & \textbf{Marija Sakota}\\
\small{\'Ecole Polytechnique F\'ed\'eralede Lausanne (EPFL)} & \small{\'Ecole Polytechnique F\'ed\'erale de Lausanne (EPFL)}\\
\small{Switzerland} & \small{Switzerland}\\
\href{mailto:panayiotis.danassis@alumni.epfl.ch}{\small{\texttt{panayiotis.danassis@alumni.epfl.ch}}} & \href{mailto:marija.sakota@epfl.ch}{\small{\texttt{marija.sakota@epfl.ch}}}\\
& \\
\textbf{Aris Filos-Ratsikas} & \textbf{Boi Faltings}\\
\small{University of Liverpool} & \small{\'Ecole Polytechnique F\'ed\'erale de Lausanne (EPFL)}\\
\small{United Kingdom} & \small{Switzerland}\\
\href{mailto:Aris.Filos-Ratsikas@liverpool.ac.uk}{\small{\texttt{aris.filos-ratsikas@liverpool.ac.uk}}} & \href{mailto:boi.faltings@epfl.ch}{\small{\texttt{boi.faltings@epfl.ch}}}\\
\end{tabular}
}

\date{}

\begin{document}
\maketitle

\begin{abstract}
	We study the optimization of large-scale, real-time ridesharing systems and propose a modular design methodology, \emph{Component Algorithms for Ridesharing} (CAR). We evaluate a diverse set of CARs (14 in total), focusing on the \emph{key algorithmic components} of ridesharing. We take a multi-objective approach, evaluating 12 metrics related to \emph{global efficiency}, \emph{complexity}, \emph{passenger}, \emph{driver}, and \emph{platform} incentives, in settings designed to \emph{closely resemble reality} in every aspect, focusing on vehicles of capacity two. To the best of our knowledge, this is the \emph{largest} and most \emph{comprehensive} evaluation to date. We (i) identify CARs that perform well on global, passenger, driver or platform metrics, (ii) demonstrate that lightweight \emph{relocation schemes} can significantly improve the Quality of Service by up to $50\%$, and (iii) highlight a \emph{practical}, \emph{scalable}, \emph{on-device} CAR that works well across all metrics.
\end{abstract}

\section{Introduction} \label{Introduction}

The emergence and widespread use of \emph{Mobility-on-Demand} systems in recent years has had a profound impact on urban transportation in a variety of ways. Amongst other advantages, these systems have the potential to mitigate congestion costs (such as commute times, fuel usage, accident propensity, etc.), enable marketplace optimization for both passengers and drivers, and provide great environmental benefits. A prominent such example is ridesharing\footnote{Throughout the paper, we use the term `ridesharing' to refer to passengers (potentially) using the same vehicle at the same time, also referred to as `ridepooling' \cite{doi:10.1080/01441647.2018.1497728}.}. Ridesharing however results in some passenger disruption as well, due to compromise in flexibility, increased travel time, and loss of privacy and convenience. Thus, in the core of any ridesharing platform lies the need for an efficient balance between the incentives\footnote{We remark that the term `incentive' is not used in the strict game-theoretic sense.} of the passengers, the drivers, and those of the platform.


Optimizing the usage of transportation resources is not an easy task, especially for cities like New York, with more than $13000$ taxis and $270$ ride requests per minute. For example, \cite{buchholz2018spatial} estimates that $45000$ customer requests remain unmet each day in New York, despite the fact that approximately $5000$ taxis are vacant at any time. In fact, on aggregate, drivers spend about $47\%$ of their time not serving any passengers \cite{buchholz2018spatial}. Moreover, up to $80\%$ of the taxi rides in Manhattan could be shared by two riders, with only a few minutes increase in travel time \cite{alonso2017demand}. A \emph{more sophisticated matching policy} could mitigate these costs by better allocating available supply to demand. As a second example, \emph{coordinated vehicle relocation} could also be employed to bridge the gap on the spatial supply/demand imbalance and improve passenger satisfaction and Quality of Service (QoS) metrics. Drivers often relocate to find passengers: $61.3\%$ of trips begin in a different neighborhood than the drop-off location of the last passenger \cite{buchholz2018spatial}, yet currently drivers move without any coordinated search behavior, resulting in spatial search frictions.

Given the importance of the problem for transportation and the economy, it is not surprising that the related literature is populated with a plethora of papers, proposing different solutions along different axes, such as efficiency \cite{santi2014quantifying,alonso2017demand,agatz2011dynamic,ashlagi2017min,huang2019dynamic,bienkowski2018primal,dickerson2018allocation,fagnant2018dynamic,LOKHANDWALA201845}, platform revenue \cite{banerjee2017pricing,chen2019dispatching}, driver incentives \cite{ma2019spatio,yuen2019beyond,10.1145/3391403.3399476}, fairness \cite{lesmanabalancing2019,suhr2019two,DBLP:conf/ijcai/0002X20,DBLP:conf/aaai/Nanda0SDS20}, reliability \cite{FIELBAUM2020102831,ALONSOGONZALEZ2020102621}, or analyzing the effects on sharing economies \cite{DBLP:conf/www/KootiGADRL17,DBLP:conf/www/JiangCMW18,10.1145/3391403.3399534,10.1145/3391403.3399462}.

It is well-documented (e.g., \cite{lesmanabalancing2019}) that all these different desiderata are often contrasting (e.g., fairness vs. revenue), and therefore we should not expect a single algorithm for ridesharing to be superior for all of them; rather, the design of such algorithms should be contingent on the goals of the designer, and which of those properties they consider to be more important for the application at hand. Thus, we want a \emph{flexible} and \emph{adaptable} design, able to work best with respect to any set of such objectives with `a few tweaks'.

To enable this, we propose a \emph{modular approach to algorithm design in ridesharing}, in which an algorithm consists of three different \emph{components}, namely (a) matching passengers with other passengers, (b) assigning rides to vehicles and (c) vehicle relocation, in which the taxis move, when they do not serve passengers, close to positions where requests are \emph{expected to appear} in the near future. Each component can then be seen as a different (sub)-algorithm, and those algorithms can be appropriately chosen to be geared towards the specific objectives of the designer. As a matter of fact, our approach draws inspiration from several successful algorithms in the ridesharing literature, such as the well-known \emph{High Capacity} algorithm of \cite{alonso2017demand}, or the recent algorithm of \cite{ijcai2020-609}, who can both be cast as examples of algorithms in this modular design setting.

\begin{figure*}[t!]
	\centering
	\begin{subfigure}[t]{0.33\textwidth}
		\centering
		\includegraphics[width = 1 \linewidth]{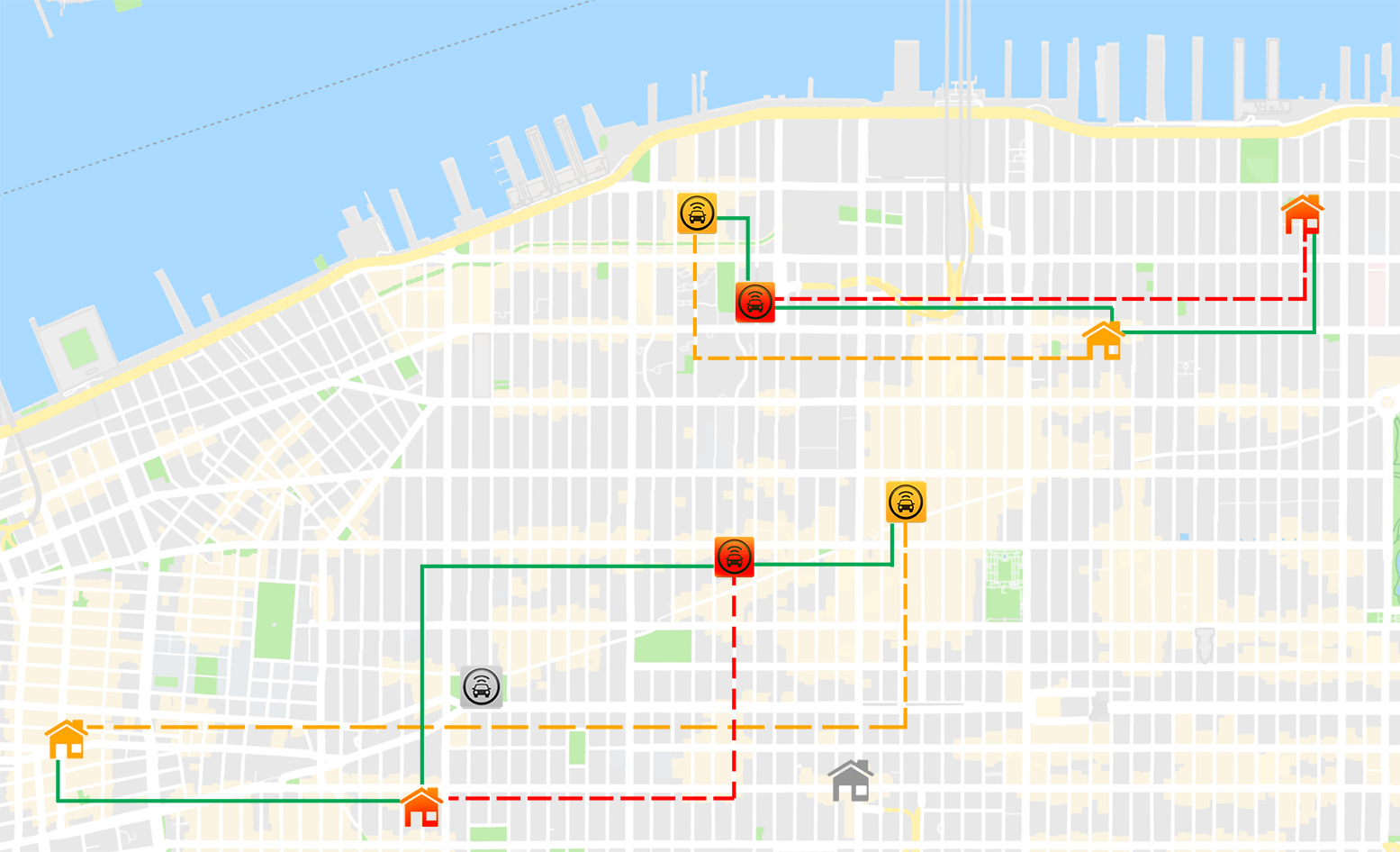}
		\caption{Request -- Request}
		\label{fig: Problem Setting (a)}
	\end{subfigure}%
	~ 
	\begin{subfigure}[t]{0.33\textwidth}
		\centering
		\includegraphics[width = 1 \linewidth]{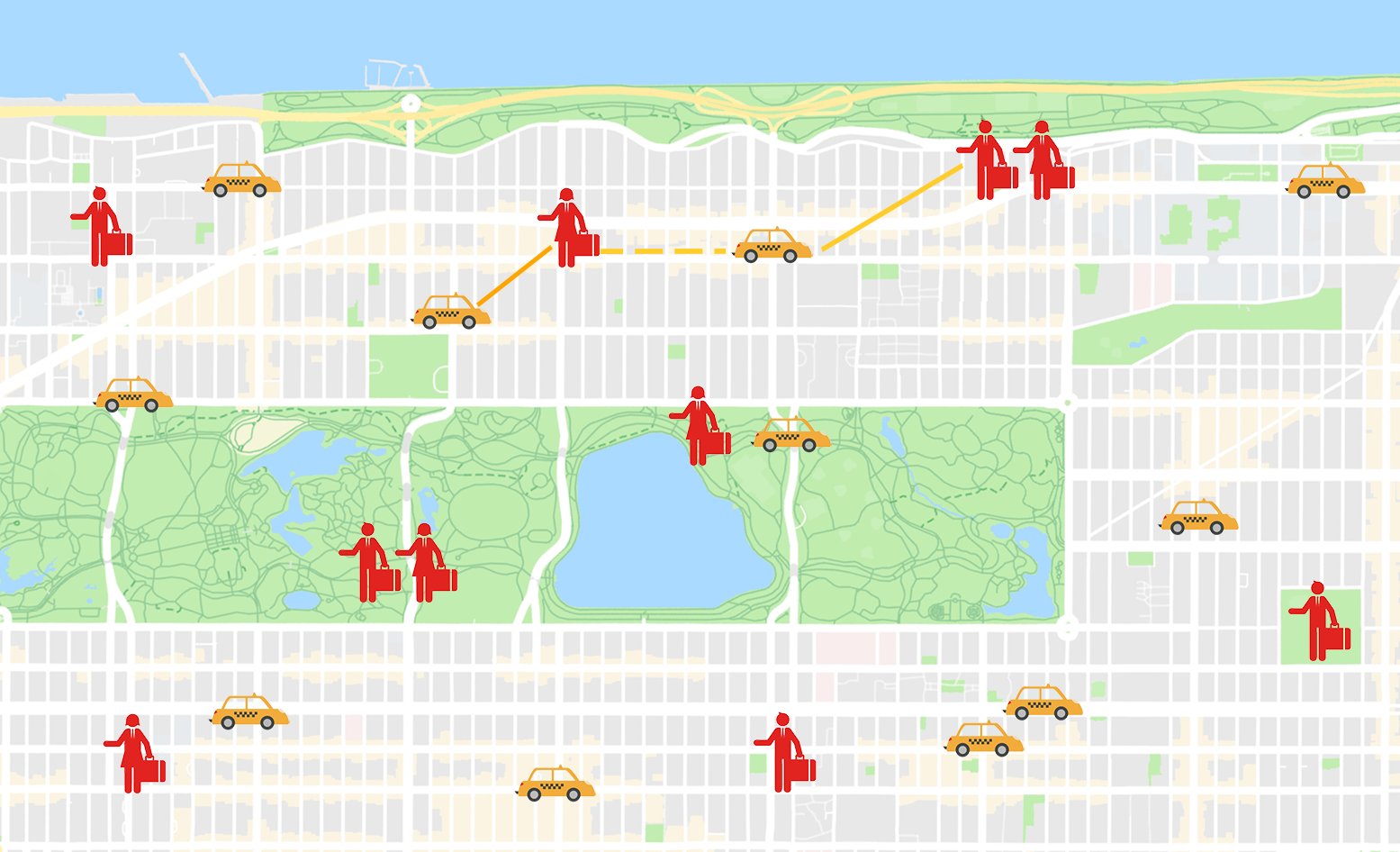}
		\caption{(Shared) Ride -- Taxi}
		\label{fig: Problem Setting (b)}
	\end{subfigure}%
	~ 
	\begin{subfigure}[t]{0.33\textwidth}
		\centering
		\includegraphics[width = 1 \linewidth]{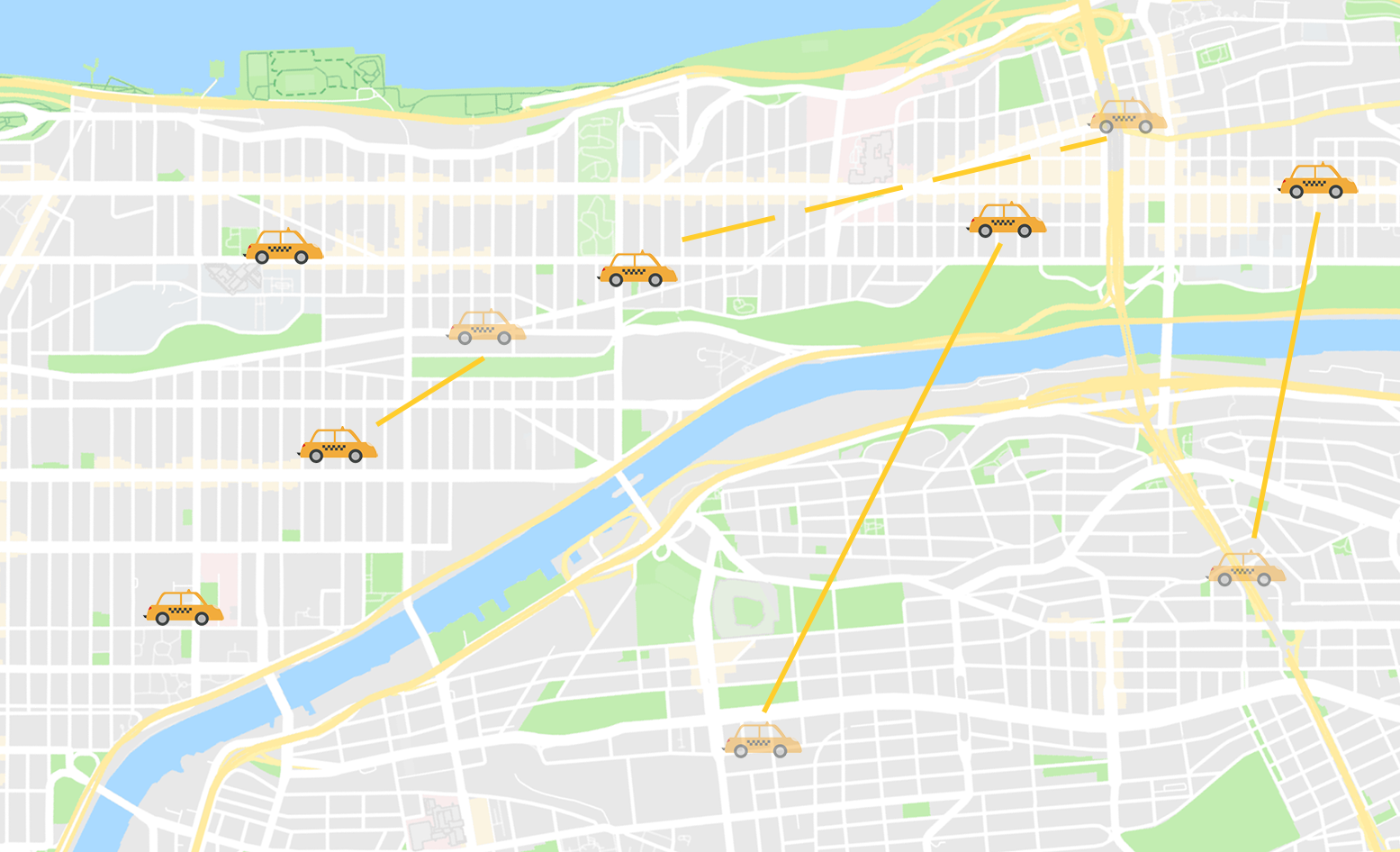}
		\caption{Vehicle Relocation}
		\label{fig: Problem Setting (c)}
	\end{subfigure}%
	\caption{The three components of a CAR.}
	\label{fig: Problem Setting}
\end{figure*}

\subsection{Our Contributions} \label{Our Contributions}

\subsubsection{CARs.} We initiate the \emph{systematic} study of \textbf{\emph{Component Algorithms for Ridesharing} (CARs)}. A CAR is an algorithm consisting of three sub-algorithms, each solving one of the following components of the ridesharing problem (Figure \ref{fig: Problem Setting}).
\begin{itemize}[leftmargin=*]
	\item[-] \emph{Matching passengers to other passengers}. For this component, the underlying algorithmic problem is that of \emph{Online Maximum Weight Matching}, where the ``online'' part stems from the fact that passenger requests appear at different points in time, and we have to \emph{account for the future} when deciding which passengers to match. As such, we have a lot of classic as well as modern matching algorithms at our disposal.
	\item[-] \emph{Assigning rides to vehicles}. For this component, the underlying algorithmic problem can either be seen as an \emph{Online Maximum Weight Bipartite Matching}, or as an instance of the \emph{$k$-Taxi Problem} and by extension as the famous \emph{$k$-Server problem} from the literature of online algorithms. Similarly to above, there is a large set of classic and modern solutions that one can plug-in as components for this part. 
	\item[-] \emph{Vehicle Relocation}. For this component, the objective is to use \emph{historical data}  to \emph{predict} the location of future requests and move idle taxis closer to those locations. From an algorithmic standpoint, this problem can be cast as either as \emph{$k$-Facility Location} problem, concerned with the optimal placement of facilities (taxis) to minimize transportation costs, or as an \emph{Online Maximum Weight Matching} problem on the \emph{history of requests}.
\end{itemize}


\subsubsection{Evaluation Platform.} While several papers in the literature provide evaluations on realistic datasets, (e.g., see \cite{ijcai2020-609,santi2014quantifying,alonso2017demand,agatz2011dynamic,santos2013dynamic,ijcai201931}), they either (a) only consider parts of the ridesharing problem and therefore do not propose end-to-end solutions, (b) only evaluate a few newly-proposed algorithms against some basic baselines, (c) only consider a limited number of performance metrics, predominantly with regard to the overall efficiency and often without regard to QoS metrics, or (d) perform evaluations on a much smaller scale, thus not capturing the real-life complexity of the problem. On the contrary, \emph{our work provides a comprehensive evaluation of a large number of proposed algorithms, over multiple different metrics, and for real-world scale, end-to-end problems}. Specifically:
\begin{itemize}[leftmargin=*]
	\item[-] We meticulously design an experimental setting to \emph{resemble reality as close as possible} in \emph{every} aspect of the problem. To the best of our knowledge, this is the \emph{first end-to-end experimental evaluation of this magnitude}, and could serve as a \emph{common-ground} for evaluating future work in a setting designed to capture real-world challenges.
	\item[-] We evaluate our CARs for a host of different objectives (12 metrics) related to \emph{global efficiency}, \emph{complexity}, \emph{passenger}, \emph{driver}, and \emph{platform} incentives (see Table~\ref{tb: Performance Metrics}). 
\end{itemize}
We focus on (shared) rides of at most two requests (i.e., vehicles of capacity two) for two reasons: complexity, and passenger satisfaction; as we explain in detail in Section \ref{Rides}.

\subsubsection{Results.} Applying the modular approach we advocated above, we design a large set of CARs, based on different classic and modern algorithms for the different components (14 in total, see Table \ref{tb: algorithms}). The main take-away is the following:
\begin{itemize}[leftmargin=*]
\item[-] CARs based on off-line, in-batches maximum-weight matching approaches perform well on global efficiency and passenger related metrics.
\item[-] CARs based on $k$-server algorithms perform well on driver/platform related metrics (e.g., the Balance algorithm \cite{manasse1990competitive}).
\item[-] Lightweight CARs perform better in real-world, large-scale settings since real-time constraints dictate short planning windows which can diminish the benefit of cumbersome optimization techniques compared to myopic approaches.
\item[-] Simple, lightweight relocation schemes can significantly improve Quality of Service metrics by up to $50\%$.
\item[-] We identify a \emph{scalable}, \emph{on-device} CAR based on ALMA \cite{ijcai201931} that performs well \emph{across the board}.
\end{itemize}
Our findings provide convincing evidence to a ridesharing platform as to which combination of components would be most suitable for a given set of objectives.

\section{Discussion and Related Work} \label{Discussion and Related Work}

The literature on ridesharing is rather extensive; here we only highlight the key algorithmic principles in our design of CARs.

The dynamic ridesharing -- and the closely related \emph{dynamic dial-a-ride} (see \cite{agatz2012optimization}) -- problem has drawn the attention of diverse disciplines over the past few years, from operations research to transportation engineering, and computer science. Solution approaches include constrained optimization \cite{qian2017optimal,simonetto2019real,agatz2011dynamic,alonso2017demand,ijcai2020-609}, weighted matching \cite{ashlagi2017min,bei2018algorithms,dickerson2018allocation,zhao2019preference,ijcai201931}, other heuristics \cite{qian2017optimal,santos2015taxi,bathla2018real,lowalekar2019zac,santos2013dynamic,7080855,8031493,Shah_Lowalekar_Varakantham_2020}, reinforcement learning \cite{gueriau2018samod,DBLP:conf/www/LiQJYWWWY19,DBLP:conf/www/HeS19}, or model predictive control \cite{chen2019optimization,ijcai2020-609,8794194}, among others. We refer the interested reader to the following surveys \cite{agatz2012optimization,silwal2019survey,furuhata2013ridesharing,ho2018survey,mourad2019survey,cordeau2007dial} for a review on the optimization challenges, various algorithmic designs adopted over the years, a classification of existing ridesharing systems, models and algorithms for shared mobility, and finally models and solution methodologies for the dial-a-ride problem, respectively.


As we mentioned in the introduction, the key algorithmic components of ridesharing are the following. First, it is an \emph{online} problem, as the decisions made at some point in time clearly affect the possible decisions in the future, and therefore the the literature of \emph{online algorithms} and competitive analysis \cite{borodin2005online,manasse1988competitive} offers clear-cut candidates for CARs. Second, all of the components can be seen as some type of \emph{matching} both for bipartite graphs (for matching passengers with taxis, or idle taxis with `future' requests) and for general graphs (for matching passengers to shared rides). In fact, several of the algorithms that have been proposed in the literature for the problem are for different variants of online matching.

Finally, ridesharing displays an inherent connection to the \emph{$k$-taxi problem} \cite{coester2018online,buchbinder2020online,fiat1994competitive,kosoresow1997design}, which, in turn, is a generalization of the well-known \emph{$k$-server problem} \cite{koutsoupias1995k,KOUTSOUPIAS2009105}\footnote{In fact the latter two problems are quite closely connected, and algorithms for the $k$-server problem can be used to solve the $k$-taxi problem. See \cite{coester2018online} for more details.}. In the $k$-taxi problem, once a request appears (with a source and a destination), one of the $k$ taxis at the platform's disposal must serve the request. Viewing shared rides (multiple passengers that have already been matched in a previous step) as requests, one can clearly apply the $k$-taxi (and $k$-server algorithms) to the ridesharing setting. Granted, the $k$-server algorithms have been designed to operate in a more challenging setting in which (a) the requests have to be served \emph{immediately}, whereas normally there is some leeway in that regard, often at the expense of customer satisfaction, and (b) the positions of requests are typically \emph{adversarially} chosen, rather than following some distribution, as is the case in reality. Despite those facts, the fundamental idea behind these algorithms is a pivotal part of ridesharing, as it aims to \emph{serve existing requests efficiently, but at the same time place the vehicles as well as possible to serve future requests}. This is also the main principle of the relocation strategies for idle taxis.

The algorithms that we consider are appropriate modifications of the most significant ones that have been proposed for the aforementioned key algorithmic primitives of the ridesharing problem, as well as heuristic approaches which are based on the same principles, but were specifically designed with the ridesharing application in mind. We emphasize that such modifications are needed, primarily because many of these algorithms were tailored for sub-problems of the ridesharing setting, and end-to-end solutions in the literature are rather scarce.

Much of the related work in the literature focuses on approaches that are inherently centralized and require knowledge of the full ridesharing network, which makes them rather computationally intensive. As an additional goal of our investigation, we would like to identify solutions that are lightweight, decentralized, and which ideally run \emph{on-device}. Of course, some hybrid and decentralized approaches for the ridesharing problem have been proposed (e.g., \cite{simonetto2019real,gueriau2018samod}), and several of the algorithms that we include in our experimental evaluation can be implemented in a decentralized manner (e.g., \cite{giordani2010distributed,7991447,zavlanos2008distributed,burger2012distributed}), but that would typically require a larger amount of communication between the agents; in this case, the vehicles. As it turns out though, the ALMA algorithm of \cite{ijcai201931}, which has been designed with precisely these objectives in mind (low computational complexity, scalability, and low communication cost), performs very well across the board with respect to our objectives.

The third component of our CARs is the relocation of idle taxis. Relocation is an important component of a successful ridesharing application. Many studies in shared mobility systems have shown that the adoption of a relocation strategy can help improve the system performance for their specific context \cite{gueriau2018samod,vosooghi2019shared,martinez2017insights,belanger2016empirical,ruch2018amodeus,alonso2017demand,buchholz2018spatial,lioris2016dynamic,Spieser2014,8794194,VANENGELEN2018110,8317908,8593743}. Strategies include using a short window of known active requests \cite{alonso2017demand}, historical demand \cite{gueriau2018samod,8206203,fielbaum2021anticipatory,6754040,articleXueRui,VANENGELEN2018110}, or techniques to predict future demand \cite{spieser2016shared}. Yet, relocation by nature increases vehicle travel distance, leading to undesirable consequences (economical, environmental, maintenance, management of human resources, etc.), thus a balance needs to be struck. Most of the employed relocation approaches are course-grained; the network is generally divided into several zones, blocks, etc. \cite{gueriau2018samod,vosooghi2019shared,martinez2017insights} and the entities (e.g., the vehicles) move between the zones. However, compared to other shared mobility systems, dynamic ridesharing posses unique challenges, meaning that such coarse-grained approaches are not appropriate: most of them are centralized -- thus computationally intensive and not scalable --, they might not take into account the actions of other vehicles, potentially leading to over-saturation of high demand areas, and, most importantly, they are \emph{slow to adapt} to the highly dynamic nature of the problem (e.g., responding to high demand generated by a concert, or the fact that vehicles remain free for only a few minutes at a time). The problem clearly calls for fine-grained solutions, yet such approaches in the literature are still rather scarce. In this paper, we employ such a fine-grained relocation scheme (similarly to \cite{alonso2017demand}), based on matching between the idle taxis and the \emph{potential} requests, which is better suited for the problem at hand.

Relocation  can be either viewed as the $k$-center or $k$-Facility Location Problem \cite{guha1999greedy}, or as an \emph{Online Maximum Weight Matching} problem on the \emph{history} of requests. Given the high complexity of the former problems (they are both NP-hard, in fact, APX-hard \cite{hsu1979easy,feder1988optimal}), we have opted for the latter interpretation.



\section{Problem Statement \& Modeling} \label{Problem Statement & Modeling}

In this section we formally present the Ridesharing problem. To avoid introducing unnecessary notation, we only present the description of the model here; precise notation and details are provided in the respective sections where they are used. 

In the Ridesharing problem there is a (potentially infinite) metric space $\mathcal{X}$ representing the topology of the environment, equipped with a distance function $\delta: \mathcal{X} \times \mathcal{X} \rightarrow \mathds{R}_{\geq 0}$. Both are known in advance. At any moment, there is a (\emph{dynamic}) set of available taxi vehicles $\mathcal{V}_t$, ready to serve customer requests (i.e., drive to the pick-up, and subsequently to the destination location). Between serving requests, vehicles can \emph{relocate} to locations of potentially higher demand, to mitigate spatial search frictions between drivers. Customer requests appear in an online manner at their respective pick-up locations, \emph{wait} to potentially be matched to a shared ride, and finally are served by a taxi to their respective destination. In order for two requests to be able to share a ride, they must satisfy \emph{spatial}, and \emph{temporal} constraints. The former dictates that requests should be matched only if there is good spatial overlap among their routes. Yet, due to the latter constraint, requests cannot be matched even if they have perfect spatial overlap, if they are not both `active' at the same time. Finally, ridesharing is an inherently \emph{online} problem, as we are unaware of the requests that will appear in the future, and need to make decisions before the requests expire, while taking into account the dynamics of the fleet of taxis.

\subsection{Performance Metrics} \label{Performance Metrics}

The goal is to \emph{minimize the cumulative distance driven} by the fleet of taxis, while maintaining \emph{high Quality of Service (QoS)}, given that we \emph{serve all requests} (service guarantee). Serving all requests improves passenger satisfaction, and, most importantly, allows us to ground our evaluation to a common scenario, ensuring a fair comparison.

\subsubsection{Global Metrics}

\noindent
\textbf{Distance Driven}: Minimize the cumulative distance driven by all vehicles for serving all the requests. We chose this objective as it directly correlates to passenger, driver, company, and environmental objectives (minimize operational cost, delay, CO$_2$ emissions, maximize the number of shared rides, improve QoS, etc.). All of the evaluated algorithms have to \emph{serve all the requests}, either as shared, or single rides. \\

\noindent
\textbf{Complexity}: Real-world time constraints dictate that the employed solution produces results in a reasonable time-frame\footnote{\label{footnote: waiting period}For example UberPool has a waiting period of at most 2 minutes until you get a match (\url{https://www.uber.com/au/en/ride/uberpool/}), thus any algorithm has to run in under that time to be applicable in real life.}.

\subsubsection{Passenger Specific Metrics -- Quality of Service (QoS)} \label{Passenger Specific Matrics} 

\noindent
\textbf{Time to Pair}: Expected time to be paired in a shared ride, i.e., $\mathds{E}[t_{\text{paired}} - t_{\text{open}}]$, where $t_{\text{open}}, t_{\text{paired}}$ denote the time the request appeared, and was paired as a shared ride, respectively. If the request is served as a single ride, then $t_{\text{paired}}$ refers to the time the algorithm chose to serve it as such. \\

\noindent
\textbf{Time to Pair with Taxi}: Expected time to be paired with a taxi, i.e., $\mathds{E}[t_{\text{taxi}} - t_{\text{paired}}]$, where $t_{\text{taxi}}$ denotes the time the (shared) ride was paired with a taxi. \\

\noindent
\textbf{Time to Pick-up}: Expected time to passenger pickup, i.e., $\mathds{E}[t_{\text{pickup}} - t_{\text{taxi}}]$, where $t_{\text{pickup}}$ denotes the time the request was picked-up. \\

\noindent
\textbf{Delay}: Additional travel time over the expected direct travel time (when served as a single ride, instead of a shared ride), i.e., $\mathds{E}[(t_{\text{dest}} - t_{\text{pickup}}) - (t'_{\text{dest}} - t_{\text{pickup}})]$. $t_{\text{dest}}$, and $t'_{\text{dest}}$ denote the time the request reaches, and would have reached as a single ride, its destination. \\

Research conducted by ridesharing companies shows that passengers' satisfaction level remains sufficiently high as long as the pick-up time is less than a certain threshold. The latter is corroborated by data on booking cancellation rate against pick-up time \cite{8259801}. In other words, passengers would rather have a short pick-up time and long detour, than vice-versa \cite{lyftblog2}. This also suggests that an effective relocation scheme can considerably improve passenger satisfaction by reducing the average pick-up time (see Section \ref{Relocation Simulation Results}).

Given the importance of short pick-up times in passengers' satisfaction, we opted to distinguish and study each segment of the waiting process separately (`Time to Pair', `Time to Pair with Taxi', and `Time to Pick-up'). To the best of our knowledge, we are the first to do so. Such analysis can provide a clear picture of sources of inefficiency to a ridesharing platform, and improve the overall satisfaction which in turn correlates to the growth of the company.

\subsubsection{Driver Specific Metrics}

\noindent
\textbf{Driver Profit}: Total revenue earned minus total travel costs. \\

\noindent
\textbf{Number of Shared Rides}: Related to the profit. By carrying more than one passenger at a time, drivers can serve more requests in a day, which consequently, increases their income \cite{widdows2017grabshare}. \\

\noindent
\textbf{Frictions}: Waiting time experienced by drivers between serving requests (i.e., time between dropping-off a ride, and getting matched with another). Search frictions occur when drivers are unable to locate rides due to spatial supply and demand imbalance. Even though in our scenario matchings are performed automatically, without any searching involved by the drivers, lower frictions indicate a better distribution of the platform's supply. \\

\subsubsection{Platform Specific Metrics}

\noindent
\textbf{Platform Profit}: Usually a commission on the driver's fee\footnote{\label{uberpayments}E.g., Uber charges partners 25\% fee on all fares (\url{https://www.uber.com/en-GH/drive/resources/payments/}).}, and passenger fees (which, given that we serve all the requests, the latter would be constant across all the employed algorithms). \\

\noindent
\textbf{Quality of Service (QoS)}: Refer to the aforementioned, passenger specific metrics. Improving the QoS to their costumers correlates to the growth of the company. \\

\noindent
\textbf{Number of Shared Rides}: The matching rate is important especially in the nascent stage of a ridesharing platform \cite{dutta2018online}. \\

We do not report separate values on the aforementioned metrics, as they directly correlate to their respective passenger, and driver specific ones.

\subsection{Modeling} \label{Modeling}

Our evaluation setting is \emph{meticulously designed to resemble reality as closely as possible}, in every aspect of the problem. We achieve this by (a) using actual data from the NYC's yellow taxi trip records\footnote{\label{TLCTripRecordData}\url{https://www1.nyc.gov/site/tlc/about/tlc-trip-record-data.page}}, both for modeling customer requests and taxis (b) following closely the pricing model employed by ridesharing platforms and (c) running our simulations to the scale of the actual problem faced by the ridesharing platforms (we run simulations with more than $390,000$ requests and $12,000$ taxis). Moreover, we have exhaustively designed every detail of the problem, such as speed of the vehicles, initial positions, distance function, pricing, etc. In what follows, we describe each design aspect in detail.

\subsubsection{Dataset}

\begin{figure}[t!]
	\centering
	\includegraphics[width = 1 \linewidth, trim={0.5em 0.8em 0.65em 0.7em}, clip]{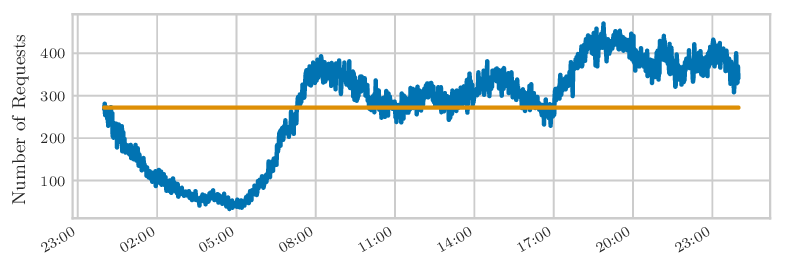}
	\caption{Request per minute on Jan. 15, 2016 (blue line). Mean = 272 requests (yellow line).}
	\label{fig: requests}
\end{figure}

We have used the yellow taxi trip records of 2016, provided by the NYC Taxi and Limousine Commission\footnoteref{TLCTripRecordData}. The dataset was cleaned to remove requests with travel time shorter than 1 minute, or invalid geo-locations (e.g., outside Manhattan, Bronx, Staten Island, Brooklyn, or Queens). For every request, the dataset provides amongst others the pick-up and drop-off times, and geo-location coordinates. Time is discrete, with granularity of 1 minute (same as the dataset). On average, there are $272$ new requests per minute, totaling to $391479$ requests in the broader NYC area ($352455$ in Manhattan) on the evaluated day (Jan, 15). Figure \ref{fig: requests} depicts the number of request per minute on the aforementioned day.

\subsubsection{Taxi Vehicles} \label{Taxi Vehicles}

A unique feature of the NYC Yellow taxis is that they may only be hailed from the street and are not authorized to conduct pre-arranged pick-ups. This provides an ideal setting for a counter-factual analysis for several reasons: (1) We can assume a realistic position of each taxi at the beginning of the simulation (last drop-off location). (2) Door-to-door service can be inefficient \cite{FIELBAUM2021103061,STIGLIC201536}, thus users may be requested to walk to/from a nearby fast street. Given that users have presumably hailed the taxis from larger streets, this results to a more accurate modeling of the origins of supply and demand. Finally, (3) all observed rides are obtained through search, thus -- assuming reasonable prices, and delays -- customers do not have nor are willing to take an alternative means of transportation. The latter validates our choice that all of the algorithms considered will have to eventually serve all the requests.

By law, there are $13,587$ taxis in NYC\footnoteref{TLCYellowCab}. The majority of the results presented in this paper use a much lower number of vehicles (what we call \emph{base number}) for three reasons: (1) to reduce the complexity of the problem, given that most of the employed algorithms can not handle such a large number of vehicles, (2) to evaluate under resource scarcity -- making the problem harder -- to better differentiate between the results, and (3) to investigate the possibility of a more efficient utilization of resources, with minimal cost to the consumers. However, we still present simulations for a wide range of vehicles, up to close to the total number.

The number, initial location, and speed of the taxi vehicles were calculated as follows:

\begin{itemize}[leftmargin=*]
	\item We calculated the \emph{base number} of taxis, as the minimum number of taxis required to serve all requests as single rides (no ridesharing). If a request appears, and all taxis are occupied serving other requests, we increase the required number of taxis by one. This resulted to around $4000 - 5000$ vehicles (depending on the size of the simulation, see Section \ref{Simulation Results}). Simulations were conducted for $\{\times 0.5, \times 0.75, \times 1.0, \times 2.0, \times 3.0\}$ the base number.
	\item Given a number of taxis, $V$, the initial position of each taxi is the drop-off location of the last $V$ requests, prior to the starting time of the simulation. To avoid cold start, we compute the drop-off time of each request, and assume the vehicle occupied until then.
	\item The vehicles' average speed is estimated to $6.2$ m/s ($22.3$ km/h), based on the trip distance and time per trip as reported in the dataset, and corroborated by the related literature (in \cite{santi2014quantifying} the speed was estimated at $5.5 - 8.5$ m/s depending on the time of day).
\end{itemize}

\subsubsection{Customer Requests} \label{Customer Requests}

A request, $r$, is a tuple $\langle t_r, s_r, d_r, k_r \rangle$. Request $r$ appears (becomes \emph{open}) at its respective pick-up time ($t_r$), and geo-location ($s_r$). Let $d_r$ denote the destination. Each request admits a willingness to wait ($k_r$) to find a match (rideshare), i.e., we assume \emph{dynamic} waiting periods per request. The rationale behind $k_r$ is that requests with longer trips are more willing to wait to find a match than requests with destinations near-by. After $k_r$ time-steps we call request $r$, \emph{critical}. If a critical request is not matched, it has to be served as a single ride. Recall that in our setting \emph{all} of the requests must be served. Let $\mathcal{R}_t^{\text{open}}, \mathcal{R}_t^{\text{critical}}$ denote the sets of open, and critical requests respectively, and let $\mathcal{R}_t = \mathcal{R}_t^{\text{open}} \cup \mathcal{R}_t^{\text{critical}}$.

We calculate $k_r$ as in related literature \cite{ijcai201931}. Let $w_{\text{min}}$, and $w_{\text{max}}$ be the minimum and maximum possible waiting time, i.e., $w_{\text{min}} \leq k_r \leq w_{\text{max}}, \forall r$. Knowing $s_r, d_r$, we can compute the expected trip time ($\mathds{E}[t_{\text{trip}}]$). Assuming people are willing to wait proportional to their trip time, let $k_r = q \times \mathds{E}[t_{\text{trip}}]$, where $q \in [0, 1]$. $w_{\text{min}}, w_{\text{max}}$, and $q$ can be set by the ridesharing company, based on customer satisfaction (following \cite{ijcai201931}, let $w_{\text{min}} = 1, w_{\text{max}} = 3$, and $q = 0.1$).

\subsubsection{Rides} \label{Rides}

A (shared)ride, $\rho$, is a pair $\langle r_1, r_2 \rangle$, composed of two requests. If a request $r$ is served as a single ride, then $r_1 = r_2 = r$. Let $\mathcal{P}_t$ denote the set of rides waiting to be matched to a taxi at time $t$. Contrary to some recent literature on high capacity ridesharing (e.g., \cite{alonso2017demand,lowalekar2019zac}), we \emph{purposefully restricted ourselves to rides of at most two requests} for two reasons: \emph{complexity}, and \emph{passenger satisfaction}. The complexity of the problem grows rapidly as the number of potential matches increases, while most of the proposed/evaluated approaches already struggle to tackle matchings of size two on the scale of a real-world application. Moreover, even though a fully utilized vehicle would ultimately be a more efficient use of resources, it diminishes passenger satisfaction (a frequent worry being that the ride will become interminable, according to internal research by ridesharing companies) \cite{widdows2017grabshare,lyftblog1}. Given that a hard constraint is the serving of all requests, we do not assume a time limit on matching rides with taxis; instead we treat it as a QoS metric.

\subsubsection{Distance Function}

The optimal choice for a distance function would be the actual driving distance. Yet, our simulations require trillions of distance calculations, which is not attainable. Given that the locations are given in latitude and longitude coordinates, it is tempting to use the Haversine formula\footnote{\url{https://en.wikipedia.org/wiki/Haversine_formula}} to estimate the Euclidean distance, as in related literature \cite{santos2013dynamic,lyftblog1}. We have opted to use the Manhattan distance, given that the simulation takes place mostly in Manhattan. To evaluate our choice, we collected more than $12$ million actual driving distances using the Open Source Routing Machine (\href{http://project-osrm.org/}{project-osrm.org}), which computes the shortest path in road networks. Manhattan distance's standard and mean squared error, compared to the actual driving distance, was $-0.5 \pm 2.9$ km, and $1.7 \pm 2.4$ km respectively, while Euclidean distance's was $-3.2 \pm 3.8$ km, and $3.2 \pm 3.8$ km respectively.

\subsubsection{Pricing}

A combination of an one-time flag drop fee\footnote{\label{uberpriceestimate}\url{https://www.uber.com/us/en/price-estimate/}} ($\beta = 2.2$ \$), distance fare\footnoteref{uberpriceestimate} ($\pi_{_I} = 0.994$ \$/km for a single ride, $\pi_{_{II}} = 0.8$ \$/km shared), fuel price\footnote{\url{https://www.eia.gov/dnav/pet/pet_pri_gnd_dcus_sny_m.htm}} ($3.2$ \$/gal), and vehicle mileage ($46.671$ km/gal \cite{buchholz2018spatial}). The aforementioned fuel price and mileage result in a cost per km $c = 0.0686$ \$/km. The revenue $M(\rho)$ of a taxi driver from serving ride $\rho$ is given by \cite{buchholz2018spatial}:

\begin{equation*} \label{Eq: Pricing}
	M(\rho) = 
	\begin{cases}
		\beta + \pi_{_I} \delta(s_{r}, d_{r}) - c \delta(s_v, s_{r}, d_{r}) & \text{, if $\rho$ single} \\
		2 \beta + \pi_{_{II}} \delta(s_{r_1}, d_{r_1} | r_2) + \pi_{_{II}} \delta(s_{r_2}, d_{r_2} | r_1) \\ \quad  - c \delta(s_v, s_{r_1}, s_{r_2}, d_{r_1}, d_{r_2}) & \text{, if $\rho$ shared}
	\end{cases}
\end{equation*}

\noindent
where, with some slight abuse of notation, $\delta(s_v, s_{r}, d_{r})$ denotes the distance from the current location of the taxi $s_v$, to the pick-up and subsequently drop-off location of the ride, $\delta(s_{r_1}, d_{r_1} | r_2)$ denotes the distance driven from the pick-up to the destination of $r_1$, given that $r_1$ will share the ride with $r_2$ (similarly $\delta(s_{r_2}, d_{r_2} | r_1)$ for $r_2$), and finally, $\delta(s_v, s_{r_1}, s_{r_2}, d_{r_1}, d_{r_2})$ denotes the total driving distance of the taxi for serving the two requests starting from $s_v$.


\subsubsection{Embedding into HSTs} \label{Embedding into HSTs}

A starting point of many of the employed $k$-server algorithms is embedding the input metric space $\mathcal{X}$ into a distribution $\mu$ over $\sigma$-hierarchically well-separated trees (HSTs), with separation $\sigma = \Theta(\log |\mathcal{X}| \log(k \log |\mathcal{X}|))$, where $|\mathcal{X}|$ denotes the number of points. It has been shown that solving the problem on HSTs suffices, as any finite metric space can be embedded into a probability distribution over HSTs with low distortion \cite{Fakcharoenphol2003}. The distortion is of order $\mathcal{O}(\sigma \log_\sigma |\mathcal{X}|)$, and the resulting HSTs have depth $\mathcal{O}(\log_\sigma \Delta)$, where $\Delta$ is the diameter of $\mathcal{X}$ \cite{bansal2015polylogarithmic}.

Given the popularity of the aforementioned method, it is worth examining the size of the resulting trees. Given that the geo-coordinate system is a discrete metric space, we could directly embed it into HSTs. Yet, the size of the space is huge, thus for better discretization we have opted to generate the graph of the street network of NYC. To do so, we used data from \href{https://www.openstreetmap.org}{openstreetmap.org}. Similarly to \cite{santi2014quantifying}, we filtered the streets selecting only primary, secondary, tertiary, residential, unclassified, road, and living street classes, using those as undirected edges and street intersections as nodes. The resulting graph for NYC contains $66543$ nodes, and $95675$ edges ($5018$, and $8086$ for Manhattan). Given that graph, we generate the HSTs \cite{santi2014quantifying}.

\begin{table*}[t!]
\centering
\caption{Evaluated CARs.}
\resizebox{\textwidth}{!}{%
\begin{tabular}{@{}lccc@{}}
\toprule
                                                                        & \textbf{Step (a)} & \textbf{Step (b)} & \textbf{Step (c)} \\ \midrule
\textbf{Maximum Weight Matching (MWM)}                                  & MWM               & MWM               & MWM/ALMA/Greedy   \\
\textbf{ALtruistic MAtching Heuristic (ALMA)} \cite{ijcai201931}        & ALMA              & ALMA              & ALMA              \\
\textbf{Greedy}                                                         & Greedy            & Greedy            & Greedy            \\
\textbf{Approximation (Appr)} \cite{bei2018algorithms}                  & Appr              & Appr              & -                 \\
\textbf{Postponed Greedy (PG)} \cite{ashlagi2018maximum}                & PG                & MWM               & -                 \\
\textbf{Greedy Dual (GD)} \cite{bienkowski2018primal}                   & GD                & MWM               & -                 \\
\textbf{Balance (Bal)} \cite{manasse1990competitive}                    & MWM               & Bal               & -                 \\
\textbf{Harmonic (Har)} \cite{raghavan1989memory}                       & MWM               & Har               & -                 \\
\textbf{Double Coverage (DC)} \cite{Chrobak90newresults}                & MWM               & DC                & -                 \\
\textbf{Work Function (WFA)} \cite{koutsoupias1995k}  & MWM               & WFA               & -                 \\
\textbf{$k$-Taxi} \cite{coester2018online}                              & MWM               & $k$-Taxi          & -                 \\
\textbf{High Capacity (HC)} \cite{alonso2017demand}                     & HC                & HC                & (HC)              \\
\textbf{Baseline: Single Ride}                                          & -                 & MWM               & -                 \\
\textbf{Baseline: Random}                                               & -                 & Random            & -                 \\ \bottomrule
\end{tabular}
}
\label{tb: algorithms}
\end{table*}

\section{Component Algorithms for Ridesharing}

In this section, we describe our design choices for developing \emph{Component Algorithms for Ridesharing (CARs)}. Each CAR is composed of three parts (Figure \ref{fig: Problem Setting}): (a) request -- request matching to create a (shared) ride, (b) ride to taxi matching, and (c) relocation of the idle fleet. Each of these components is a significant problem in its own right. Complexity issues make the simultaneous consideration of all three problems impractical. Instead, a more realistic approach is to tackle each component individually, under minimum consideration of the remaining two\footnote{To have a comprehensive analysis, we have also evaluated the HC algorithm, a highly cited approach that tackles steps (a), and (b) simultaneously. Yet, this results in a prohibitively large ILP (see Sections \ref{High Capacity} and \ref{Challenges}).}. The algorithms that we consider are appropriate modifications of the most significant ones that have been proposed for the key algorithmic primitives of the ridesharing problem (see Sections \ref{Our Contributions} and \ref{Discussion and Related Work}), i.e., \emph{online and offline matching algorithms, with or without delays} for steps (a), (b), and (c), \emph{$k$-taxi/server algorithms} for step (b), as well as \emph{heuristic approaches} that were specifically designed with the ridesharing application in mind. 

A list of all the CARs that we designed and evaluated (14 in total) can be found in Table \ref{tb: algorithms}, while in the following sections we provide a detailed description of each CAR component.

\subsection{CAR components} \label{algorithms}

We have evaluated a variety of approaches ranging from offline maximum weight matching (MWM), and greedy solutions, to online MWM, $k$-Taxi/Server algorithms, and linear programming. Offline algorithms (e.g., MWM, ALMA, Greedy) can be run either in a just-in-time (JiT) manner -- i.e., when a request becomes critical -- or in batches, i.e., every x minutes (given that our dataset has granularity of 1 minute, we run in batches of 1, and 2 minutes).

\paragraph{Matching Graphs:}
At time $t$, let $\mathcal{G}_a = (\mathcal{R}_t, \mathcal{E}^a_t)$, where $\mathcal{E}^a_t$ denotes the weighted edges between requests. With a slight abuse of notation, let $\delta(s_{r_1}, s_{r_2}, d_{r_1}, d_{r_2})$ denote the minimum distance required to serve both $r_1$, and $r_2$ (as a shared ride, i.e., excluding the case of first serving one of them and then the other) with a single taxi located either in $s_1$, or $s_2$. The weight $w_{r_1, r_2}$ of an edge $(r_1, r_2) \in \mathcal{E}^a_t$ is defined as $w_{r_1, r_2} = \delta(s_1, d_1) + \delta(s_2, d_2) - \delta(s_{r_1}, s_{r_2}, d_{r_1}, d_{r_2})$ (similarly to \cite{ijcai201931,alonso2017demand}). If $r_1 = r_2$, let $w_{r_1, r_2} = 0$ (single passenger ride). Intuitively, this number represents an approximation (given that it is impossible to know in advance the location of the taxi that will serve the ride) on the travel distance saved by matching requests $r_1$, and $r_2$\footnote{It also ensures that the shared ride will cost less than the single ride option.}.

Similarly, at time $t$, let $\mathcal{G}_b = (\mathcal{V}_t \cup \mathcal{P}_t, \mathcal{E}^b_t)$, where $\mathcal{E}^b_t$ denotes the weighted edges between rides and taxis. With a slight abuse of notation, let $\delta(s_v, s_{r_1}, s_{r_2}, d_{r_1}, d_{r_2})$ denote the minimum distance required (out of all the possible pick-up and drop-off combinations) to serve both requests $r_1$, and $r_2$ (that compose the (shared) ride $\rho$) with a single taxi located at $s_v$. The weight $w_{v, \rho}$ of an edge $(v, \rho) \in \mathcal{E}^b_t$ is defined as $w_{v, \rho} = 1 / \delta(s_v, s_{r_1}, s_{r_2}, d_{r_1}, d_{r_2})$. If $r_1 = r_2$ (single passenger ride), let $\allowbreak\delta(s_v, s_{r_1}, s_{r_2}, d_{r_1}, d_{r_2}) = \delta(s_v, s_{r_1}, d_{r_1})$. For the step (b) of the Ridesharing problem, we run the offline algorithms every time the set of rides ($\mathcal{P}_t$) is not empty.

\subsubsection{Maximum Weight Matching (MWM)}

The maximum weight matching algorithm finds a matching with maximum total edge weight in a graph. We use a maximum wieght matching algorithm to 
\begin{itemize}
	\item[-] match requests into shared rides (step (a) of the Ridesharing problem), i.e., find a matching on $\mathcal{G}^a$ that maximizes the quantity $\sum_{(r_1,r_2) \in \mathcal{E}^a_t} w_{r_1,r_2}$.
	\item[-] match rides with taxis (step (b) of the Ridesharing problem), i.e., find a matching on $\mathcal{G}_b$ that maximizes the quantity $\sum_{(v, \rho) \in \mathcal{E}^b_t} w_{v, \rho}$.
\end{itemize}
In both cases we use the well-known \emph{blossom algorithm} of \citet{edmonds1965maximum}. Not surprisingly, MWM results in high quality allocations, but that comes with an overhead in running time, compared to simpler, `local' solutions (see Section \ref{Simulation Results}). This is because blossom's worst-case time complexity -- on a graph $(V, E)$ -- is $\mathcal{O}(|E| |V|^2)$, and we have to run it three times, one for each step of the Ridesharing problem. Additionally, the MWM algorithm inherently requires a global view of the whole request set in a time window, and is therefore not a good candidate for the fast, decentralized solutions that are more appealing for real-life applications.

\subsubsection{ALtruistic MAtching Heuristic (ALMA), \cite{ijcai201931,Danassis2021,ijcai2021-18,danassis2022Scalable,DanassisEcai2020}}

ALMA is a recently proposed lightweight heuristic for weighted matching. A distinctive characteristic of ALMA is that agents (in our context: requests / rides) make decisions locally, based solely on their own utilities. In particular, while contesting for a resource (in our context: request / taxi), each agent will back-off with probability that depends on their own utility loss of switching to their next most preferred resource. E.g., for step (b) of the Ridesharing problem, suppose that for the agent representing ride $\rho$, the next most preferred taxi to $v$ is $v'$, then $loss = w_{v, \rho} -  w_{v', \rho}$. The back-off probability ($P(\cdot)$) is computed individually and locally, based on Equation\footnote{The parameter $\epsilon$ places a threshold on the minimum / maximum back-off probability.} \ref{Eq: backoff}.
\begin{equation} \label{Eq: backoff}
	P(loss) =
	\begin{cases}
		1 - \epsilon, & \text{ if } loss \leq \epsilon \\
		\epsilon, & \text{ if } 1 - loss \leq \epsilon \\
		1 - loss, & \text{ otherwise}
	\end{cases}
\end{equation}

Intuitively, agents that do not have good alternatives will be less likely to back-off and vice versa. The algorithm is inherently decentralized, requires only a 1-bit partial feedback from the resource (indicating whether the resource is free or not), and has constant in the total problem size running time, under reasonable assumptions on the preference domain of the agents. Thus, it is an ideal candidate for an \emph{on-device} solution. Moreover, in \cite{ijcai201931} it was shown to achieve high quality results on a simpler version of step (a) of the Ridesharing problem, and in \cite{Danassis2021} it was shown that it can be adapted to protect the privacy of the agents.

\subsubsection{Greedy}

Greedy is a very simple algorithm, which selects a node $i \in V$ of a graph $G=(V,E)$ uniformly at random, considers all the edges $(i,j)$ with endpoint $i$, and matches $i$ with a node $j^{*}$ that is the endpoint of the edge with the largest weight among those, i.e.,  $(i,j^{*}) \in \arg\max(w_{i, j})$. Greedy approaches are appealing\footnote{\label{AppealOfGreedy}\cite{widdows2017grabshare} reports that GrabShare's scheduling component has used an entirely greedy approach to allocate bookings to drivers. Lyft also started with a greedy system \cite{lyftblog1}.}, not only due to their low complexity, but also because real-time constraints dictate short planning windows which diminish the benefit of batch optimization solutions compared to myopic approaches \cite{widdows2017grabshare}.

\subsubsection{Approximation (Appr), \cite{bei2018algorithms}}

Approximation (Appr) is a recently-proposed offline algorithm due to \citet{bei2018algorithms} which can be used to solve steps (a), and (b) of the Ridesharing problem. The algorithm takes a two-phase approach which is also based on maximum weight matchings (or more accurately, the equivalent notion of minimum cost matchings), but on a set of different weights (to the ones we defined for the MWM algorithm). In particular:
\begin{itemize}
	\item[-] First, it matches requests to shared rides using minimum cost matching based on the shortest distance to serve any request pair but on the worst pickup choice. Formally, the algorithm defines the quantities:
	\[
	w_{ij} = \min\{\delta(s_1,s_2)+\delta(s_2,d_1)+\delta(d_1,d_2), \delta(s_1,s_2)+\delta(s_2,d_2)+\delta(d_2,d_1)\} 
	\]
	\[
	w_{ji} = \min\{\delta(s_2,s_1)+\delta(s_1,d_1)+\delta(d_1,d_2), \delta(s_2,s_1)+\delta(s_1,d_2)+\delta(d_2,d_1)\} 
	\]
and then chooses $w^1(i,j) = \max\{w_{ij},w_{ji}\}$. Intuitively, $w_{ij}$ is the distance of the shortest path that picks up request $r_1$ first (at its source location $s_1$), and similarly, $w_{ji}$ is the distance of the shortest path that picks up request $r_2$ first.  
\item[-] Then it matches rides to taxis using again minimum cost matching, and assuming the weight to be the distance of the closest pick-up location of the two. Formally, let $w^2(v,\langle r_i, r_j \rangle) = \min\{\delta(s_v, s_i), \delta(s_v,s_j)\}$, where $s_v$ is the position of taxi $v$, and compute a minimum cost matching in the bipartite graph defined by pairs $\langle r_i, r_j \rangle$ matched in the previous step and taxis, with weights defined by $u^2$.
\end{itemize}
\citet{bei2018algorithms} prove a worst-case approximation guarantee of $2.5$ for the algorithm.

\subsubsection{Postponed Greedy (PG), \cite{ashlagi2018maximum}}

Postponed Greedy (PG) is another very recently proposed, algorithm for the maximum weight online matching problem with deadlines (step (a) of the Ridesharing problem). The algorithm is online, meaning that it considers the potential requests that might appear in the future when making decisions about the present; its competitive ratio was proven to be $1/4$ by \citet{ashlagi2018maximum}. Contrary to our setting, the algorithm was designed for fixed deadlines, i.e., $k_r = c, \forall r \in \mathcal{R}$. 

The algorithm is best described in terms of an \emph{auction} environment \cite{ashlagi2018maximum} as follows. Let $S_t$ and $B_t$ be the sets of \emph{virtual sellers} and \emph{virtual buyers} at time $t$ respectively. When a request $r$ appears at time $t$, the algorithm creates a virtual seller $s_r$ and a virtual buyer $b_r$ for that request, and adds them to the aforementioned sets, i.e., $S_t \leftarrow S_{t-1} \cup \{s_r\}$ and  $B_t \leftarrow B_{t-1} \cup \{b_r\}$. In other words, every request has two copies: a buyer and a seller. These are then placed in a virtual weighted bipartite graph $G=(S_t,B_t,E_t)$, where the edge weights are defined in the same manner as the weights of $\mathcal{G}_a$ (see `Matching Graphs' in Section \ref{algorithms}). The algorithm proceeds to match the newly added buyer $b_r$ with a seller $s_{r^*}$ in a greedy manner, i.e.,  $(b_r,s_{r^*}) \in \underset{r' \in S_{t-1}}{\arg\max}(w_{r, r'})$. This choice remains fixed for subsequent time steps. When the request $r$ becomes critical (i.e., the deadline is about to be met), the `role' of the request as either a seller or a buyer is conclusively chosen (uniformly at random). If $r$ is a seller, and a subsequent buyer was matched with $r$, the match is finalized and is included in the output matching.


The major difference between the setting consider by \citet{ashlagi2018maximum} and our setting is that for us, requests become critical out-of-order, and a critical request cannot be matched later. Thus, we apply the following modification: when a request becomes critical, if determined to be a seller, the match is finalized (if one has been found), otherwise the request is treated as a single ride.

\subsubsection{Greedy Dual (GD), \cite{bienkowski2018primal}}

Greedy Dual is an online algorithm for solving the minimum cost (bipartite) perfect matching with delays, i.e., both steps (a), and (b) of the Ridesharing problem, which is based on the popular primal-dual technique \cite{goemans1997primal}. The weight (cost) of an edge in this setting includes arrival times as well, specifically: 
\[
w_{r_1, r_2} = \frac{(\delta(s_1, s_2) + \delta(d_1, d_2))}{u_{\text{average}}} + |t_1 - t_2|,\] 
where $u_{\text{average}}$ is the average speed (see Section \ref{Taxi Vehicles}). The algorithm partitions all the requests into \emph{active sets}, starting with the singleton $\{r\}$ for a newly arrived request $r$. As is typical in the primal-dual approach, at every time-step $t$ these actives sets `grow', until the weight of the edges of different active sets make the dual constraints of the problem tight (i.e., satisfied with equality). At this point the active sets merge, and the algorithm matches as many pairs of free requests in these sets as possible. 

The algorithm has a competitive ratio of $\mathcal{O}(|\mathcal{R}|)$ and works with infinite metric spaces, potentially making the algorithm better suited for applications like the Ridesharing problem. Yet, in terms of our setup, it does not take into account the willingness to wait ($k_r$), thus missing matches of requests that became critical. Despite being designed for bipartite matchings as well, we opted out from using it for step (b) since it would require to create a new node every time a taxi vehicle drops-off a ride and becomes available.

\subsubsection{Balance (Bal), \cite{manasse1990competitive}}

Balance is a simple and classic algorithm for the $k$-server problem from the literature of competitive analysis. The rationale behind the algorithm is that it tries to balance out the distance traveled by taxis over the course of their operation, trying to maintain the workload as equal as possible. In particular, a ride is served by the taxi that has the minimum sum of the distance traveled so far plus its distance to the source of the ride (chosen uniformly at random between the sources of the two requests composing the ride). Specifically, ride $\rho$ will be matched to taxi $v$:
\begin{equation}
	(v, \rho) = \underset{v \in \mathcal{V}_t}{\arg\min}(\text{driven}(v) + \delta(s_v, s_{\rho}))
\end{equation}
\noindent
where $\text{driven}(v)$ denotes the distance driven by taxi $v$ so far, and $s_{\rho}$ is selected equiprobably among $s_1$ and $s_2$. The algorithm is min-max fair, i.e., it greedily minimizes the maximum accumulated distance among the taxis. The competitive ratio of the algorithm is $|\mathcal{X}|-1$ in arbitrary metric spaces with $|\mathcal{X}|$ points \cite{manasse1990competitive}.

\subsubsection{Harmonic (Har), \cite{raghavan1989memory}}

The Harmonic algorithm (Har) is another classic randomized algorithm from the $k$-server problem literature, which is simple and memoryless (i.e., it does not need to `remember' the decisions that it took in previous steps). The algorithm matches a taxi with a ride with probability inversely proportional to the distance from its source (chosen uniformly at random between the sources of the two requests composing the ride). Specifically, ride $\rho$ will be matched to taxi $v$ with probability:
\begin{equation}
	P(v, \rho) = \frac{\frac{1}{\delta(s_v, s_{\rho})}}{\underset{\rho' \in \mathcal{P}_t}{\sum}{\frac{1}{\delta(s_v, s_{\rho'})}}}
\end{equation}

\noindent
where $s_{\rho}$ and $s_{\rho'}$ are both selected equiprobably among $s_1$, $s_2$ and $s_{1'}$, $s_{2'}$, respectively. The trade-off for its simplicity is the high competitive ratio, which is $\mathcal{O}(2^{|\mathcal{V}|} \log |\mathcal{V}|)$ \cite{bartal2000harmonic}.

\subsubsection{Double Coverage (DC), \cite{Chrobak90newresults}}

Double Coverage (DC) is one of the two most famous $k$-server algorithms in the literature. The algorithm is designed to run on a specific type of metric space called an HST (Hierarchical Separated Tree, see Section \ref{Embedding into HSTs}). 
For a general metric spaces $\mathcal{X}$, the algorithm can be applied by first embedding $\mathcal{X}$ to an HST (a process which is referred to as an `HST embedding'). This process `simulates' the general space $\mathcal{X}$ by an HST, in the sense that the HST approximately captures the properties of the original space $\mathcal{X}$. The points of $\mathcal{X}$ are the leaves of the HST. 

Given an HST, the algorithm works as follows. To determine which taxi will serve a ride, all \emph{unobstructed} taxis move towards its source, i.e., a leaf of the HST (chosen randomly between the sources of the two requests sharing the ride) with equal speed. Initially, all taxis are unobstructed. During this movement process, a taxi becomes \emph{obstructed} when its path from its current location to the leaf corresponding to the ride is `blocked' by another taxi, meaning that it would have to move through the same position in the tree that another taxi has already been at, to reach the leaf. In this case, the taxi stops (as the `blocking' taxi is closer to serving the ride), while the remaining taxis keep moving as before. When some taxi reaches the leaf corresponding to the ride, the process stops, and each taxi maintains its current position on the HST.

To implement the algorithm, we first appropriately discretize our metric space and then perform the HST embedding as described in \cite{bartal1996probabilistic,fakcharoenphol2004tight} (see Section~\ref{Embedding into HSTs} for more details).  Given that only leaves correspond to locations on $\mathcal{X}$, we chose to implement the \emph{lazy} version of the algorithm (which is worst-case equivalent to the original definition e.g., see \cite{KOUTSOUPIAS2009105}), i.e., only the taxi serving the ride will move on $\mathcal{X}$; one can envision a process in which the taxis `virtually' move as described above, but once the ride has been served, all taxis are restored to their original positions. This is also on par with the main goal of minimizing the distance driven. The algorithm is $k$-competitive on all tree metrics~\cite{chrobak1991optimal}.

\subsubsection{Work Function (WFA), \cite{chrobak1991server,koutsoupias1995k}}

The Work Function algorithm (WFA) is perhaps the most important $k$-server algorithm, as it provides the best competitive ratio to date, due to the celebrated result of \cite{koutsoupias1995k}. Intuitively, to decide which taxi (or server) will be the one to serve a ride that just appeared at time $t$, and, more generally, the movement of the other taxis, the algorithm:
\begin{itemize}
	\item[-] computes the (offline) optimal solution until time $t - 1$, meaning the best possible allocation of rides to taxis using the information from the beginning of the algorithm until the appearance of the ride at time $t$,
	\item[-] computes a \emph{greedy cost} for switching between configurations,
	\item[-] chooses the new taxi positions that minimize the sum of the two aforementioned costs.
\end{itemize}
More formally, let $L^t = (l^t_1, l^t_2, \dots, l^t_{|\mathcal{V}|})$ denote the \emph{configuration} of the fleet of taxis $\mathcal{V}$ at time-step $t$, i.e., a vector of taxi locations, where $l^t_v$ specifies the location of taxi $v$. Let $\text{OPT}_t(L)$ be the optimal (total distance-minimizing) way of serving rides that appear at times $1$ through $t$, such that the taxis end up at configuration $L$. To choose configuration $L^{t}$, it uses the following rule:
\[
L^{t} = \arg\min_{L}\{\text{OPT}_t(L)+ d(L^{t-1},L)\}
\] 
The WFA serves ride $\rho_t$ at time-step $t$ by switching from the current taxi configuration $L^{t-1}$, to a new configuration $L^{t}$. Specifically, it selects $L^{t}$ which minimizes (a) the minimum total cost of starting from $L^{0}$, serving in turn $\rho_0, \rho_1, \dots, \rho_{t-1}$, and ending up in $L^{t}$, plus (b) the distance traveled by a taxi to move from its position in $L^{t-1}$ to that in $L^{t}$.

An obvious obstacle that makes the algorithm intractable in practice is that the complexity increases from step to step, resulting in computation and/or memory issues. To circumvent this obstacle, we implemented an efficient variant using network flows, as described in \cite{Rudec2013}. Yet, as the authors of \cite{Rudec2013} state as well, the only practical way of using the WFA is switching to its window version $w$-WFA, where we only optimize for the last $w$ rides. Even though the complexity of $w$-WFA does not change between time-steps, it does change with the number of taxis. The resulting network has $2|\mathcal{P}| + 2|\mathcal{V}| + 2$ nodes, and we have to run the Bellman–Ford algorithm \cite{bellman1958routing} at least once to compute the potential of nodes and make the costs positive (Bellman–Ford runs in $\mathcal{O}(|\mathcal{P}||\mathcal{V}|)$. We refer the reader to \cite{bertsekas1998network} for more details on network optimization. As before, the source of the ride is chosen randomly between the sources of the two requests composing the ride.


\subsubsection{k-Taxi, \cite{coester2018online}}

This is a very recent algorithm for the $k$-taxi problem, which provides the best possible competitive ratio. The algorithm operates on HSTs, where the rides and taxis at any time are placed at its leaves. First, it generates a Steiner tree that spans the leaves that have taxis or rides, and then uses this tree to schedule rides, by simulating an electrical circuit. In particular, whenever a ride appears at a leaf, the algorithm interprets the edges of the tree with length $R$ as resistors with resistance $R$, which determine the fraction of the current flow that will be routed from the node corresponding to the taxi towards the ride. These fractions are then interpreted as probabilities which determine which taxi will be chosen to pick up the ride.

\subsubsection{High Capacity (HC), \cite{alonso2017demand}} \label{High Capacity}

This algorithm comes from a highly-cited paper, and is the only one in our evaluated approaches that addresses vehicle relocation (step (c)). Contrary to our approach, it tackles steps (a), and (b) simultaneously, leaving step (c) as a separate sub-problem. The algorithm consists of five steps:

\begin{enumerate}[label=(\roman*)]
	\item Computing a pairwise request-vehicle shareability graph (RV-graph) \cite{santi2014quantifying}. The RV-graph represents which requests and vehicles might be pairwise-shared, with edges connecting all possible requests to pair and all possible vehicles to serve a request.
	\item Computing a graph consisting of feasible (candidate) trips and the set of vehicles that can execute them (RTV-graph). This is a tripartite graph with edges connecting requests to trips (a request is connected to a trip if it is part of it), and edges connecting trips to vehicles (an edge between vehicle and a trip exists if the vehicle is able to serve it).
	\item Computing a greedy solution for the RTV-graph. In this step, rides are assigned to vehicles iteratively in decreasing size of the trip (in our case, we first assign shared rides (two requests), and then single rides) and increasing cost (e.g., delay).
	\item Solving an ILP to compute the best assignment of vehicles to trips, using the previously computed greedy solution as an initial solution.
	\item (optional) Rebalancing of free vehicles. If there remain any unassigned requests, it solves an ILP to optimally assign them to idle vehicles based on travel times.
\end{enumerate}

We use CPLEX \cite{bliek1u2014solving} to solve the ILPs.


\subsubsection{Baseline: Single Ride} \label{single ride baseline}

Uses MWM to schedule the serving of single rides to taxis (there is no ridesharing, i.e., we omit step (a) of the Ridesharing problem).

\subsubsection{Baseline: Random} \label{random baseline}

Makes random matches, provided that the edge weight is non-negative. \\

While our evaluation contains many recently proposed algorithms for matching, the observant reader might notice that, with the exception of $k$-taxi, our $k$-server algorithms are from the classical literature. We did consider more recent $k$-server algorithms (e.g., \cite{dehghani2017stochastic,lee2018fusible,bansal2015polylogarithmic}), but their complexity turns out to be prohibitive. This is mainly because they proceed via an `online rounding' of an LP-relaxation of the problem, which maintains a variable for every (time-step, point in the metric space) pair. Even for one hour (3600 time-steps) and our discretization of Manhattan (5018 nodes), we need more than 18 million variables (230 million for NYC).

\section{Scalability Challenges} \label{Challenges}

To highlight the challenges in the design of CARs, we will be referring to our evaluation setting (see Section \ref{Modeling}), which accurately models a real-world application, in terms of both \emph{scale} and \emph{detail}. Let $\mathcal{V}$, $\mathcal{R}$ denote the set of vehicles / requests, respectively. Recall that in our setting, which involves real data from NYC taxi records, there are $272$ new requests per minute on average, totaling to $391479$ requests in the broader NYC area ($352455$ in Manhattan) on the evaluated day (Jan, 15, 2006). By law, there are $13,587$ taxis in NYC\footnote{\label{TLCYellowCab}\url{https://www1.nyc.gov/site/tlc/businesses/yellow-cab.page}}.

\subsection{ILP Approaches}

A natural approach would be to try to use Integer Linear Programs (ILPs) for matching passengers to other passengers or rides, under spatial and temporal constraints, similarly to the High Capacity algorithm of \cite{alonso2017demand} (which can be seen as a CAR with steps (a) and (b) intertwined). As is commonly the case with ILPs, the problem is scalability; the number of variables can be as large as $\mathcal{O}(|\mathcal{V}| |\mathcal{R}|^2)$ -- which results in $27$ - $216$ million variables, given that every time-step we have approximately $300$ - $600$ requests, and as many taxis -- and the number of constraints is $|\mathcal{V}| + |\mathcal{R}|$. This makes ILP approaches prohibitive as components in CARs.  The latter make hard to even compute the initial greedy solution in real-time. \citeauthor{alonso2017demand} circumvent this issue by enforcing delay constraints, specifically they ignore requests that are not matched to any vehicle within a maximum waiting time. This is not possible in our model since we have to serve all requests (service guarantee).\footnote{For the sake of completeness we have evaluated the High Capacity algorithm on much smaller test cases; see Appendix \ref{appendix: Simulation Results}.}


\subsection{MWM Approaches}

Given that all three parts of the ridesharing problem can be viewed as matching problems, a natural approach would be to run maximum-weight matching (MWM) \emph{in batches} (e.g., \cite{bei2018algorithms}), meaning that we serve the requests that have accumulated over a pre-specified time window. The MWM problem can be solved via the classic \emph{blossom algorithm} \cite{edmonds1965maximum} with run time -- on a graph $(V, E)$ -- of $\mathcal{O}(|E| |V|^2)$.

\subsection{\texorpdfstring{$k$}-server/taxi Algorithms}

Many of these algorithms operate by embedding the input metric space $\mathcal{X}$ into a distribution $\mu$ over Hierarchical Separated Trees (HSTs) (e.g., the classic \emph{double-coverage} \cite{Chrobak90newresults}), and thus to apply them in practice, it is necessary to examine the size of these trees. Given that the geo-coordinate system is a discrete metric space, we could directly embed it into HSTs. Yet, the size of the space is huge, and hence for better discretization we have opted to generate the graph of the street network of NYC (see Section \ref{Embedding into HSTs}). The resulting graph for NYC contains $66543$ nodes, and $95675$ edges ($5018$, and $8086$ for Manhattan). Here, there is an obvious interplay between the accuracy of the embedding and the algorithm's complexity.


More recent $k$-server algorithms (e.g., \cite{dehghani2017stochastic,lee2018fusible,bansal2015polylogarithmic}) use sophisticated `online rounding' techniques; these however require maintaining variables for every (time-step, point in the metric space) pair, which makes them prohibitive for any large-scale real-world application; even for one hour (3600 time-steps) and our discretization of Manhattan (5018 nodes), we would need more than 18 million variables (230 million for NYC).

\subsection{Observability}

Most approaches are centralized, and require a \emph{global} view of the \emph{entire window}, which is hard to scale. As autonomous agents proliferate, a practical and applicable CAR must be distributed and ideally run \emph{on-device}.

\section{Vehicle Relocation Challenges} \label{Relocation}

There are two ways to enforce relocation: \emph{passive}, and \emph{active}. Ridesharing platforms, like Uber and Lyft, have implemented market-driven pricing as a passive form of relocation. Counterfactual analysis performed in \cite{buchholz2018spatial} shows that implementing pricing rules can result in daily net surplus gains of up to $232000$ and $93000$ additional daily taxi-passenger matches. While the gains are substantial, the market might be slow to adapt, and drivers and passengers do not always follow equilibrium policies. Contrary to that, our approach is \emph{active}, in the sense that we directly enforce relocation. Moreover, we adopt a more \emph{anthropocentric} approach: in our setting, the demand is fixed, thus the goal is not to increase revenue as a result of serving more rides, but rather to improve the QoS\footnote{Decreased delays can also in turn improve revenue by serving more requests in a fixed time window.}.

There are many ways to approach dynamic relocation. Most of the employed relocation approaches are course-grained; the network is generally divided into several zones, blocks, etc. \cite{gueriau2018samod,vosooghi2019shared,martinez2017insights} and the entities (e.g., the vehicles) move between the zones. However, compared to other shared mobility systems, dynamic ridesharing posses unique challenges, meaning that such coarse-grained approaches are not appropriate\footnote{As a matter of fact, we tried zone based relocation (generating zones based on historical data using the OPTICS clustering algorithm \cite{Xu2015}, or using pre-defined clusters based on population density according to the NYC census data (\url{https://guides.newman.baruch.cuny.edu/nyc_data/nbhoods})). Due to the vast number of requests, the only discernible clusters were of large regions (Manhattan, Bronx, Staten Island, Brooklyn, or Queens), which does not allow for fine-grained relocation. As a result, we achieved significantly inferior results.}: most of them are centralized -- thus computationally intensive and not scalable --, they might not take into account the actions of other vehicles, potentially leading to over-saturation of high demand areas, and, most importantly, they are \emph{slow to adapt} to the highly dynamic nature of the problem (e.g., responding to high demand generated by a concert, or the fact that vehicles remain free for only a few minutes at a time). The problem clearly calls for fine-grained solutions, yet such approaches in the literature are still rather scarce. High Capacity (HC) employs fine-grained relocation. HC solves an ILP, which could reach high quality results, but it is not scalable nor practical. Ideally, we would like a solution that can run \emph{on-device}. The $k$-server algorithms perform an implicit relocation, yet they are primarily developed for adversarial scenarios, and do not utilize the plethora of historic data\footnote{NYC TLC has been proving data on yellow taxi trips since 2009.}. In reality, requests follow patterns that emerge due to human habituality (e.g., during the first half of the day in Manhattan, there are many more drop-offs in Midtown compared to pickups \cite{buchholz2018spatial}).

\subsection{Patterns in Customer Requests}

\begin{figure}[t!]
	\centering
	\includegraphics[width = 1 \linewidth, trim={0.5em 0.8em 0.65em 0.7em}, clip]{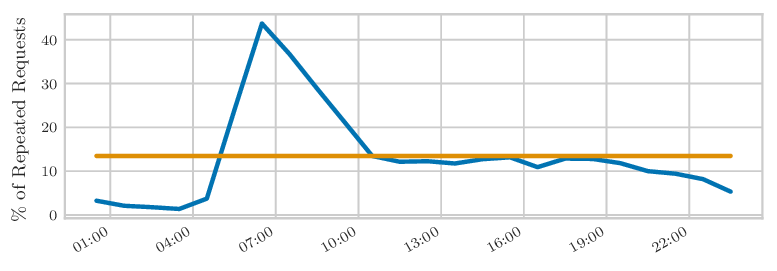}
	\caption{Percentage of similar trips per hour in Manhattan, January 15, 2016 (blue line). Mean value = $13.3\%$ (yellow line).}
	\label{fig: repeatedRequests}
\end{figure}

To confirm the existence of transportation patterns, we performed the following analysis: For each request $r$ on January 15\footnote{January 15, 2016 was selected as a representative date for our simulations since it is not a holiday, and it is a Friday thus sampling for past requests results in a representative pattern (contrary to sampling on a weekend for example).}, we searched the past three days for requests $r'$ such that $|t_{r} - t_{r'}| \leq 10$, $\delta(s_{r}, s_{r'}) \leq 250$, and $\delta(d_{r}, d_{r'}) \leq 250$. The results are depicted in Figure \ref{fig: repeatedRequests}. On average, $13.3\%$ of the trips are repeated across all three previous days, peaking at $43.7\%$ on rush hours (e.g., 6-8 in the morning). Note that predicting transport demand based on historic data is not an easy task; $13.3\%$ is about $47000$ trips, which is rather significant in raw numbers.

\subsection{Relocation Matching Graph}

Given the high density of the requests, and the low frictions of the taxis (i.e., taxis remain free for relocation only for a short time window), we opted for a simple, fine-grained, matching approach. We use the history to predict a set of \emph{expected future requests}. Specifically, let $D$, and $T$ be the sampling windows, in days and minutes respectively (we used $D = 3$, and $T = 2$). Let $t$ denote the current time-step. The set of past requests on our sampling window is $\mathcal{R}_{\text{past}} = \{r: t_{r} - t \leq T\}$, as long as $r$ appeared at most $D$ number of days prior to $t$. The set of expected future requests $\mathcal{R}_{\text{future}}$ is generated by sampling from $\mathcal{R}_{\text{past}}$. Relocation is performed in a just-in-time manner, every time the set of idle vehicles is not empty. We generate similar matching graphs as in Section \ref{algorithms}, and then we proceed to match requests to shared rides, and rides to idle taxis. The difference being that now the set of nodes of $\mathcal{G}_a$ is $\mathcal{R}_{\text{future}} \cup \mathcal{R}_{t}$. Finally, each idle taxi starts moving towards the source of its match (given that these are expected rides, the source is picked at random between the sources of the two requests composing the ride).

\section{Evaluation} \label{Evaluation}

\subsection{Employed CARs} \label{Employed CARs}

Evaluating all of the possible combinations of CAR components is infeasible. To make the evaluation tractable, we first consider only the first two steps of the ridesharing problem (i.e., no relocation). When possible, we use the same component for both steps (a) and (b). $k$-Taxi/Server algorithms, though, can not solve step (a), thus we opted to use the best performing component for step (a) (namely the offline maximum-weight matching (MWM) run in batches). Then, we move to evaluate step (c), testing only the most promising components (namely the MWM and ALMA, plus the Greedy as a baseline). We begin by isolating step (c); we fix the component for (a) and (b) to MWM, to have a common-ground for evaluating relocation. Finally, we present results on \emph{end-to-end} solutions. A list of all the evaluated CARs can be found in Table \ref{tb: algorithms}, while Table \ref{tb: Performance Metrics} contains a summary of all the evaluated metrics.

\begin{table*}[t!]
\centering
\caption{Evaluated performance metrics (global, passenger (Quality of Service), driver, and platform specific).}
\label{tb: Performance Metrics}
\begin{tabularx}{\textwidth}{@{}rX@{}}
\toprule
\textbf{Distance Driven} & Minimize the cumulative distance driven by all vehicles for serving all the requests. We chose this objective as it directly correlates to passenger, driver, company, and environmental objectives. \\
\textbf{Complexity} & Real-world time constraints dictate that the employed solution produces results in a reasonable time-frame\footnoteref{footnote: waiting period}. \\
\midrule
\textbf{Time to Pair} & Expected time to be paired in a shared ride. \\
\textbf{Time to Pair with Taxi} & Expected time to be paired with a taxi. \\
\textbf{Time to Pick-up} & Expected time to passenger pickup. \\
\textbf{Delay} & Additional travel time over the expected direct travel time (when served as a single, instead of a shared ride). \\
\midrule
\textbf{Driver Profit} & Total revenue earned minus total travel costs. \\
\textbf{Number of Shared Rides} & Related to the profit. By carrying more than one passenger at a time, drivers can serve more requests in a day. \\
\textbf{Frictions} & Waiting time experienced by drivers between serving requests (i.e., time between dropping-off a ride, and getting matched with another). Lower frictions indicate lower regret by the drivers. \\
\midrule
\textbf{Platform Profit} & A commission on the driver's fee, and passenger fees. \\
\textbf{Quality of Service (QoS)} & Refer to the passenger metrics. Improving the QoS to their customers correlates to the growth of the company. \\
\textbf{Number of Shared Rides} & The matching rate is important especially in the nascent stage of the platform \cite{dutta2018online}. \\ \bottomrule
\end{tabularx}%
\end{table*}

\subsection{Simulation Results} \label{Simulation Results}

In this section we present the results of our evaluation. For every metric we report the average value out of 8 runs. In what follows we shortly detail only the most relevant results. Please refer to Appendix \ref{appendix: Simulation Results} for the complete results including larger test-cases on the broader NYC area and \emph{omitted metrics, standard deviation values, algorithms} (e.g., WFA, and HC had to be evaluated in smaller test-cases), etc.

\begin{figure*}[t!]
	\centering
	\begin{subfigure}[t]{0.5\textwidth}
		\centering
		\includegraphics[width = 1 \linewidth, trim={0.6em 0.6em 0.6em 1.75em}, clip]{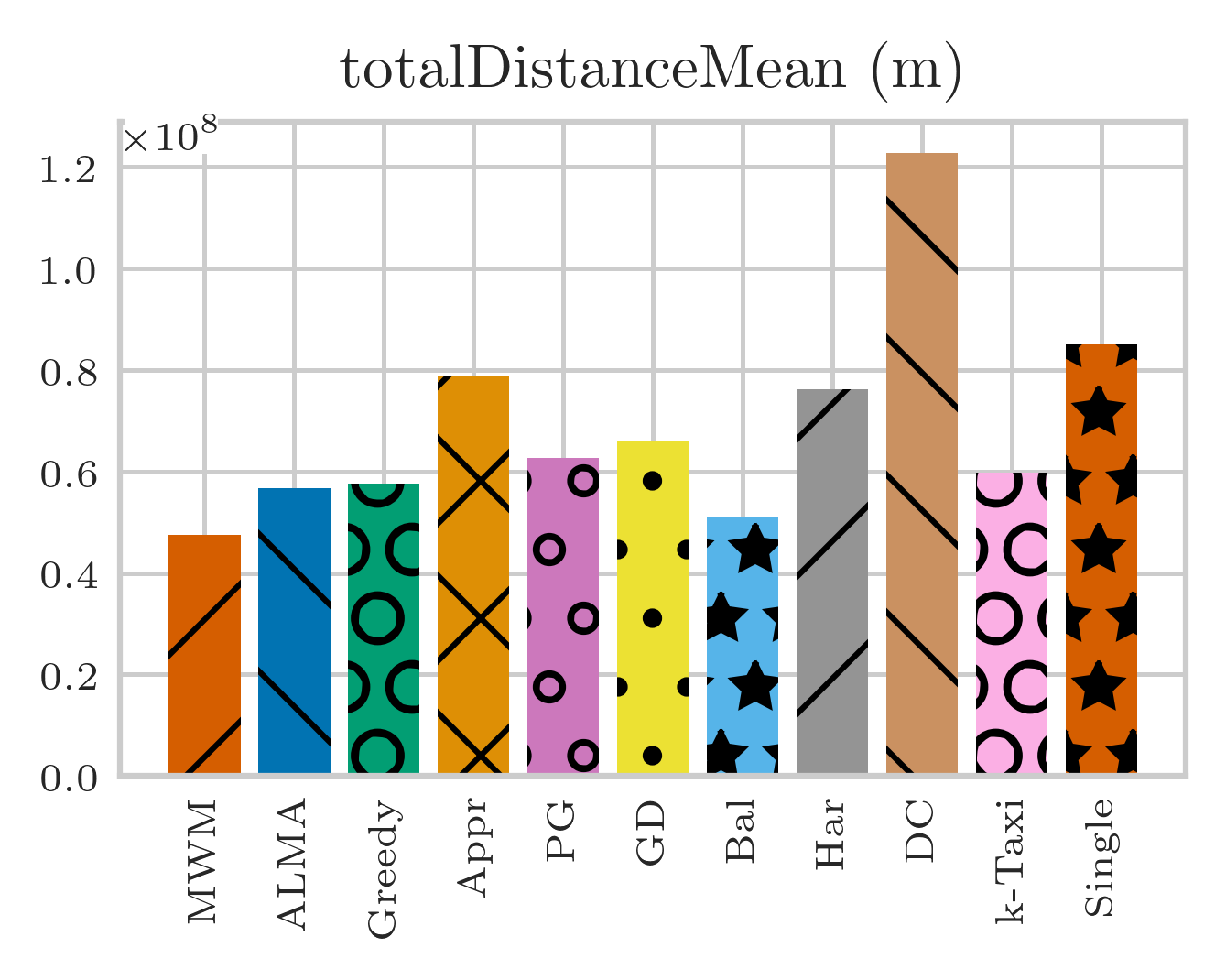}
		\caption{Distance Driven (m)}
		\label{fig: Jan15Hour8to9Manhattan_totalDistanceMean}
	\end{subfigure}%
	~ 
	\begin{subfigure}[t]{0.5\textwidth}
		\centering
		\includegraphics[width = 1 \linewidth, trim={0.6em 0.6em 0.6em 1.85em}, clip]{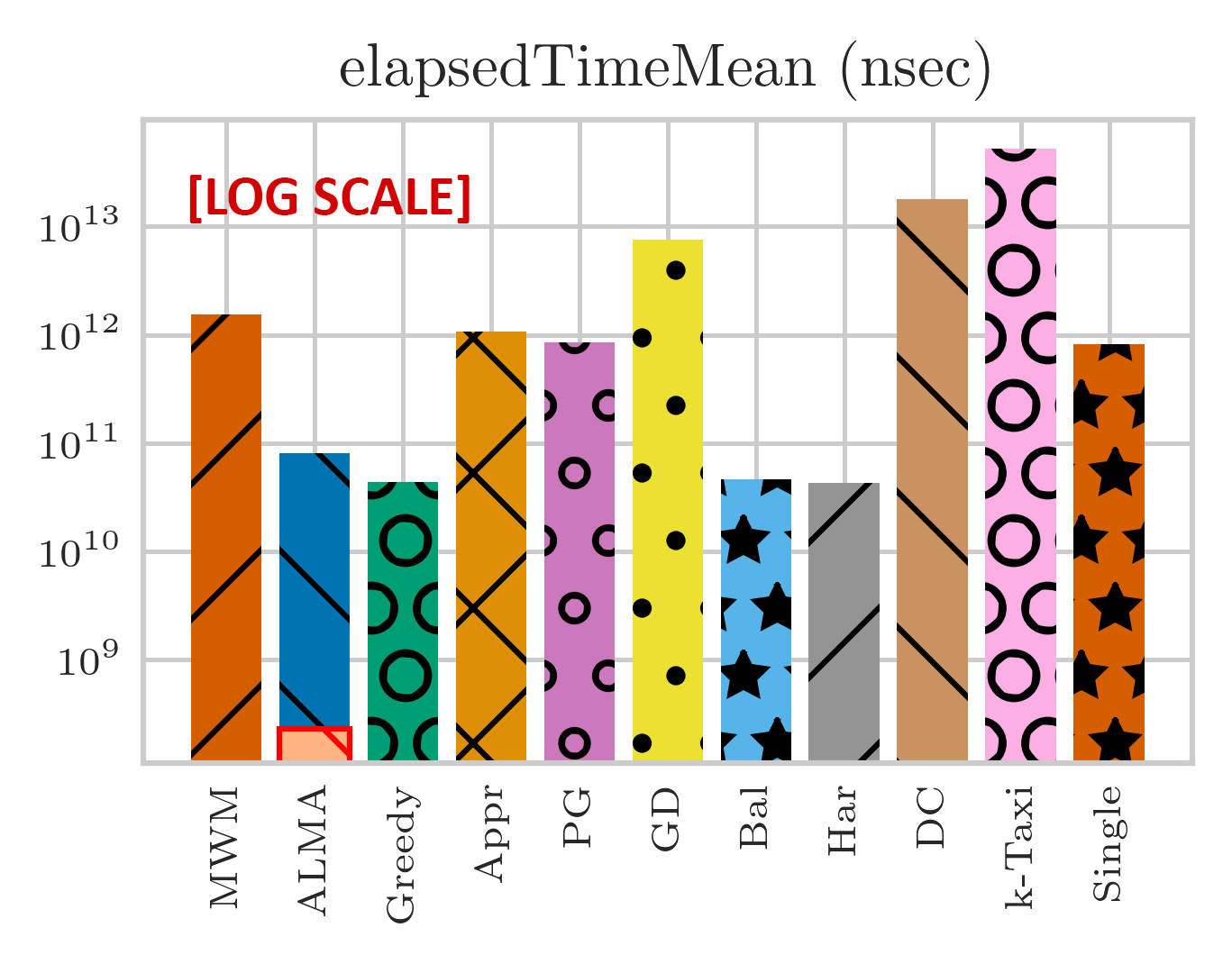}
		\caption{Elapsed Time (ns)}
		\label{fig: Jan15Hour8to9Manhattan_elapsedTimeMean}
	\end{subfigure}%
	
	\begin{subfigure}[t]{0.5\textwidth}
		\centering
		\includegraphics[width = 1 \linewidth, trim={0.6em 0.6em 0.6em 1.85em}, clip]{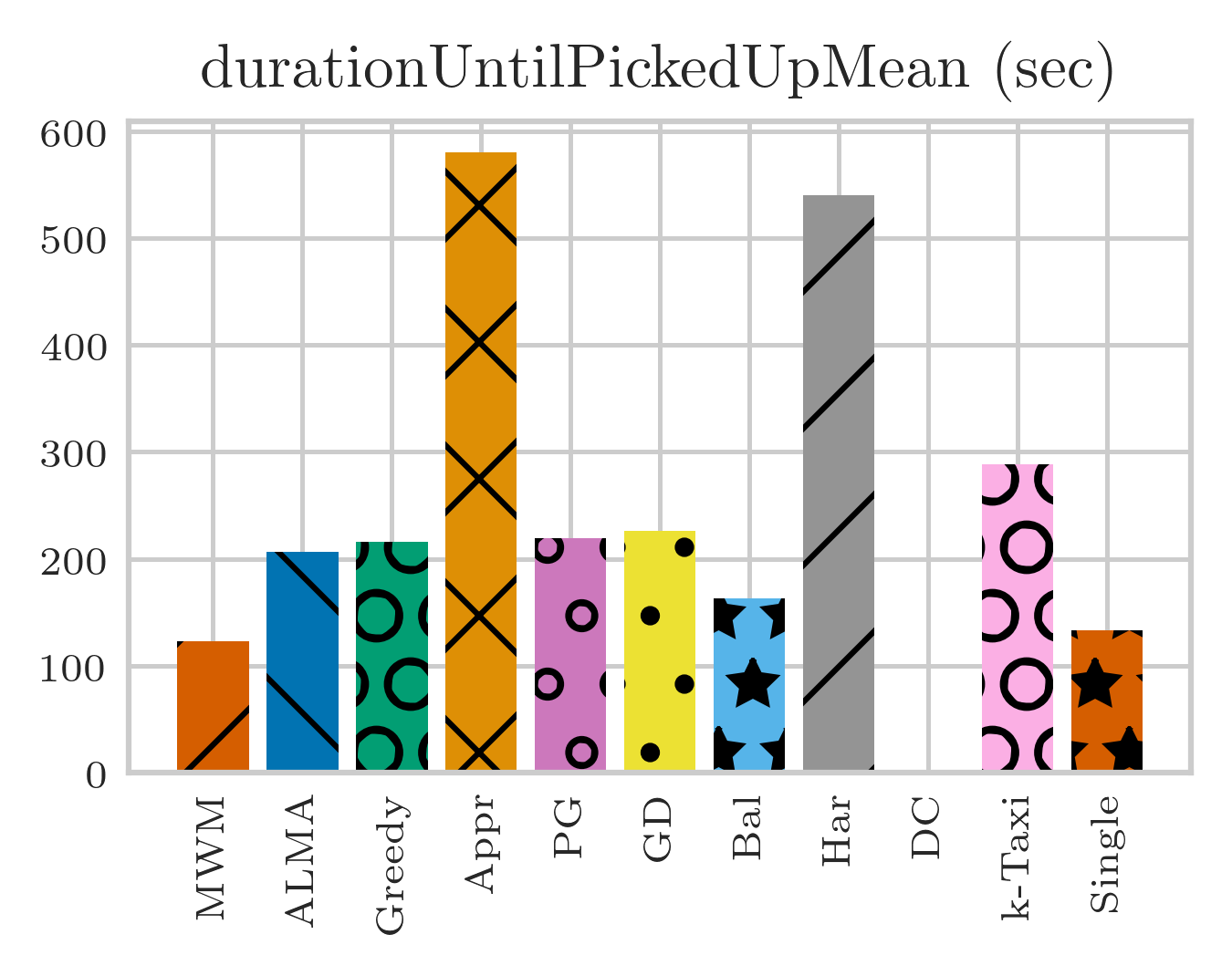}
		\caption{Time to Pick-up (s)}
		\label{fig: Jan15Hour8to9Manhattan_durationUntilPickedUpMean}
	\end{subfigure}%
	~ 
	\begin{subfigure}[t]{0.5\textwidth}
		\centering
		\includegraphics[width = 1 \linewidth, trim={0.6em 0.em 0.6em 1.85em}, clip]{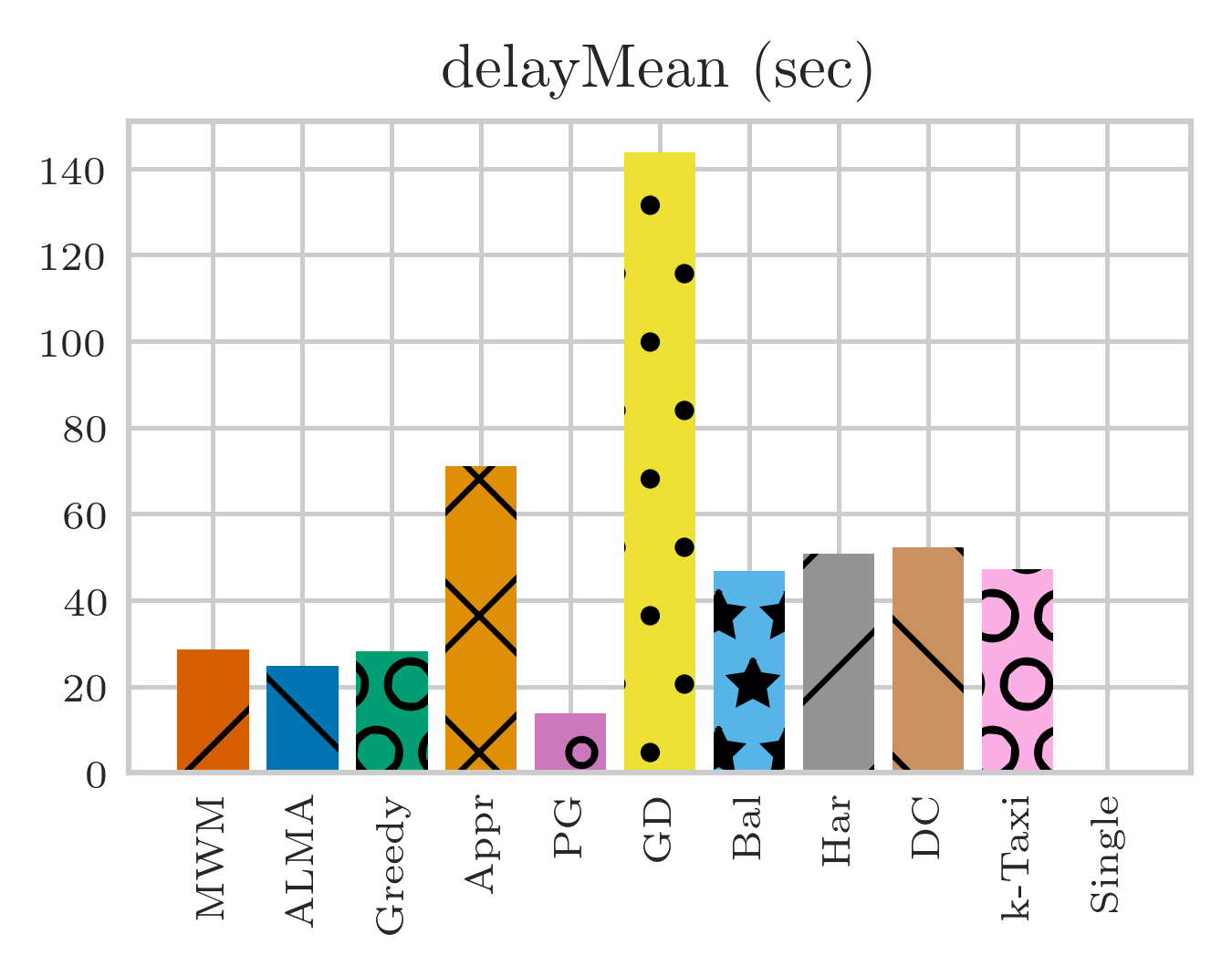}
		\caption{Delay (s)}
		\label{fig: Jan15Hour8to9Manhattan_delayMean}
	\end{subfigure}%
	\caption{08:00 - 09:00, \#Taxis = 4276 (base number). Manhattan, January 15, 2016}
	\label{fig: results one hour}
\end{figure*}

\begin{figure*}[t!]
	\centering
	\begin{subfigure}[t]{0.5\textwidth}
		\centering
		\includegraphics[width = 1 \linewidth, trim={0.6em 0.6em 0.6em 1.6em}, clip]{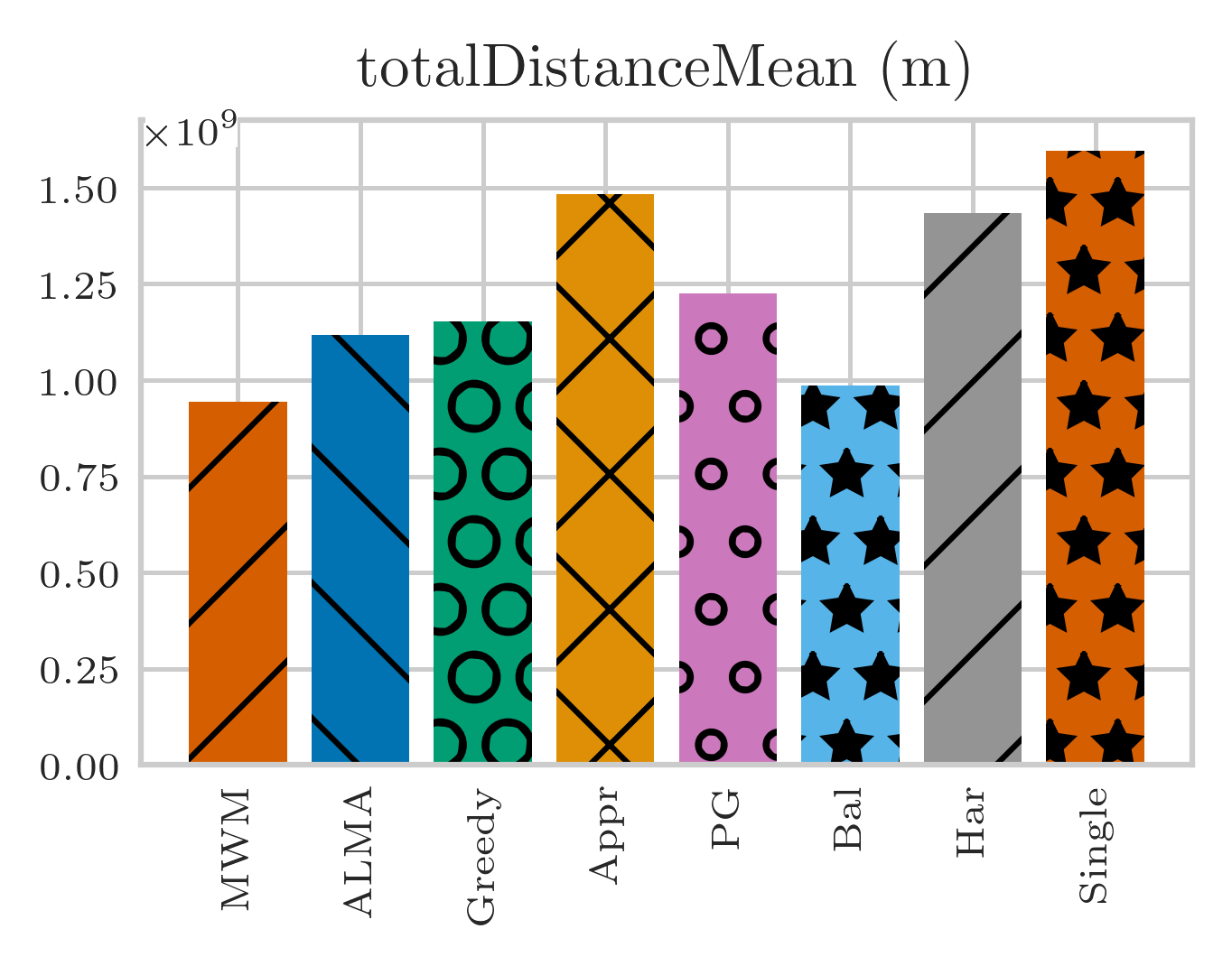}
		\caption{Distance Driven (m)}
		\label{fig: Jan15Manhattan_totalDistanceMean}
	\end{subfigure}%
	~ 
	\begin{subfigure}[t]{0.5\textwidth}
		\centering
		\includegraphics[width = 1 \linewidth, trim={0.6em 0.6em 0.6em 1.85em}, clip]{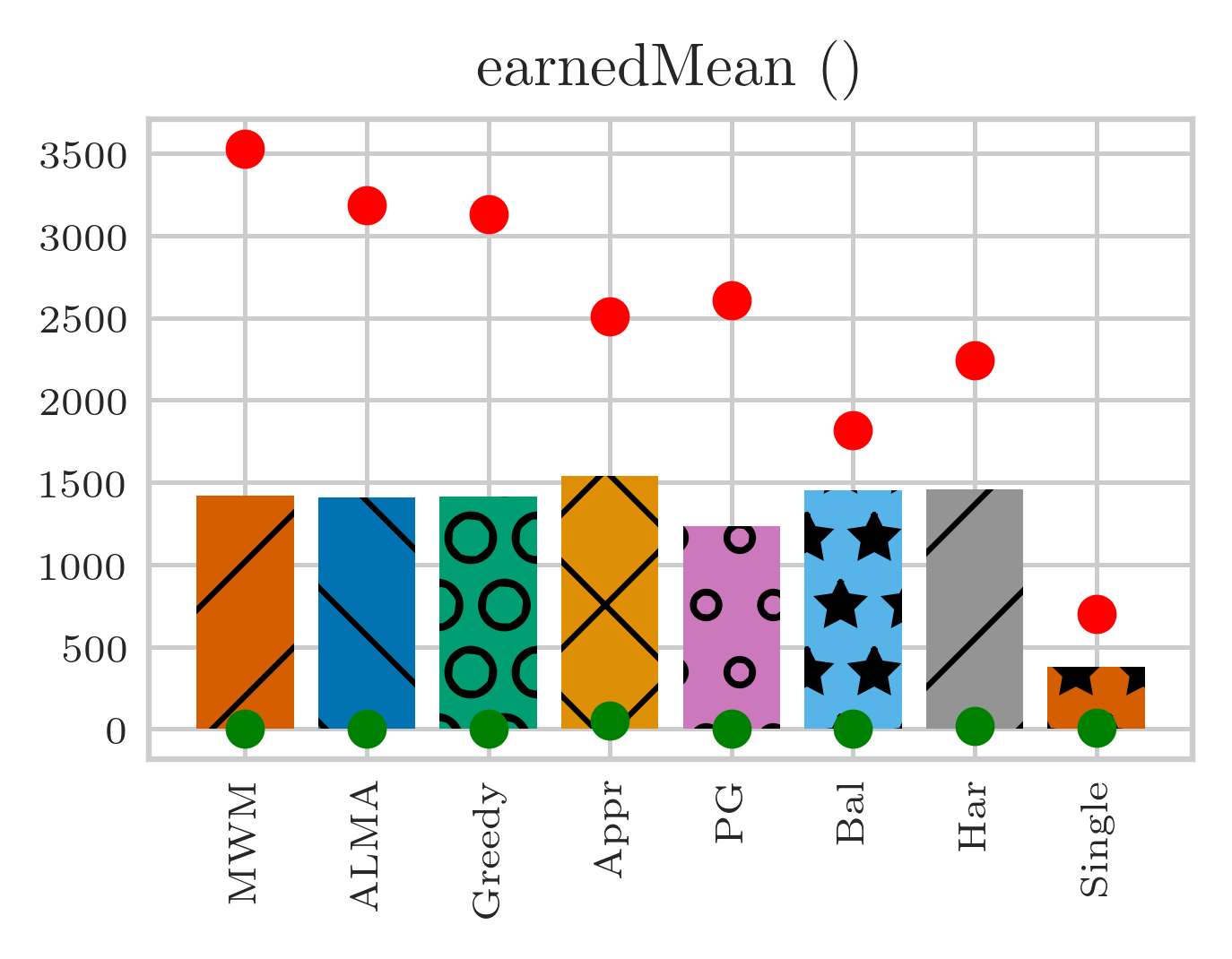}
		\caption{Driver Profit (\$)}
		\label{fig: Jan15Manhattan_earnedMeanMaxMin}
	\end{subfigure}%
	
	\begin{subfigure}[t]{0.5\textwidth}
		\centering
		\includegraphics[width = 1 \linewidth, trim={0.6em 0.6em 0.6em 1.85em}, clip]{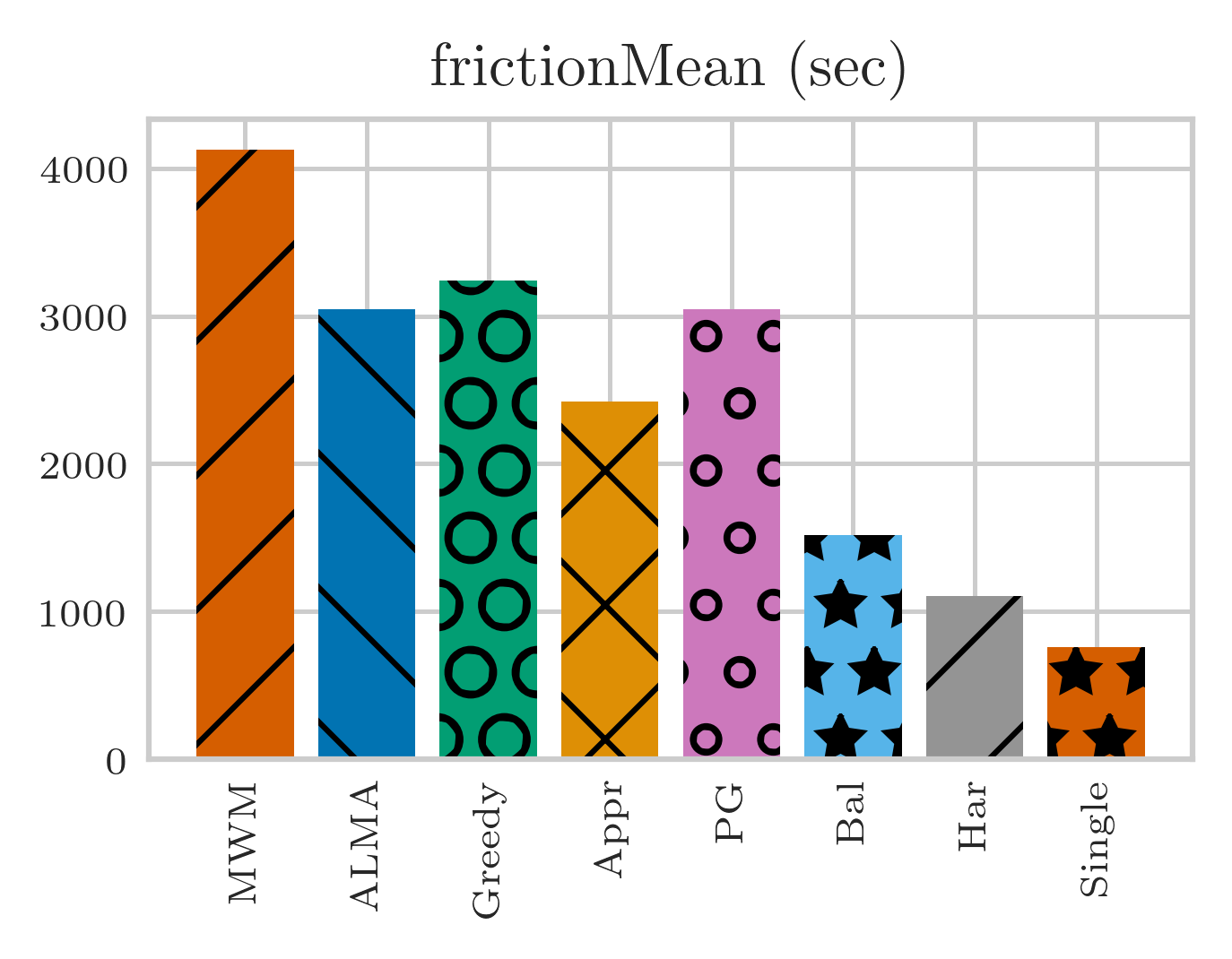}
		\caption{Frictions (s)}
		\label{fig: Jan15Manhattan_frictionMean}
	\end{subfigure}%
	~ 
	\begin{subfigure}[t]{0.5\textwidth}
		\centering
		\includegraphics[width = 1 \linewidth, trim={0.6em 0.6em 0.6em 1.85em}, clip]{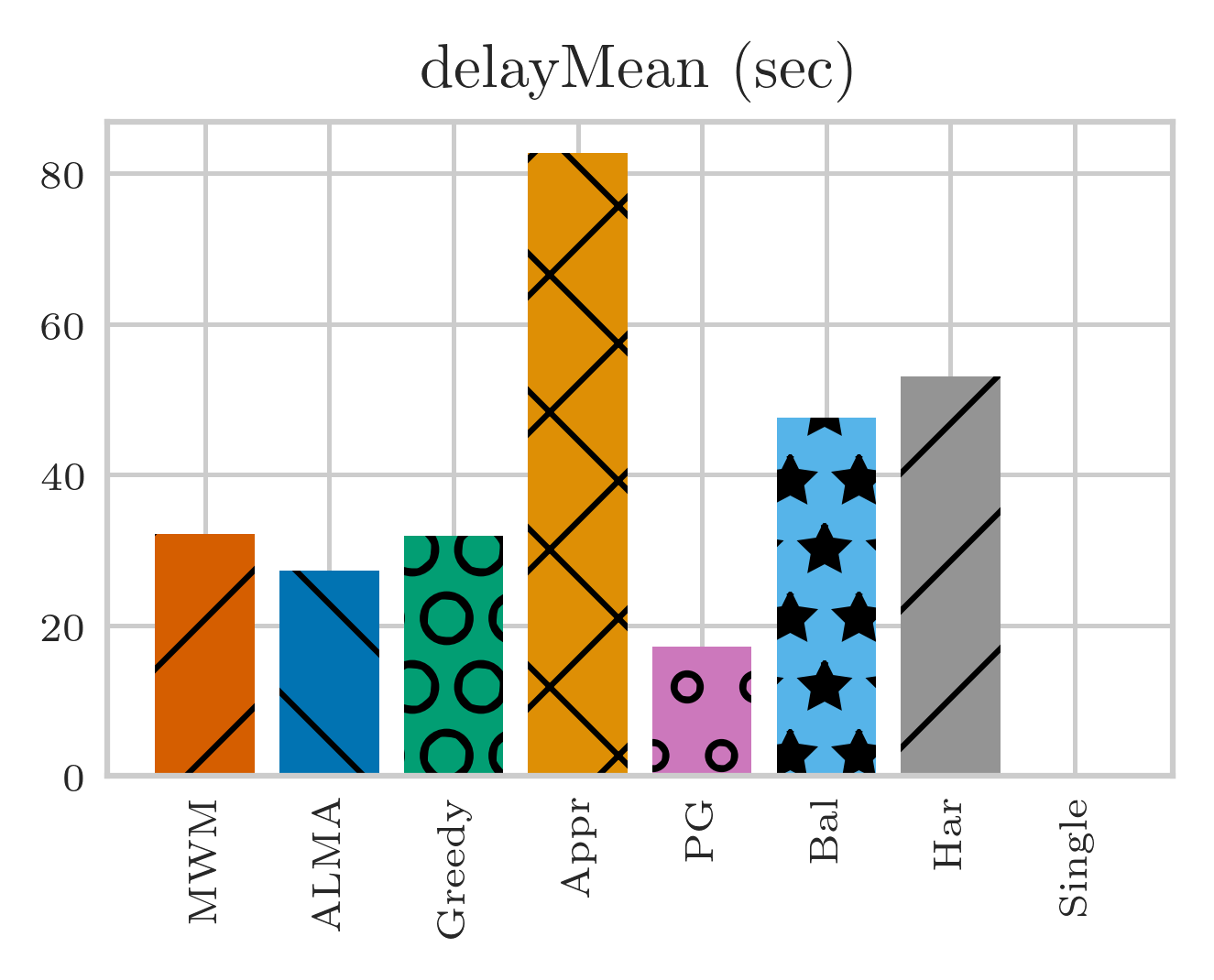}
		\caption{Delay (s)}
		\label{fig: Jan15Manhattan_delayMean}
	\end{subfigure}%
	\caption{00:00 - 23:59 (full day), \#Taxis = 5081 (base number). Manhattan, January 15, 2016}
	\label{fig: results full day}
\end{figure*}

\begin{figure*}[t!]
	\centering
	\begin{subfigure}[t]{0.5\textwidth}
		\centering
		\includegraphics[width = 1 \linewidth, trim={1.85em 0.7em 0.6em 1.2em}, clip]{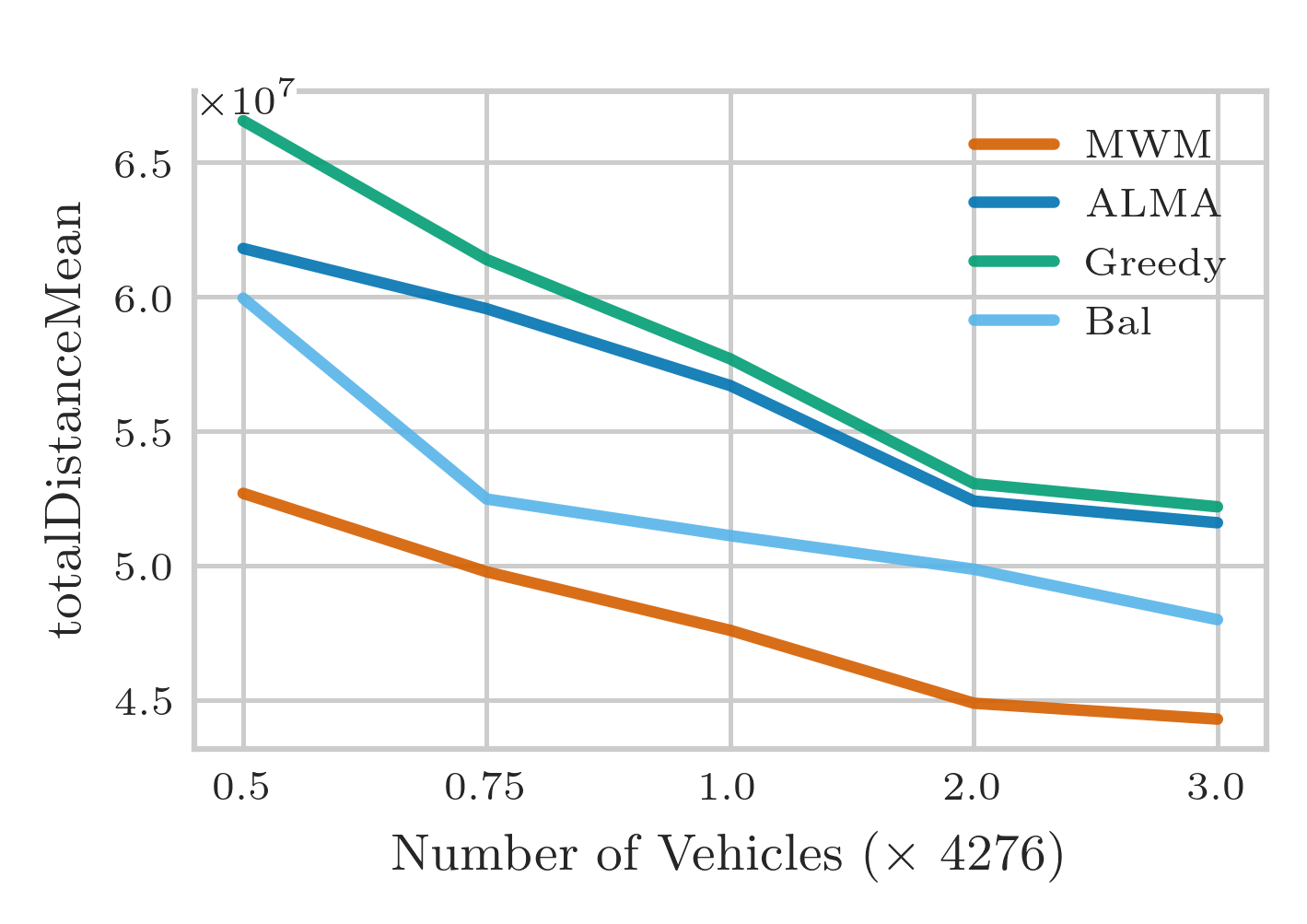}
		\caption{Distance Driven (m)}
		\label{fig: Jan15Hour8to9Manhattan_LinePlots_totalDistanceMean}
	\end{subfigure}%
	~ 
	\begin{subfigure}[t]{0.5\textwidth}
		\centering
		\includegraphics[width = 1 \linewidth, trim={1.85em 0.7em 0.6em 0.6em}, clip]{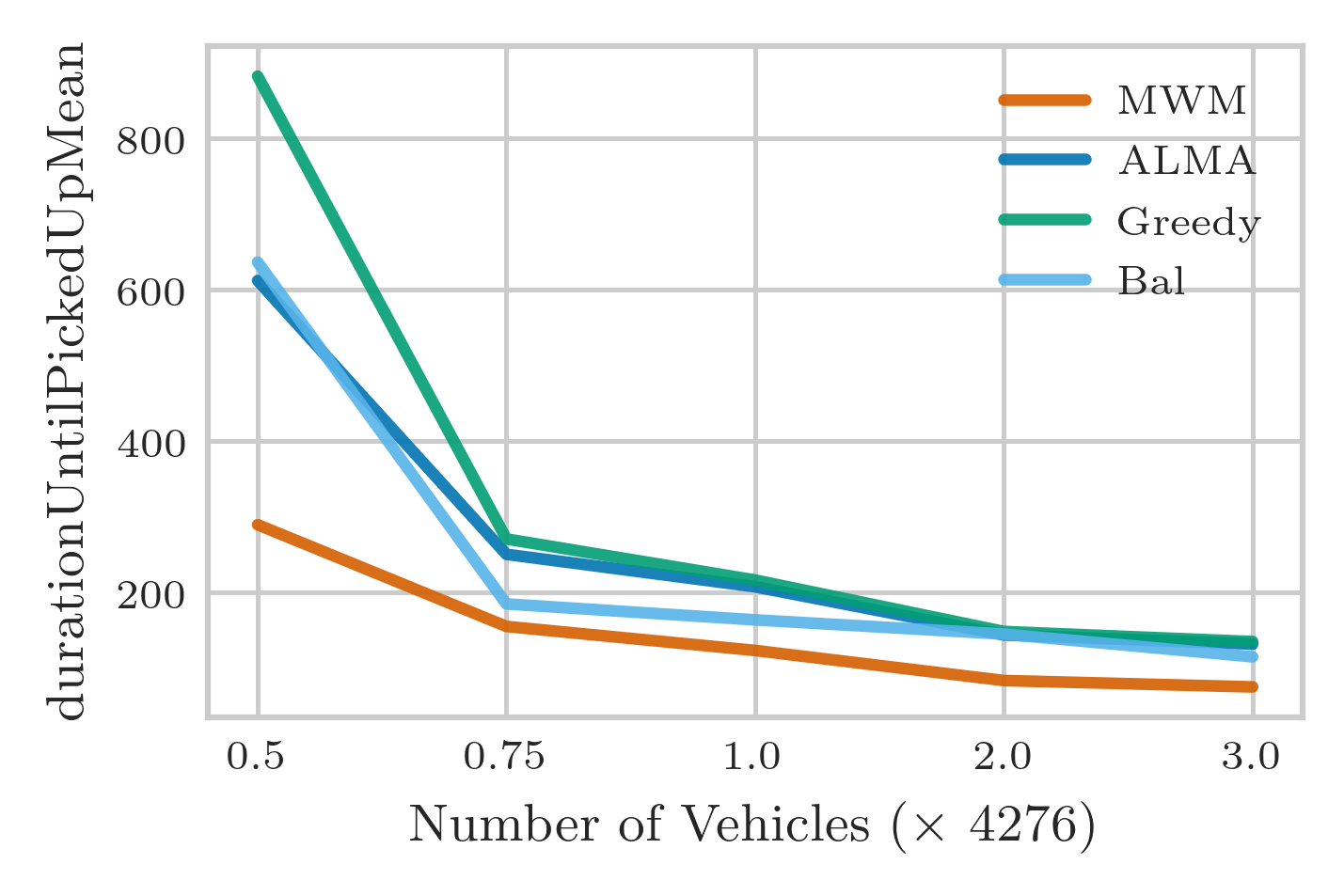}
		\caption{Time to Pick-up (s)}
		\label{fig: Jan15Hour8to9Manhattan_LinePlots_durationUntilPickedUpMean}
	\end{subfigure}%
	
	\begin{subfigure}[t]{0.5\textwidth}
		\centering
		\includegraphics[width = 1 \linewidth, trim={1.85em 0.7em 0.6em 0.6em}, clip]{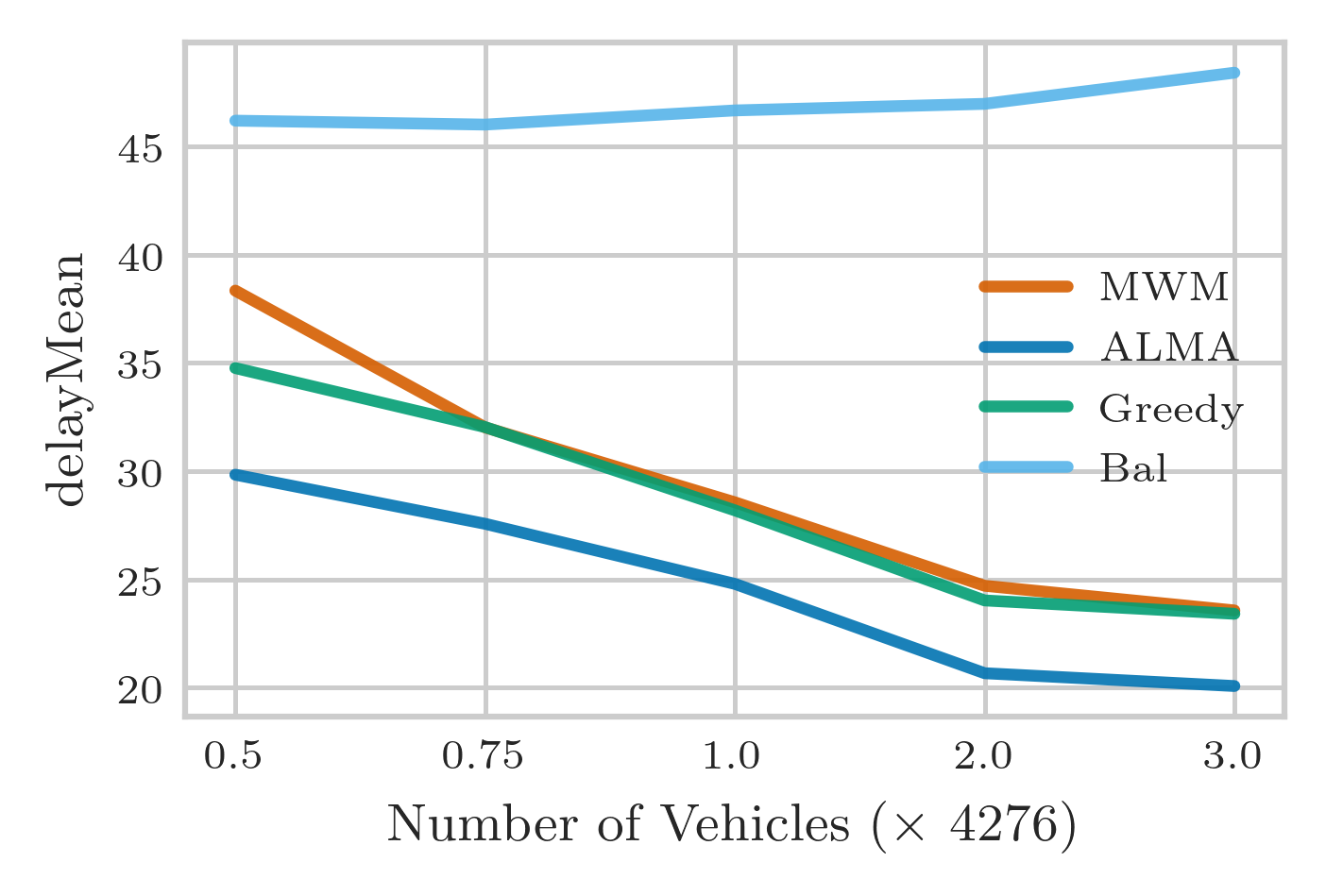}
		\caption{Delay (s)}
		\label{fig: Jan15Hour8to9Manhattan_LinePlots_delayMean}
	\end{subfigure}%
	~ 
	\begin{subfigure}[t]{0.5\textwidth}
		\centering
		\includegraphics[width = 1 \linewidth, trim={1.85em 0.7em 0.6em 0.6em}, clip]{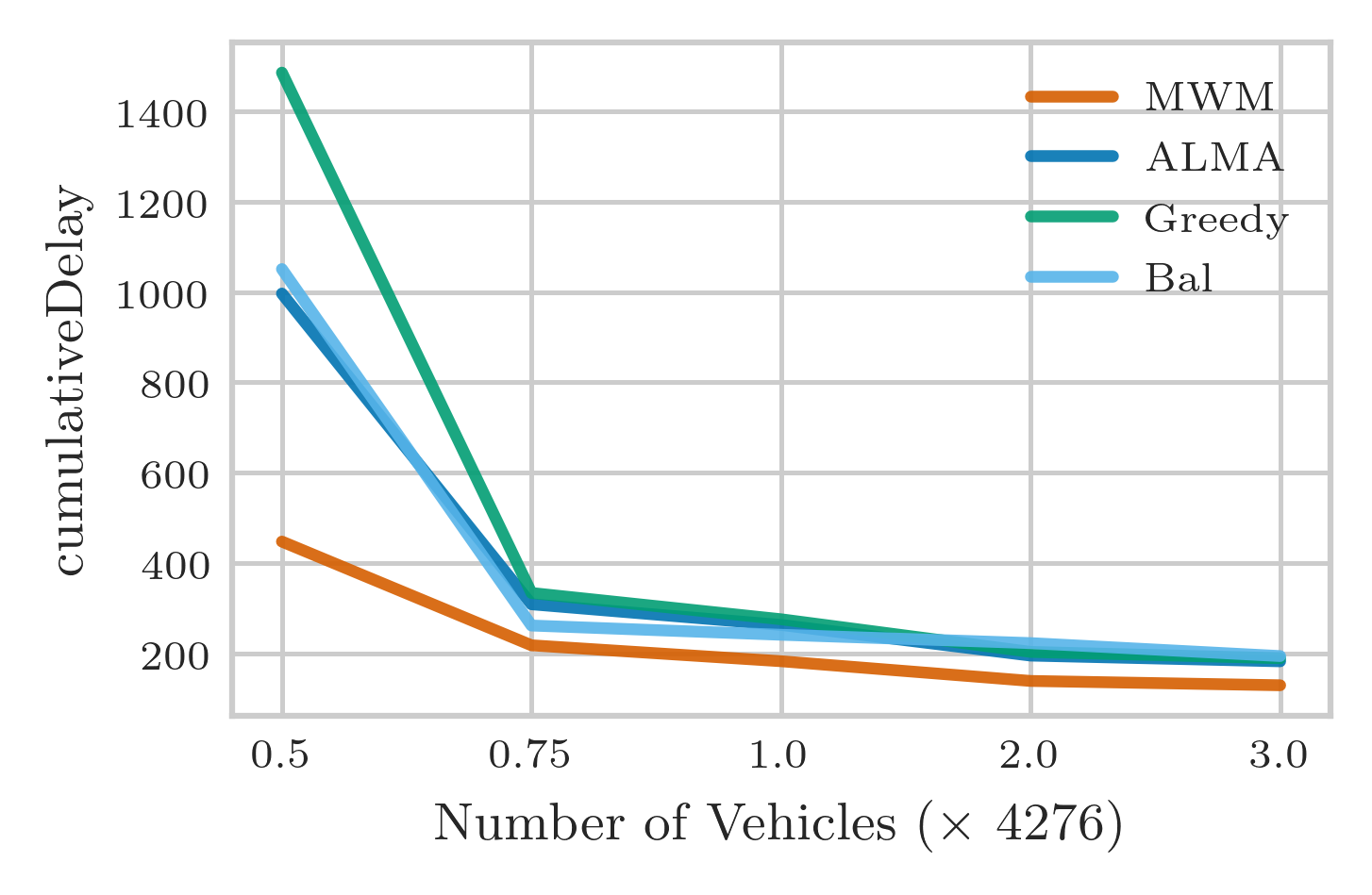}
		\caption{Cumul. Delay (s)}
		\label{fig: Jan15Hour8to9Manhattan_LinePlots_cumulativeDelay}
	\end{subfigure}%
	\caption{08:00 - 09:00, \#Taxis = $\{2138, 3207, 4276, 8552, 12828\}$. Manhattan, January 15, 2016}
	\label{fig: results increasing number of taxis}
\end{figure*}

\begin{figure*}[t!]
	\centering
	\begin{subfigure}[t]{1\textwidth}
		\centering
		\includegraphics[width = 1 \linewidth, trim={0.7em 0.7em 0.6em 1.8em}, clip]{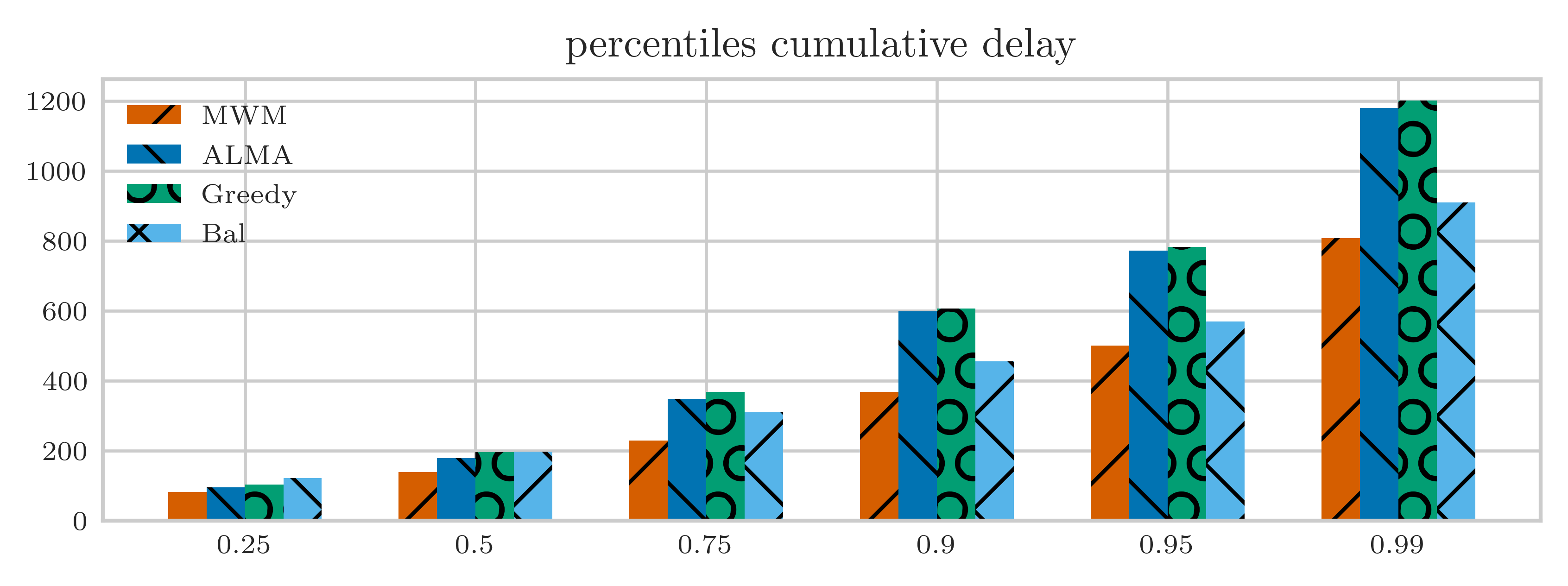}
		\caption{Sequence of Percentiles for Cumulative Delay (s)}
		\label{fig: Jan15Hour8to9Manhattan_percentiles_cumulative_delay}
	\end{subfigure}
	\caption{08:00 - 09:00, \#Taxis = 4276 (base number). Manhattan, January 15, 2016}
	\label{fig: results percentiles}
\end{figure*}

Figures \ref{fig: results one hour}, \ref{fig: results percentiles}, \ref{fig: results full day}, and \ref{fig: results increasing number of taxis} present the results without relocation. We first present results on one hour (Figures \ref{fig: results one hour} and \ref{fig: results percentiles}) and base number of taxis (see Section \ref{Taxi Vehicles}). Then, we show that the results are robust at a larger time-scale\footnote{Missing components were too computationally expensive to simulate for an entire day.} (Figure \ref{fig: results full day}), and varying number of vehicles\footnote{We only present the most promising solutions.} ($2138$ - $12828$) (Figure \ref{fig: results increasing number of taxis}). Finally, we present results on the step (c) of the Ridesharing problem: dynamic relocation (Table \ref{tb: relocation gains}, Figure \ref{fig: Jan15ManhattanRelocation_durationUntilPickedUpMean}).

\subsubsection{Distance Driven:} In the small test-case (Figure \ref{fig: Jan15Hour8to9Manhattan_totalDistanceMean}) MWM performs the best, followed by Bal ($+7\%$). ALMA comes third ($+19\%$), and then Greedy ($+21\%$). The high performance of Bal in this metric is because it uses MWM for step (a), which has a more significant impact on the distance driven. Similar results are observed for the whole day (Figure \ref{fig: Jan15Manhattan_totalDistanceMean}), with Bal, ALMA, and Greedy achieving $+4\%$, $+18\%$, and $+22\%$ compared to MWM, respectively. Figure \ref{fig: Jan15Hour8to9Manhattan_LinePlots_totalDistanceMean} shows that as we decrease the number of taxis, Bal loses its advantage, Greedy is pulling away from ALMA ($9\%$ worse than ALMA), while ALMA closes the gap to MWM ($+17\%$).

\subsubsection{Complexity:} To estimate the complexity, we measured the elapsed time of each algorithm. Greedy is the fastest one (Figure \ref{fig: Jan15Hour8to9Manhattan_elapsedTimeMean}), closely followed by Har, Bal, and ALMA. ALMA is inherently decentralized. The red overlay denotes the parallel time for ALMA, which is $2.5$ orders of magnitude faster than Greedy.

\subsubsection{Time to Pick-up:} MWM exhibits exceptionally low time to pick-up (Figure \ref{fig: Jan15Hour8to9Manhattan_durationUntilPickedUpMean}), lower than the single ride baseline. ALMA, Greedy, and Bal have $+69\%$, $+76\%$, and $+33\%$ compared to MWM, respectively. As before, Figure \ref{fig: Jan15Hour8to9Manhattan_LinePlots_durationUntilPickedUpMean} shows that as we decrease the number of taxis, Bal loses its advantage, and Greedy is pulling further away from ALMA. Note that to improve visualization, we removed DC's pick-up time as it was one order of magnitude larger than Appr.

\subsubsection{Delay:} \label{Delay}
PG exhibits the lowest delay (Figure \ref{fig: Jan15Hour8to9Manhattan_delayMean}), but this is because it makes $26\%$ fewer shared rides than the rest of the high performing algorithms. ALMA has the smallest delay ($-13\%$ compared to MWM), with Greedy following at $-1\%$, while Bal has $+63\%$ (both compared to MWM). As the number of taxis decrease (Figure \ref{fig: Jan15Hour8to9Manhattan_LinePlots_delayMean}), ALMA's gains increase further ($-22\%$ compared to MWM).

Figure \ref{fig: Jan15Hour8to9Manhattan_LinePlots_cumulativeDelay} depicts the cumulative delay, which is the sum of all delays described in Section \ref{Passenger Specific Matrics}, namely the time to pair, time to pair with taxi, time to pick-up, and delay. An interesting observation is that reducing the fleet size from $12828$ ($\times 3.0$ of the base number) to just $3207$ ($\times 0.75$ of the base number) vehicles ($75\%$ reduction) results in only approximately 2 minutes of additional delay. This goes to show the great potential for efficiency gains such technologies have to offer.

Finally, we wanted to investigate the distribution of the achieved QoS metrics and, consequently, the reliability/fairness of each CAR. As such, we plotted in Figure \ref{fig: Jan15Hour8to9Manhattan_percentiles_cumulative_delay} the sequence of percentiles\footnote{Given a vector $V$ of cumulative delays per request, the $q$-th percentile of $V$ is the value $q/100$ of the way from the minimum to the maximum in a sorted copy of $V$.} for the cumulative delay. As shown, the vast majority of the users ($75\%$) experience cumulative delay close to the average value (only $46$, $85$, $92$, $69$ additional seconds of cumulative delay than the average value for MWM, ALMA, Greedy, and BAL, respectively). Of course, some of the users experiences high cumulative delay, but this is a small percentage of them. Specifically, less than $5\%$ of requests experience a delay of more than $8.5$, $13$, $13$, and $9.5$ minutes for MWM, ALMA, Greedy, and BAL, respectively. Given the size and the average speed of taxi vehicles in Manhattan, such delays could be expected and, thus, acceptable; ultimately, it is up to the ridesharing platform to impose hard constraints and reject requests with potentially high delay.

\begin{table}[t!]
\centering
\caption{Fairness of the Drivers' Profit.\\08:00 - 09:00, \#Taxis = 4276 (base number). Manhattan, January 15, 2016}
\label{tb: results profit jain index}
\begin{tabular}{@{}lcc@{}}
\toprule
                & \textbf{Jain Index} & \textbf{Relative Diff. to MWM} \\ \midrule
\textbf{MWM}    & 0.71                & 0.0\%                           \\
\textbf{ALMA}   & 0.75                & 6.0\%                           \\
\textbf{Greedy} & 0.75                & 6.9\%                           \\
\textbf{Appr}   & 0.86                & 21.4\%                          \\
\textbf{PG}     & 0.75                & 7.0\%                           \\
\textbf{GD}     & 0.77                & 9.0\%                           \\
\textbf{Bal}    & 0.90                & 28.2\%                          \\
\textbf{Har}    & 0.84                & 19.1\%                          \\
\textbf{DC}     & 0.19                & -73.0\%                         \\
\textbf{k-Taxi} & 0.71                & 0.3\%                           \\
\textbf{Single} & 0.92                & 30.5\%                          \\ \bottomrule
\end{tabular}%
\end{table}

\subsubsection{Profit \& Frictions:} \label{Profit & Frictions}
Contrary to their performance in QoS metrics, GD, and Appr achieve the highest driver profit, $12\%$ and $8\%$ higher than MWM, respectively (although the low QoS and increased distance driven suggest low quality matchings, which can explain the higher revenue, yet deems them undesirable). Bal, and Har follow with $+2-3\%$. ALMA and Greedy achieve the similar profit to MWM. PG exhibits significantly worse results ($-13\%$), due to the lower number of shared rides it matches.

Small differences in driver profit can have a significant impact on the platform's profit. There are $13587$ taxis in NYC\footnoteref{TLCYellowCab}, $67-85\%$ of which are on the road at one time (i.e., $9103$ - $11549$ taxis). The additional $2\%$ profit of Bal translates to $\$32.3$ additional revenue in a day. Multiplied by the total number of taxis, and assuming that the platform keeps $25\%$ as commission\footnoteref{uberpayments}, this results in $\$73506$ - $\$93258$ additional revenue per day for the platform.

Figure \ref{fig: Jan15Manhattan_earnedMeanMaxMin} also depicts the maximum (red dot), and minimum (green dot) value of a driver's profit. Closer to the mean maximum value suggests a fairer algorithm for the drivers. Moreover, it is worth noting that the minimum value for all the algorithms is zero, meaning that there are taxis which remain unutilized (in spite of the fact that the number of taxis -- in this scenario $5081$ -- is considerably lower than the current fleet size of yellow taxis).

In order to investigate the fairness of the distribution of profits amongst drivers, we calculated the Jain index \cite{DBLP:journals/corr/cs-NI-9809099}, a well-established fairness metric. The Jain index exhibits a lot of desirable properties such as population size independence, continuity, scale and metric independence, and boundedness. For an allocation between $N$ agents, such that the $n^{\text{th}}$ agent is alloted $x_n$, the Jain index is given by Equation \ref{eq: Jain Index}. $\mathds{J}(\mathbf{x}) \in [0, 1]$. An allocation $\mathbf{x} = (x_1, \dots, x_N) ^\top$ is considered fair, iff $\mathds{J}(\mathbf{x}) = 1$.
\begin{equation} \label{eq: Jain Index}
	\mathds{J}(\mathbf{x}) = \frac{\left(\underset{n = 1}{\overset{N}{\sum}} x_n\right) ^ 2}{N \underset{n = 1}{\overset{N}{\sum}}  x_n ^ 2}
\end{equation}


Table \ref{tb: results profit jain index} shows the Jain index, and the relative difference compared to MWM. Bal achieves the most fair allocation (excluding the single ride baseline\footnote{It is expected that the single ride baseline would result in a fair allocation of the profit, as it constantly utilizes the entire fleet of vehicles (see the definition of the base number of taxis in Section \ref{Taxi Vehicles}).}), with a Jain index of 0.9, closely followed by Appr with 0.86. MWM, ALMA, and Greedy all achieve relatively fair allocations, with the latter two achieving a $6\%$ and $7\%$ improvement over MWM.

Figure \ref{fig: Jan15Manhattan_frictionMean} shows the driver frictions. Just like with the profit, $k$-server algorithms seem to outperform matching algorithms by far. Compared to MWM, Bal and Har achieve a $63\%$ and $73\%$ decrease, respectively, while ALMA and Greedy achieve a $26\%$, and $21\%$ decrease, respectively. Given that we have a fixed supply, lower frictions indicate a more even distribution of rides amongst taxis.

It is important to note that while the results for all the other metrics are consistent when moving from the one hour test-case to the full day test-case, this is not true for the frictions (see Figures \ref{fig_appendix: Jan15Hour8to9Manhattan_frictionMean} and \ref{fig_appendix: Jan15Manhattan_frictionMean} and Tables \ref{tb_appendix: Jan15Hour8to9ManhattanPercentages} and \ref{tb_appendix: Jan15Manhattan} in the Appendix). This is because taxis that serve zero or one rides are assumed to have zero friction by definition. Algorithms like Bal -- which attempts to balance the distance driven by each taxi -- will utilize each vehicle multiple times, even for the short time window of one hour. This results to a deceivingly high number in the frictions in the one hour test-case. As a matter of fact, the number of taxis that served less than two rides (and, thus, had zero friction) in the one hour test-case for Bal were 483. For MWM this number is 1368 (almost 3 times larger), for ALMA it is 1181, and for Greedy 1120. This is why we opted to present the frictions for the full day test-case in Figure \ref{fig: Jan15Manhattan_frictionMean}.

\subsubsection{Time to Pair with Taxi \& Number of Shared Rides:} \label{Time to Pair with Taxi & Number of Shared Rides}
Excluding the test-case with the smallest taxi fleet ($\times 0.5$ the base number), the time to pair with taxi was zero, or close to zero, for all the evaluated algorithms. The latter comes to show the potential for efficiency gains and better utilization of resources using smart technologies. The reason for the low time to pair with a taxi is that, for the step (b) of the ridesharing problem (matching (shared) rides to taxis), we run the offline algorithms in a just-in-time (JiT) manner, i.e., every time the set of rides ($\mathcal{P}_t$) is not empty (see Section \ref{algorithms}). We opted to do so for simplicity -- the alternative would require to run all combinations of batch sizes for both steps (a) and (b). Results from step (a), though, suggest that running in batches is more beneficial (running in batch size of two minutes consistently outperformed the JiT version, see Appendix \ref{appendix: Simulation Results}). There is a clear trade-off: match with a taxi as soon as possible (JiT), and have a vehicle moving to pick-up the ride earlier, or wait (match in batches every $x$ minutes), potentially allowing for better matches? Answering this question remains open for future work.


The number of shared rides is approximately the same for all the employed algorithms, with notable exception the PG which makes $26\%$ fewer shared rides.

\subsubsection{Relocation} \label{Relocation Simulation Results}

The aim of any relocation strategy is to improve the spatial allocation of supply. Serving requests redistributes the taxis, resulting in an inefficient allocation. One can assume a `lazy' approach, relocating vehicles only to serve requests. While this minimizes the cost of serving a request (e.g., distance driven, fuel, etc.), it results in sub-optimal QoS. Improving the QoS (especially the time to pick-up, since it highly correlates to passenger satisfaction, see Section \ref{Passenger Specific Matrics}) plays a vital role in the growth of a company. Thus, \emph{a crucial trade-off of any relocation scheme is improving the QoS metrics, while minimizing the excess distance driven.}

\begin{table}[t!]
\centering
\caption{Relocation Gains.}
\label{tb: relocation gains}
\begin{tabular}{@{}rccc@{}}
\toprule
                            & \textbf{MWM}  & \textbf{ALMA}  & \textbf{Greedy}  \\ \midrule
\textbf{Time to Pick-up}    & -48.95\%      & -55.18\%       & -55.03\%         \\
\textbf{Time to Pick-up SD} & -52.97\%      & -58.22\%       & -58.21\%         \\
\textbf{Delay}              & -15.95\%      & -17.79\%       & -17.73\%         \\
\textbf{Delay SD}           & -19.25\%      & -20.96\%       & -20.98\%         \\
\textbf{Cumulative Delay}   & -38.37\%      & -43.23\%       & -43.11\%         \\
\textbf{Total Distance}     & 5.48\%        & 6.25\%         & 6.24\%           \\ \bottomrule
\end{tabular}
\end{table}

\begin{figure}[t!]
	\centering
	\includegraphics[width = 1 \linewidth, trim={0.6em 0.75em 0.6em 1.8em}, clip]{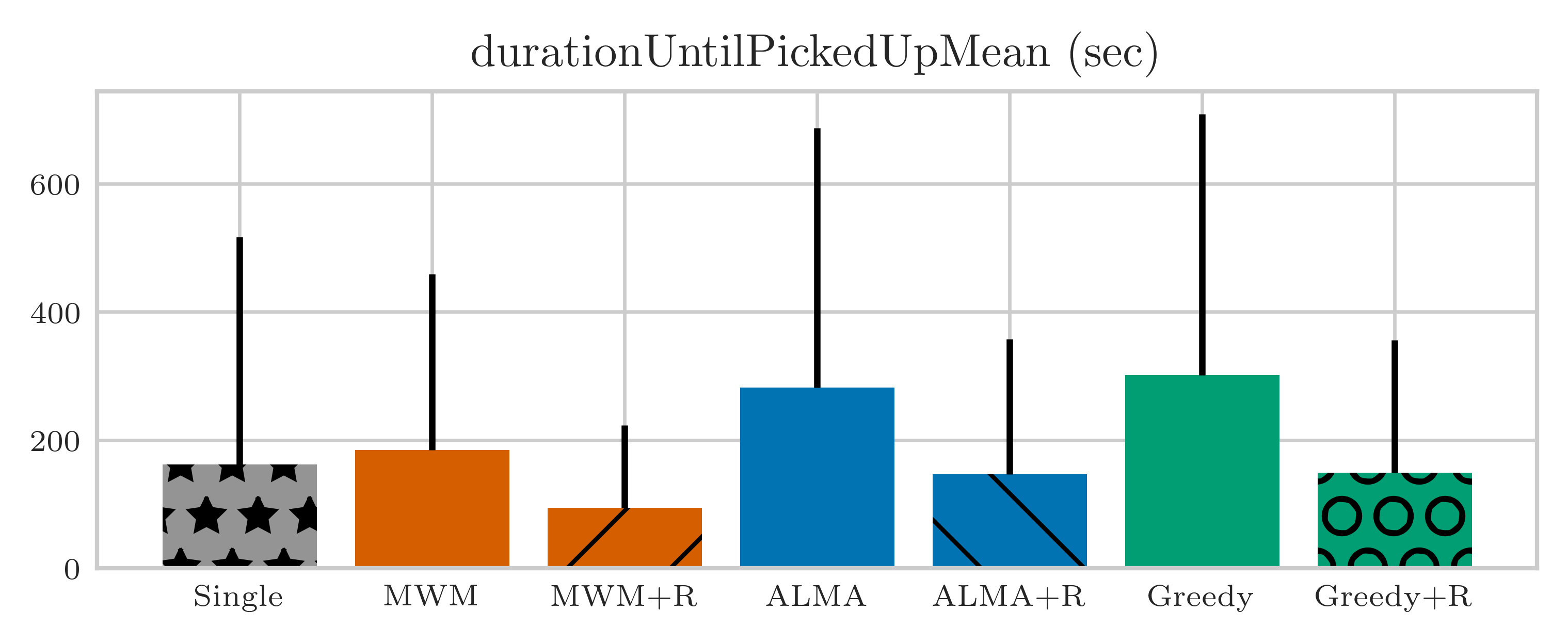}
	\caption{Time to Pick-up (s) -- End-To-End Solution. January 15, 2016 -- 00:00 - 23:59 -- Manhattan -- \#Taxis = 5081}
	\label{fig: Jan15ManhattanRelocation_durationUntilPickedUpMean}
\end{figure}

CARs with relocation successfully balance this trade-off (Table \ref{tb: relocation gains}). In particular, ALMA -- the best performing overall -- radically improves the QoS metrics by more than $50\%$ (e.g., it decreases the pick-up time by $55\%$, and its standard deviation (SD) by $58\%$), while increasing the driving distance by only $6\%$. The cumulative delay is decreased by $43\%$.

As a final step, we evaluate \emph{end-to-end} solutions, using MWM, ALMA, and Greedy to solve all three steps of the ridesharing problem. Figure \ref{fig: Jan15ManhattanRelocation_durationUntilPickedUpMean} depicts the time to pick-up (error bars denote one SD of uncertainty), a metric highly correlated to passenger satisfaction level \cite{8259801,lyftblog2}. We compare against the single ride baseline (no delay due to sharing a ride, see Section \ref{single ride baseline}). Once more, the proposed relocation scheme results in radical improvements, as the time to pick-up drops (compared to the single ride) from $+14.09\%$ to $-41.76\%$ for MWM, from $+74.14\%$ to $-9.33\%$ for ALMA, and from $+86.10\%$ to $-7.97\%$ for Greedy. This comes to show that \emph{simple relocation schemes can eliminate the negative effects of ridesharing on the QoS}.

\subsubsection{ALMA as an end-to-end CAR}

While MWM seems to perform the best in the total distance driven, and most QoS metrics -- which is reasonable since it makes optimal matches amongst passengers -- it hard to scale and requires a centralized solution. In contrast, greedy approaches are appealing\footnoteref{AppealOfGreedy} not only due to their low complexity, but also because real-time constraints dictate short planning windows which can diminish the benefit of batch optimization solutions compared to myopic approaches \cite{widdows2017grabshare}.

In fact, ALMA is of a greedy nature as well, albeit it utilizes a more intelligent backing-off scheme, thus there are scenarios where ALMA significantly outperforms the greedy, as proven by the simulation results. For example, in more challenging scenarios (smaller taxi fleet, or potentially different types of taxis) the smarter back off mechanism results in a more profound difference.

Most importantly, ALMA was inherently developed for multi-agent applications. Agents make decisions locally, using completely uncoupled learning rules, and require only a 1-bit partial feedback \cite{ijcai201931}, making it an ideal candidate for an \emph{on-device} implementation. This is fundamentally different than a decentralized implementation of the Greedy algorithm for example. Even in decentralized algorithms, the number of communication rounds required grows with the size of the problem. However, in practice the real-time constraints impose a limit on the number of rounds, and thus on the size of the problem that can be solved within them.

\begin{table}[p!]
\centering
\caption{High level (qualitative) ranking of the evaluated CARs.\\
For each metric, the best performing CAR receives four stars ($\star\star\star\star$). Then, for the rest of the CARs, we compute the relative difference to the best performing one, i.e., $r = (x - OPT) / OPT$, where $x$ is the value for the CAR considered, and $OPT$ is the value achieved by the best performing CAR in the specific metric. If the relative difference is $< 0.1$, this CAR also receives four stars ($\star\star\star\star$), if it is $0.1 \leq r < 0.5$, the CAR receives three stars ($\star\star\star$), if it is $0.5 \leq r < 1$, the CAR receives two stars ($\star\star$), and finally, if $r \geq 1$, the CAR receives one star ($\star$).\\
The ranking is primarily based on the one hour test-case and base number of taxis (08:00 - 09:00, Manhattan, \#Taxis = 4276, see Tables \ref{tb_appendix: Jan15Hour8to9Manhattan}, and \ref{tb_appendix: Jan15Hour8to9ManhattanPercentages} in the Appendix).\\
The same ranking holds in most cases for the full day test-case and base number of taxis (00:00 - 23:59, Manhattan, \#Taxis = 5081, see Tables \ref{tb_appendix: Jan15Manhattan}, and \ref{tb_appendix: Jan15ManhattanPercentages} in the Appendix). The only exceptions are the ones awarded one additional star in the full day test-case, which is denote inside a parenthesis when relevant.\\
As mentioned in Section \ref{Profit & Frictions}, reporting frictions for the one hour test-case can be deceiving, thus we report the ranking based on the full day test-case. To make the distinction clear, the awarded stars are also inside a parenthesis. Algorithms that were too computationally heavy to run for a full day lack a ranking for the frictions.\\
The ranking for the WFA and HC (which we were able to run only in much smaller test-cases, see Tables \ref{tb_appendix: Jan15Hour8to9ManhattanFull}, and \ref{tb_appendix: Jan15Hour8to9ManhattanFullPercentages} for WFA, and Tables \ref{tb_appendix: Jan15Hour8to810Manhattan}, and \ref{tb_appendix: Jan15Hour8to810ManhattanPercentages} for HC in the Appendix) was extrapolated based on their performance against the baseline CARs that were common in all test-cases (MWM, ALMA, and Greedy).\\
The Time to Pair with a Taxi is not included as it was zero for most CARs (see Section \ref{Time to Pair with Taxi & Number of Shared Rides}).\\
It is important to note that since the ranking is based on the relative difference in performance compared to the best performing CAR in each metric, it can be misleading in cases where the change is insignificant in terms of absolute values. Such a case is the Time to Pick-up, where the difference between MWM vs. Bal, or ALMA, or Greedy is only 1-2 minutes.\\
\ddag ALMA run in a decentralized manner is orders of magnitude faster than any other CAR. In order to allow for a clear ranking between the rest of the CARs, we performed the ranking based on the second best CAR (i.e., Greedy).\\
\dag Again, to allow for a proper ranking between the CARs, PG was not included in the delay metric because its notably low delay is only a result of making significantly fewer shared rides ($26\%$ less, see Section \ref{Delay}).
}
\label{tb: results high level analysis}
\resizebox{\textwidth}{!}{%
\begin{tabular}{@{}lccccccccc@{}}
\toprule
                                              & \multicolumn{3}{c}{\textbf{Operational Efficiency}}                            & \multicolumn{4}{c}{\textbf{Quality of Service}}                                                     & \multicolumn{2}{c}{\textbf{Drivers' Metrics}}     \\ \midrule
                                              & \textbf{Distance}      & \textbf{Computational}       & \textbf{Number of}     & \textbf{Time to}       & \textbf{Time to}         &                        & \textbf{Cumulative}    &                        &                          \\
                                              & \textbf{Driven}        & \textbf{Complexity}          & \textbf{Shared Rides}  & \textbf{Pair}          & \textbf{Pick-up}         & \textbf{Delay}         & \textbf{Delay}         & \textbf{Profit}        & \textbf{Frictions}       \\
\textbf{Maximum Weight Matching (MWM)}        & $\star\star\star\star$ & $\star$                      & $\star\star\star\star$ & $\star\star\star\star$ & $\star\star\star\star$   & $\star\star\star$      & $\star\star\star\star$ & $\star\star\star$      & $(\star)$                \\
\textbf{ALtruistic MAtching Heuristic (ALMA)} & $\star\star\star$      & $\star\star\star\star$ \ddag & $\star\star\star\star$ & $\star\star\star\star$ & $\star\star$             & $\star\star\star\star$ & $\star\star\star$      & $\star\star\star$      & $(\star)$                \\
\textbf{Greedy}                               & $\star\star\star$      & $\star\star\star\star$       & $\star\star\star\star$ & $\star\star\star\star$ & $\star\star$             & $\star\star\star$      & $\star\star(\star)$    & $\star\star\star$      & $(\star)$                \\
\textbf{Approximation (Appr)}                 & $\star\star$           & $\star$                      & $\star\star\star\star$ & $\star\star\star\star$ & $\star$                  & $\star$                & $\star$                & $\star\star\star\star$ & $(\star)$                \\
\textbf{Postponed Greedy (PG)}                & $\star\star\star$      & $\star$                      & $\star\star\star$      & $\star$                & $\star\star$             & \dag                   & $\star\star(\star)$    & $\star\star\star$      & $(\star)$                \\
\textbf{Greedy Dual (GD)}                     & $\star\star\star$      & $\star$                      & $\star\star\star\star$ & $\star\star$           & $\star\star$             & $\star$                & $\star$                & $\star\star\star\star$ & -                        \\
\textbf{Balance (Bal)}                        & $\star\star\star\star$ & $\star\star\star\star$       & $\star\star\star\star$ & $\star\star\star\star$ & $\star\star\star(\star)$ & $\star\star$           & $\star\star\star$      & $\star\star\star\star$ & $(\star\star\star)$      \\
\textbf{Harmonic (Har)}                       & $\star\star$           & $\star\star\star\star$       & $\star\star\star\star$ & $\star\star\star\star$ & $\star$                  & $\star(\star)$         & $\star$                & $\star\star\star\star$ & $(\star\star\star\star)$ \\
\textbf{Double Coverage (DC)}                 & $\star$                & $\star$                      & $\star\star\star\star$ & $\star\star\star\star$ & $\star$                  & $\star$                & $\star$                & $\star\star\star\star$ & -                        \\
\textbf{Work Function (WFA)}                  & $\star\star\star$      & $\star$                      & $\star\star\star\star$ & $\star$                & $\star$                  & $\star\star\star$      & $\star$                & $\star\star\star$      & -                        \\
\textbf{$k$-Taxi}                             & $\star\star\star$      & $\star$                      & $\star\star\star\star$ & $\star\star\star\star$ & $\star$                  & $\star\star$           & $\star$                & $\star\star\star\star$ & -                        \\
\textbf{High Capacity (HC)}                   & $\star\star\star\star$ & $\star$                      & $\star\star\star\star$ & $\star$                & $\star\star\star$        & $\star\star$           & $\star\star\star\star$ & $\star\star\star$      & -                        \\ \bottomrule
\end{tabular}%
}
\end{table}

\subsection{High-level Analysis}

Applying the modular approach we advocate, allowed us to thoroughly test a wide variety of state-of-the-art algorithms for ridesharing. When dealing with a multi-objective optimization problem, it is unreasonable to expect to identify an approach that outperforms the competition across the board. Nevertheless, our findings provide convincing evidence to a ridesharing platform as to which CARs would be most suitable for a given set of objectives. Specifically: (i) CARs that rely on off-line (in-batches) maximum-weight matching solutions perform well on global efficiency and passenger related metrics, (ii) CARs based on $k$-server algorithms perform well on driver/platform related metrics (e.g., Bal), (iii) lightweight CARs perform better in real-world, large-scale settings due to short planning windows imposed by the requirement to run in real-time, (iv) a simple, fine-grained relocation scheme based on the history of requests can significantly improve Quality of Service metrics by up to $50\%$, and finally, (v) we identify a scalable, on-device CAR based on ALMA that performs well across the board. A summary of the results can be found in Table \ref{tb: results high level analysis}.

\section{Conclusion} \label{Conclusion}

Managing transportation resources on a large scale remains a critical open problem. We initiate the \emph{systematic} study of \emph{Component Algorithms for Ridesharing} (CARs), a modular design methodology for ridesharing. To gain insight into the intricate dynamics of the problem, it is highly important to evaluate a diverse set of candidate solutions in settings designed to closely resemble reality. We evaluate a diverse set of candidate CARs (14 in total) -- focused on the \emph{key algorithmic components} of ridesharing -- over 12 metrics, in settings designed to \emph{closely resemble reality} in every aspect of the problem. To the best of our knowledge, this is the \emph{first end-to-end evaluation of this magnitude}. We show the capacity of \emph{simple relocation schemes} to improve QoS metrics radically, eliminating the negative effects of ridesharing, and identify an ALMA-based CAR that offers an \emph{efficient} (across all metrics), \emph{scalable}, \emph{on-device}, end-to-end solution.

\clearpage

\appendix

\section*{APPENDIX}

\section{Simulation Results in Detail} \label{appendix: Simulation Results}

We present in detail the results of Section \ref{Simulation Results} including, but not limited to, larger test-cases (broader NYC area), and the omitted algorithms, graphs, and tables. For every metric we report the average value out of 8 runs. \\

\noindent
\textbf{Section \ref{Appendix Jan15Hour8to9Manhattan} 08:00 - 09:00 -- Manhattan:} We begin with our small test-case: one hour (08:00 - 09:00), base number of taxis (i.e., $4276$, see Section \ref{Taxi Vehicles}), limited to Manhattan. Figure \ref{fig_appendix: Jan15Hour8to9Manhattan_BarPlots}, and Table \ref{tb_appendix: Jan15Hour8to9Manhattan} depict all the evaluated metrics, while the latter also includes the standard deviation of each value. Finally, Table \ref{tb_appendix: Jan15Hour8to9ManhattanPercentages} presents the relative difference (percentage of gain or loss) compared to MWM (first line of the table). In what follows, we will adhere to the same pattern, i.e., presenting two tables for the same evaluation, one containing the absolute values, and one presenting the relative difference compared to the algorithm in the first line of the table. We were able to run most of the algorithms in this test-case, except for WFA which we run only for $\{\times 0.5, \times 0.75\}$ the base number of taxis, and HC which is so computationally heavy, that we had to run a separate test-case of only 10 minutes (see Section \ref{Appendix Jan15Hour8to810Manhattan}).

Offline algorithms (e.g., MWM, ALMA, Greedy) can be run either in a just-in-time (JiT) manner -- i.e., when a request becomes critical -- or in batches. The following two tables (Tables \ref{tb_appendix: Jan15Hour8to9ManhattanFull}, and \ref{tb_appendix: Jan15Hour8to9ManhattanFullPercentages}) evaluate the performance of each algorithm for each option. Given that our dataset has granularity of one minute, we run in batches of one, and two minutes. Moreover, due to the large number of requests, at least one request turns critical in every time-step. Thus, JiT and in batches of one minute produced the exact same results. To allow for the evaluation of every algorithm (except HC), we run the evaluation in a smaller scale, i.e., $2138$ taxis ($\{\times 0.5\}$ the base number of taxis). These tables also include the results for the WFA algorithm. Every other result presented in this paper assumes the best performing option for each of the algorithms (usually batch size of two minutes).

Figure \ref{fig_appendix: Jan15Hour8to9Manhattan_Percentiles} shows the sequence of percentiles for the various delays introduced in Section \ref{Passenger Specific Matrics}, while Table \ref{tb_appendix: Jan15Hour8to9Manhattan_Percentiles} presents the complete results.

Finally, Figure \ref{fig_appendix: Jan15Hour8to9Manhattan_LinePlots} shows that our results are robust to a varying number of vehicles ($2138$ - $12828$). \\

\noindent
\textbf{Section \ref{Appendix Jan15Manhattan} 00:00 - 23:59 (full day) -- Manhattan:} We continue to show that the results are robust to a larger time-scale. As before, Figure \ref{fig_appendix: Jan15Manhattan}, and Tables \ref{tb_appendix: Jan15Manhattan}, and \ref{tb_appendix: Jan15ManhattanPercentages} depict all the evaluated metrics. \\

\noindent
\textbf{Sections \ref{Appendix Jan15Hour8to9} 08:00 - 09:00, and \ref{Appendix Jan15} 00:00 - 23:59 (full day) -- Broader NYC Area:} In the following two sections, we show that our results are robust to larger geographic areas, specifically in the broader NYC Area, including Manhattan, Bronx, Staten Island, Brooklyn, and Queens. Figure \ref{fig_appendix: Jan15Hour8to9}, and Tables \ref{tb_appendix: Jan15Hour8to9}, and \ref{tb_appendix: Jan15Hour8to9Percentages}, and Figure \ref{fig_appendix: Jan15}, and Tables \ref{tb_appendix: Jan15}, and \ref{tb_appendix: Jan15Percentages} depict all the evaluated metrics, for one hour, and one day respectively. \\

\noindent
\textbf{Section \ref{Appendix Jan15Hour8to810Manhattan} 08:00 - 08:10 -- Manhattan:} This is a limited test-case aimed to evaluate the HC algorithm, due to its high computational complexity. Figure \ref{fig_appendix: Jan15Hour8to810Manhattan}, and Tables \ref{tb_appendix: Jan15Hour8to810Manhattan}, and \ref{tb_appendix: Jan15Hour8to810ManhattanPercentages} depict all the evaluated metrics. \\

\noindent
\textbf{Section \ref{Appendix Jan15ManhattanRelocation} Dynamic Vehicle Relocation -- 00:00 - 23:59 (full day) -- Manhattan:} In this section, we present results on the step (c) of the Ridesharing problem: dynamic relocation. We fix an algorithm for steps (a), and (b) -- specifically MWM -- to allow for a common ground and a fair comparison, focused only on the relocation part. Figure \ref{fig_appendix: Jan15ManhattanRelocation}, and Tables \ref{tb_appendix: Jan15ManhattanRelocation}, and \ref{tb_appendix: Jan15ManhattanRelocationPercentages} depict all the evaluated metrics. \\

\noindent
\textbf{Section \ref{Appendix Jan15ManhattanRelocationEnd2End} End-To-End Solution -- 00:00 - 23:59 (full day) -- Manhattan:} As a final step, we evaluate end-to-end solutions, using MWM, ALMA, and Greedy to solve all three of the steps of the Ridesharing problem. Figure \ref{fig_appendix: Jan15ManhattanRelocation_end2end}, and Tables \ref{tb_appendix: Jan15ManhattanRelocation_end2end}, and \ref{tb_appendix: Jan15ManhattanRelocation_end2endPercentages} present all the evaluated metrics.

\clearpage

\input{arXiv_ridesharing_supplement.tex}

\clearpage

\bibliographystyle{plainnat}
\bibliography{arXiv_ridesharing_bibliography}

\end{document}

%% file: arXiv_ridesharing_supplement.tex

\begin{figure*}[t!]
\subsection{08:00 - 09:00 -- Manhattan} \label{Appendix Jan15Hour8to9Manhattan}
\end{figure*}

\begin{figure*}[t!]
	\centering
	\begin{subfigure}[t]{0.23\textwidth}
		\centering
		\includegraphics[width = 1 \linewidth, trim={0.6em 0.6em 0.5em 1.8em}, clip]{./Published/Jan15Hour8to9Manhattan_BarPlots_totalDistanceMean.png}
		\caption{Total Distance Driven (m)}
		\label{fig_appendix: Jan15Hour8to9Manhattan_totalDistanceMean}
	\end{subfigure}
	~ 
	\begin{subfigure}[t]{0.23\textwidth}
		\centering
		\includegraphics[width = 1 \linewidth, trim={0.6em 0.6em 0.5em 1.8em}, clip]{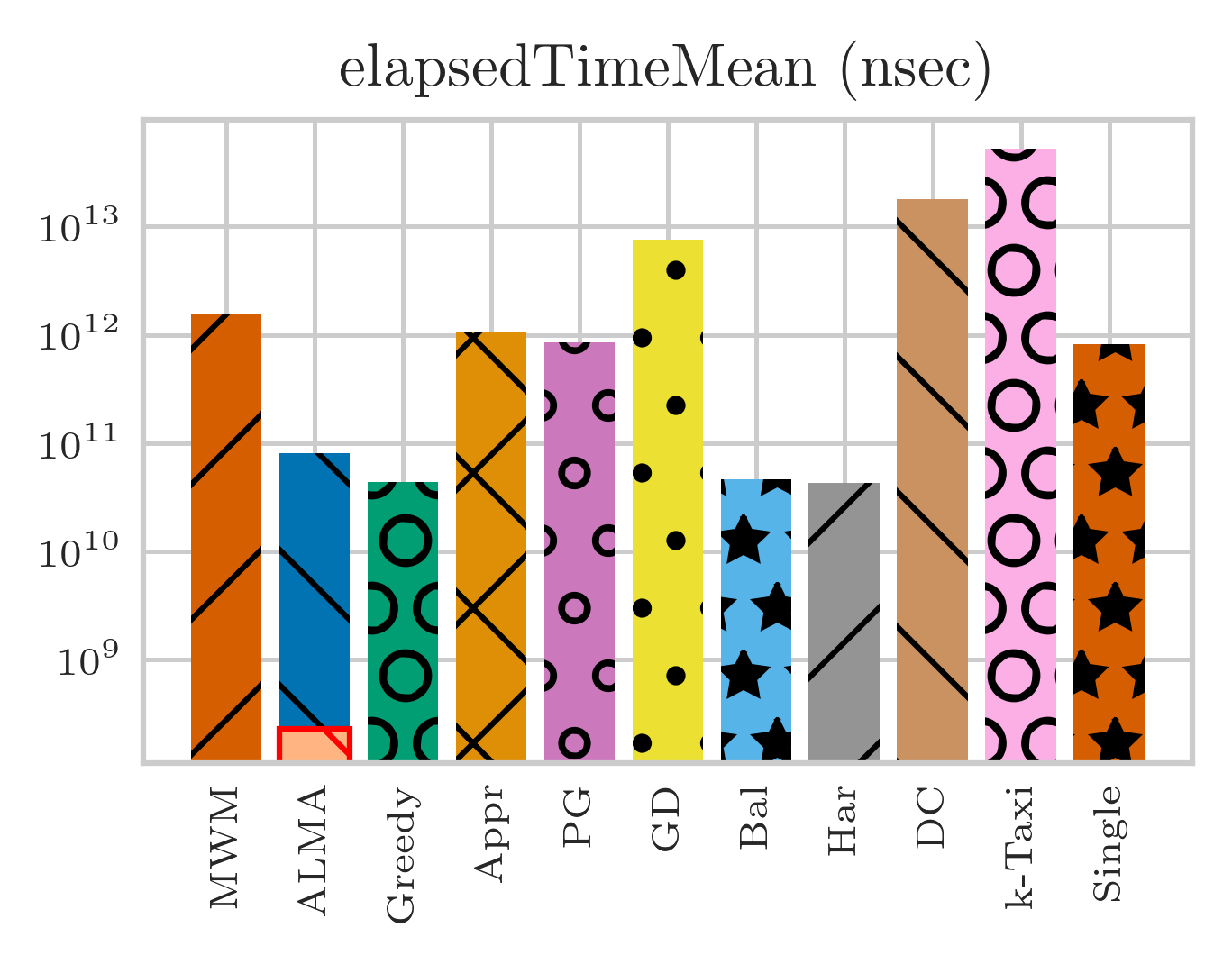}
		\caption{Elapsed Time (ns) [LOG]}
		\label{fig_appendix: Jan15Hour8to9Manhattan_elapsedTimeMean}
	\end{subfigure}
	~
	\begin{subfigure}[t]{0.23\textwidth}
		\centering
		\includegraphics[width = 1 \linewidth, trim={0.6em 0.6em 0.5em 1.8em}, clip]{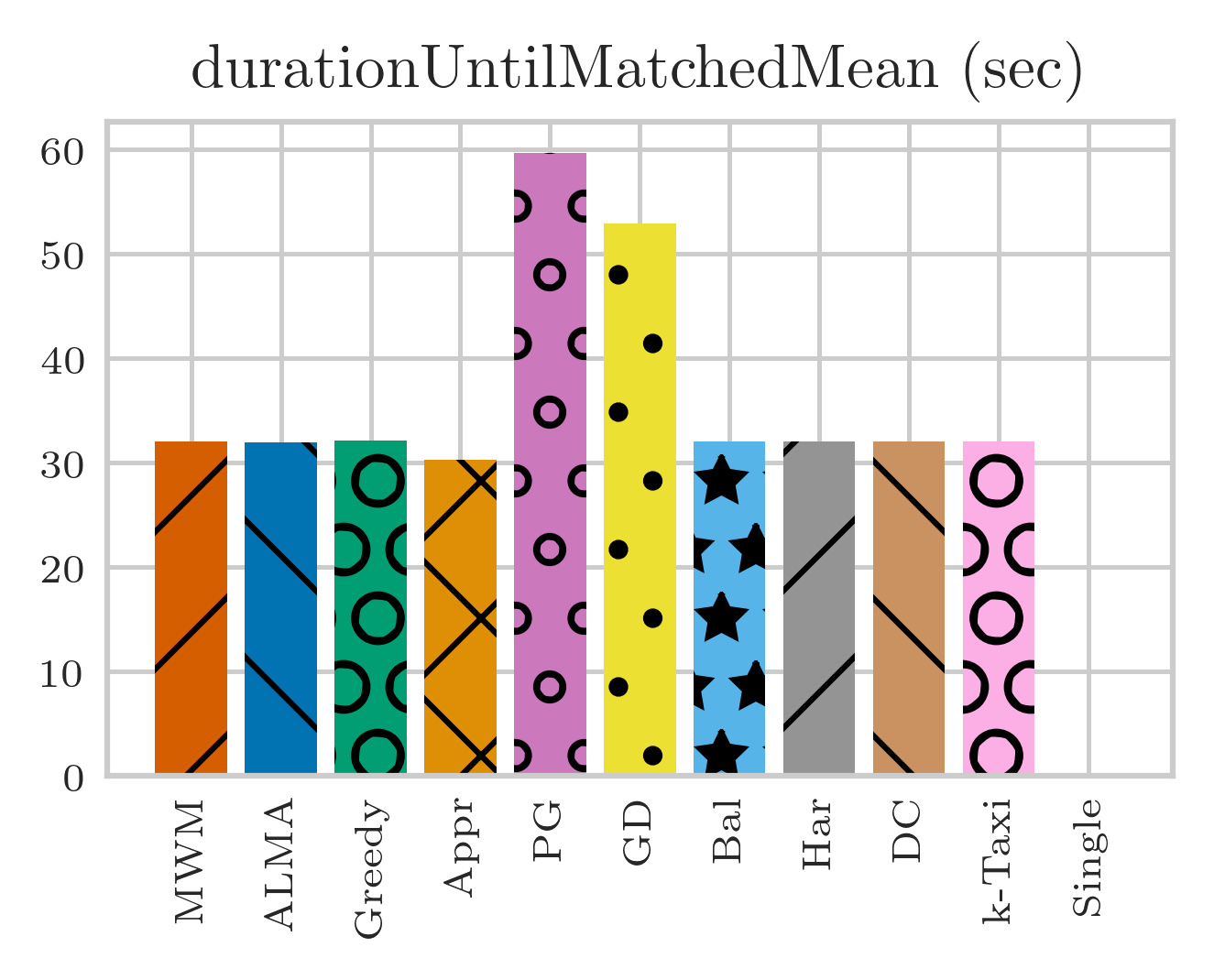}
		\caption{Time to Pair (s)}
		\label{fig_appendix: Jan15Hour8to9Manhattan_durationUntilMatchedMean}
	\end{subfigure}
	~
	\begin{subfigure}[t]{0.23\textwidth}
		\centering
		\includegraphics[width = 1 \linewidth, trim={0.6em 0.6em 0.5em 1.8em}, clip]{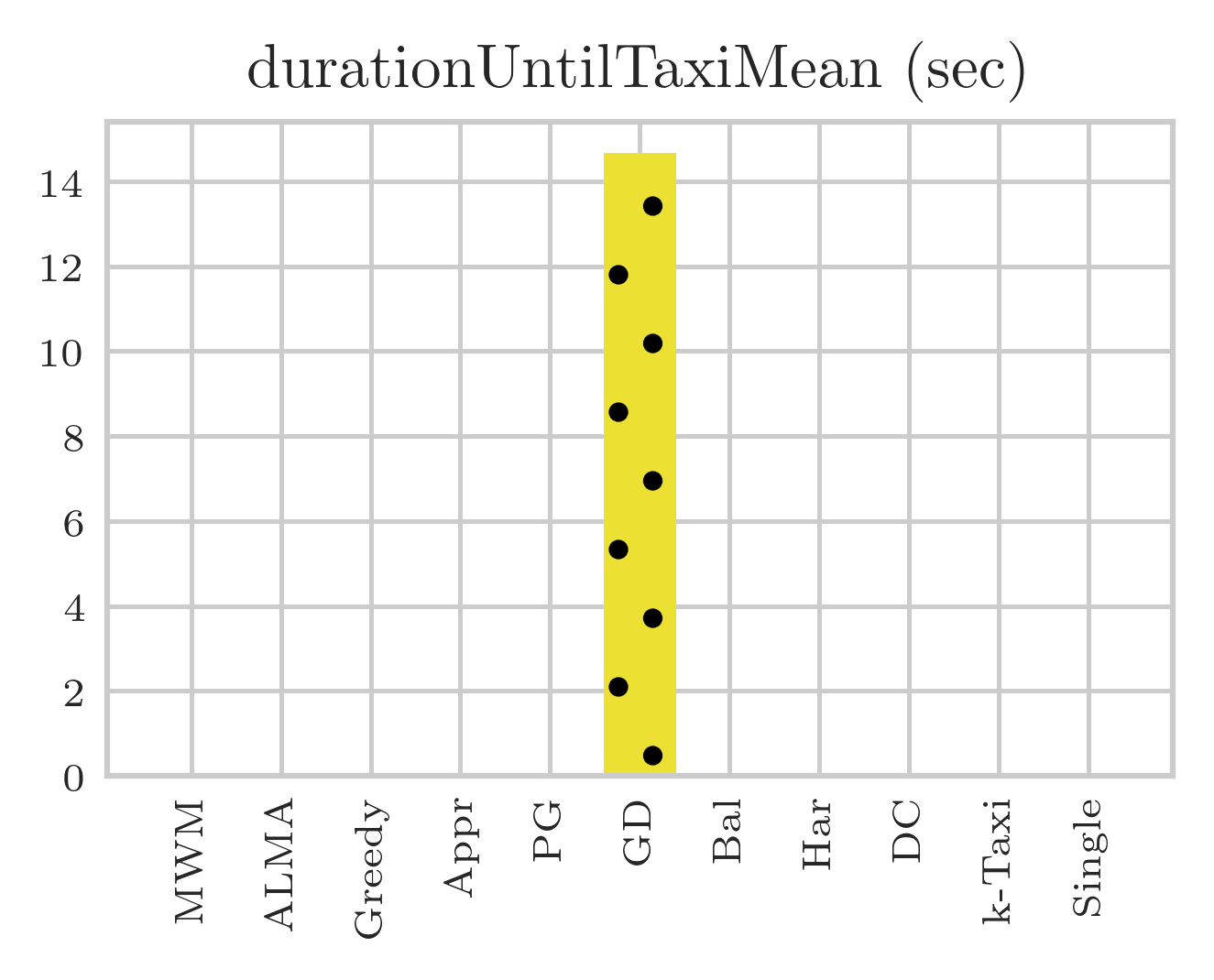}
		\caption{Time to Pair with Taxi (s)}
		\label{fig_appendix: Jan15Hour8to9Manhattan_durationUntilTaxiMean}
	\end{subfigure}

	\begin{subfigure}[t]{0.23\textwidth}
		\centering
		\includegraphics[width = 1 \linewidth, trim={0.6em 0.6em 0.5em 1.8em}, clip]{./Published/Jan15Hour8to9Manhattan_BarPlots_durationUntilPickedUpMean.png}
		\caption{Time to Pick-up (s)}
		\label{fig_appendix: Jan15Hour8to9Manhattan_durationUntilPickedUpMean}
	\end{subfigure}
	~
	\begin{subfigure}[t]{0.23\textwidth}
		\centering
		\includegraphics[width = 1 \linewidth, trim={0.6em 0.6em 0.5em 1.8em}, clip]{./Published/Jan15Hour8to9Manhattan_BarPlots_delayMean.png}
		\caption{Delay (s)}
		\label{fig_appendix: Jan15Hour8to9Manhattan_delayMean}
	\end{subfigure}
	~
	\begin{subfigure}[t]{0.23\textwidth}
		\centering
		\includegraphics[width = 1 \linewidth, trim={0.6em 0.6em 0.5em 1.8em}, clip]{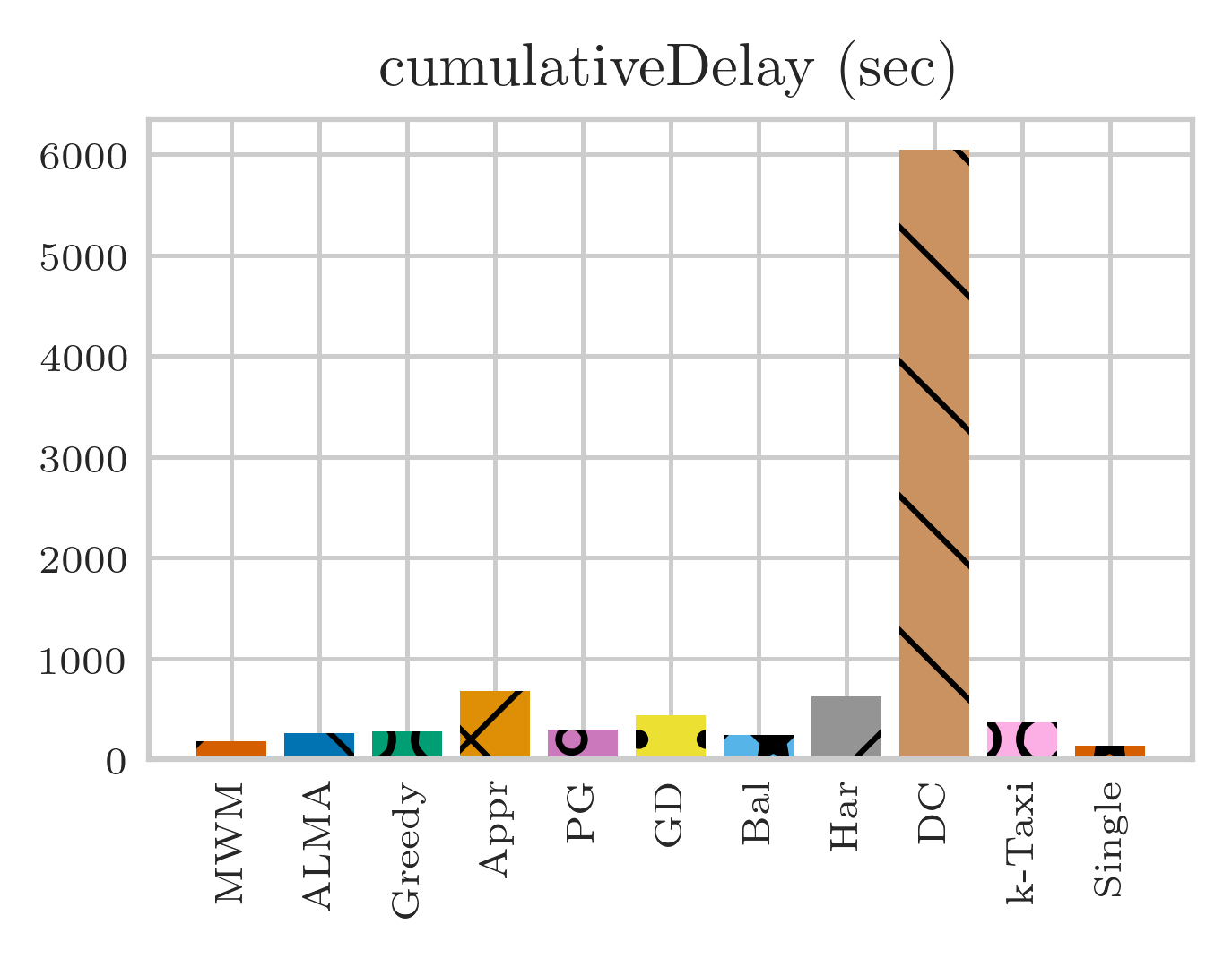}
		\caption{Cumulative Delay (s)}
		\label{fig_appendix: Jan15Hour8to9Manhattan_cumulativeDelay}
	\end{subfigure}
	~
	\begin{subfigure}[t]{0.23\textwidth}
		\centering
		\includegraphics[width = 1 \linewidth, trim={0.6em 0.6em 0.5em 1.8em}, clip]{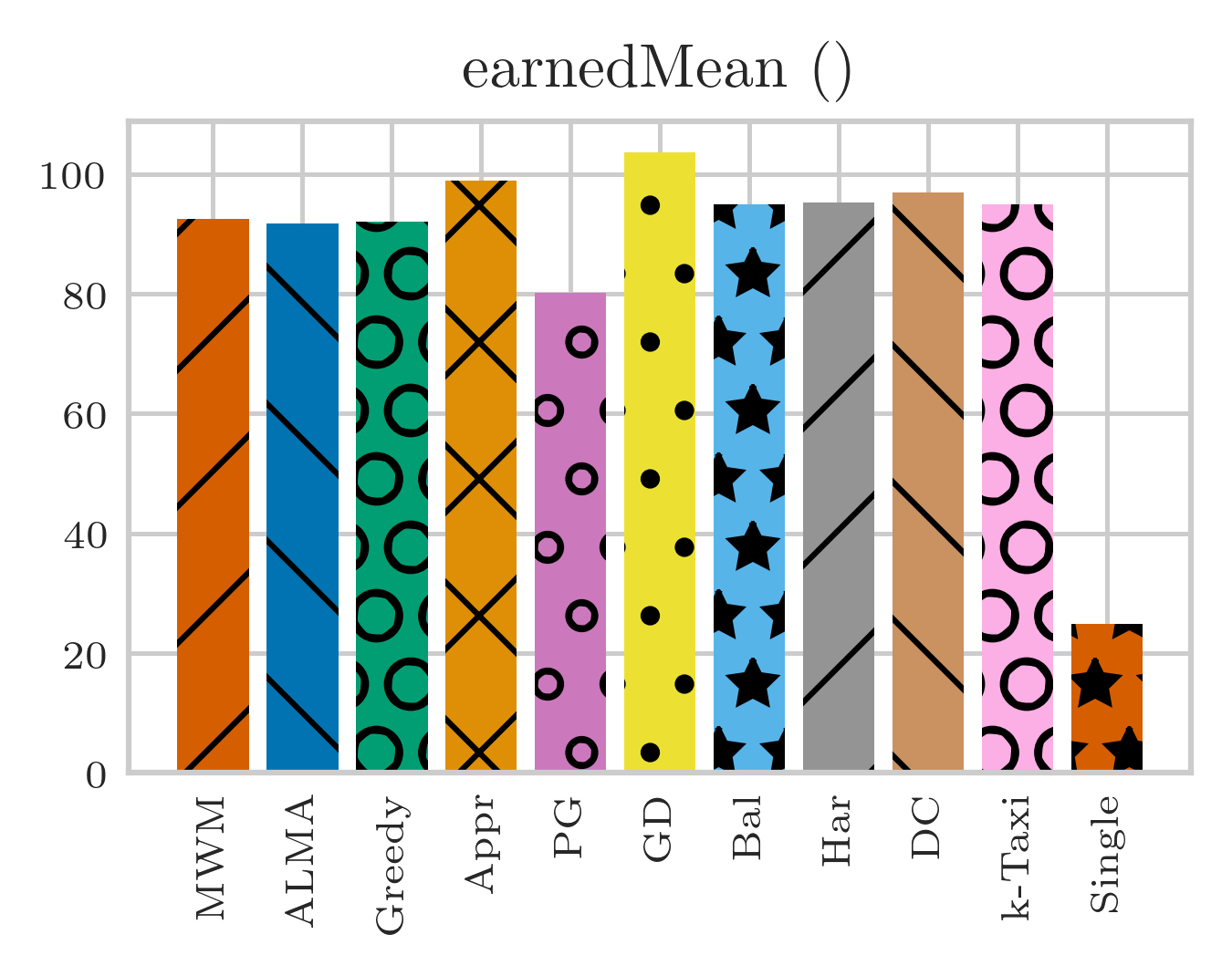}
		\caption{Driver Profit (\$)}
		\label{fig_appendix: Jan15Hour8to9Manhattan_earnedMean}
	\end{subfigure}

	\begin{subfigure}[t]{0.23\textwidth}
		\centering
		\includegraphics[width = 1 \linewidth, trim={0.6em 0.6em 0.5em 1.8em}, clip]{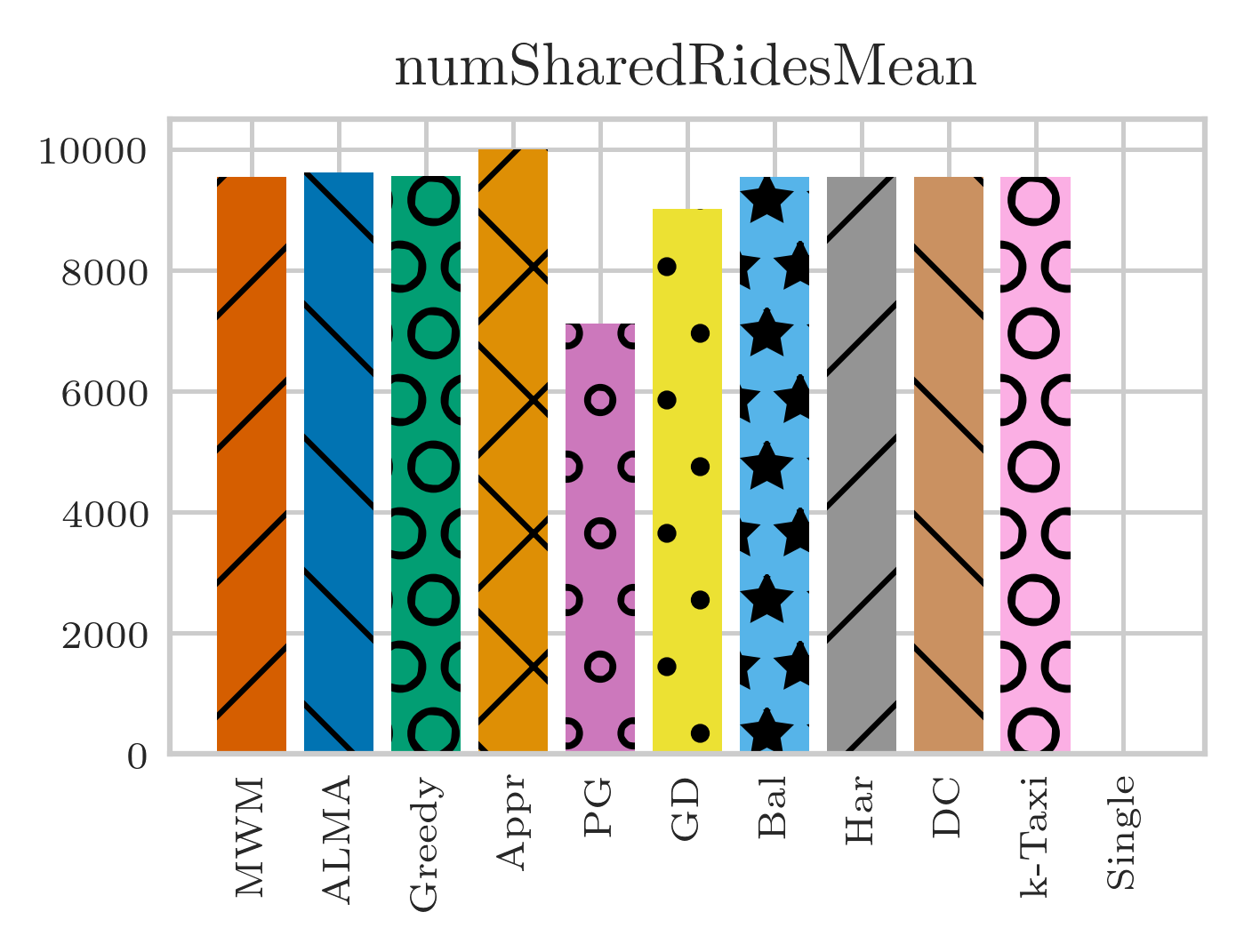}
		\caption{Number of Shared Rides}
		\label{fig_appendix: Jan15Hour8to9Manhattan_numSharedRidesMean}
	\end{subfigure}
	~
	\begin{subfigure}[t]{0.23\textwidth}
		\centering
		\includegraphics[width = 1 \linewidth, trim={0.6em 0.6em 0.5em 1.8em}, clip]{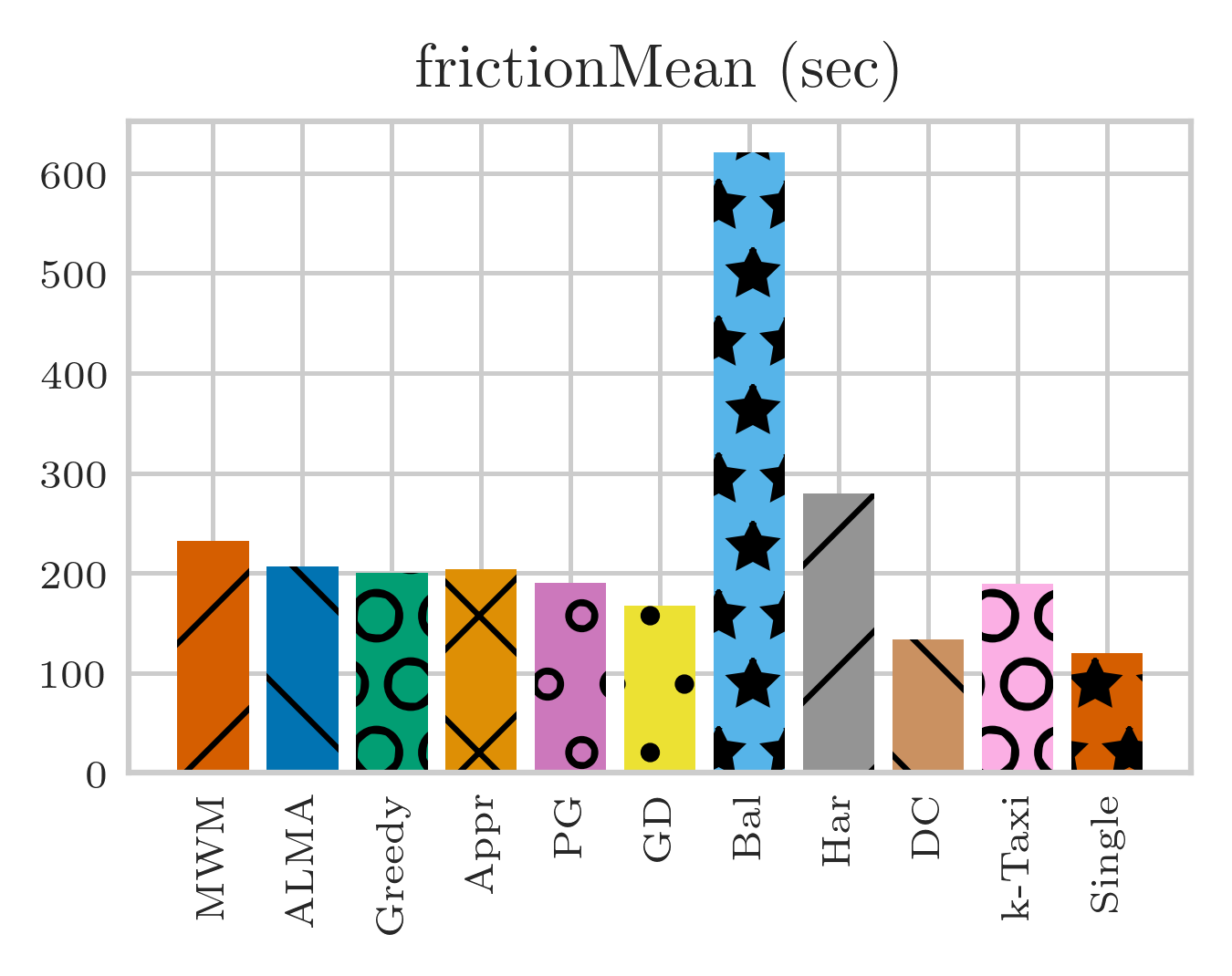}
		\caption{Frictions (s)}
		\label{fig_appendix: Jan15Hour8to9Manhattan_frictionMean}
	\end{subfigure}%
	\caption{January 15, 2016 -- 08:00 - 09:00 -- Manhattan -- \#Taxis = 4276 (base number).}
	\label{fig_appendix: Jan15Hour8to9Manhattan_BarPlots}
\end{figure*}

\begin{table*}[b!]
\centering
\caption{January 15, 2016 -- 08:00 - 09:00 -- Manhattan -- \#Taxis = 4276 (base number).}
\vspace{-0.25em}
\label{tb_appendix: Jan15Hour8to9Manhattan}
\resizebox{\textwidth}{!}{%
\begin{tabular}{@{}lccccccccccccccccccc@{}}
\toprule
\textbf{} & \textbf{\begin{tabular}[c]{@{}c@{}}Distance\\ Driven (m)\end{tabular}} & \textbf{SD} & \textbf{\begin{tabular}[c]{@{}c@{}}Elapsed\\ Time (ns)\end{tabular}} & \textbf{SD} & \textbf{\begin{tabular}[c]{@{}c@{}}Time to\\ Pair (s)\end{tabular}} & \textbf{SD} & \textbf{\begin{tabular}[c]{@{}c@{}}Time to Pair\\ with Taxi (s)\end{tabular}} & \textbf{SD} & \textbf{\begin{tabular}[c]{@{}c@{}}Time to\\ Pick-up (s)\end{tabular}} & \textbf{SD} & \textbf{Delay (s)} & \textbf{SD} & \textbf{\begin{tabular}[c]{@{}c@{}}Cumulative\\ Delay (s)\end{tabular}} & \textbf{\begin{tabular}[c]{@{}c@{}}Driver\\ Profit (\$)\end{tabular}} & \textbf{SD} & \textbf{\begin{tabular}[c]{@{}c@{}}Number of \\ Shared Rides\end{tabular}} & \textbf{SD} & \textbf{Frictions (s)} & \textbf{SD} \\ \midrule
\textbf{MWM}             & 4.76E+07 & 0.00E+00   & 1.54E+12 & 0.00E+00     & 32.05 & 30.90                  & 0.00  & 0.00            & 122.78 & 146.36   & 28.56  & 76.94            & 183.39        & 92.42  & 59.75                  & 9.54E+03 & 0.00      & 232.30 & 420.13 \\
\textbf{ALMA}            & 5.67E+07 & 1.14E+05   & 8.09E+10 & 4.36E+09     & 31.95 & 30.99                  & 0.00  & 0.00            & 206.93 & 246.41   & 24.79  & 77.61            & 263.67        & 91.65  & 53.22                  & 9.61E+03 & 10.46     & 206.59 & 387.78 \\
\textbf{Greedy}          & 5.77E+07 & 8.97E+04   & 4.37E+10 & 2.73E+09     & 32.16 & 31.02                  & 0.00  & 0.00            & 215.82 & 249.18   & 28.19  & 79.89            & 276.17        & 92.05  & 52.55                  & 9.55E+03 & 18.52     & 200.17 & 381.71 \\
\textbf{Appr}            & 7.90E+07 & 0.00E+00   & 1.06E+12 & 0.00E+00     & 30.29 & 30.19                  & 0.00  & 0.00            & 580.48 & 427.45   & 71.13  & 133.34           & 681.90        & 98.89  & 40.52                  & 1.00E+04 & 0.00      & 203.84 & 315.85 \\
\textbf{PG}              & 6.27E+07 & 1.05E+05   & 8.55E+11 & 2.18E+10     & 59.69 & 43.21                  & 0.00  & 0.00            & 219.16 & 282.64   & 13.77  & 61.12            & 292.62        & 80.08  & 45.69                  & 7.11E+03 & 28.92     & 190.24 & 384.91 \\
\textbf{GD}              & 6.62E+07 & 0.00E+00   & 7.54E+12 & 9.01E+10     & 52.91 & 32.09                  & 14.67 & 18.27           & 225.96 & 267.47   & 143.82 & 313.95           & 437.36        & 103.65 & 56.84                  & 9.01E+03 & 0.00      & 166.78 & 348.97 \\
\textbf{Bal}             & 5.11E+07 & 5.16E+04   & 4.60E+10 & 2.09E+09     & 32.05 & 30.89                  & 0.00  & 0.00            & 163.20 & 156.45   & 46.67  & 120.62           & 241.91        & 94.98  & 30.95                  & 9.54E+03 & 0.00      & 621.14 & 490.55 \\
\textbf{Har}             & 7.61E+07 & 2.47E+05   & 4.28E+10 & 2.41E+09     & 32.05 & 30.89                  & 0.00  & 0.00            & 540.38 & 479.35   & 50.84  & 129.03           & 623.27        & 95.18  & 41.59                  & 9.54E+03 & 0.00      & 279.55 & 282.88 \\
\textbf{DC}              & 1.23E+08 & 5.41E+06   & 1.79E+13 & 1.05E+12     & 32.05 & 30.89                  & 0.00  & 0.00            & 0.00   & 10458.42 & 52.29  & 125.51           & 6051.60       & 96.94  & 196.20                 & 9.54E+03 & 0.00      & 133.54 & 349.94 \\
\textbf{k-Taxi}          & 5.97E+07 & 2.31E+05   & 5.23E+13 & 3.00E+12     & 32.05 & 30.89                  & 0.00  & 0.00            & 288.89 & 372.65   & 47.09  & 120.88           & 368.02        & 94.90  & 62.00                  & 9.54E+03 & 0.00      & 188.68 & 343.92 \\
\textbf{Single}          & 8.51E+07 & 0.00E+00   & 8.12E+11 & 0.00E+00     & 0.00  & 0.00                   & 0.02  & 1.04            & 133.36 & 201.19   & 0.00   & 0.00             & 133.38        & 24.88  & 7.30                   & 0.00E+00 & 0.00      & 119.80 & 291.01 \\ \bottomrule
\end{tabular}%
}
\end{table*}

\begin{table*}[b!]
\centering
\caption{January 15, 2016 -- 08:00 - 09:00 -- Manhattan -- \#Taxis = 4276 (base number). Each column presents the relative difference compared to the first line, i.e., the MWM (algorithm - MWM) / MWM, for each metric.}
\vspace{-0.25em}
\label{tb_appendix: Jan15Hour8to9ManhattanPercentages}
\resizebox{\textwidth}{!}{%
\begin{tabular}{@{}lccccccccccccccccccc@{}}
\toprule
\textbf{} & \textbf{\begin{tabular}[c]{@{}c@{}}Distance\\ Driven (m)\end{tabular}} & \textbf{SD} & \textbf{\begin{tabular}[c]{@{}c@{}}Elapsed\\ Time (ns)\end{tabular}} & \textbf{SD} & \textbf{\begin{tabular}[c]{@{}c@{}}Time to\\ Pair (s)\end{tabular}} & \textbf{SD} & \textbf{\begin{tabular}[c]{@{}c@{}}Time to Pair\\ with Taxi (s)\end{tabular}} & \textbf{SD} & \textbf{\begin{tabular}[c]{@{}c@{}}Time to\\ Pick-up (s)\end{tabular}} & \textbf{SD} & \textbf{Delay (s)} & \textbf{SD} & \textbf{\begin{tabular}[c]{@{}c@{}}Cumulative\\ Delay (s)\end{tabular}} & \textbf{\begin{tabular}[c]{@{}c@{}}Driver\\ Profit (\$)\end{tabular}} & \textbf{SD} & \textbf{\begin{tabular}[c]{@{}c@{}}Number of \\ Shared Rides\end{tabular}} & \textbf{SD} & \textbf{Frictions (s)} & \textbf{SD} \\ \midrule
\textbf{MWM}             & 0.00\%   & --         & 0.00\%    & --           & 0.00\%    & 0.00\%                 & -- & --              & 0.00\%    & 0.00\%    & 0.00\%    & 0.00\%           & 0.00\%        & 0.00\%   & 0.00\%                 & 0.00\%    & --        & 0.00\%   & 0.00\%   \\
\textbf{ALMA}            & 19.15\%  & --         & -94.75\%  & --           & -0.29\%   & 0.30\%                 & -- & --              & 68.54\%   & 68.36\%   & -13.19\%  & 0.88\%           & 43.78\%       & -0.84\%  & -10.93\%               & 0.72\%    & --        & -11.07\% & -7.70\%  \\
\textbf{Greedy}          & 21.24\%  & --         & -97.16\%  & --           & 0.35\%    & 0.40\%                 & -- & --              & 75.78\%   & 70.25\%   & -1.30\%   & 3.84\%           & 50.59\%       & -0.41\%  & -12.06\%               & 0.08\%    & --        & -13.83\% & -9.14\%  \\
\textbf{Appr}            & 65.95\%  & --         & -30.85\%  & --           & -5.47\%   & -2.28\%                & -- & --              & 372.79\%  & 192.06\%  & 149.01\%  & 73.30\%          & 271.83\%      & 6.99\%   & -32.18\%               & 4.81\%    & --        & -12.25\% & -24.82\% \\
\textbf{PG}              & 31.69\%  & --         & -44.50\%  & --           & 86.27\%   & 39.86\%                & -- & --              & 78.51\%   & 93.12\%   & -51.81\%  & -20.56\%         & 59.57\%       & -13.35\% & -23.54\%               & -25.48\%  & --        & -18.10\% & -8.38\%  \\
\textbf{GD}              & 39.03\%  & --         & 389.79\%  & --           & 65.11\%   & 3.87\%                 & -- & --              & 84.04\%   & 82.75\%   & 403.50\%  & 308.04\%         & 138.49\%      & 12.15\%  & -4.87\%                & -5.63\%   & --        & -28.21\% & -16.94\% \\
\textbf{Bal}             & 7.41\%   & --         & -97.01\%  & --           & 0.00\%    & 0.00\%                 & -- & --              & 32.92\%   & 6.90\%    & 63.39\%   & 56.77\%          & 31.91\%       & 2.76\%   & -48.19\%               & 0.00\%    & --        & 167.39\% & 16.76\%  \\
\textbf{Har}             & 59.98\%  & --         & -97.22\%  & --           & 0.00\%    & 0.00\%                 & -- & --              & 340.13\%  & 227.52\%  & 78.00\%   & 67.70\%          & 239.87\%      & 2.98\%   & -30.39\%               & 0.00\%    & --        & 20.34\%  & -32.67\% \\
\textbf{DC}              & 158.13\% & --         & 1061.10\% & --           & 0.00\%    & 0.00\%                 & -- & --              & -100.00\% & 7045.82\% & 83.09\%   & 63.13\%          & 3199.91\%     & 4.88\%   & 228.37\%               & 0.00\%    & --        & -42.51\% & -16.71\% \\
\textbf{k-Taxi}          & 25.48\%  & --         & 3293.85\% & --           & 0.00\%    & 0.00\%                 & -- & --              & 135.29\%  & 154.62\%  & 64.85\%   & 57.11\%          & 100.68\%      & 2.68\%   & 3.76\%                 & 0.00\%    & --        & -18.78\% & -18.14\% \\
\textbf{Single}          & 78.81\%  & --         & -47.28\%  & --           & -100.00\% & -100.00\%              & -- & --              & 8.62\%    & 37.46\%   & -100.00\% & -100.00\%        & -27.27\%      & -73.08\% & -87.78\%               & -100.00\% & --        & -48.43\% & -30.73\% \\ \bottomrule
\end{tabular}%
}
\end{table*}

\clearpage

\begin{landscape}
\begin{table*}[t!]
\centering
\caption{January 15, 2016 -- 08:00 - 09:00 -- Manhattan -- \#Taxis = 2138. \\
Offline algorithms are run either in Just-in-Time (JiT) manner, or in batches (with batch size 1, or 2 min). Because of the density of the dataset, requests become critical every time-step, thus JiT is the same as in batches with batch size 1.}
\label{tb_appendix: Jan15Hour8to9ManhattanFull}
\resizebox{\textwidth}{!}{%
\begin{tabular}{@{}lccccccccccccccccccc@{}}
\toprule
\textbf{} & \textbf{\begin{tabular}[c]{@{}c@{}}Distance\\ Driven (m)\end{tabular}} & \textbf{SD} & \textbf{\begin{tabular}[c]{@{}c@{}}Elapsed\\ Time (ns)\end{tabular}} & \textbf{SD} & \textbf{\begin{tabular}[c]{@{}c@{}}Time to\\ Pair (s)\end{tabular}} & \textbf{SD} & \textbf{\begin{tabular}[c]{@{}c@{}}Time to Pair\\ with Taxi (s)\end{tabular}} & \textbf{SD} & \textbf{\begin{tabular}[c]{@{}c@{}}Time to\\ Pick-up (s)\end{tabular}} & \textbf{SD} & \textbf{Delay (s)} & \textbf{SD} & \textbf{\begin{tabular}[c]{@{}c@{}}Cumulative\\ Delay (s)\end{tabular}} & \textbf{\begin{tabular}[c]{@{}c@{}}Driver\\ Profit (\$)\end{tabular}} & \textbf{SD} & \textbf{\begin{tabular}[c]{@{}c@{}}Number of \\ Shared Rides\end{tabular}} & \textbf{SD} & \textbf{Frictions (s)} & \textbf{SD} \\ \midrule
\textbf{MWM (1)}         & 5.46E+07 & 0.00E+00   & 9.91E+10 & 0.00E+00     & 5.17  & 18.08                  & 102.26  & 407.94          & 318.23   & 542.59   & 45.92  & 123.45           & 471.57        & 190.44 & 52.46                  & 9.72E+03 & 0.00      & 29.68 & 12.38  \\
\textbf{MWM (2)}         & 5.27E+07 & 0.00E+00   & 1.45E+11 & 0.00E+00     & 32.05 & 30.90                  & 88.64   & 308.08          & 288.94   & 441.85   & 38.34  & 112.20           & 447.97        & 187.50 & 49.91                  & 9.54E+03 & 0.00      & 34.32 & 16.06  \\
\textbf{ALMA (1)}        & 6.30E+07 & 1.38E+05   & 3.96E+10 & 4.12E+09     & 3.97  & 15.71                  & 356.07  & 614.16          & 654.41   & 806.25   & 39.48  & 119.40           & 1053.93       & 188.79 & 36.01                  & 9.85E+03 & 8.36      & 29.43 & 9.91   \\
\textbf{ALMA (2)}        & 6.18E+07 & 1.02E+05   & 6.02E+10 & 7.57E+09     & 31.91 & 30.92                  & 323.09  & 566.04          & 611.59   & 758.26   & 29.84  & 102.38           & 996.43        & 184.66 & 35.59                  & 9.62E+03 & 6.88      & 30.00 & 10.28  \\
\textbf{Greedy (1)}      & 6.76E+07 & 2.28E+05   & 6.78E+09 & 1.68E+09     & 4.41  & 16.52                  & 577.85  & 706.95          & 932.92   & 813.98   & 44.64  & 121.06           & 1559.82       & 189.99 & 36.36                  & 9.82E+03 & 6.13      & 29.54 & 9.76   \\
\textbf{Greedy (2)}      & 6.66E+07 & 1.01E+05   & 1.31E+10 & 3.75E+09     & 32.14 & 31.04                  & 536.36  & 668.99          & 881.67   & 783.43   & 34.77  & 106.58           & 1484.94       & 185.72 & 36.99                  & 9.55E+03 & 17.72     & 29.97 & 10.00  \\
\textbf{Appr (1)}        & 8.04E+07 & 0.00E+00   & 5.41E+11 & 0.00E+00     & 27.94 & 30.07                  & 852.44  & 1048.80         & 1454.91  & 1185.15  & 71.59  & 137.37           & 2406.88       & 198.00 & 36.05                  & 1.00E+04 & 0.00      & 45.28 & 18.08  \\
\textbf{Appr (2)}        & 8.05E+07 & 0.00E+00   & 5.42E+11 & 0.00E+00     & 30.29 & 30.19                  & 804.67  & 995.86          & 1410.40  & 1145.77  & 69.81  & 128.89           & 2315.17       & 197.35 & 35.61                  & 1.00E+04 & 0.00      & 29.69 & 10.13  \\
\textbf{PG}              & 6.74E+07 & 1.17E+05   & 5.20E+11 & 1.81E+10     & 59.53 & 43.27                  & 297.99  & 764.61          & 664.30   & 1053.38  & 19.90  & 107.57           & 1041.73       & 161.90 & 39.97                  & 7.12E+03 & 31.39     & 29.56 & 8.29   \\
\textbf{GD}              & 6.92E+07 & 0.00E+00   & 7.11E+12 & 3.28E+11     & 52.91 & 32.09                  & 358.38  & 801.79          & 626.96   & 945.13   & 153.49 & 333.21           & 1191.73       & 210.00 & 54.47                  & 9.01E+03 & 0.00      & 30.21 & 9.70   \\
\textbf{Bal (1)}         & 6.22E+07 & 9.71E+04   & 1.35E+10 & 4.56E+09     & 5.17  & 18.08                  & 403.38  & 535.51          & 725.18   & 637.49   & 55.53  & 134.36           & 1189.26       & 192.96 & 41.38                  & 9.72E+03 & 0.00      & 30.23 & 11.05  \\
\textbf{Bal (2)}         & 5.99E+07 & 1.38E+05   & 3.42E+10 & 7.09E+09     & 32.05 & 30.89                  & 336.45  & 466.17          & 635.90   & 569.04   & 46.20  & 120.74           & 1050.59       & 189.53 & 41.81                  & 9.54E+03 & 0.00      & 31.76 & 12.20  \\
\textbf{Har (1)}         & 8.10E+07 & 2.36E+05   & 1.37E+10 & 4.70E+09     & 5.17  & 18.08                  & 946.56  & 1058.46         & 1555.42  & 1190.40  & 61.98  & 146.65           & 2569.12       & 194.22 & 49.97                  & 9.72E+03 & 0.00      & 29.45 & 9.67   \\
\textbf{Har (2)}         & 7.96E+07 & 2.87E+05   & 3.36E+10 & 6.90E+09     & 32.05 & 30.89                  & 906.49  & 1024.29         & 1501.74  & 1153.03  & 51.28  & 130.04           & 2491.55       & 190.37 & 51.20                  & 9.54E+03 & 0.00      & 29.65 & 9.65   \\
\textbf{DC (1)}          & 1.41E+08 & 1.90E+06   & 1.16E+13 & 5.48E+11     & 5.17  & 18.08                  & 0.00    & 0.00            & 7660.58  & 10890.62 & 62.70  & 142.71           & 7728.45       & 192.45 & 280.85                 & 9.72E+03 & 0.00      & 85.76 & 269.59 \\
\textbf{DC (2)}          & 1.38E+08 & 1.71E+06   & 8.99E+12 & 3.74E+11     & 32.05 & 30.89                  & 0.00    & 0.00            & 7290.07  & 10196.66 & 51.77  & 124.76           & 7373.89       & 188.37 & 272.89                 & 9.54E+03 & 0.00      & 93.37 & 287.47 \\
\textbf{k-Taxi (1)}      & 7.10E+07 & 3.00E+05   & 3.41E+12 & 2.12E+11     & 5.17  & 18.08                  & 646.48  & 528.54          & 1099.19  & 702.35   & 58.32  & 141.50           & 1809.15       & 193.48 & 46.97                  & 9.72E+03 & 0.00      & 29.33 & 9.74   \\
\textbf{k-Taxi (2)}      & 6.92E+07 & 2.04E+05   & 4.32E+12 & 4.05E+11     & 32.05 & 30.89                  & 586.22  & 491.05          & 1020.87  & 670.76   & 48.51  & 126.93           & 1687.65       & 189.90 & 48.16                  & 9.54E+03 & 0.00      & 30.14 & 10.10  \\
\textbf{WFA (1)}         & 8.48E+07 & 3.25E+05   & 1.28E+14 & 6.77E+12     & 5.17  & 18.08                  & 0.00    & 0.00            & 30966.54 & 40847.98 & 64.04  & 145.43           & 31035.75      & 194.69 & 597.69                 & 9.72E+03 & 0.00      & 63.79 & 254.48 \\
\textbf{WFA (2)}         & 8.09E+07 & 0.00E+00   & 9.69E+13 & 0.00E+00     & 32.05 & 30.90                  & 0.00    & 0.00            & 28964.94 & 39066.93 & 51.85  & 127.05           & 29048.84      & 190.49 & 593.79                 & 9.54E+03 & 0.00      & 69.68 & 264.10 \\
\textbf{Single}          & 8.63E+07 & 0.00E+00   & 2.02E+12 & 0.00E+00     & 0.00  & 0.00                   & 700.50  & 1317.51         & 843.80   & 1444.94  & 0.00   & 0.00             & 1544.30       & 49.72  & 6.90                   & 0.00E+00 & 0.00      & 29.41 & 6.11   \\
\textbf{Random}          & 1.14E+08 & 2.56E+05   & 6.26E+09 & 1.42E+09     & 3.61  & 15.05                  & 1963.08 & 1457.54         & 2903.86  & 1580.37  & 161.07 & 328.92           & 5031.62       & 222.47 & 47.27                  & 9.88E+03 & 6.54      & 29.56 & 9.37   \\ \bottomrule
\end{tabular}%
}
\end{table*}

\begin{table*}[t!]
\centering
\caption{January 15, 2016 -- 08:00 - 09:00 -- Manhattan -- \#Taxis = 2138. \\
Offline algorithms are run either in Just-in-Time (JiT) manner, or in batches (with batch size 1, or 2 min). Because of the density of the dataset, requests become critical every time-step, thus JiT is the same as in batches with batch size 1.\\Each column presents the relative difference compared to the first line, i.e., MWM of batch size one (algorithm - MWM(1)) / MWM(1), for each metric.}
\label{tb_appendix: Jan15Hour8to9ManhattanFullPercentages}
\resizebox{\textwidth}{!}{%
\begin{tabular}{@{}lccccccccccccccccccc@{}}
\toprule
\textbf{} & \textbf{\begin{tabular}[c]{@{}c@{}}Distance\\ Driven (m)\end{tabular}} & \textbf{SD} & \textbf{\begin{tabular}[c]{@{}c@{}}Elapsed\\ Time (ns)\end{tabular}} & \textbf{SD} & \textbf{\begin{tabular}[c]{@{}c@{}}Time to\\ Pair (s)\end{tabular}} & \textbf{SD} & \textbf{\begin{tabular}[c]{@{}c@{}}Time to Pair\\ with Taxi (s)\end{tabular}} & \textbf{SD} & \textbf{\begin{tabular}[c]{@{}c@{}}Time to\\ Pick-up (s)\end{tabular}} & \textbf{SD} & \textbf{Delay (s)} & \textbf{SD} & \textbf{\begin{tabular}[c]{@{}c@{}}Cumulative\\ Delay (s)\end{tabular}} & \textbf{\begin{tabular}[c]{@{}c@{}}Driver\\ Profit (\$)\end{tabular}} & \textbf{SD} & \textbf{\begin{tabular}[c]{@{}c@{}}Number of \\ Shared Rides\end{tabular}} & \textbf{SD} & \textbf{Frictions (s)} & \textbf{SD} \\ \midrule
\textbf{MWM (1)}         & 0.00\%   & --         & 0.00\%      & --           & 0.00\%    & 0.00\%                 & 0.00\%    & 0.00\%          & 0.00\%    & 0.00\%    & 0.00\%    & 0.00\%           & 0.00\%        & 0.00\%   & 0.00\%                 & 0.00\%    & --        & 0.00\%   & 0.00\%    \\
\textbf{MWM (2)}         & -3.44\%  & --         & 46.61\%     & --           & 520.26\%  & 70.89\%                & -13.32\%  & -24.48\%        & -9.20\%   & -18.57\%  & -16.49\%  & -9.12\%          & -5.00\%       & -1.54\%  & -4.86\%                & -1.82\%   & --        & 15.65\%  & 29.68\%   \\
\textbf{ALMA (1)}        & 15.50\%  & --         & -60.03\%    & --           & -23.20\%  & -13.08\%               & 248.20\%  & 50.55\%         & 105.64\%  & 48.59\%   & -14.01\%  & -3.29\%          & 123.49\%      & -0.87\%  & -31.36\%               & 1.36\%    & --        & -0.84\%  & -20.00\%  \\
\textbf{ALMA (2)}        & 13.27\%  & --         & -39.29\%    & --           & 517.65\%  & 71.03\%                & 215.95\%  & 38.76\%         & 92.19\%   & 39.75\%   & -35.02\%  & -17.07\%         & 111.30\%      & -3.04\%  & -32.16\%               & -1.03\%   & --        & 1.08\%   & -16.96\%  \\
\textbf{Greedy (1)}      & 23.91\%  & --         & -93.16\%    & --           & -14.58\%  & -8.61\%                & 465.07\%  & 73.30\%         & 193.16\%  & 50.02\%   & -2.78\%   & -1.94\%          & 230.77\%      & -0.24\%  & -30.70\%               & 1.01\%    & --        & -0.47\%  & -21.15\%  \\
\textbf{Greedy (2)}      & 21.98\%  & --         & -86.81\%    & --           & 522.14\%  & 71.70\%                & 424.50\%  & 63.99\%         & 177.06\%  & 44.39\%   & -24.29\%  & -13.67\%         & 214.89\%      & -2.48\%  & -29.48\%               & -1.72\%   & --        & 0.99\%   & -19.22\%  \\
\textbf{Appr (1)}        & 47.32\%  & --         & 445.49\%    & --           & 440.80\%  & 66.34\%                & 733.59\%  & 157.10\%        & 357.19\%  & 118.42\%  & 55.92\%   & 11.28\%          & 410.39\%      & 3.97\%   & -31.28\%               & 2.92\%    & --        & 52.57\%  & 46.03\%   \\
\textbf{Appr (2)}        & 47.52\%  & --         & 447.16\%    & --           & 486.30\%  & 66.99\%                & 686.88\%  & 144.12\%        & 343.20\%  & 111.17\%  & 52.04\%   & 4.40\%           & 390.95\%      & 3.63\%   & -32.13\%               & 2.90\%    & --        & 0.02\%   & -18.16\%  \\
\textbf{PG}              & 23.56\%  & --         & 424.86\%    & --           & 1052.23\% & 139.31\%               & 191.40\%  & 87.43\%         & 108.75\%  & 94.14\%   & -56.65\%  & -12.87\%         & 120.90\%      & -14.99\% & -23.81\%               & -26.70\%  & --        & -0.41\%  & -33.08\%  \\
\textbf{GD}              & 26.84\%  & --         & 7071.20\%   & --           & 924.12\%  & 77.50\%                & 250.45\%  & 96.55\%         & 97.01\%   & 74.19\%   & 234.27\%  & 169.91\%         & 152.71\%      & 10.27\%  & 3.83\%                 & -7.35\%   & --        & 1.79\%   & -21.63\%  \\
\textbf{Bal (1)}         & 14.02\%  & --         & -86.42\%    & --           & 0.00\%    & 0.00\%                 & 294.47\%  & 31.27\%         & 127.88\%  & 17.49\%   & 20.94\%   & 8.83\%           & 152.19\%      & 1.33\%   & -21.13\%               & 0.00\%    & --        & 1.87\%   & -10.75\%  \\
\textbf{Bal (2)}         & 9.88\%   & --         & -65.49\%    & --           & 520.26\%  & 70.88\%                & 229.01\%  & 14.28\%         & 99.83\%   & 4.87\%    & 0.61\%    & -2.20\%          & 122.78\%      & -0.48\%  & -20.31\%               & -1.82\%   & --        & 7.00\%   & -1.50\%   \\
\textbf{Har (1)}         & 48.42\%  & --         & -86.16\%    & --           & 0.00\%    & 0.00\%                 & 825.64\%  & 159.47\%        & 388.77\%  & 119.39\%  & 34.98\%   & 18.79\%          & 444.80\%      & 1.98\%   & -4.75\%                & 0.00\%    & --        & -0.76\%  & -21.90\%  \\
\textbf{Har (2)}         & 45.93\%  & --         & -66.14\%    & --           & 520.26\%  & 70.88\%                & 786.45\%  & 151.09\%        & 371.90\%  & 112.50\%  & 11.68\%   & 5.34\%           & 428.35\%      & -0.04\%  & -2.41\%                & -1.82\%   & --        & -0.08\%  & -22.10\%  \\
\textbf{DC (1)}          & 159.34\% & --         & 11636.41\%  & --           & 0.00\%    & 0.00\%                 & -100.00\% & -100.00\%       & 2307.25\% & 1907.15\% & 36.55\%   & 15.59\%          & 1538.87\%     & 1.05\%   & 435.35\%               & 0.00\%    & --        & 188.95\% & 2076.91\% \\
\textbf{DC (2)}          & 152.49\% & --         & 8969.76\%   & --           & 520.26\%  & 70.88\%                & -100.00\% & -100.00\%       & 2190.83\% & 1779.25\% & 12.75\%   & 1.06\%           & 1463.68\%     & -1.09\%  & 420.17\%               & -1.82\%   & --        & 214.60\% & 2221.34\% \\
\textbf{k-Taxi (1)}      & 30.13\%  & --         & 3339.41\%   & --           & 0.00\%    & 0.00\%                 & 532.19\%  & 29.56\%         & 245.41\%  & 29.44\%   & 27.00\%   & 14.61\%          & 283.64\%      & 1.60\%   & -10.47\%               & 0.00\%    & --        & -1.18\%  & -21.38\%  \\
\textbf{k-Taxi (2)}      & 26.82\%  & --         & 4258.51\%   & --           & 520.26\%  & 70.88\%                & 473.26\%  & 20.37\%         & 220.80\%  & 23.62\%   & 5.65\%    & 2.81\%           & 257.88\%      & -0.28\%  & -8.20\%                & -1.82\%   & --        & 1.54\%   & -18.41\%  \\
\textbf{WFA (1)}         & 55.42\%  & --         & 129083.20\% & --           & 0.00\%    & 0.00\%                 & -100.00\% & -100.00\%       & 9630.90\% & 7428.32\% & 39.48\%   & 17.80\%          & 6481.32\%     & 2.23\%   & 1039.29\%              & 0.00\%    & --        & 114.92\% & 1954.90\% \\
\textbf{WFA (2)}         & 48.26\%  & --         & 97654.76\%  & --           & 520.26\%  & 70.89\%                & -100.00\% & -100.00\%       & 9001.91\% & 7100.07\% & 12.93\%   & 2.91\%           & 6059.99\%     & 0.03\%   & 1031.86\%              & -1.82\%   & --        & 134.77\% & 2032.61\% \\
\textbf{Single}          & 58.20\%  & --         & 1935.09\%   & --           & -100.00\% & -100.00\%              & 585.01\%  & 222.97\%        & 165.16\%  & 166.30\%  & -100.00\% & -100.00\%        & 227.48\%      & -73.89\% & -86.85\%               & -100.00\% & --        & -0.91\%  & -50.64\%  \\
\textbf{Random}          & 108.44\% & --         & -93.69\%    & --           & -30.10\%  & -16.76\%               & 1819.69\% & 257.30\%        & 812.51\%  & 191.26\%  & 250.78\%  & 166.43\%         & 966.99\%      & 16.82\%  & -9.89\%                & 1.60\%    & --        & -0.40\%  & -24.38\%  \\ \bottomrule
\end{tabular}%
}
\end{table*}
\end{landscape}

\clearpage

\begin{figure*}[p!]
	\centering
	\begin{subfigure}[t]{1\textwidth}
		\centering
		\includegraphics[width = 1 \linewidth, trim={0.5em 0.7em 0.6em 1.8em}, clip]{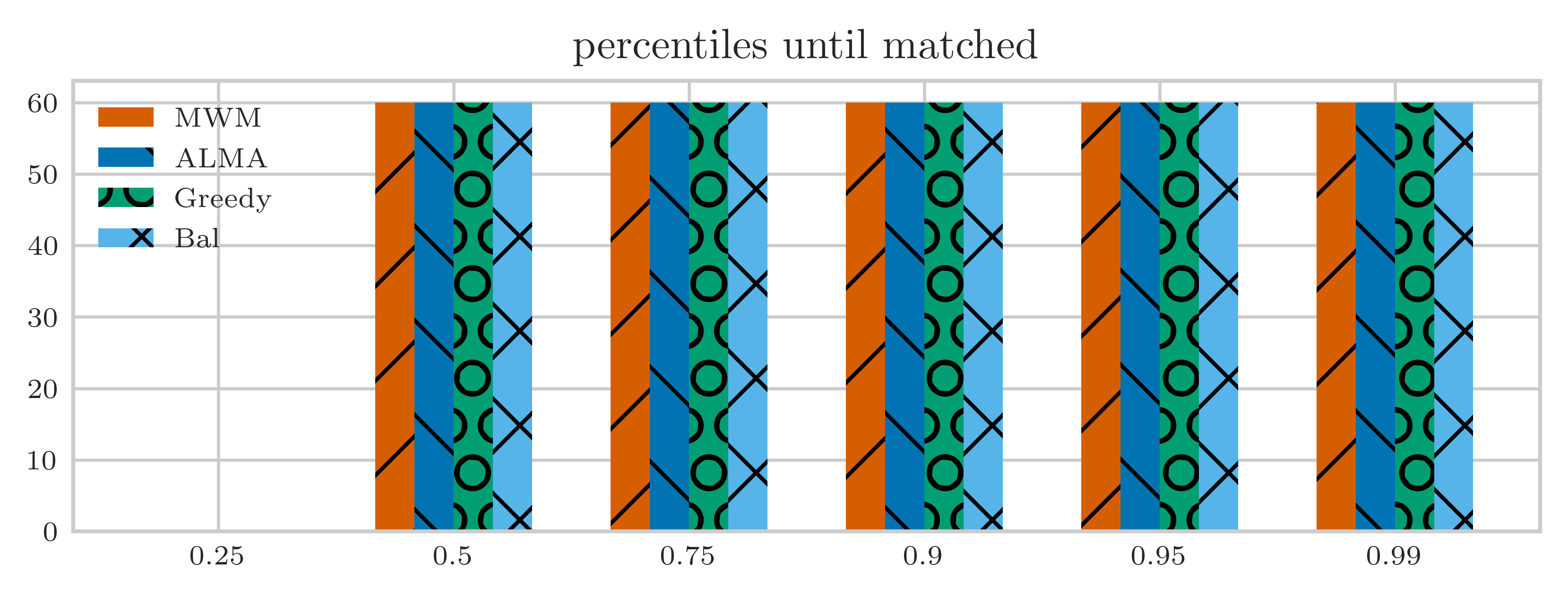}
		\caption{Sequence of Percentiles for Time to Pair (s)}
		\label{fig_appendix: Jan15Hour8to9Manhattan_percentiles_until_matched}
	\end{subfigure}
	~ 
	\begin{subfigure}[t]{1\textwidth}
		\centering
		\includegraphics[width = 1 \linewidth, trim={0.5em 0.7em 0.6em 1.8em}, clip]{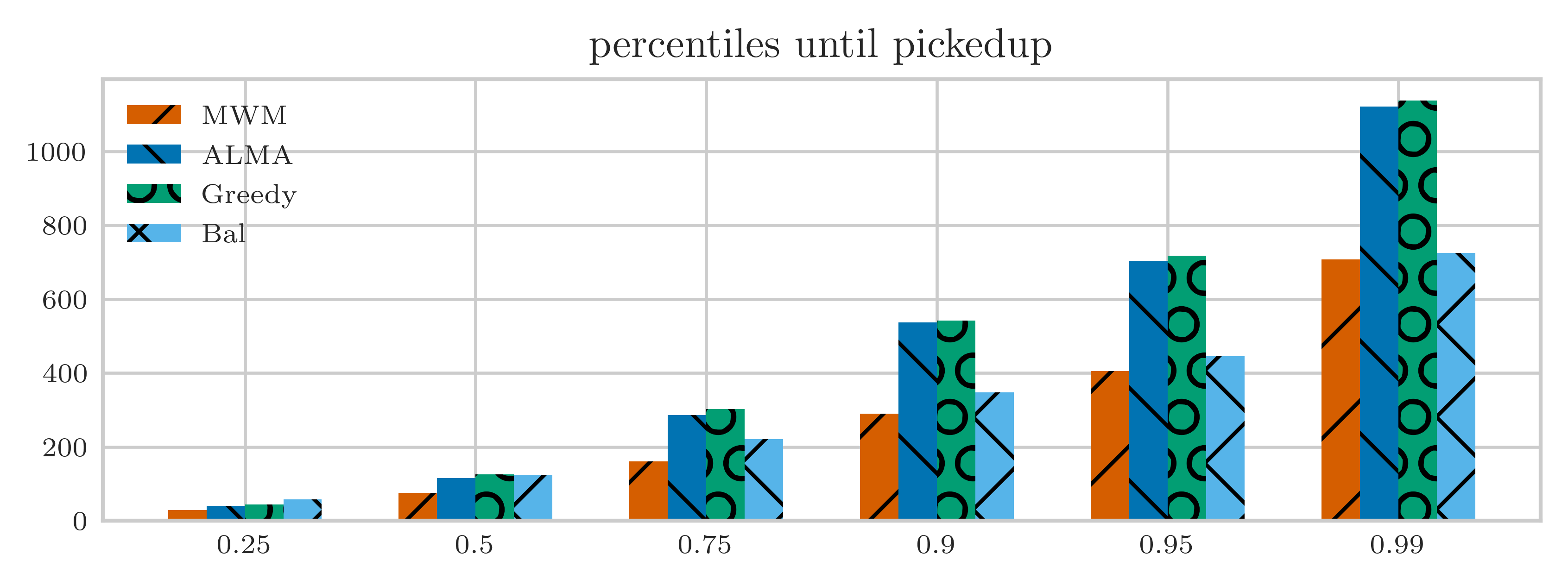}
		\caption{Sequence of Percentiles for Time to Pick-up (s)}
		\label{fig_appendix: Jan15Hour8to9Manhattan_percentiles_until_pickedup}
	\end{subfigure}
	~ 
	\begin{subfigure}[t]{1\textwidth}
		\centering
		\includegraphics[width = 1 \linewidth, trim={0.5em 0.7em 0.6em 1.8em}, clip]{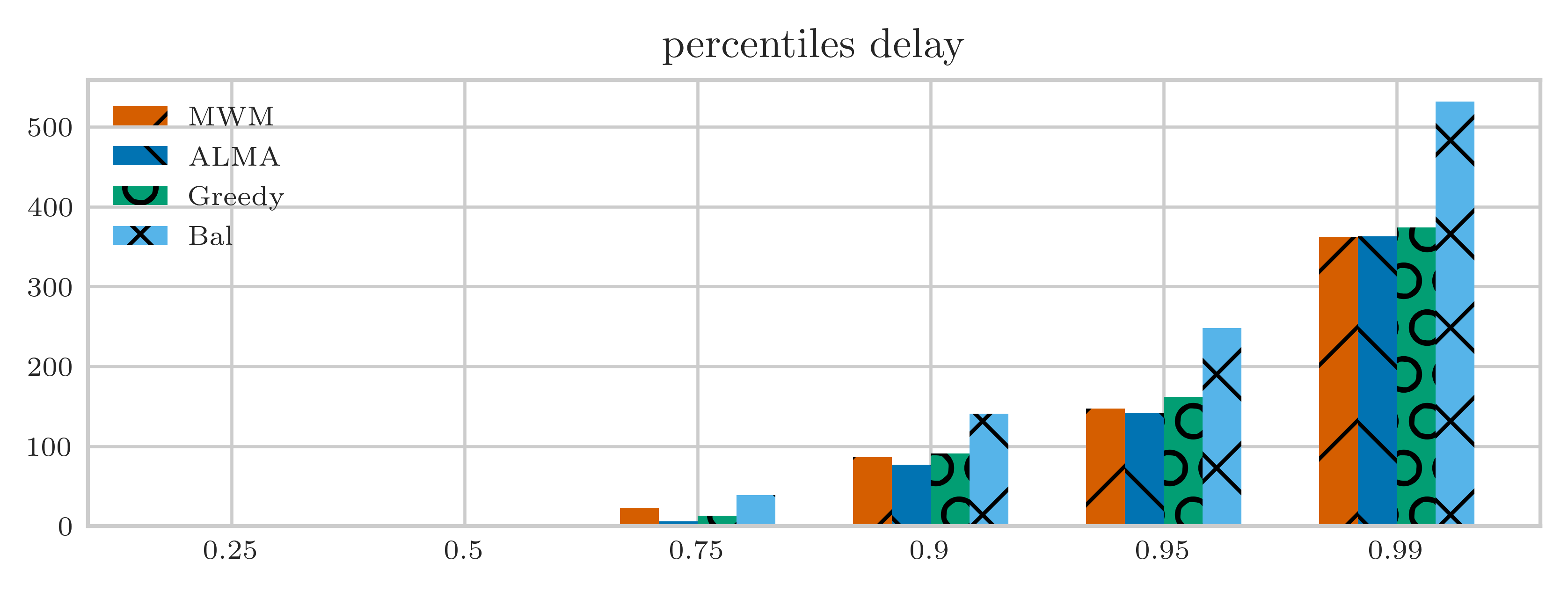}
		\caption{Sequence of Percentiles Delay (s)}
		\label{fig_appendix: Jan15Hour8to9Manhattan_percentiles_delay}
	\end{subfigure}
	~ 
	\begin{subfigure}[t]{1\textwidth}
		\centering
		\includegraphics[width = 1 \linewidth, trim={0.5em 0.7em 0.6em 1.8em}, clip]{./Published/Jan15Hour8to9Manhattan_percentiles_cumulative_delay.png}
		\caption{Sequence of Percentiles for Cumulative Delay (s)}
		\label{fig_appendix: Jan15Hour8to9Manhattan_percentiles_cumulative_delay}
	\end{subfigure}%
	\caption{08:00 - 09:00, \#Taxis = 4276 (base number). Manhattan, January 15, 2016}
	\label{fig_appendix: Jan15Hour8to9Manhattan_Percentiles}
\end{figure*}

\clearpage

\begin{table*}[t!]
\centering
\caption{January 15, 2016 -- 08:00 - 09:00 -- Manhattan -- \#Taxis = 4276 (base number).\\(Given a vector $V$, the $q$-th percentile of $V$ is the value $q/100$ of the way from the minimum to the maximum in a sorted copy of $V$.)}
\vspace{-0.25em}
\label{tb_appendix: Jan15Hour8to9Manhattan_Percentiles}

(a) Sequence of Percentiles for Time to Pair (s).\\
\vspace{0.5em}
\resizebox{0.3\textwidth}{!}{%
\begin{tabular}{lcccccc}
\hline
\textbf{}       & \textbf{0.25} & \textbf{0.5} & \textbf{0.75} & \textbf{0.9} & \textbf{0.95} & \textbf{0.99} \\ \hline
\textbf{MWM}    & 0             & 60           & 60            & 60           & 60            & 60            \\
\textbf{ALMA}   & 0             & 60           & 60            & 60           & 60            & 60            \\
\textbf{Greedy} & 0             & 60           & 60            & 60           & 60            & 60            \\
\textbf{Appr}   & 0             & 60           & 60            & 60           & 60            & 60            \\
\textbf{PG}     & 60            & 60           & 60            & 120          & 120           & 180           \\
\textbf{GD}     & 40            & 52           & 59            & 105          & 118           & 173           \\
\textbf{Bal}    & 0             & 60           & 60            & 60           & 60            & 60            \\
\textbf{Har}    & 0             & 60           & 60            & 60           & 60            & 60            \\
\textbf{DC}     & 0             & 60           & 60            & 60           & 60            & 60            \\
\textbf{k-Taxi} & 0             & 60           & 60            & 60           & 60            & 60            \\
\textbf{Single} & 0             & 0            & 0             & 0            & 0             & 0             \\ \hline
\end{tabular}%
}

\vspace{0.5em}
(b) Sequence of Percentiles for Time to Pair with Taxi (s).\\
\vspace{0.5em}
\resizebox{0.3\textwidth}{!}{%
\begin{tabular}{lcccccc}
\hline
\textbf{}       & \textbf{0.25} & \textbf{0.5} & \textbf{0.75} & \textbf{0.9} & \textbf{0.95} & \textbf{0.99} \\ \hline
\textbf{MWM}    & 0             & 0            & 0             & 0            & 0             & 0             \\
\textbf{ALMA}   & 0             & 0            & 0             & 0            & 0             & 0             \\
\textbf{Greedy} & 0             & 0            & 0             & 0            & 0             & 0             \\
\textbf{Appr}   & 0             & 0            & 0             & 0            & 0             & 0             \\
\textbf{PG}     & 0             & 0            & 0             & 0            & 0             & 0             \\
\textbf{GD}     & 2             & 8            & 20            & 39           & 55            & 84            \\
\textbf{Bal}    & 0             & 0            & 0             & 0            & 0             & 0             \\
\textbf{Har}    & 0             & 0            & 0             & 0            & 0             & 0             \\
\textbf{DC}     & 0             & 0            & 0             & 0            & 0             & 0             \\
\textbf{k-Taxi} & 0             & 0            & 0             & 0            & 0             & 0             \\
\textbf{Single} & 0             & 0            & 0             & 0            & 0             & 0             \\ \hline
\end{tabular}%
}

\vspace{0.5em}
(c) Sequence of Percentiles for Time to Pick-up (s).\\
\vspace{0.5em}
\resizebox{0.3\textwidth}{!}{%
\begin{tabular}{lcccccc}
\hline
\textbf{}       & \textbf{0.25} & \textbf{0.5} & \textbf{0.75} & \textbf{0.9} & \textbf{0.95} & \textbf{0.99} \\ \hline
\textbf{MWM}    & 29            & 76           & 161           & 290          & 406           & 707           \\
\textbf{ALMA}   & 40            & 115          & 286           & 537          & 704           & 1121          \\
\textbf{Greedy} & 44            & 126          & 302           & 542          & 718           & 1138          \\
\textbf{Appr}   & 288           & 454          & 737           & 1148         & 1470          & 2126          \\
\textbf{PG}     & 38            & 101          & 298           & 594          & 803           & 1300          \\
\textbf{GD}     & 56            & 133          & 296           & 539          & 749           & 1311          \\
\textbf{Bal}    & 58            & 124          & 221           & 347          & 445           & 725           \\
\textbf{Har}    & 197           & 406          & 734           & 1185         & 1521          & 2223          \\
\textbf{DC}     & 714           & 2811         & 7017          & 13954        & 21352         & 57870         \\
\textbf{k-Taxi} & 65            & 162          & 348           & 676          & 939           & 1705          \\
\textbf{Single} & 27            & 59           & 137           & 373          & 575           & 910           \\ \hline
\end{tabular}%
}

\vspace{0.5em}
(d) Sequence of Percentiles for Delay (s).\\
\vspace{0.5em}
\resizebox{0.3\textwidth}{!}{%
\begin{tabular}{lcccccc}
\hline
\textbf{}       & \textbf{0.25} & \textbf{0.5} & \textbf{0.75} & \textbf{0.9} & \textbf{0.95} & \textbf{0.99} \\ \hline
\textbf{MWM}    & 0             & 0            & 23            & 86           & 147           & 362           \\
\textbf{ALMA}   & 0             & 0            & 6             & 77           & 142           & 363           \\
\textbf{Greedy} & 0             & 0            & 13            & 91           & 162           & 374           \\
\textbf{Appr}   & 0             & 17           & 100           & 197          & 280           & 568           \\
\textbf{PG}     & 0             & 0            & 0             & 27           & 82            & 279           \\
\textbf{GD}     & 0             & 0            & 146           & 413          & 743           & 1608          \\
\textbf{Bal}    & 0             & 0            & 39            & 141          & 248           & 532           \\
\textbf{Har}    & 0             & 0            & 45            & 157          & 269           & 556           \\
\textbf{DC}     & 0             & 0            & 49            & 166          & 274           & 537           \\
\textbf{k-Taxi} & 0             & 0            & 39            & 142          & 250           & 537           \\
\textbf{Single} & 0             & 0            & 0             & 0            & 0             & 0             \\ \hline
\end{tabular}%
}

\vspace{0.5em}
(e) Sequence of Percentiles for Cumulative Delay (s).\\
\vspace{0.5em}
\resizebox{0.3\textwidth}{!}{%
\begin{tabular}{lcccccc}
\hline
\textbf{}       & \textbf{0.25} & \textbf{0.5} & \textbf{0.75} & \textbf{0.9} & \textbf{0.95} & \textbf{0.99} \\ \hline
\textbf{MWM}    & 82            & 139          & 229           & 369          & 501           & 808           \\
\textbf{ALMA}   & 95            & 179          & 348           & 599          & 772           & 1181          \\
\textbf{Greedy} & 103           & 196          & 368           & 607          & 783           & 1202          \\
\textbf{Appr}   & 372           & 547          & 842           & 1288         & 1626          & 2345          \\
\textbf{PG}     & 99            & 180          & 386           & 690          & 906           & 1398          \\
\textbf{GD}     & 187           & 318          & 538           & 911          & 1249          & 1971          \\
\textbf{Bal}    & 122           & 198          & 310           & 456          & 570           & 910           \\
\textbf{Har}    & 266           & 487          & 826           & 1296         & 1644          & 2364          \\
\textbf{DC}     & 803           & 2892         & 7099          & 14036        & 21446         & 57946         \\
\textbf{k-Taxi} & 133           & 240          & 440           & 775          & 1052          & 1846          \\
\textbf{Single} & 27            & 59           & 137           & 373          & 575           & 910           \\ \hline
\end{tabular}%
}
\end{table*}

\clearpage

\begin{figure*}[t!]
	\centering
	\begin{subfigure}[t]{0.23\textwidth}
		\centering
		\includegraphics[width = 1 \linewidth, trim={1.8em 0.7em 0.5em 0em}, clip]{./Published/Jan15Hour8to9Manhattan_LinePlots_totalDistanceMean.png}
		\caption{Total Distance Driven (m)}
		\label{fig_appendix: Jan15Hour8to9Manhattan_LinePlots_totalDistanceMean}
	\end{subfigure}
	~ 
	\begin{subfigure}[t]{0.23\textwidth}
		\centering
		\includegraphics[width = 1 \linewidth, trim={1.8em 0.7em 0.5em 0em}, clip]{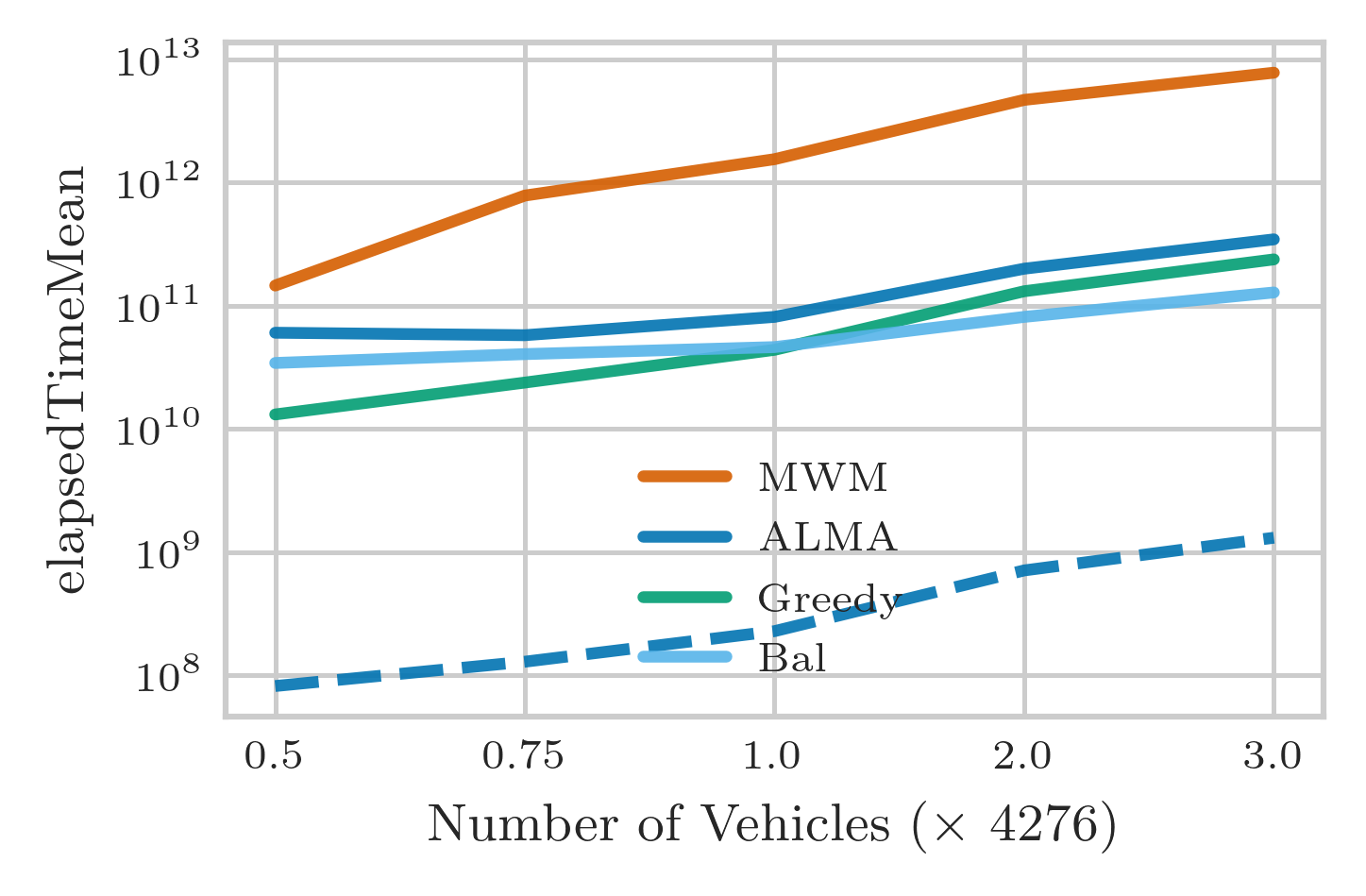}
		\caption{Elapsed Time (ns) [LOG]}
		\label{fig_appendix: Jan15Hour8to9Manhattan_LinePlots_elapsedTimeMean}
	\end{subfigure}
	~
	\begin{subfigure}[t]{0.23\textwidth}
		\centering
		\includegraphics[width = 1 \linewidth, trim={1.8em 0.7em 0.5em 0em}, clip]{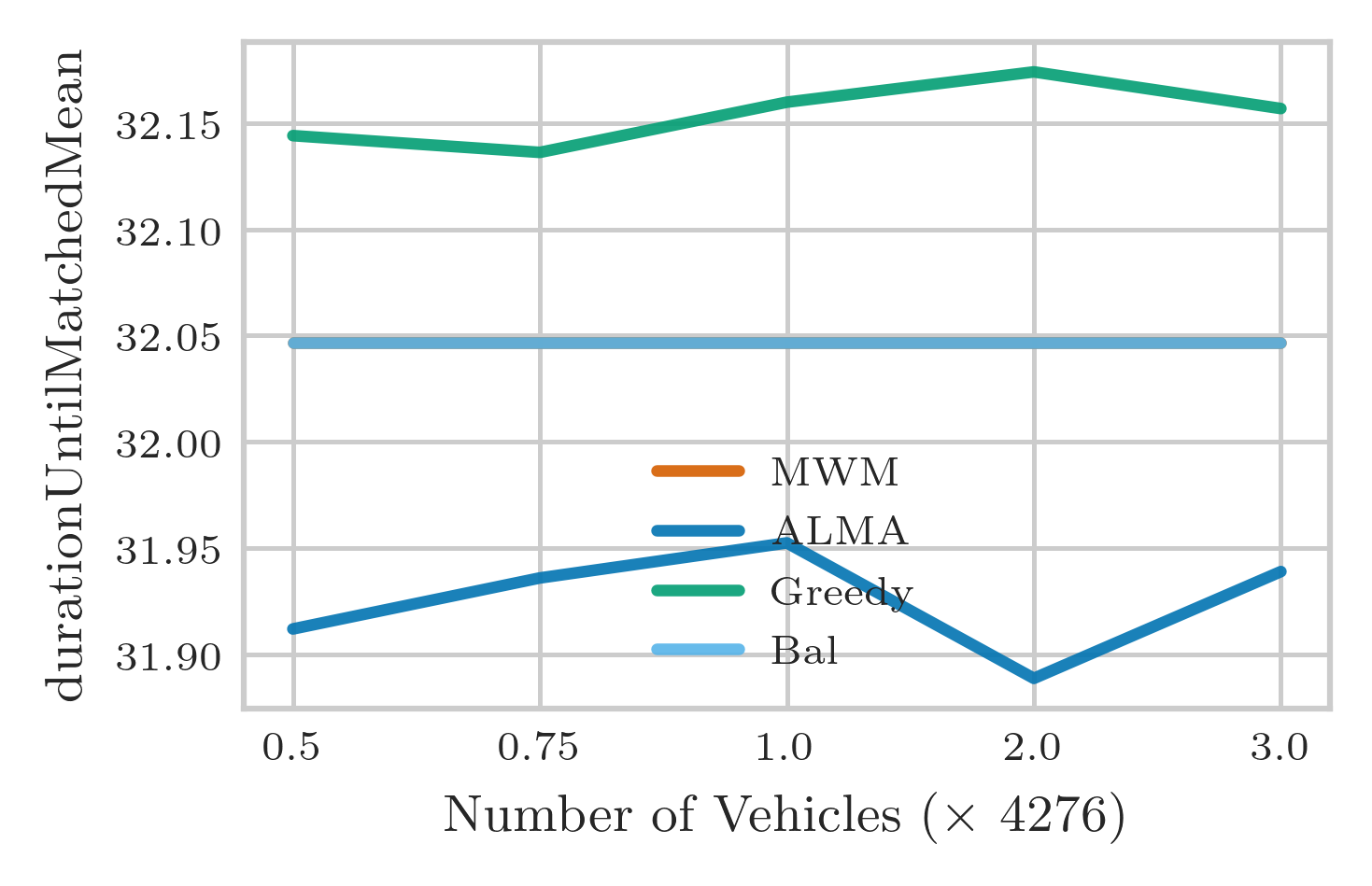}
		\caption{Time to Pair (s)}
		\label{fig_appendix: Jan15Hour8to9Manhattan_LinePlots_durationUntilMatchedMean}
	\end{subfigure}
	~
	\begin{subfigure}[t]{0.23\textwidth}
		\centering
		\includegraphics[width = 1 \linewidth, trim={1.8em 0.7em 0.5em 0em}, clip]{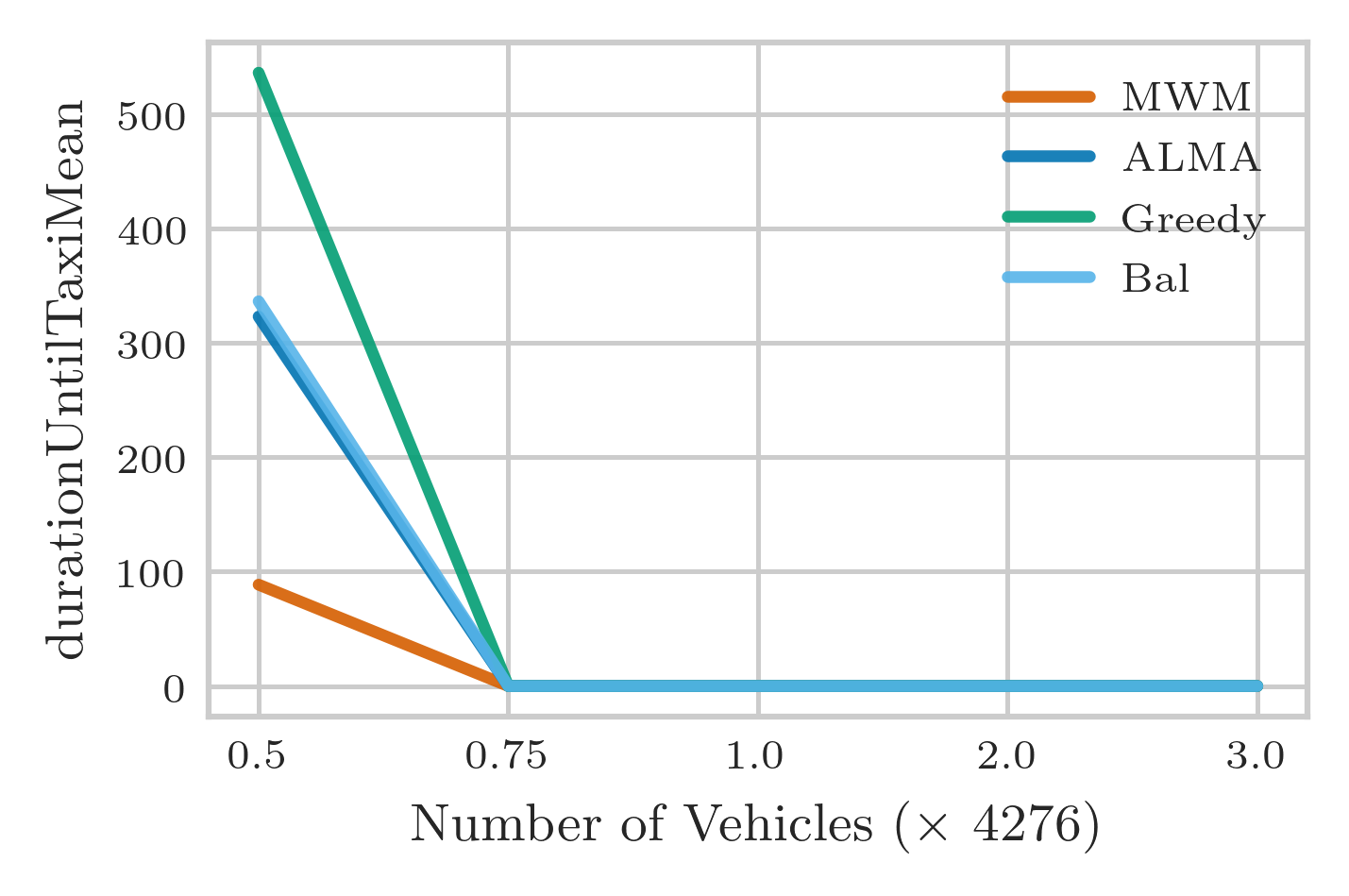}
		\caption{Time to Pair with Taxi (s)}
		\label{fig_appendix: Jan15Hour8to9Manhattan_LinePlots_durationUntilTaxiMean}
	\end{subfigure}

	\begin{subfigure}[t]{0.23\textwidth}
		\centering
		\includegraphics[width = 1 \linewidth, trim={1.8em 0.7em 0.5em 0em}, clip]{./Published/Jan15Hour8to9Manhattan_LinePlots_durationUntilPickedUpMean.png}
		\caption{Time to Pick-up (s)}
		\label{fig_appendix: Jan15Hour8to9Manhattan_LinePlots_durationUntilPickedUpMean}
	\end{subfigure}
	~
	\begin{subfigure}[t]{0.23\textwidth}
		\centering
		\includegraphics[width = 1 \linewidth, trim={1.8em 0.7em 0.5em 0em}, clip]{./Published/Jan15Hour8to9Manhattan_LinePlots_delayMean.png}
		\caption{Delay (s)}
		\label{fig_appendix: Jan15Hour8to9Manhattan_LinePlots_delayMean}
	\end{subfigure}
	~
	\begin{subfigure}[t]{0.23\textwidth}
		\centering
		\includegraphics[width = 1 \linewidth, trim={1.8em 0.7em 0.5em 0em}, clip]{./Published/Jan15Hour8to9Manhattan_LinePlots_cumulativeDelay.png}
		\caption{Cumulative Delay (s)}
		\label{fig_appendix: Jan15Hour8to9Manhattan_LinePlots_cumulativeDelay}
	\end{subfigure}
	~
	\begin{subfigure}[t]{0.23\textwidth}
		\centering
		\includegraphics[width = 1 \linewidth, trim={1.8em 0.7em 0.5em 0em}, clip]{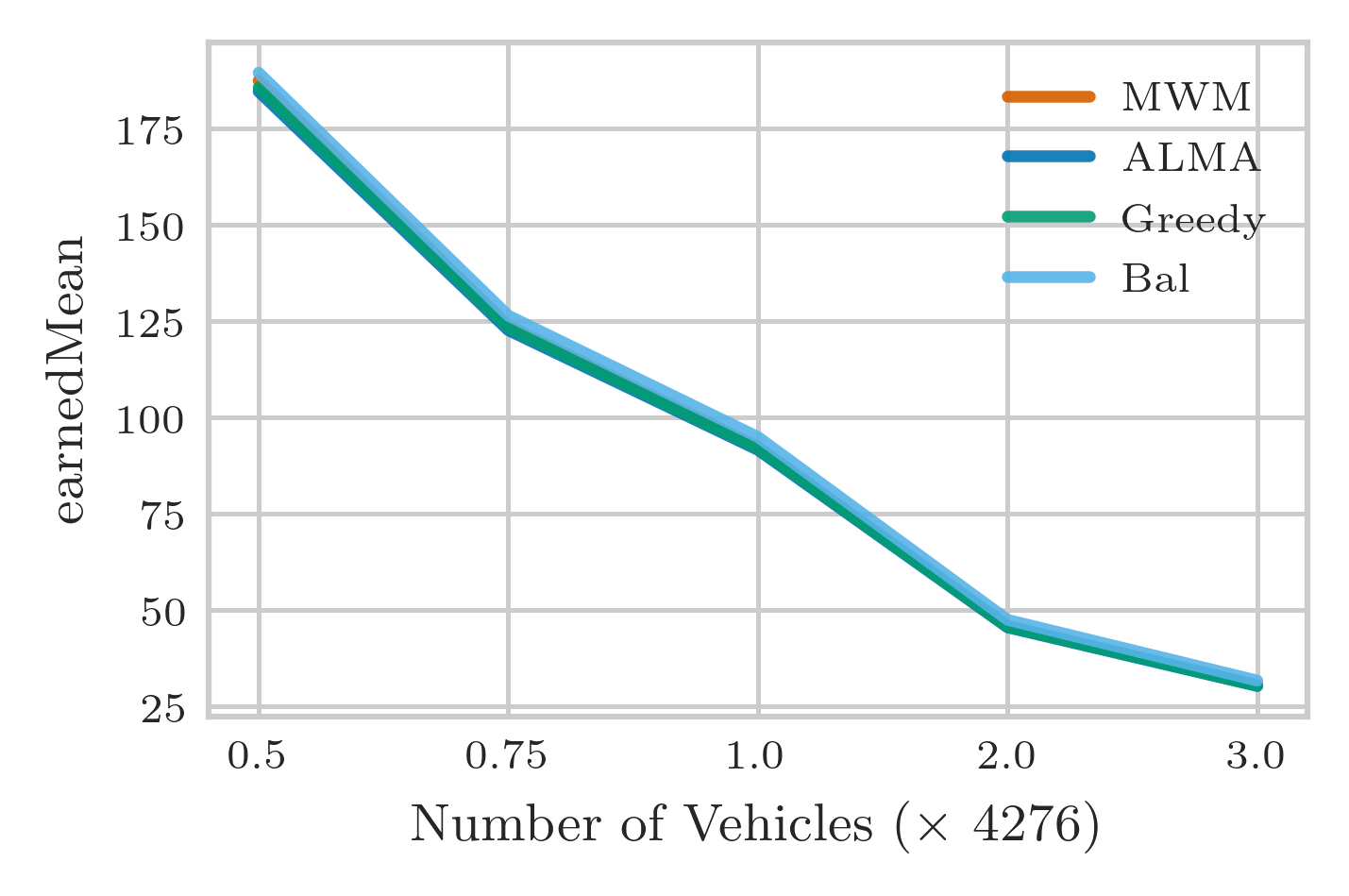}
		\caption{Driver Profit (\$)}
		\label{fig_appendix: Jan15Hour8to9Manhattan_LinePlots_earnedMean}
	\end{subfigure}

	\begin{subfigure}[t]{0.23\textwidth}
		\centering
		\includegraphics[width = 1 \linewidth, trim={1.8em 0.7em 0.5em 0em}, clip]{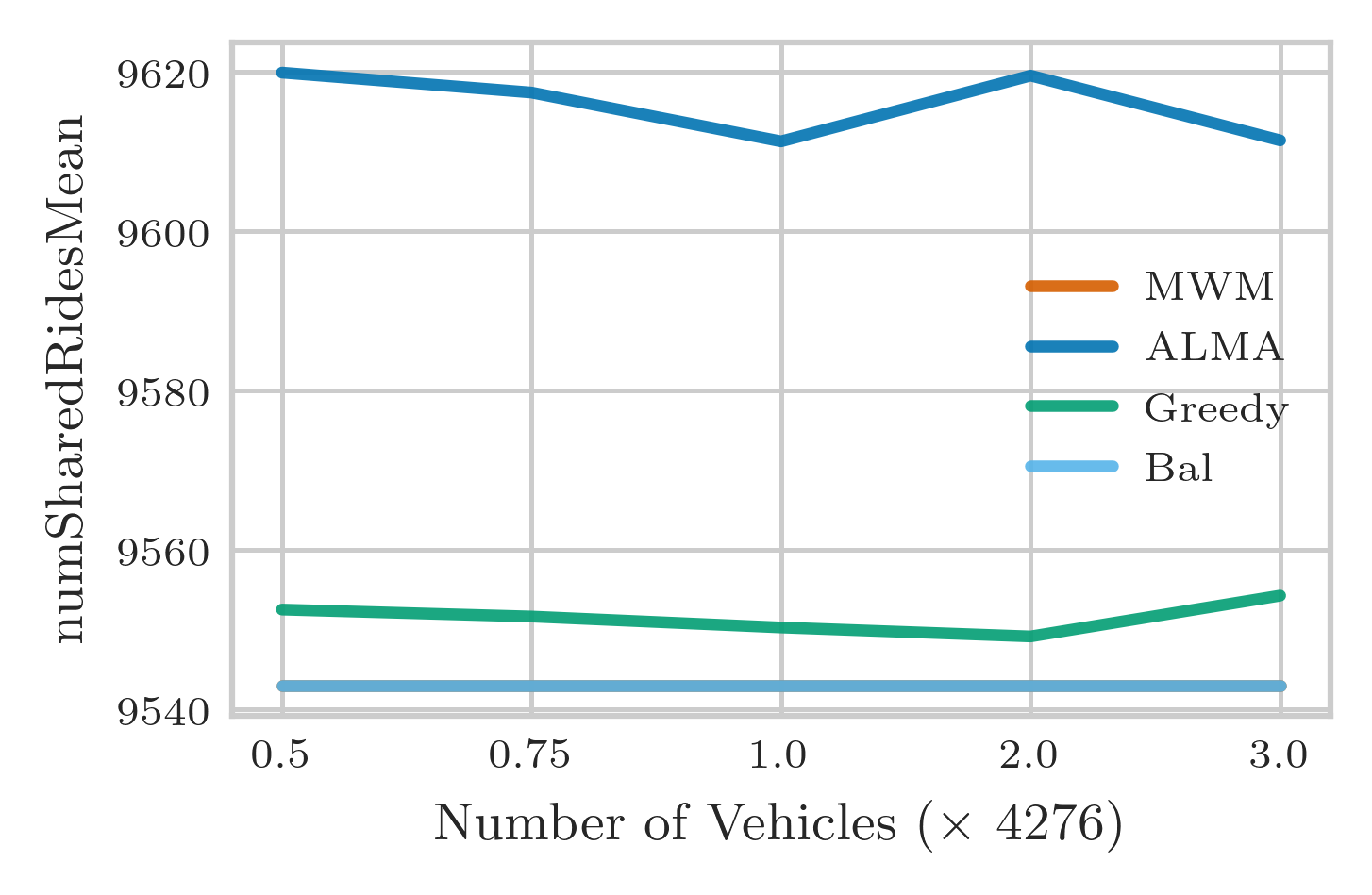}
		\caption{Number of Shared Rides}
		\label{fig_appendix: Jan15Hour8to9Manhattan_LinePlots_numSharedRidesMean}
	\end{subfigure}
	~
	\begin{subfigure}[t]{0.23\textwidth}
		\centering
		\includegraphics[width = 1 \linewidth, trim={1.8em 0.7em 0.5em 0em}, clip]{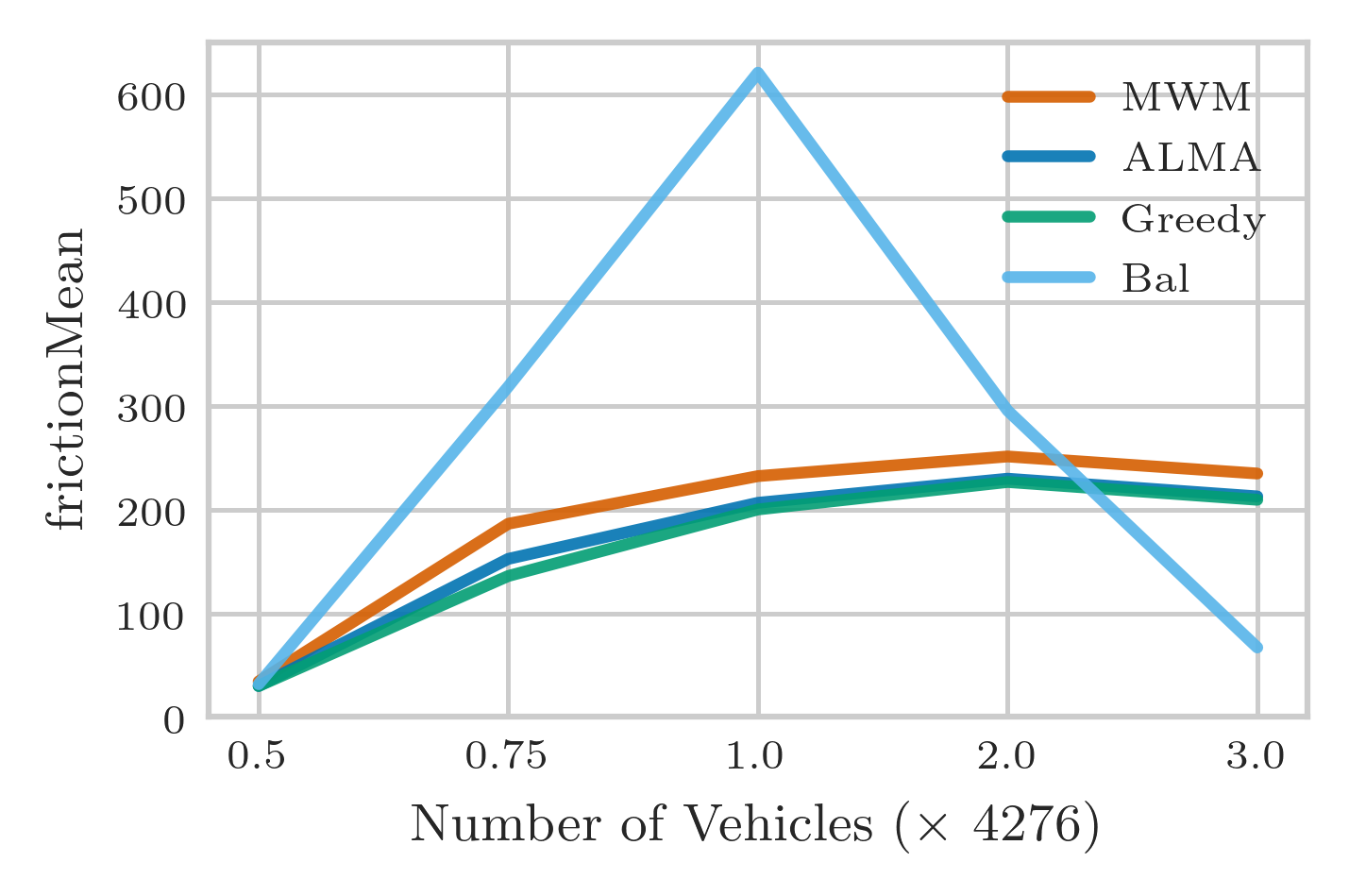}
		\caption{Frictions (s)}
		\label{fig_appendix: Jan15Hour8to9Manhattan_LinePlots_frictionMean}
	\end{subfigure}%
	\caption{January 15, 2016 -- 08:00 - 09:00 -- Manhattan -- Varying \#Taxis = $\{2138, 3207, 4276, 8552, 12828\}$.}
	\label{fig_appendix: Jan15Hour8to9Manhattan_LinePlots}
\end{figure*}

\clearpage


\begin{figure*}[t!]
\subsection{00:00 - 23:59 (full day) -- Manhattan} \label{Appendix Jan15Manhattan}
\end{figure*}

\begin{figure*}[t!]
	\centering
	\begin{subfigure}[t]{0.23\textwidth}
		\centering
		\includegraphics[width = 1 \linewidth, trim={0.6em 0.6em 0.5em 1.7em}, clip]{./Published/Jan15Manhattan_totalDistanceMean.png}
		\caption{Total Distance Driven (m)}
		\label{fig_appendix: Jan15Manhattan_totalDistanceMean}
	\end{subfigure}
	~ 
	\begin{subfigure}[t]{0.23\textwidth}
		\centering
		\includegraphics[width = 1 \linewidth, trim={0.6em 0.6em 0.5em 1.8em}, clip]{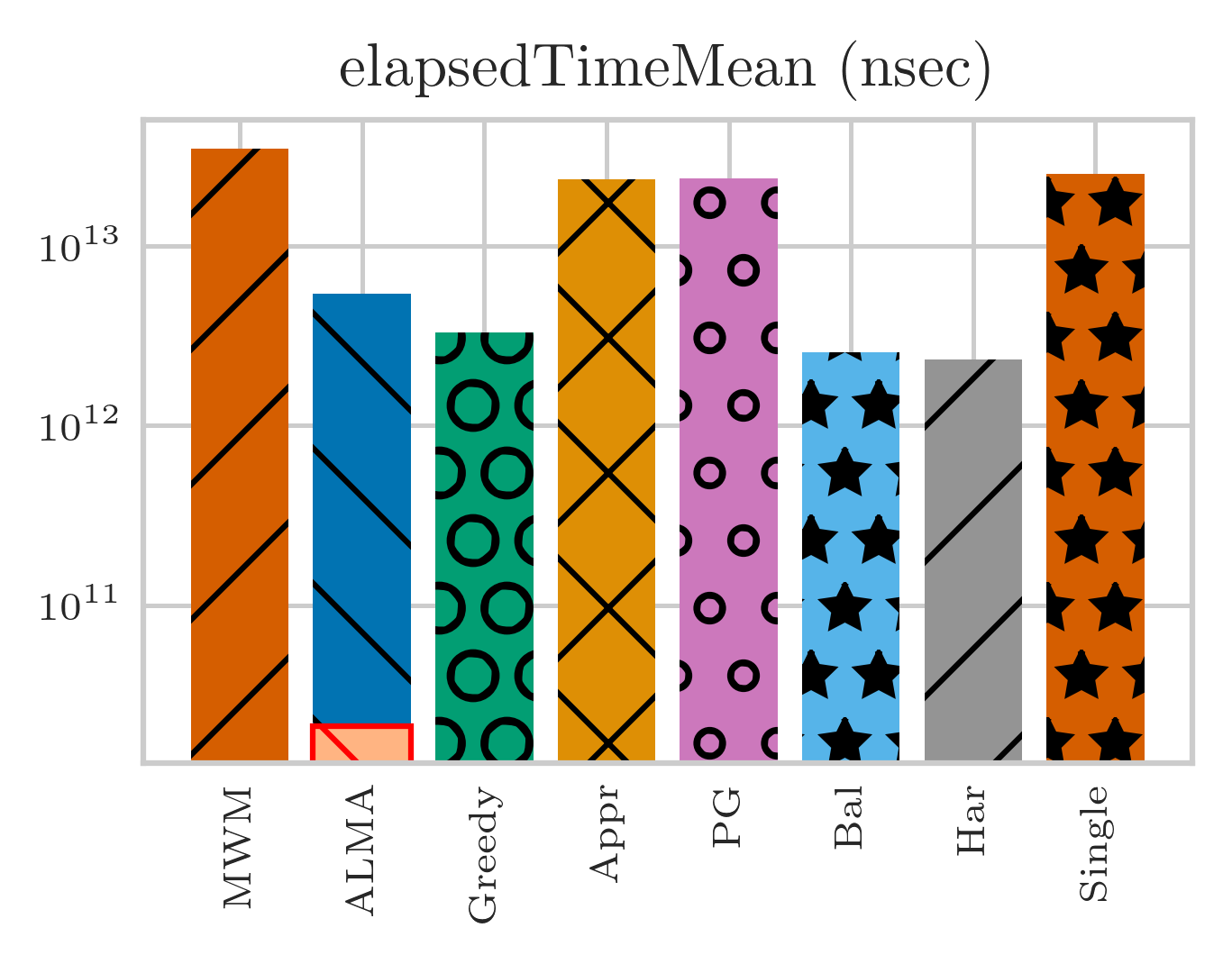}
		\caption{Elapsed Time (ns) [LOG]}
		\label{fig_appendix: Jan15Manhattan_elapsedTimeMean}
	\end{subfigure}
	~
	\begin{subfigure}[t]{0.23\textwidth}
		\centering
		\includegraphics[width = 1 \linewidth, trim={0.6em 0.6em 0.5em 1.8em}, clip]{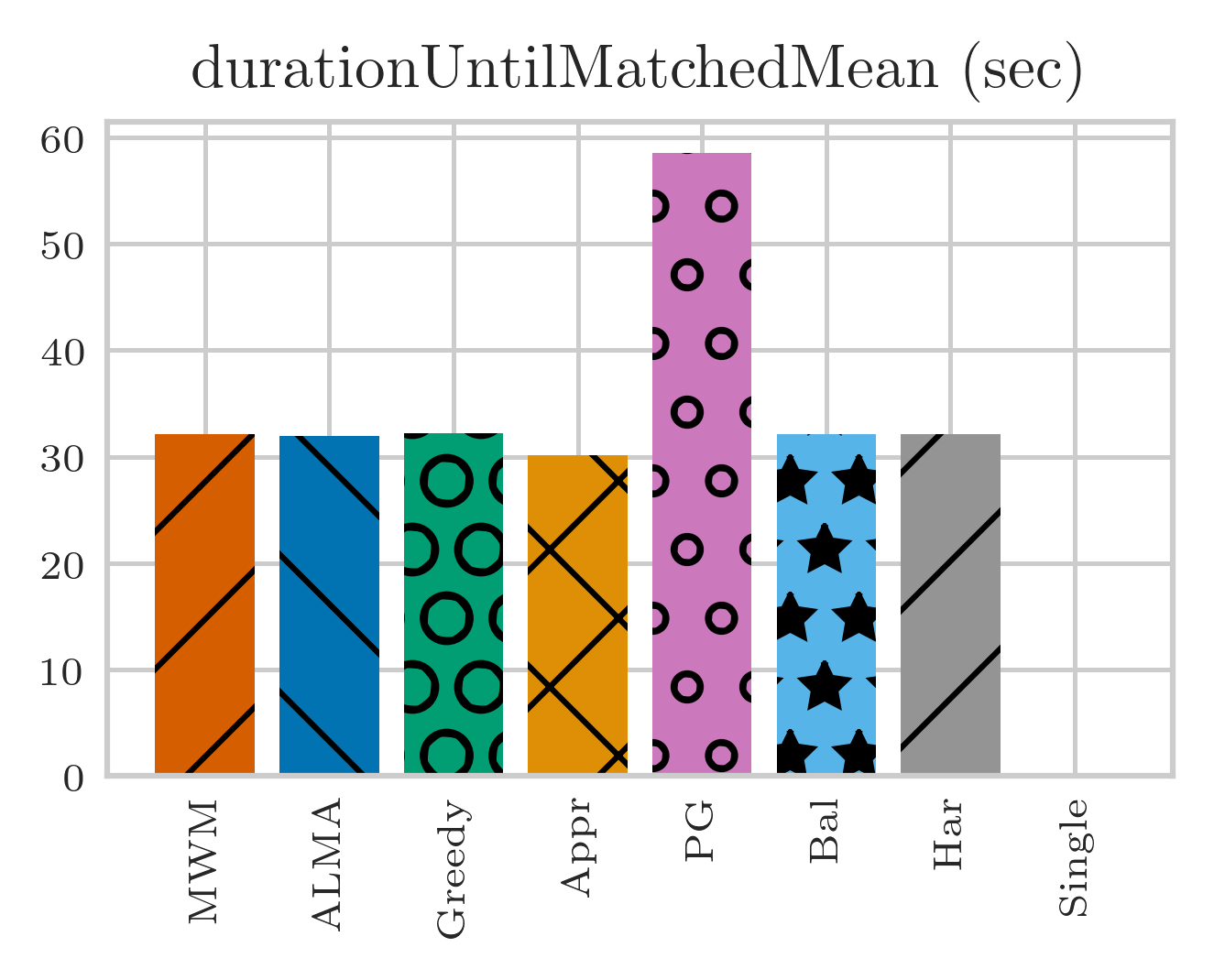}
		\caption{Time to Pair (s)}
		\label{fig_appendix: Jan15Manhattan_durationUntilMatchedMean}
	\end{subfigure}
	~
	\begin{subfigure}[t]{0.23\textwidth}
		\centering
		\includegraphics[width = 1 \linewidth, trim={0.6em 0.6em 0.5em 1.8em}, clip]{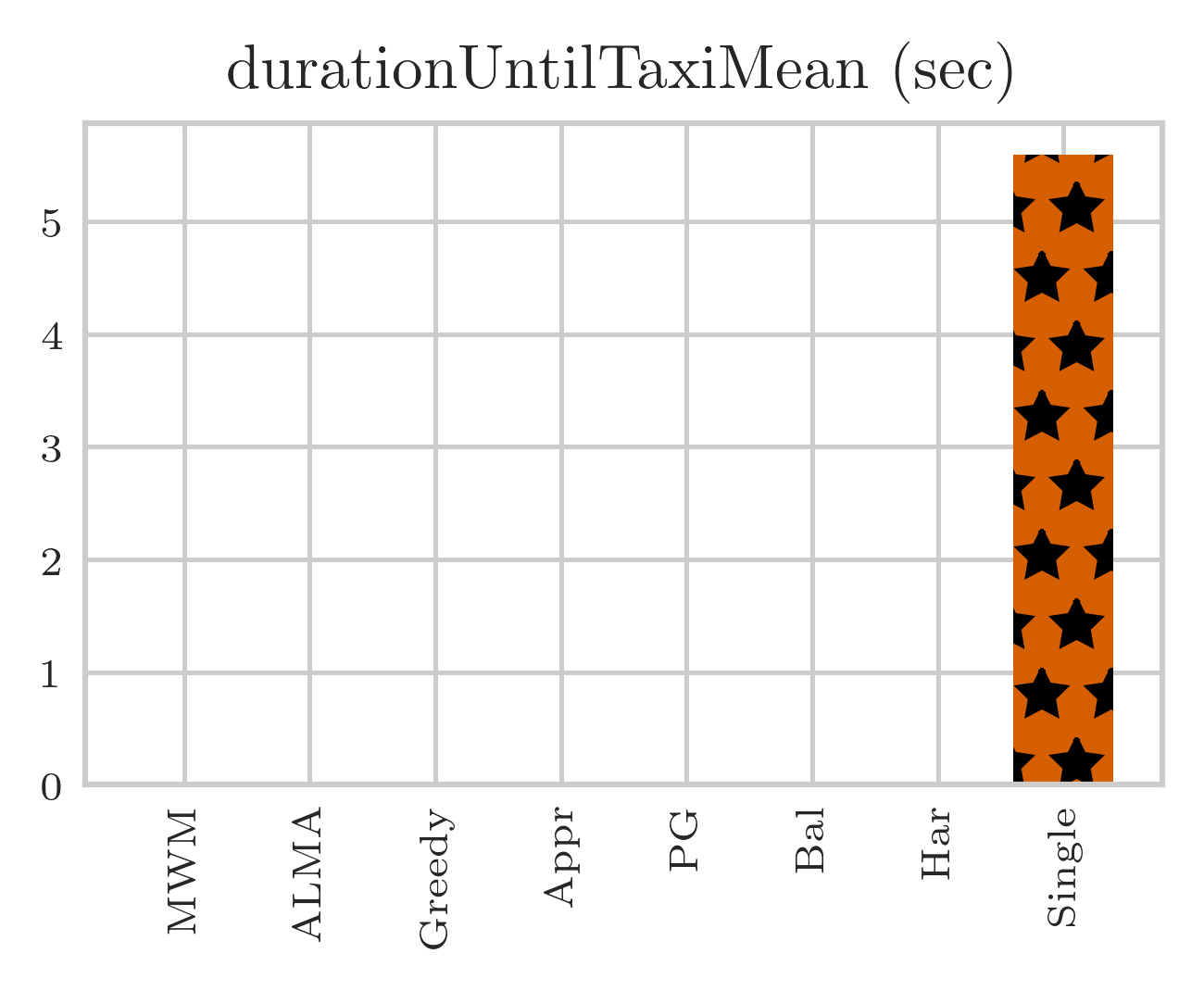}
		\caption{Time to Pair with Taxi (s)}
		\label{fig_appendix: Jan15Manhattan_durationUntilTaxiMean}
	\end{subfigure}

	\begin{subfigure}[t]{0.23\textwidth}
		\centering
		\includegraphics[width = 1 \linewidth, trim={0.6em 0.6em 0.5em 1.8em}, clip]{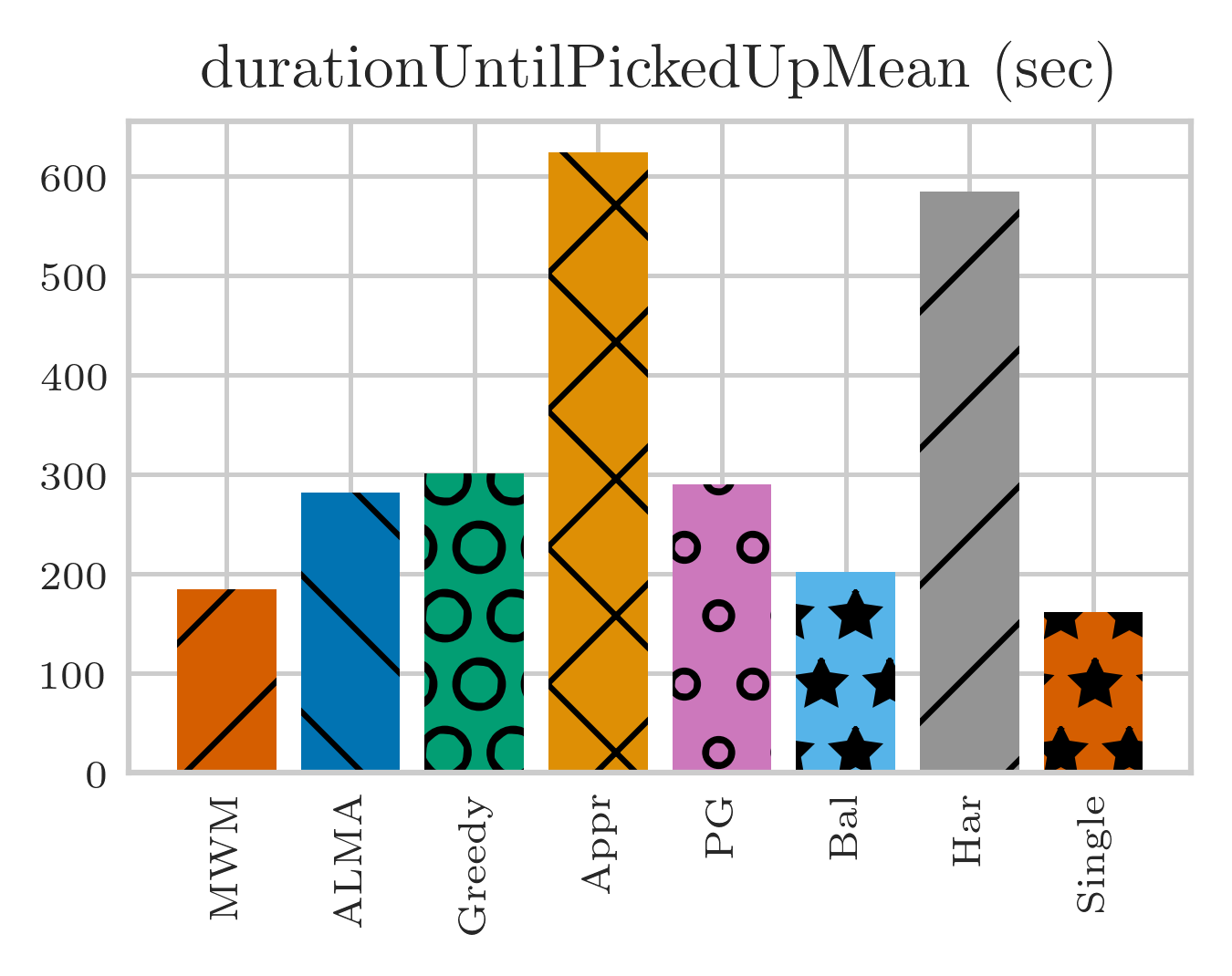}
		\caption{Time to Pick-up (s)}
		\label{fig_appendix: Jan15Manhattan_durationUntilPickedUpMean}
	\end{subfigure}
	~
	\begin{subfigure}[t]{0.23\textwidth}
		\centering
		\includegraphics[width = 1 \linewidth, trim={0.6em 0.6em 0.5em 1.8em}, clip]{./Published/Jan15Manhattan_delayMean.png}
		\caption{Delay (s)}
		\label{fig_appendix: Jan15Manhattan_delayMean}
	\end{subfigure}
	~
	\begin{subfigure}[t]{0.23\textwidth}
		\centering
		\includegraphics[width = 1 \linewidth, trim={0.6em 0.6em 0.5em 1.8em}, clip]{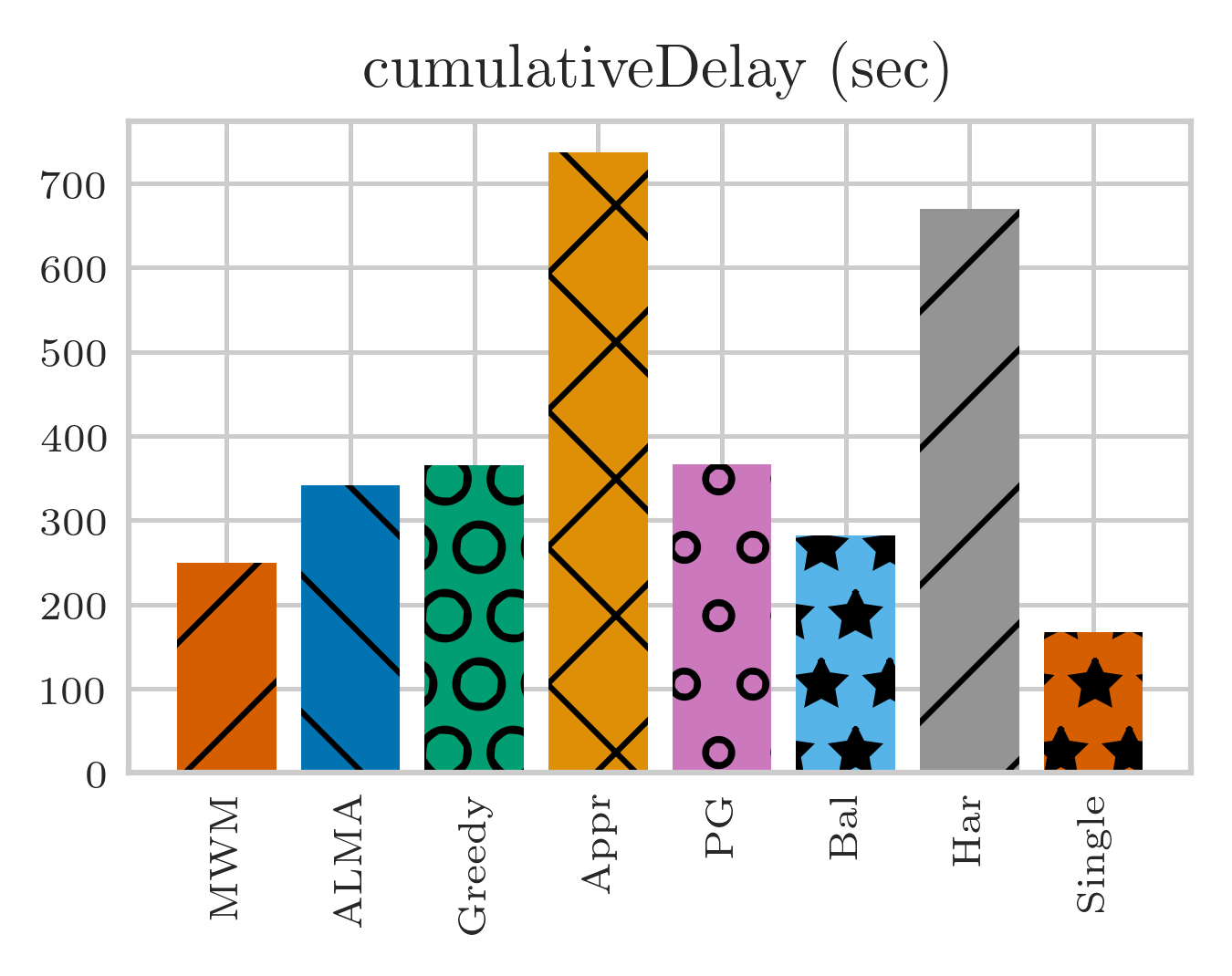}
		\caption{Cumulative Delay (s)}
		\label{fig_appendix: Jan15Manhattan_cumulativeDelay}
	\end{subfigure}
	~
	\begin{subfigure}[t]{0.23\textwidth}
		\centering
		\includegraphics[width = 1 \linewidth, trim={0.6em 0.6em 0.5em 1.8em}, clip]{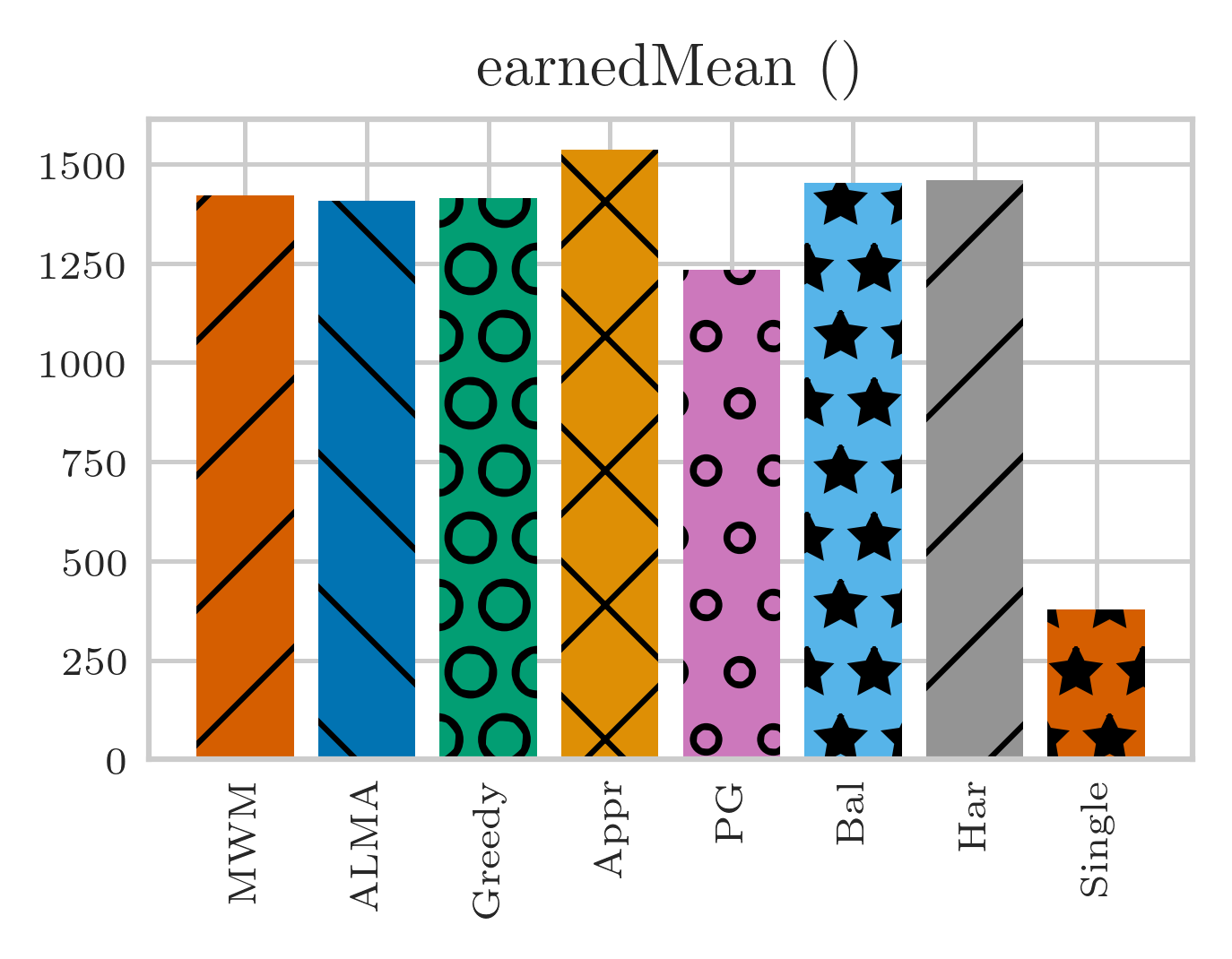}
		\caption{Driver Profit (\$)}
		\label{fig_appendix: Jan15Manhattan_earnedMean}
	\end{subfigure}

	\begin{subfigure}[t]{0.23\textwidth}
		\centering
		\includegraphics[width = 1 \linewidth, trim={0.6em 0.6em 0.5em 1.8em}, clip]{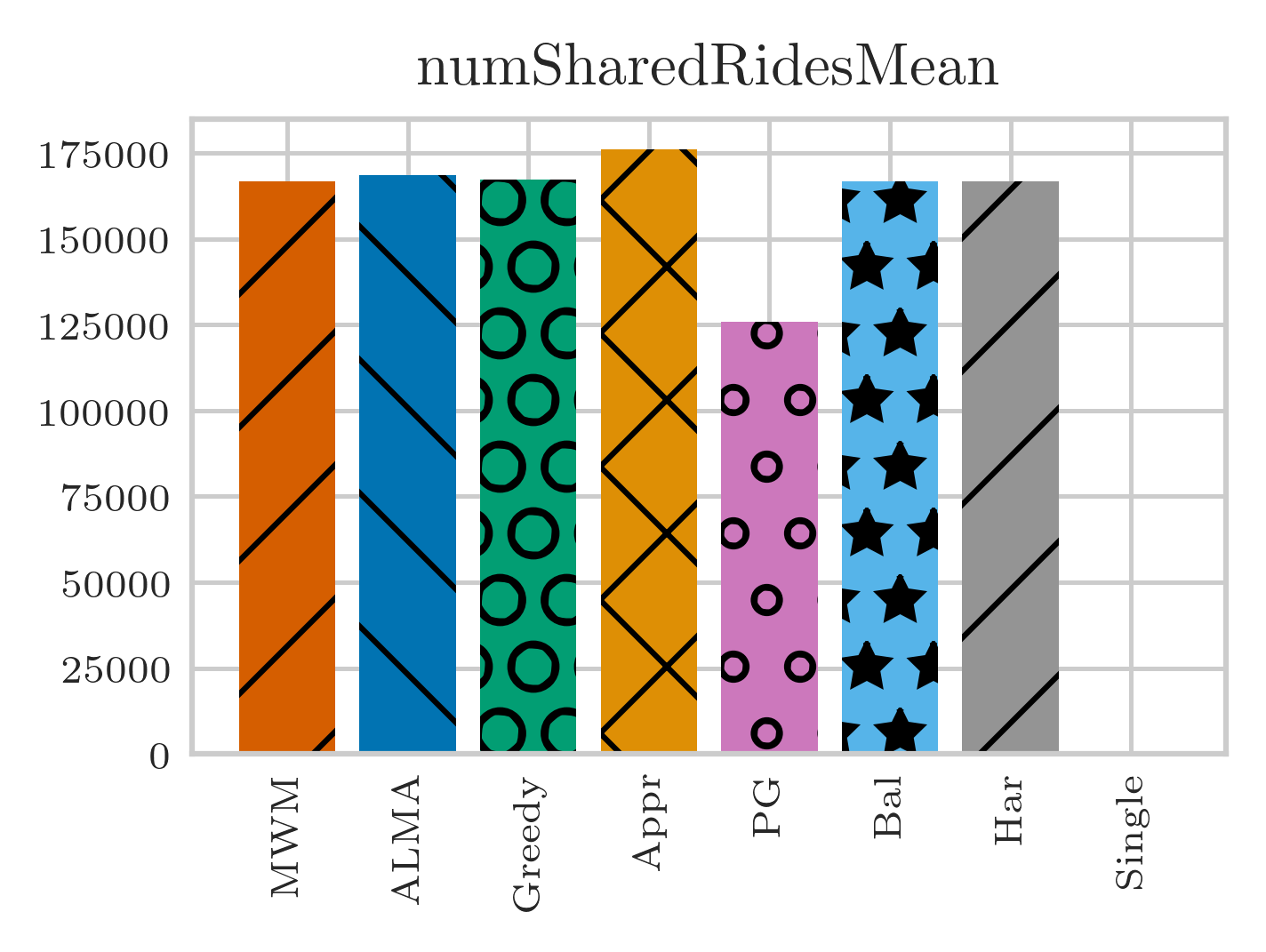}
		\caption{Number of Shared Rides}
		\label{fig_appendix: Jan15Manhattan_numSharedRidesMean}
	\end{subfigure}
	~
	\begin{subfigure}[t]{0.23\textwidth}
		\centering
		\includegraphics[width = 1 \linewidth, trim={0.6em 0.6em 0.5em 1.8em}, clip]{./Published/Jan15Manhattan_frictionMean.png}
		\caption{Frictions (s)}
		\label{fig_appendix: Jan15Manhattan_frictionMean}
	\end{subfigure}%
	\caption{January 15, 2016 -- 00:00 - 23:59 (full day) -- Manhattan -- \#Taxis = 5081 (base number).}
	\label{fig_appendix: Jan15Manhattan}
\end{figure*}

\begin{table*}[b!]
\centering
\caption{January 15, 2016 -- 00:00 - 23:59 (full day) -- Manhattan -- \#Taxis = 5081 (base number).}
\label{tb_appendix: Jan15Manhattan}
\resizebox{\textwidth}{!}{%
\begin{tabular}{@{}lccccccccccccccccccc@{}}
\toprule
\textbf{} & \textbf{\begin{tabular}[c]{@{}c@{}}Distance\\ Driven (m)\end{tabular}} & \textbf{SD} & \textbf{\begin{tabular}[c]{@{}c@{}}Elapsed\\ Time (ns)\end{tabular}} & \textbf{SD} & \textbf{\begin{tabular}[c]{@{}c@{}}Time to\\ Pair (s)\end{tabular}} & \textbf{SD} & \textbf{\begin{tabular}[c]{@{}c@{}}Time to Pair\\ with Taxi (s)\end{tabular}} & \textbf{SD} & \textbf{\begin{tabular}[c]{@{}c@{}}Time to\\ Pick-up (s)\end{tabular}} & \textbf{SD} & \textbf{Delay (s)} & \textbf{SD} & \textbf{\begin{tabular}[c]{@{}c@{}}Cumulative\\ Delay (s)\end{tabular}} & \textbf{\begin{tabular}[c]{@{}c@{}}Driver\\ Profit (\$)\end{tabular}} & \textbf{SD} & \textbf{\begin{tabular}[c]{@{}c@{}}Number of \\ Shared Rides\end{tabular}} & \textbf{SD} & \textbf{Frictions (s)} & \textbf{SD} \\ \midrule
\textbf{MWM}             & 9.45E+08 & 0.00E+00   & 3.48E+13 & 0.00E+00     & 32.10 & 30.84                  & 0.00 & 0.00            & 184.55 & 274.34 & 32.14 & 87.69            & 248.79        & 1420.37 & 895.19                 & 1.67E+05 & 0.00      & 4127.47 & 10597.84 \\
\textbf{ALMA}            & 1.12E+09 & 2.73E+05   & 5.42E+12 & 3.31E+11     & 31.98 & 31.01                  & 0.00 & 0.00            & 281.70 & 405.36 & 27.32 & 86.78            & 341.00        & 1406.70 & 736.46                 & 1.68E+05 & 29.39     & 3047.45 & 7264.56  \\
\textbf{Greedy}          & 1.15E+09 & 5.93E+05   & 3.30E+12 & 3.86E+11     & 32.18 & 31.05                  & 0.00 & 0.00            & 301.04 & 407.70 & 31.93 & 90.56            & 365.15        & 1414.66 & 719.44                 & 1.67E+05 & 26.29     & 3242.70 & 8167.00  \\
\textbf{Appr}            & 1.48E+09 & 0.00E+00   & 2.37E+13 & 0.00E+00     & 30.14 & 30.18                  & 0.00 & 0.00            & 624.13 & 473.31 & 82.75 & 156.13           & 737.02        & 1536.97 & 478.35                 & 1.76E+05 & 0.00      & 2421.75 & 4505.68  \\
\textbf{PG}              & 1.22E+09 & 4.64E+05   & 2.39E+13 & 1.50E+12     & 58.58 & 42.99                  & 0.00 & 0.00            & 290.18 & 431.80 & 17.16 & 84.29            & 365.91        & 1234.17 & 603.66                 & 1.26E+05 & 95.74     & 3044.98 & 8561.98  \\
\textbf{Bal}             & 9.85E+08 & 2.42E+05   & 2.56E+12 & 1.39E+11     & 32.10 & 30.84                  & 0.00 & 0.00            & 201.85 & 225.85 & 47.51 & 126.16           & 281.47        & 1452.66 & 107.31                 & 1.67E+05 & 0.00      & 1516.33 & 221.62   \\
\textbf{Har}             & 1.43E+09 & 1.46E+06   & 2.34E+12 & 3.85E+11     & 32.10 & 30.84                  & 0.00 & 0.00            & 584.05 & 533.66 & 53.08 & 132.72           & 669.23        & 1458.51 & 187.05                 & 1.67E+05 & 0.00      & 1106.16 & 246.76   \\
\textbf{Single}          & 1.60E+09 & 0.00E+00   & 2.54E+13 & 0.00E+00     & 0.00  & 0.00                   & 5.59 & 99.05           & 161.77 & 355.37 & 0.00  & 0.00             & 167.36        & 376.90  & 116.72                 & 0.00E+00 & 0.00      & 756.43  & 1216.88  \\ \bottomrule
\end{tabular}%
}
\end{table*}

\begin{table*}[b!]
\centering
\caption{January 15, 2016 -- 00:00 - 23:59 (full day) -- Manhattan -- \#Taxis = 5081 (base number). Each column presents the relative difference compared to the first line, i.e., the MWM (algorithm - MWM) / MWM, for each metric.}
\label{tb_appendix: Jan15ManhattanPercentages}
\resizebox{\textwidth}{!}{%
\begin{tabular}{@{}lccccccccccccccccccc@{}}
\toprule
\textbf{} & \textbf{\begin{tabular}[c]{@{}c@{}}Distance\\ Driven (m)\end{tabular}} & \textbf{SD} & \textbf{\begin{tabular}[c]{@{}c@{}}Elapsed\\ Time (ns)\end{tabular}} & \textbf{SD} & \textbf{\begin{tabular}[c]{@{}c@{}}Time to\\ Pair (s)\end{tabular}} & \textbf{SD} & \textbf{\begin{tabular}[c]{@{}c@{}}Time to Pair\\ with Taxi (s)\end{tabular}} & \textbf{SD} & \textbf{\begin{tabular}[c]{@{}c@{}}Time to\\ Pick-up (s)\end{tabular}} & \textbf{SD} & \textbf{Delay (s)} & \textbf{SD} & \textbf{\begin{tabular}[c]{@{}c@{}}Cumulative\\ Delay (s)\end{tabular}} & \textbf{\begin{tabular}[c]{@{}c@{}}Driver\\ Profit (\$)\end{tabular}} & \textbf{SD} & \textbf{\begin{tabular}[c]{@{}c@{}}Number of \\ Shared Rides\end{tabular}} & \textbf{SD} & \textbf{Frictions (s)} & \textbf{SD} \\ \midrule
\textbf{MWM}             & 0.00\%  & --         & 0.00\%   & --           & 0.00\%    & 0.00\%                 & -- & --              & 0.00\%   & 0.00\%   & 0.00\%    & 0.00\%           & 0.00\%        & 0.00\%   & 0.00\%                 & 0.00\%    & --        & 0.00\%   & 0.00\%   \\
\textbf{ALMA}            & 18.29\% & --         & -84.43\% & --           & -0.39\%   & 0.56\%                 & -- & --              & 52.64\%  & 47.76\%  & -15.00\%  & -1.04\%          & 37.06\%       & -0.96\%  & -17.73\%               & 1.08\%    & --        & -26.17\% & -31.45\% \\
\textbf{Greedy}          & 21.92\% & --         & -90.52\% & --           & 0.23\%    & 0.68\%                 & -- & --              & 63.12\%  & 48.61\%  & -0.65\%   & 3.27\%           & 46.77\%       & -0.40\%  & -19.63\%               & 0.41\%    & --        & -21.44\% & -22.94\% \\
\textbf{Appr}            & 57.08\% & --         & -32.07\% & --           & -6.12\%   & -2.16\%                & -- & --              & 238.19\% & 72.53\%  & 157.50\%  & 78.05\%          & 196.24\%      & 8.21\%   & -46.56\%               & 5.69\%    & --        & -41.33\% & -57.48\% \\
\textbf{PG}              & 29.57\% & --         & -31.48\% & --           & 82.46\%   & 39.38\%                & -- & --              & 57.23\%  & 57.40\%  & -46.61\%  & -3.87\%          & 47.08\%       & -13.11\% & -32.57\%               & -24.49\%  & --        & -26.23\% & -19.21\% \\
\textbf{Bal}             & 4.24\%  & --         & -92.64\% & --           & 0.00\%    & 0.00\%                 & -- & --              & 9.37\%   & -17.67\% & 47.85\%   & 43.87\%          & 13.13\%       & 2.27\%   & -88.01\%               & 0.00\%    & --        & -63.26\% & -97.91\% \\
\textbf{Har}             & 51.67\% & --         & -93.28\% & --           & 0.00\%    & 0.00\%                 & -- & --              & 216.47\% & 94.53\%  & 65.19\%   & 51.36\%          & 168.99\%      & 2.68\%   & -79.10\%               & 0.00\%    & --        & -73.20\% & -97.67\% \\
\textbf{Single}          & 69.00\% & --         & -27.23\% & --           & -100.00\% & -100.00\%              & -- & --              & -12.35\% & 29.54\%  & -100.00\% & -100.00\%        & -32.73\%      & -73.46\% & -86.96\%               & -100.00\% & --        & -81.67\% & -88.52\% \\ \bottomrule
\end{tabular}%
}
\end{table*}

\clearpage


\begin{figure*}[t!]
\subsection{08:00 - 09:00 -- Broader NYC Area (Manhattan, Bronx, Staten Island, Brooklyn, Queens)} \label{Appendix Jan15Hour8to9}
\end{figure*}

\begin{figure*}[t!]
	\centering
	\begin{subfigure}[t]{0.23\textwidth}
		\centering
		\includegraphics[width = 1 \linewidth, trim={0.6em 0.6em 0.5em 1.8em}, clip]{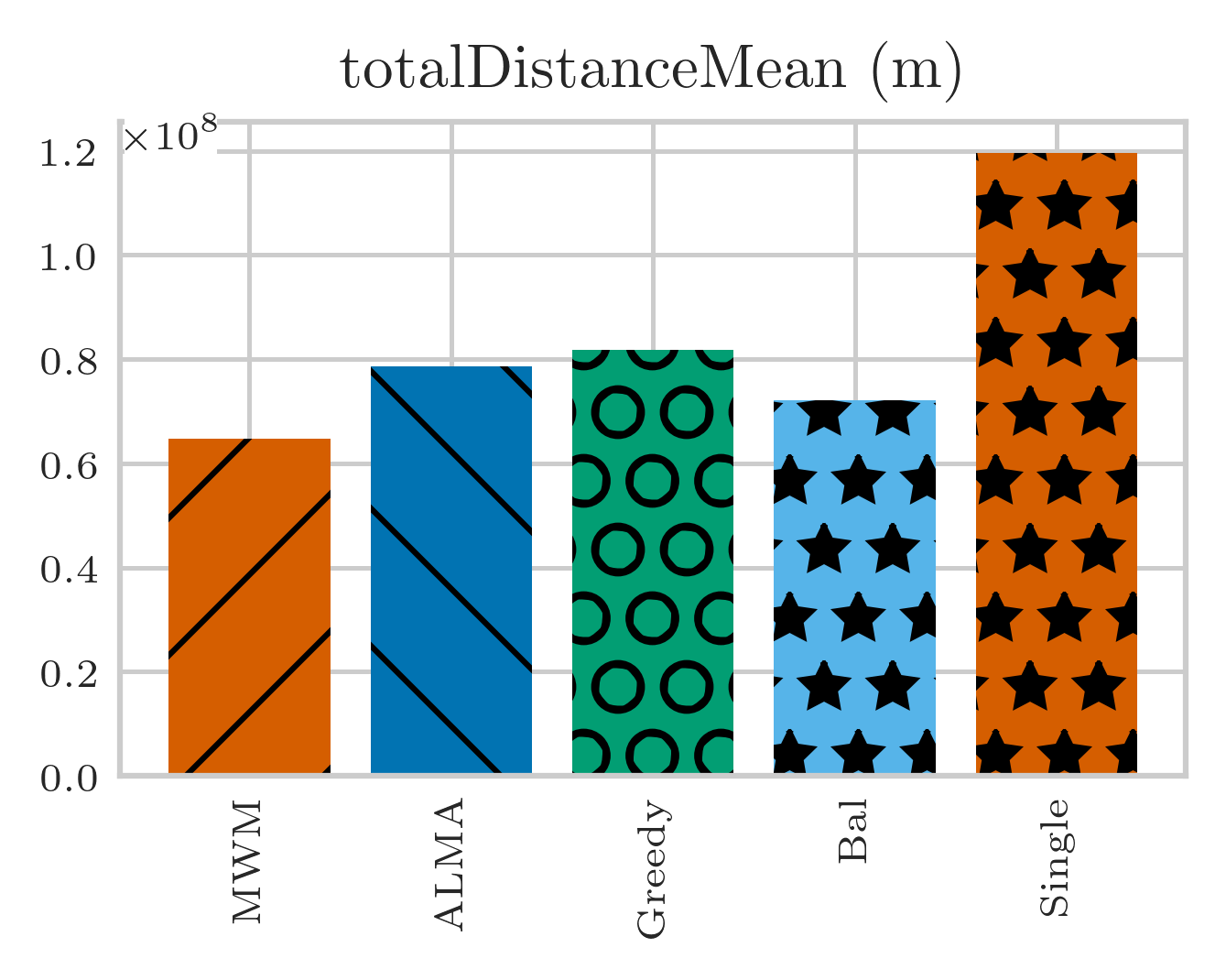}
		\caption{Total Distance Driven (m)}
		\label{fig_appendix: Jan15Hour8to9_totalDistanceMean}
	\end{subfigure}
	~ 
	\begin{subfigure}[t]{0.23\textwidth}
		\centering
		\includegraphics[width = 1 \linewidth, trim={0.6em 0.6em 0.5em 1.8em}, clip]{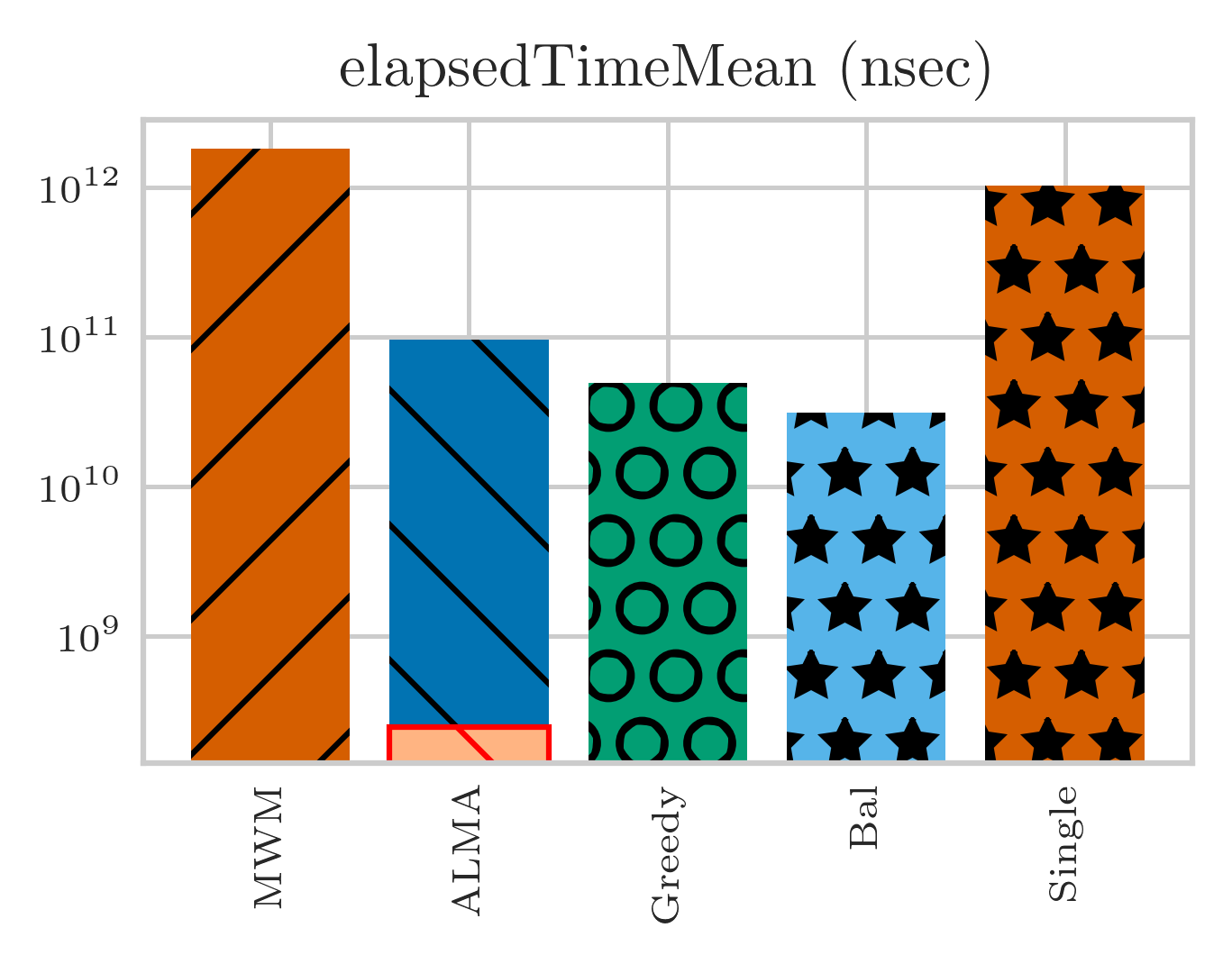}
		\caption{Elapsed Time (ns) [LOG]}
		\label{fig_appendix: Jan15Hour8to9_elapsedTimeMean}
	\end{subfigure}
	~
	\begin{subfigure}[t]{0.23\textwidth}
		\centering
		\includegraphics[width = 1 \linewidth, trim={0.6em 0.6em 0.5em 1.8em}, clip]{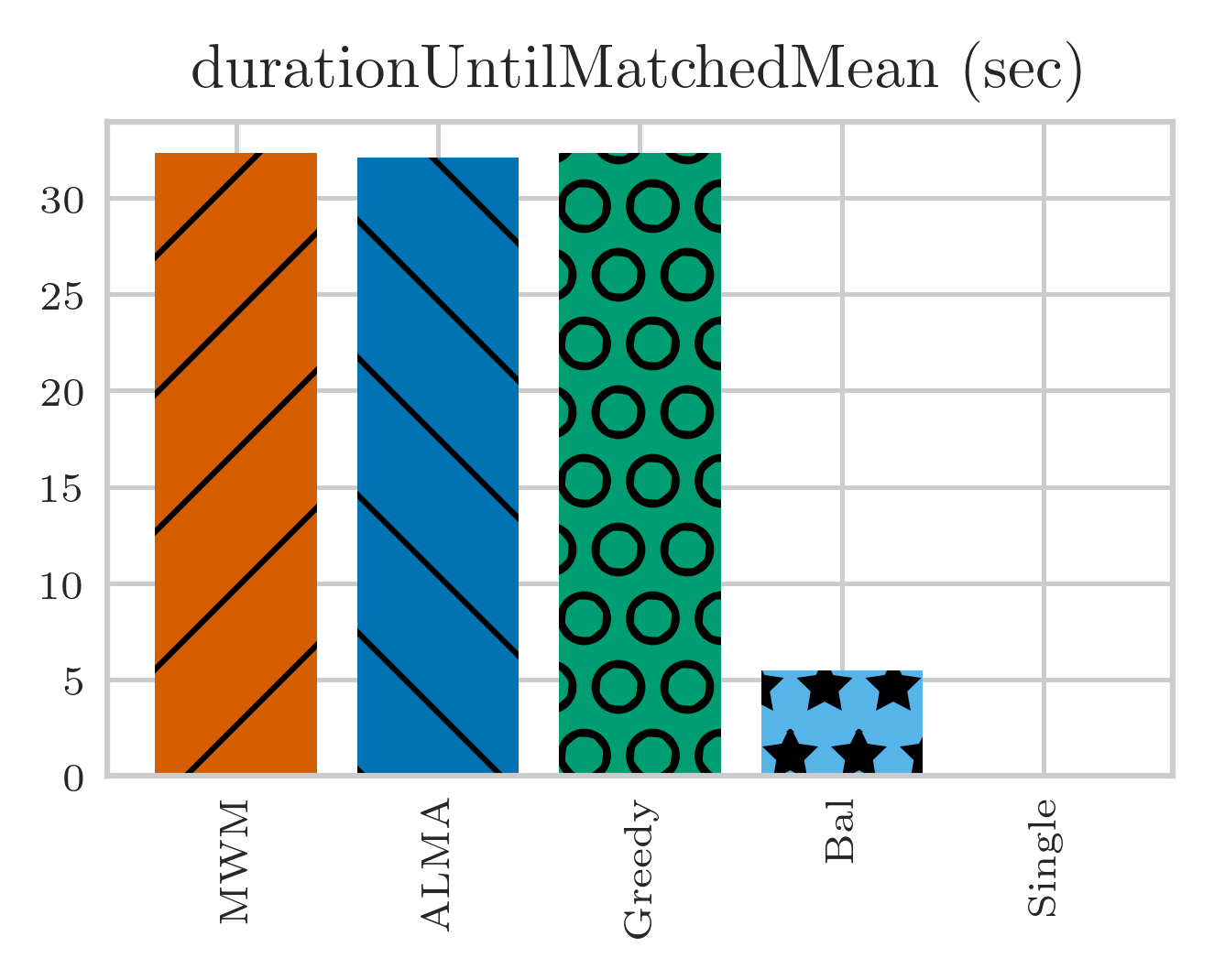}
		\caption{Time to Pair (s)}
		\label{fig_appendix: Jan15Hour8to9_durationUntilMatchedMean}
	\end{subfigure}
	~
	\begin{subfigure}[t]{0.23\textwidth}
		\centering
		\includegraphics[width = 1 \linewidth, trim={0.6em 0.6em 0.5em 1.8em}, clip]{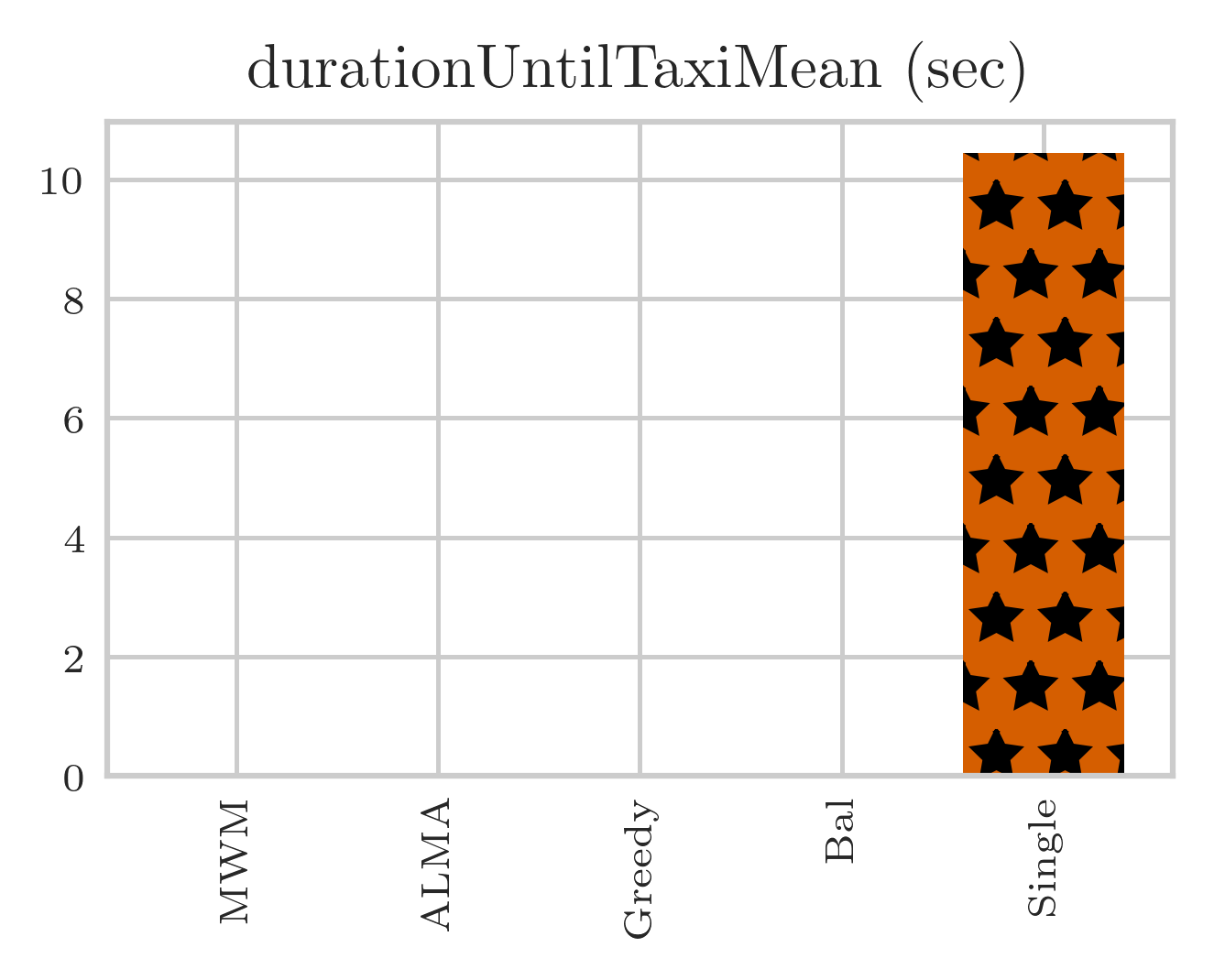}
		\caption{Time to Pair with Taxi (s)}
		\label{fig_appendix: Jan15Hour8to9_durationUntilTaxiMean}
	\end{subfigure}

	\begin{subfigure}[t]{0.23\textwidth}
		\centering
		\includegraphics[width = 1 \linewidth, trim={0.6em 0.6em 0.5em 1.8em}, clip]{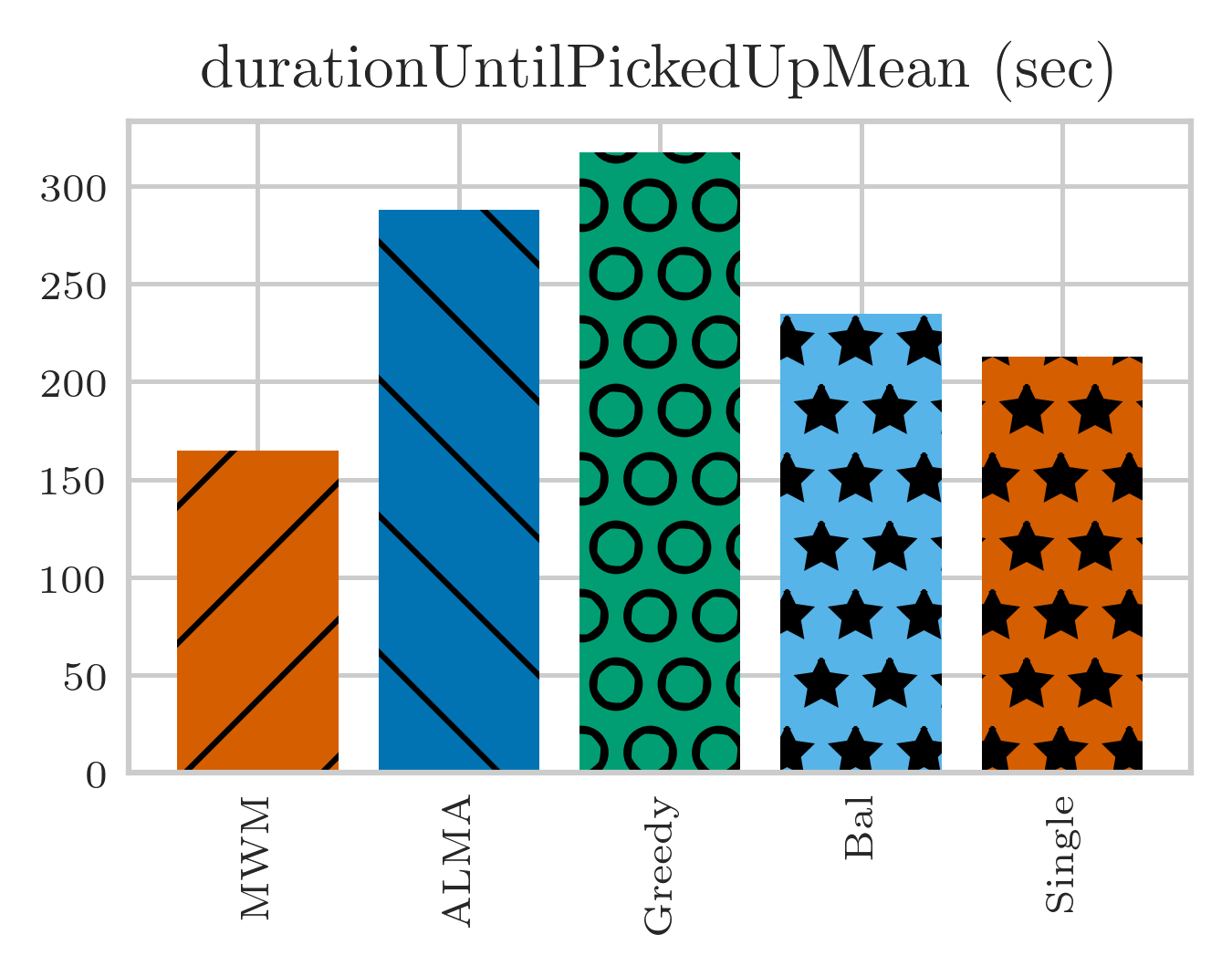}
		\caption{Time to Pick-up (s)}
		\label{fig_appendix: Jan15Hour8to9_durationUntilPickedUpMean}
	\end{subfigure}
	~
	\begin{subfigure}[t]{0.23\textwidth}
		\centering
		\includegraphics[width = 1 \linewidth, trim={0.6em 0.6em 0.5em 1.8em}, clip]{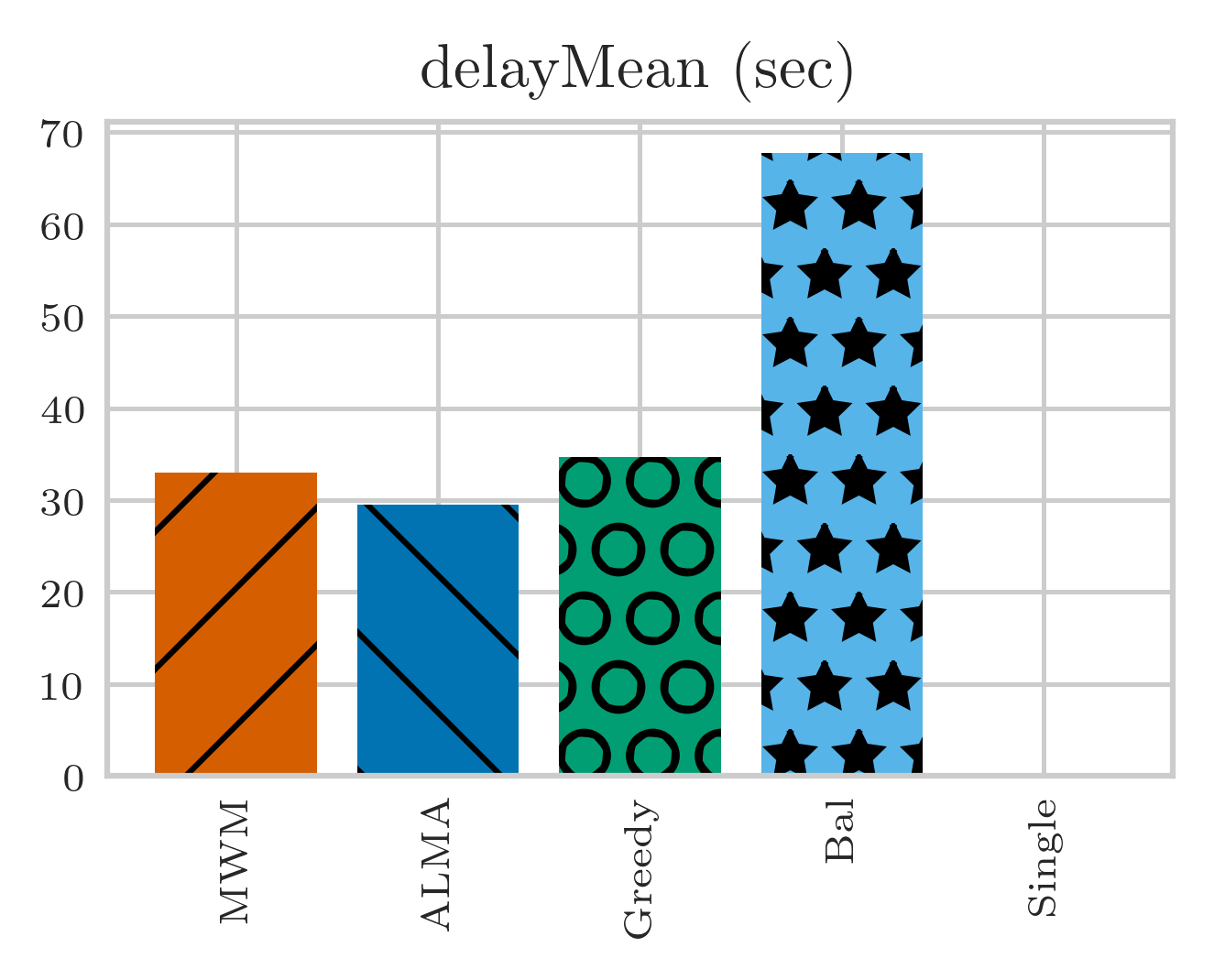}
		\caption{Delay (s)}
		\label{fig_appendix: Jan15Hour8to9_delayMean}
	\end{subfigure}
	~
	\begin{subfigure}[t]{0.23\textwidth}
		\centering
		\includegraphics[width = 1 \linewidth, trim={0.6em 0.6em 0.5em 1.8em}, clip]{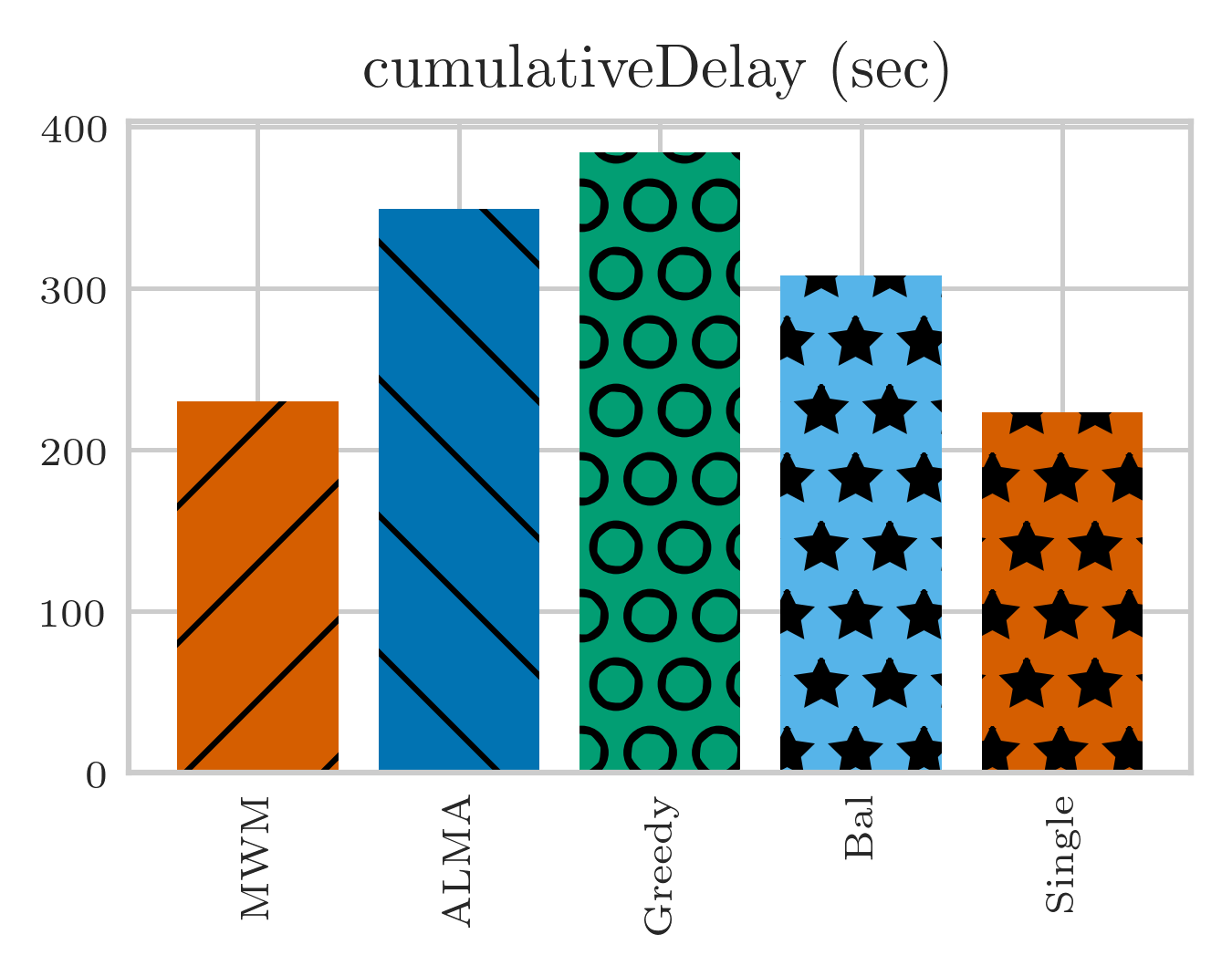}
		\caption{Cumulative Delay (s)}
		\label{fig_appendix: Jan15Hour8to9_cumulativeDelay}
	\end{subfigure}
	~
	\begin{subfigure}[t]{0.23\textwidth}
		\centering
		\includegraphics[width = 1 \linewidth, trim={0.6em 0.6em 0.5em 1.8em}, clip]{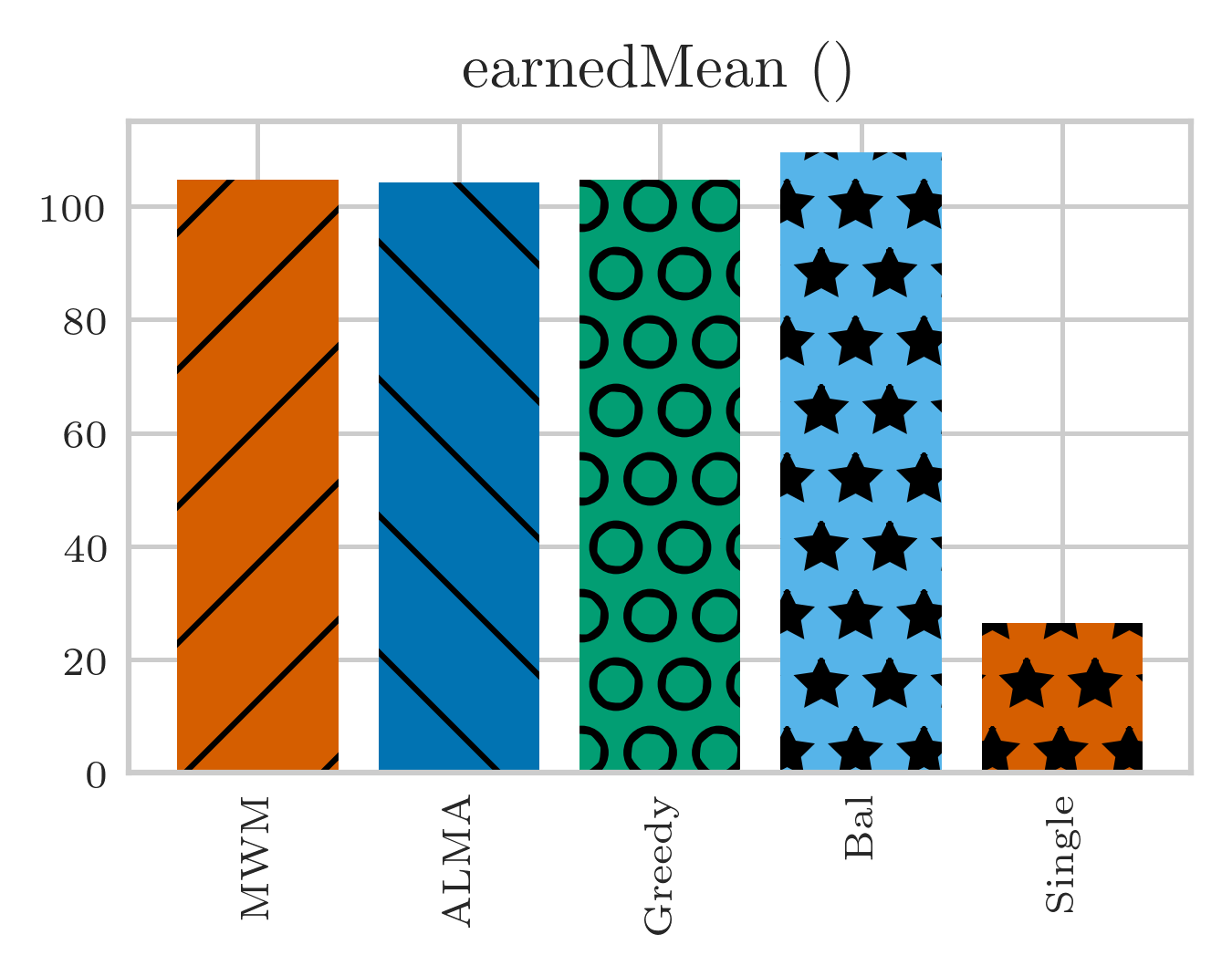}
		\caption{Driver Profit (\$)}
		\label{fig_appendix: Jan15Hour8to9_earnedMean}
	\end{subfigure}

	\begin{subfigure}[t]{0.23\textwidth}
		\centering
		\includegraphics[width = 1 \linewidth, trim={0.6em 0.6em 0.5em 1.8em}, clip]{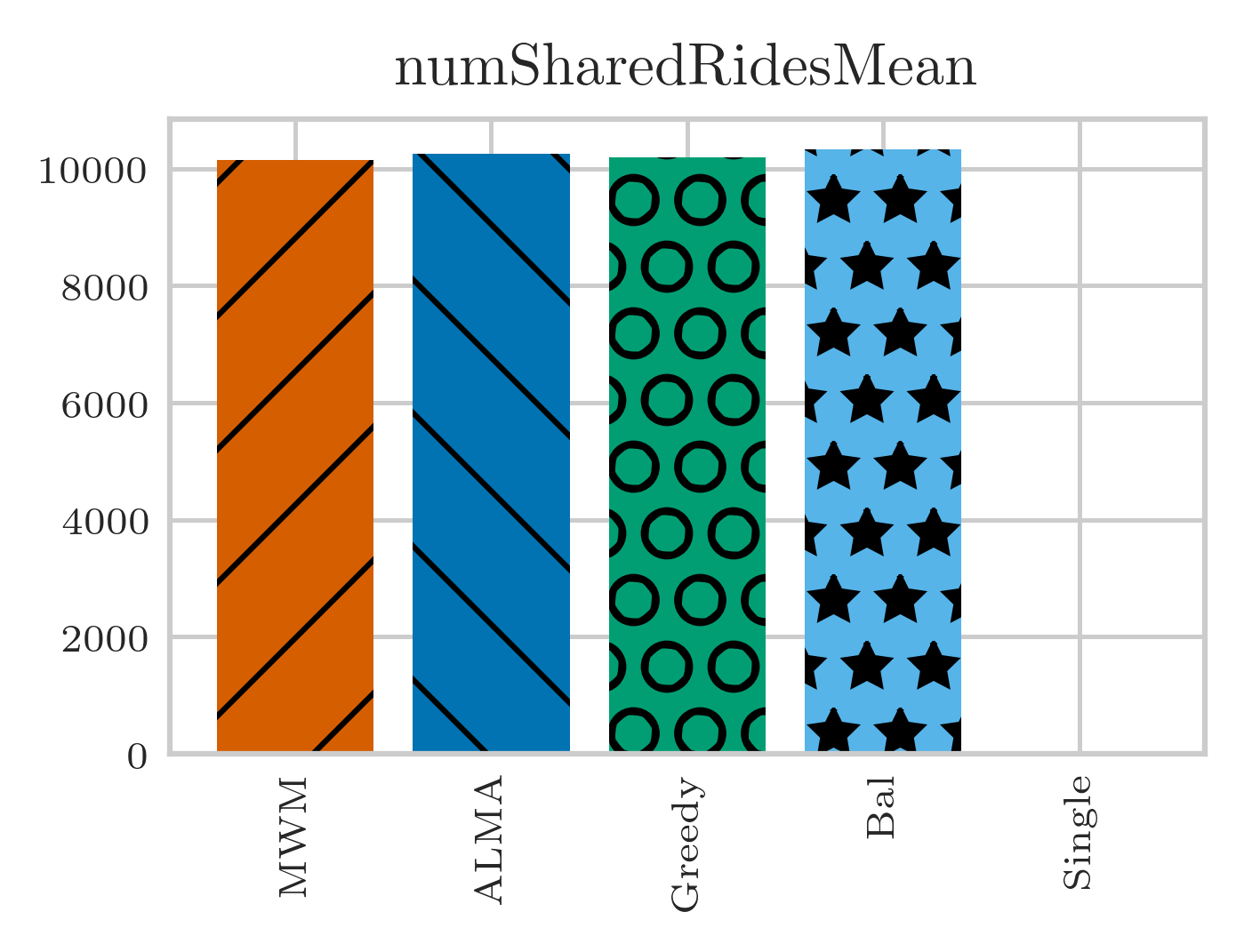}
		\caption{Number of Shared Rides}
		\label{fig_appendix: Jan15Hour8to9_numSharedRidesMean}
	\end{subfigure}
	~
	\begin{subfigure}[t]{0.23\textwidth}
		\centering
		\includegraphics[width = 1 \linewidth, trim={0.6em 0.6em 0.5em 1.8em}, clip]{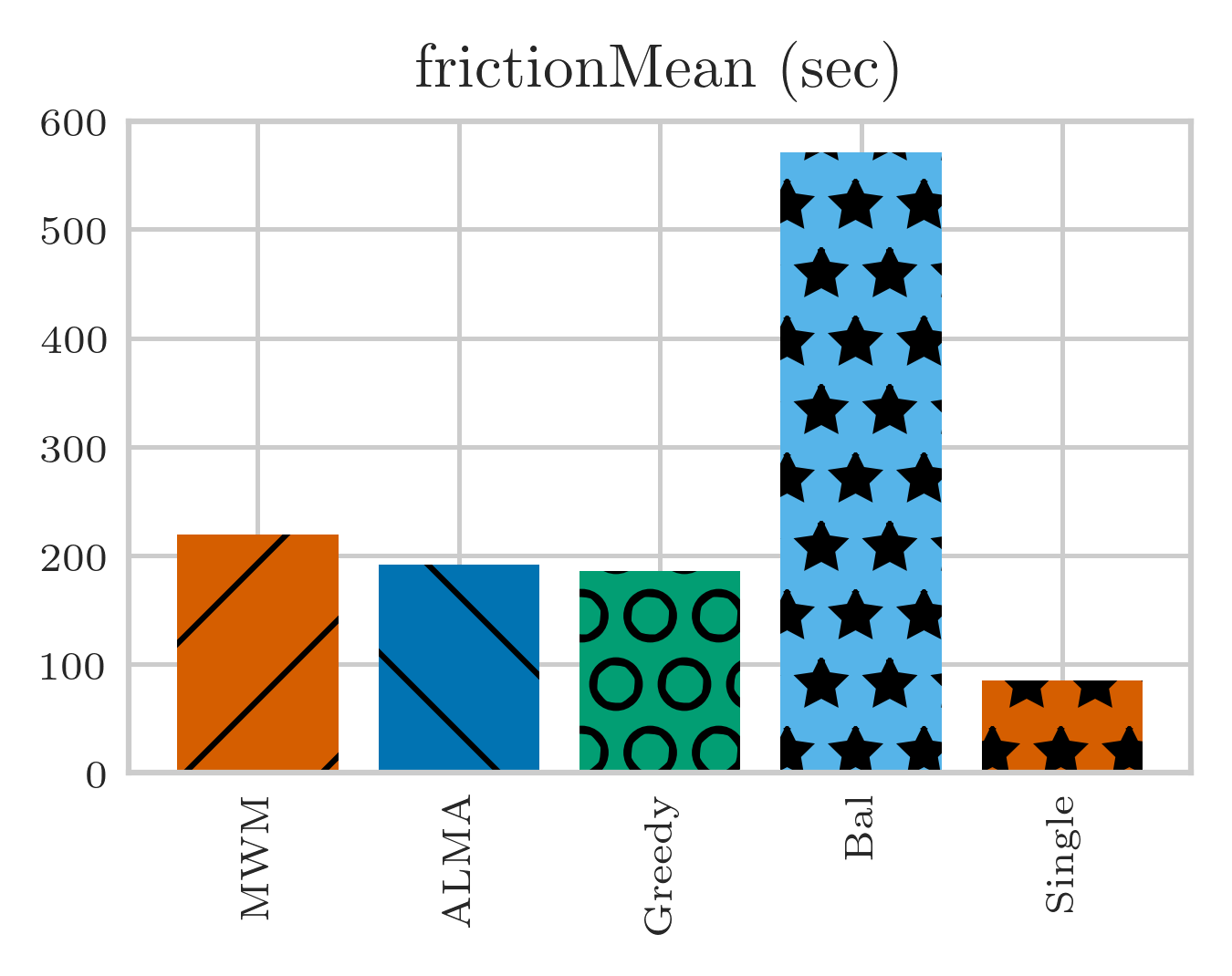}
		\caption{Frictions (s)}
		\label{fig_appendix: Jan15Hour8to9_frictionMean}
	\end{subfigure}%
	\caption{January 15, 2016 -- 08:00 - 09:00 -- Broader NYC Area -- \#Taxis = 4972 (base number).}
	\label{fig_appendix: Jan15Hour8to9}
\end{figure*}

\begin{table*}[b!]
\centering
\caption{January 15, 2016 -- 08:00 - 09:00 -- Broader NYC Area -- \#Taxis = 4972 (base number).}
\label{tb_appendix: Jan15Hour8to9}
\resizebox{\textwidth}{!}{%
\begin{tabular}{@{}lccccccccccccccccccc@{}}
\toprule
\textbf{} & \textbf{\begin{tabular}[c]{@{}c@{}}Distance\\ Driven (m)\end{tabular}} & \textbf{SD} & \textbf{\begin{tabular}[c]{@{}c@{}}Elapsed\\ Time (ns)\end{tabular}} & \textbf{SD} & \textbf{\begin{tabular}[c]{@{}c@{}}Time to\\ Pair (s)\end{tabular}} & \textbf{SD} & \textbf{\begin{tabular}[c]{@{}c@{}}Time to Pair\\ with Taxi (s)\end{tabular}} & \textbf{SD} & \textbf{\begin{tabular}[c]{@{}c@{}}Time to\\ Pick-up (s)\end{tabular}} & \textbf{SD} & \textbf{Delay (s)} & \textbf{SD} & \textbf{\begin{tabular}[c]{@{}c@{}}Cumulative\\ Delay (s)\end{tabular}} & \textbf{\begin{tabular}[c]{@{}c@{}}Driver\\ Profit (\$)\end{tabular}} & \textbf{SD} & \textbf{\begin{tabular}[c]{@{}c@{}}Number of \\ Shared Rides\end{tabular}} & \textbf{SD} & \textbf{Frictions (s)} & \textbf{SD} \\ \midrule
\textbf{MWM}             & 6.48E+07 & 0.00E+00   & 1.81E+12 & 0.00E+00     & 32.34 & 31.34                  & 0.00  & 0.00            & 164.59 & 401.29 & 33.01 & 104.44           & 229.94        & 104.67 & 81.54                  & 1.02E+04 & 0.00      & 219.19 & 415.07 \\
\textbf{ALMA}            & 7.86E+07 & 2.48E+05   & 9.61E+10 & 1.02E+10     & 32.09 & 31.32                  & 0.00  & 0.00            & 287.93 & 646.99 & 29.55 & 167.88           & 349.58        & 104.13 & 72.74                  & 1.03E+04 & 5.88      & 191.41 & 379.85 \\
\textbf{Greedy}          & 8.18E+07 & 2.96E+05   & 4.88E+10 & 7.00E+09     & 32.32 & 31.40                  & 0.00  & 0.00            & 317.48 & 720.24 & 34.73 & 170.53           & 384.53        & 104.68 & 69.88                  & 1.02E+04 & 12.28     & 185.35 & 374.04 \\
\textbf{Bal}             & 7.22E+07 & 1.15E+05   & 3.09E+10 & 6.01E+09     & 5.49  & 18.97                  & 0.00  & 0.00            & 234.85 & 428.34 & 67.79 & 219.01           & 308.14        & 109.57 & 65.24                  & 1.03E+04 & 0.00      & 571.11 & 516.22 \\
\textbf{Single}          & 1.20E+08 & 0.00E+00   & 1.03E+12 & 0.00E+00     & 0.00  & 0.00                   & 10.44 & 83.19           & 212.61 & 577.74 & 0.00  & 0.00             & 223.06        & 26.37  & 9.21                   & 0.00E+00 & 0.00      & 85.07  & 211.52 \\ \bottomrule
\end{tabular}%
}
\end{table*}

\begin{table*}[b!]
\centering
\caption{January 15, 2016 -- 08:00 - 09:00 -- Broader NYC Area -- \#Taxis = 4972 (base number). Each column presents the relative difference compared to the first line, i.e., the MWM (algorithm - MWM) / MWM, for each metric.}
\label{tb_appendix: Jan15Hour8to9Percentages}
\resizebox{\textwidth}{!}{%
\begin{tabular}{@{}lccccccccccccccccccc@{}}
\toprule
\textbf{} & \textbf{\begin{tabular}[c]{@{}c@{}}Distance\\ Driven (m)\end{tabular}} & \textbf{SD} & \textbf{\begin{tabular}[c]{@{}c@{}}Elapsed\\ Time (ns)\end{tabular}} & \textbf{SD} & \textbf{\begin{tabular}[c]{@{}c@{}}Time to\\ Pair (s)\end{tabular}} & \textbf{SD} & \textbf{\begin{tabular}[c]{@{}c@{}}Time to Pair\\ with Taxi (s)\end{tabular}} & \textbf{SD} & \textbf{\begin{tabular}[c]{@{}c@{}}Time to\\ Pick-up (s)\end{tabular}} & \textbf{SD} & \textbf{Delay (s)} & \textbf{SD} & \textbf{\begin{tabular}[c]{@{}c@{}}Cumulative\\ Delay (s)\end{tabular}} & \textbf{\begin{tabular}[c]{@{}c@{}}Driver\\ Profit (\$)\end{tabular}} & \textbf{SD} & \textbf{\begin{tabular}[c]{@{}c@{}}Number of \\ Shared Rides\end{tabular}} & \textbf{SD} & \textbf{Frictions (s)} & \textbf{SD} \\ \midrule
\textbf{MWM}             & 0.00\%  & --         & 0.00\%   & --           & 0.00\%    & 0.00\%                 & -- & --              & 0.00\%  & 0.00\%  & 0.00\%    & 0.00\%           & 0.00\%        & 0.00\%   & 0.00\%                 & 0.00\%    & --        & 0.00\%   & 0.00\%   \\
\textbf{ALMA}            & 21.37\% & --         & -94.68\% & --           & -0.75\%   & -0.08\%                & -- & --              & 74.94\% & 61.23\% & -10.48\%  & 60.74\%          & 52.03\%       & -0.52\%  & -10.79\%               & 1.02\%    & --        & -12.67\% & -8.49\%  \\
\textbf{Greedy}          & 26.34\% & --         & -97.30\% & --           & -0.04\%   & 0.20\%                 & -- & --              & 92.90\% & 79.48\% & 5.20\%    & 63.27\%          & 67.24\%       & 0.01\%   & -14.31\%               & 0.42\%    & --        & -15.44\% & -9.88\%  \\
\textbf{Bal}             & 11.44\% & --         & -98.29\% & --           & -83.01\%  & -39.48\%               & -- & --              & 42.69\% & 6.74\%  & 105.35\%  & 109.70\%         & 34.01\%       & 4.68\%   & -20.00\%               & 1.77\%    & --        & 160.56\% & 24.37\%  \\
\textbf{Single}          & 84.67\% & --         & -42.94\% & --           & -100.00\% & -100.00\%              & -- & --              & 29.18\% & 43.97\% & -100.00\% & -100.00\%        & -2.99\%       & -74.80\% & -88.71\%               & -100.00\% & --        & -61.19\% & -49.04\% \\ \bottomrule
\end{tabular}%
}
\end{table*}

\clearpage


\begin{figure*}[t!]
\subsection{00:00 - 23:59 (full day) -- Broader NYC Area  (Manhattan, Bronx, Staten Island, Brooklyn, Queens)} \label{Appendix Jan15}
\end{figure*}

\begin{figure*}[t!]
	\centering
	\begin{subfigure}[t]{0.23\textwidth}
		\centering
		\includegraphics[width = 1 \linewidth, trim={0.6em 0.6em 0.5em 1.8em}, clip]{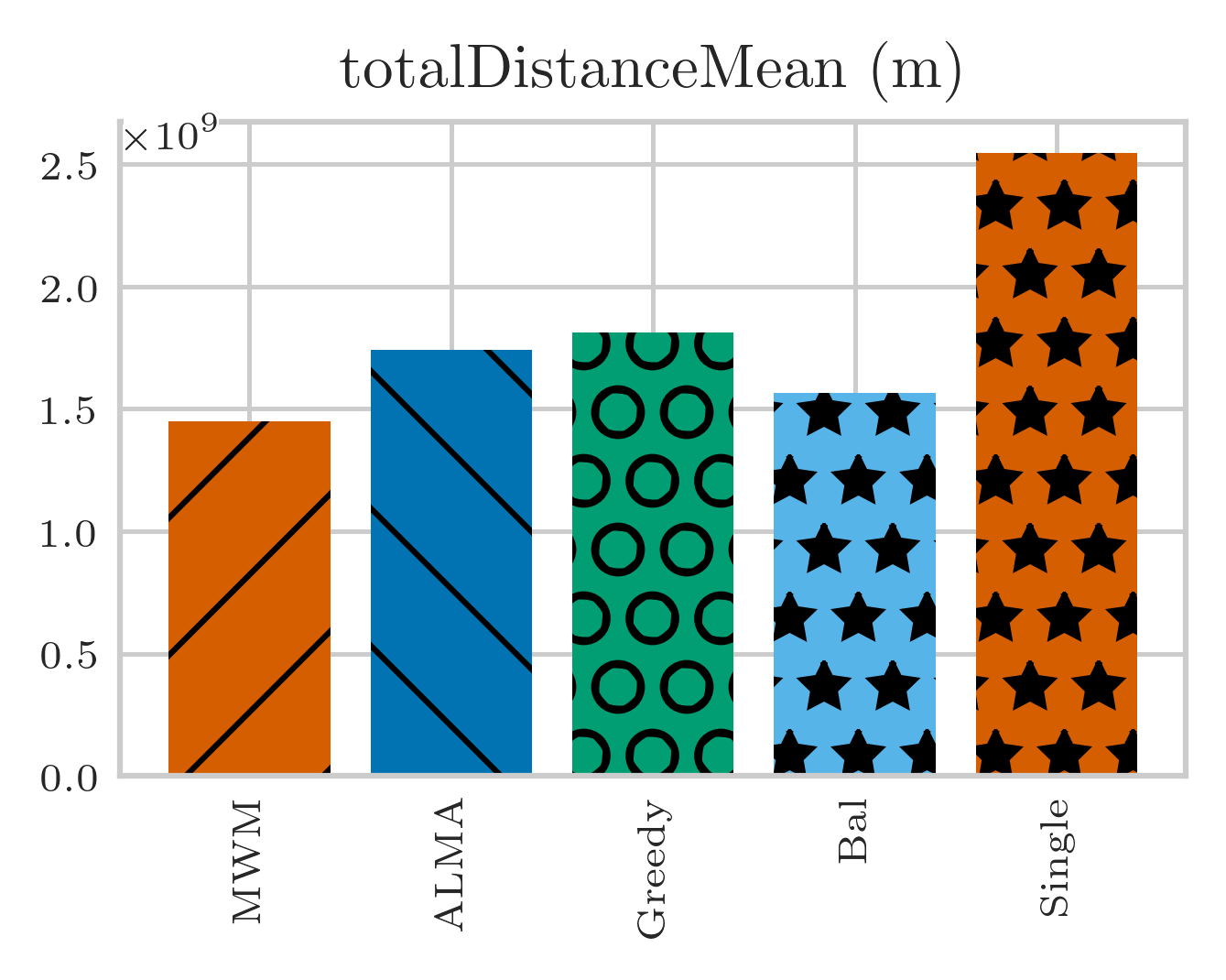}
		\caption{Total Distance Driven (m)}
		\label{fig_appendix: Jan15_totalDistanceMean}
	\end{subfigure}
	~ 
	\begin{subfigure}[t]{0.23\textwidth}
		\centering
		\includegraphics[width = 1 \linewidth, trim={0.6em 0.6em 0.5em 1.8em}, clip]{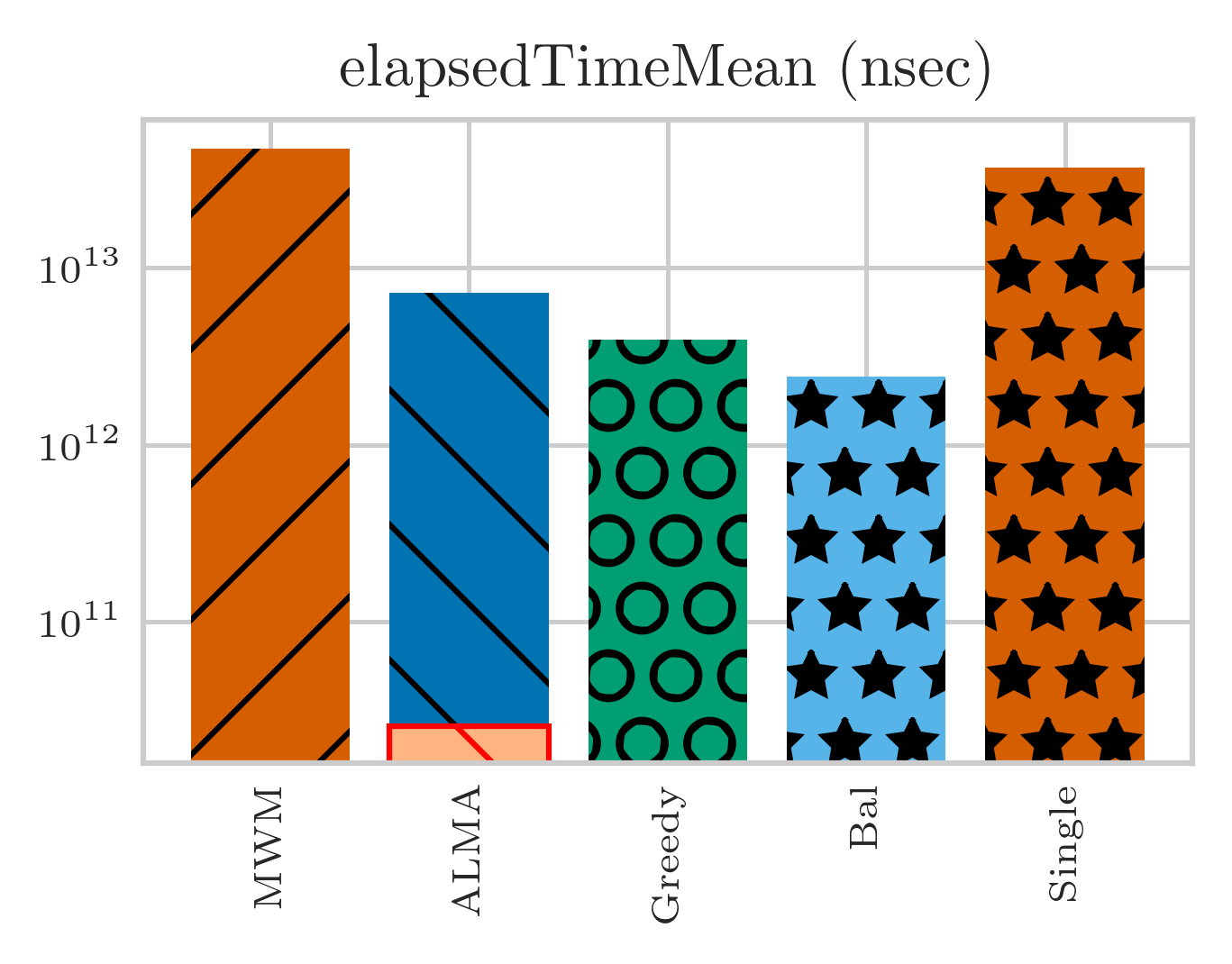}
		\caption{Elapsed Time (ns) [LOG]}
		\label{fig_appendix: Jan15_elapsedTimeMean}
	\end{subfigure}
	~
	\begin{subfigure}[t]{0.23\textwidth}
		\centering
		\includegraphics[width = 1 \linewidth, trim={0.6em 0.6em 0.5em 1.8em}, clip]{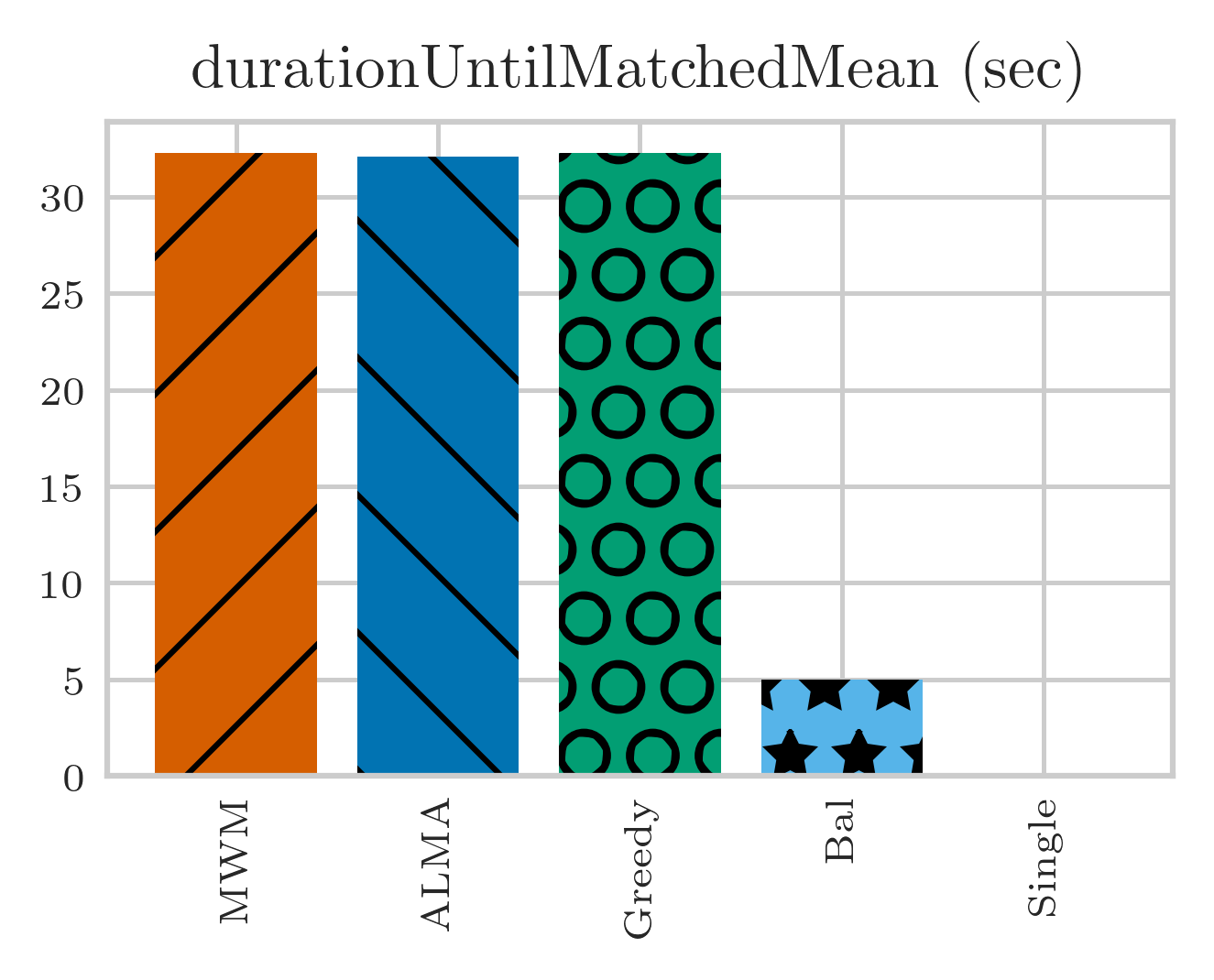}
		\caption{Time to Pair (s)}
		\label{fig_appendix: Jan15_durationUntilMatchedMean}
	\end{subfigure}
	~
	\begin{subfigure}[t]{0.23\textwidth}
		\centering
		\includegraphics[width = 1 \linewidth, trim={0.6em 0.6em 0.5em 1.8em}, clip]{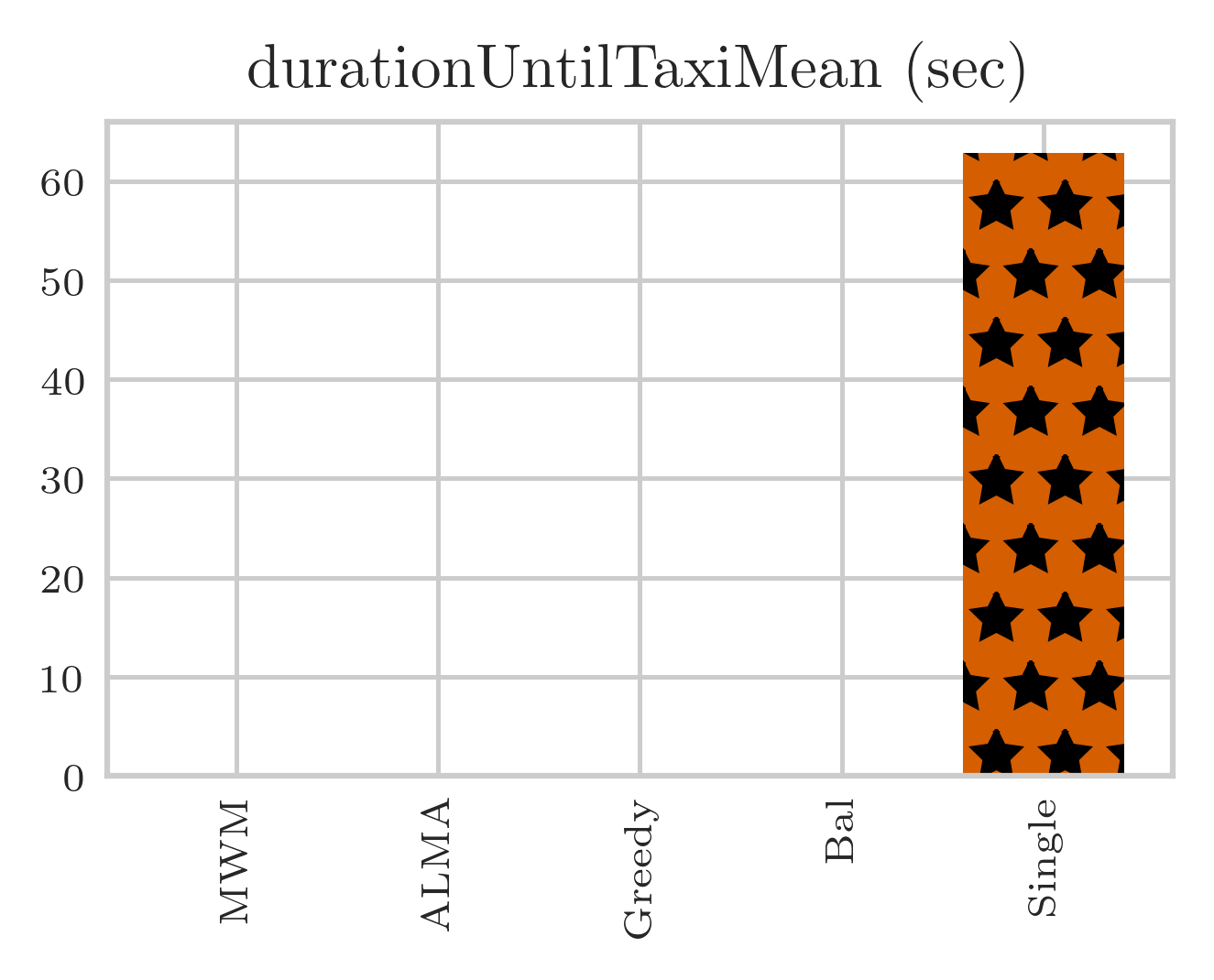}
		\caption{Time to Pair with Taxi (s)}
		\label{fig_appendix: Jan15_durationUntilTaxiMean}
	\end{subfigure}

	\begin{subfigure}[t]{0.23\textwidth}
		\centering
		\includegraphics[width = 1 \linewidth, trim={0.6em 0.6em 0.5em 1.8em}, clip]{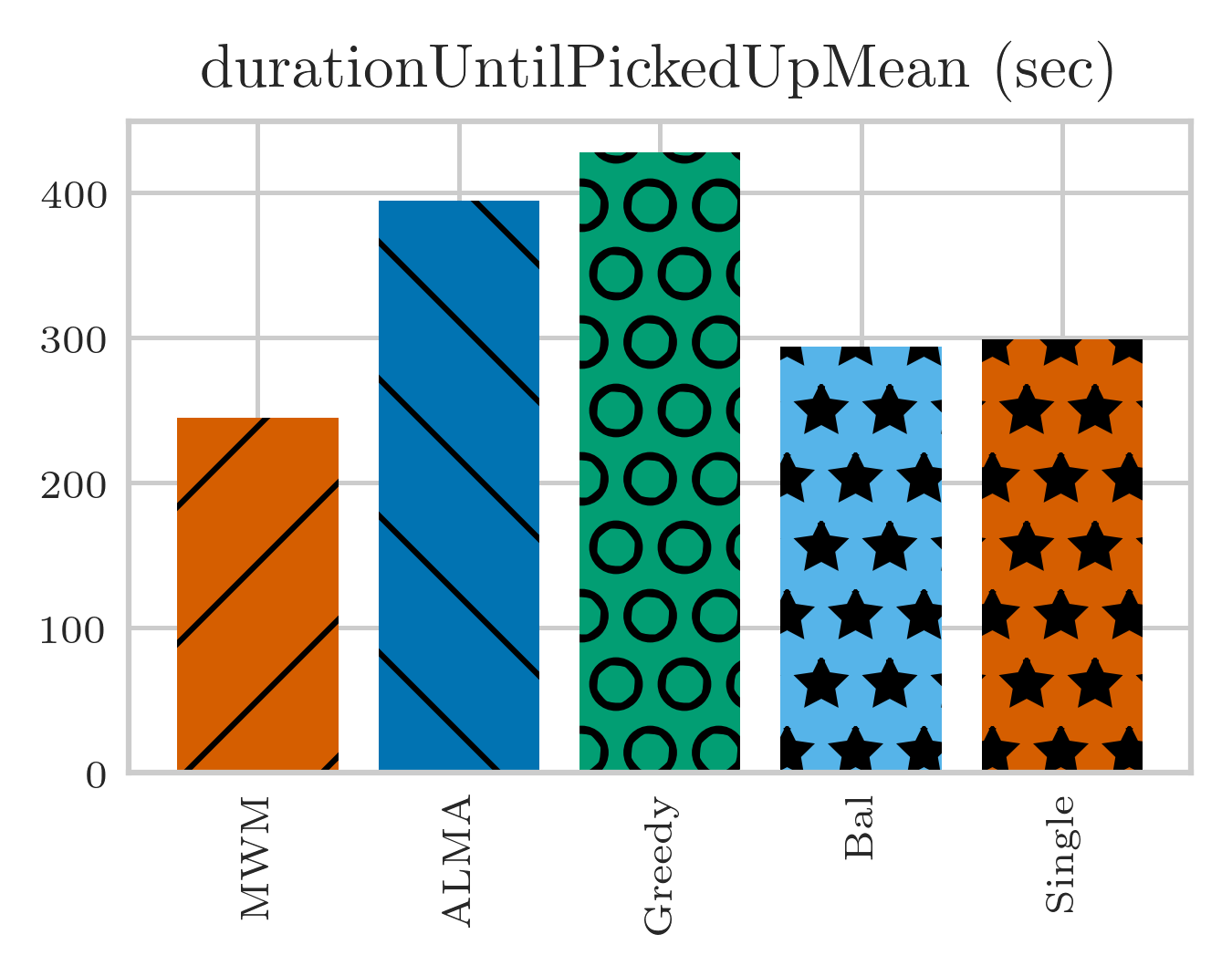}
		\caption{Time to Pick-up (s)}
		\label{fig_appendix: Jan15_durationUntilPickedUpMean}
	\end{subfigure}
	~
	\begin{subfigure}[t]{0.23\textwidth}
		\centering
		\includegraphics[width = 1 \linewidth, trim={0.6em 0.6em 0.5em 1.8em}, clip]{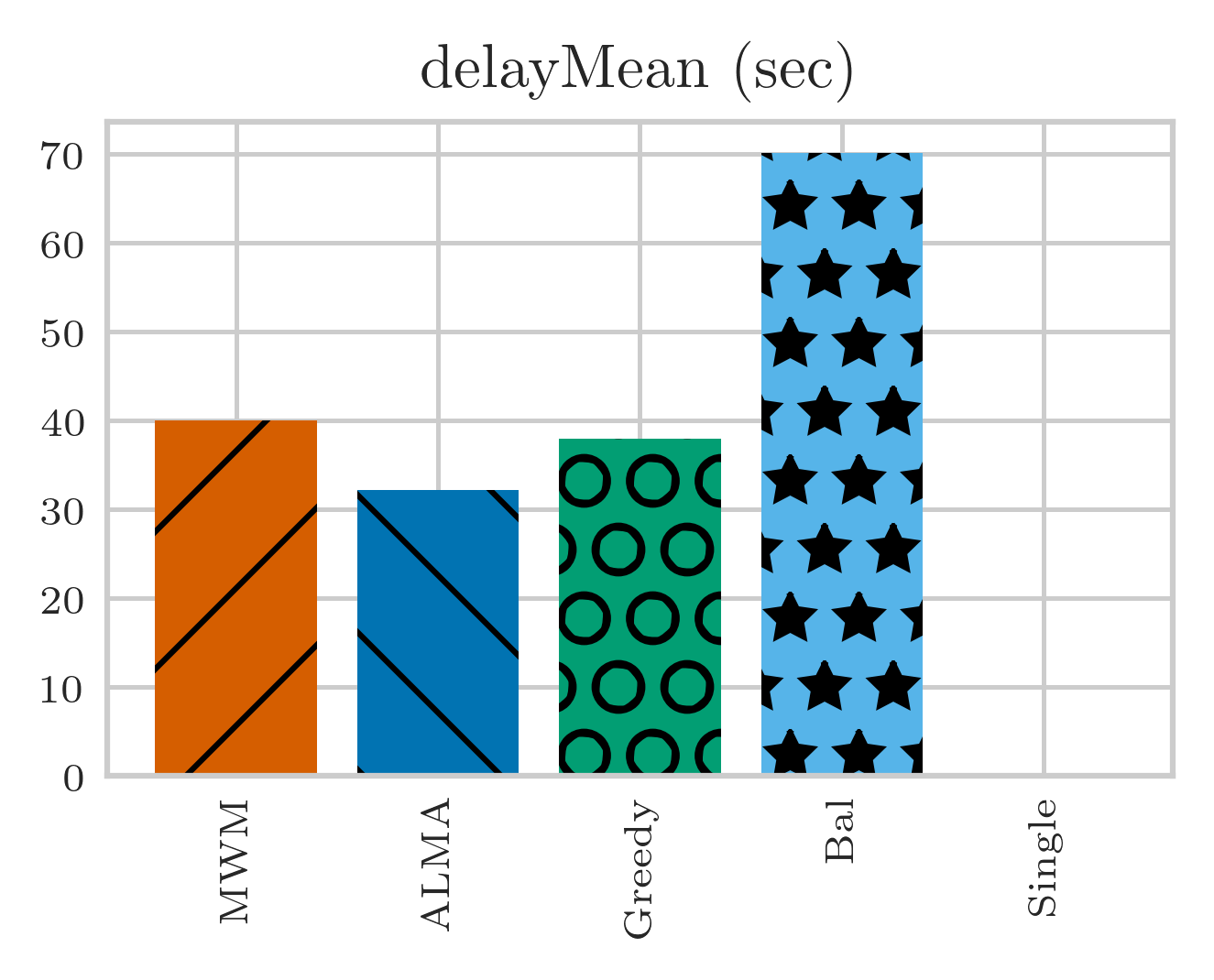}
		\caption{Delay (s)}
		\label{fig_appendix: Jan15_delayMean}
	\end{subfigure}
	~
	\begin{subfigure}[t]{0.23\textwidth}
		\centering
		\includegraphics[width = 1 \linewidth, trim={0.6em 0.6em 0.5em 1.8em}, clip]{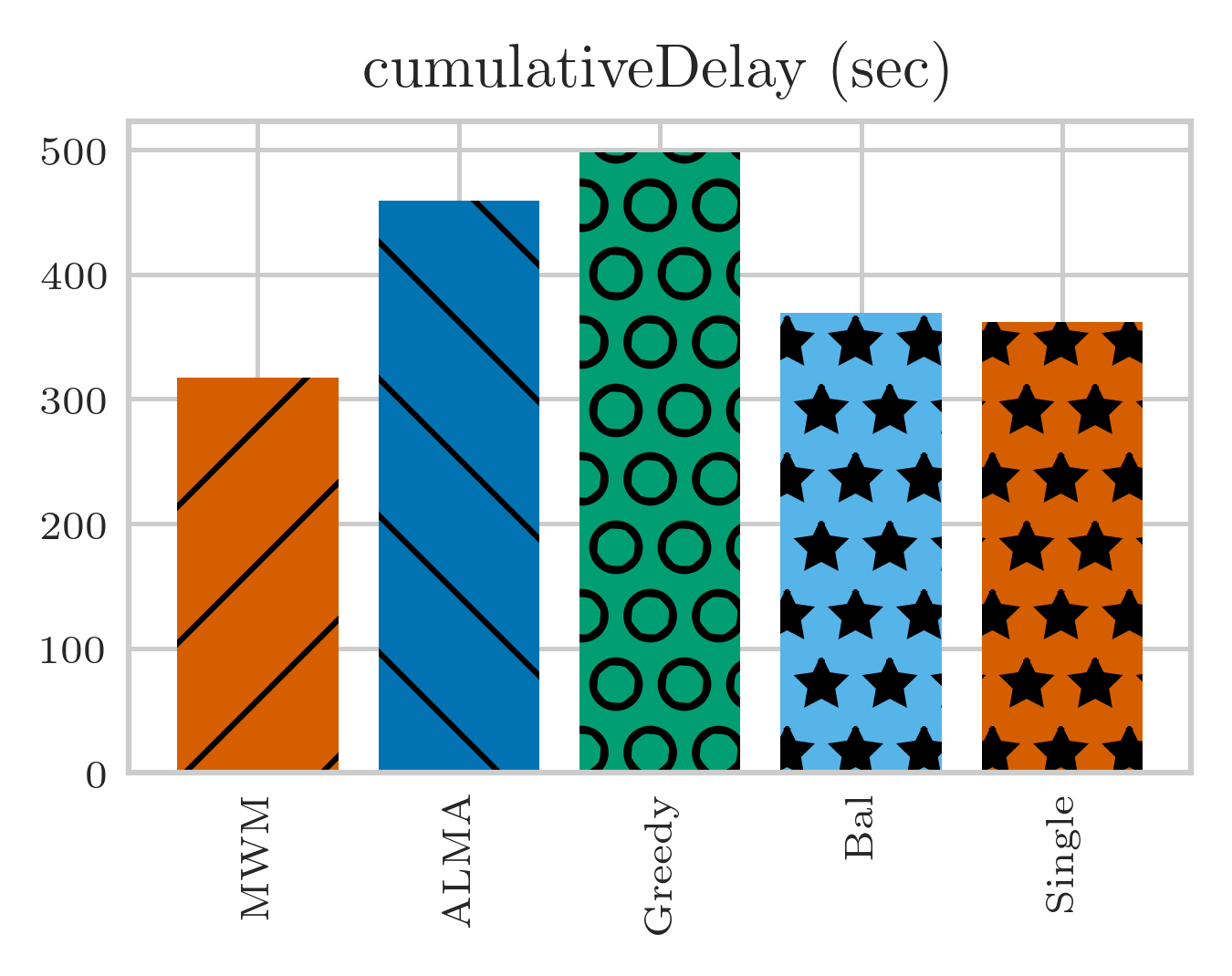}
		\caption{Cumulative Delay (s)}
		\label{fig_appendix: Jan15_cumulativeDelay}
	\end{subfigure}
	~
	\begin{subfigure}[t]{0.23\textwidth}
		\centering
		\includegraphics[width = 1 \linewidth, trim={0.6em 0.6em 0.5em 1.8em}, clip]{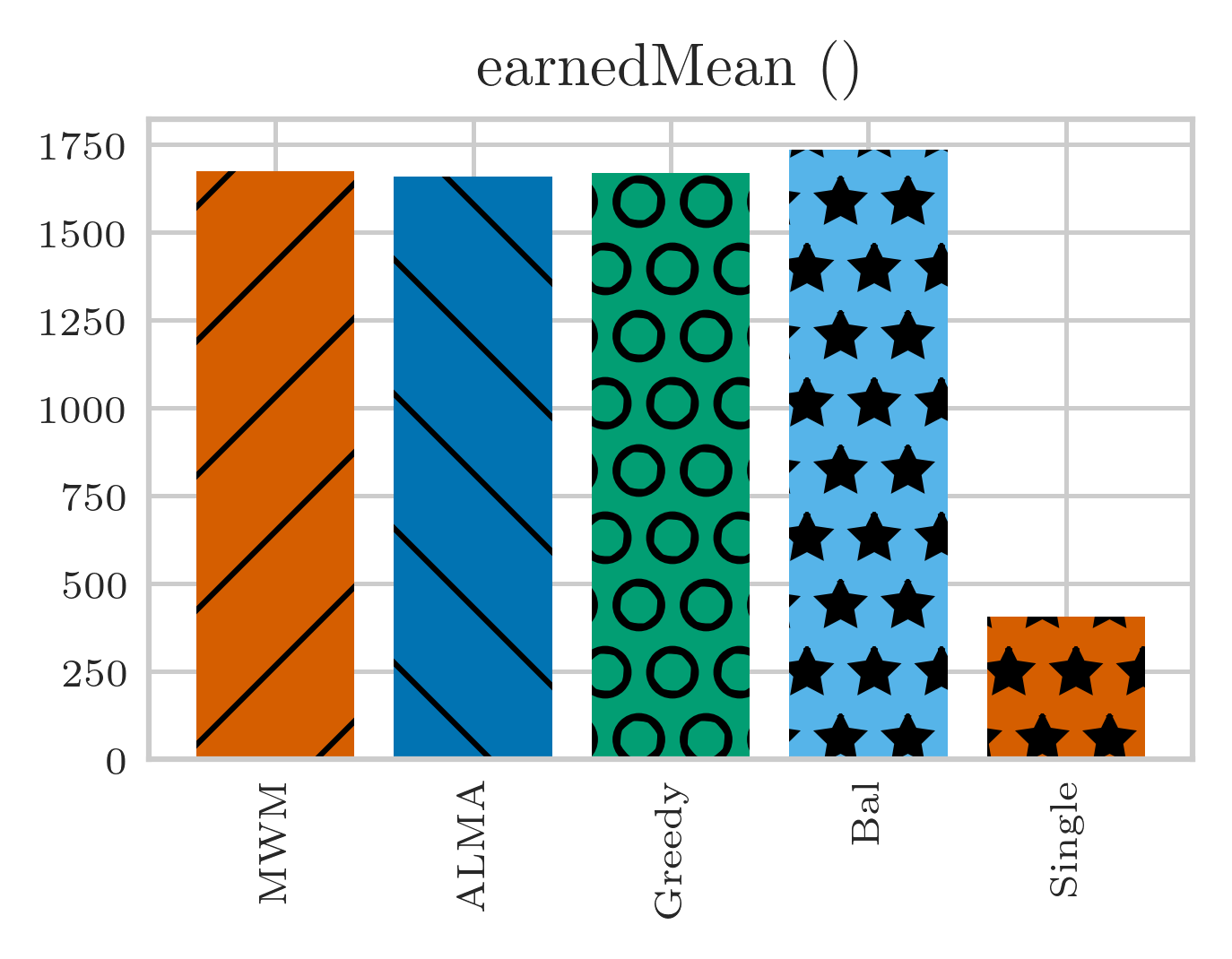}
		\caption{Driver Profit (\$)}
		\label{fig_appendix: Jan15_earnedMean}
	\end{subfigure}

	\begin{subfigure}[t]{0.23\textwidth}
		\centering
		\includegraphics[width = 1 \linewidth, trim={0.6em 0.6em 0.5em 1.8em}, clip]{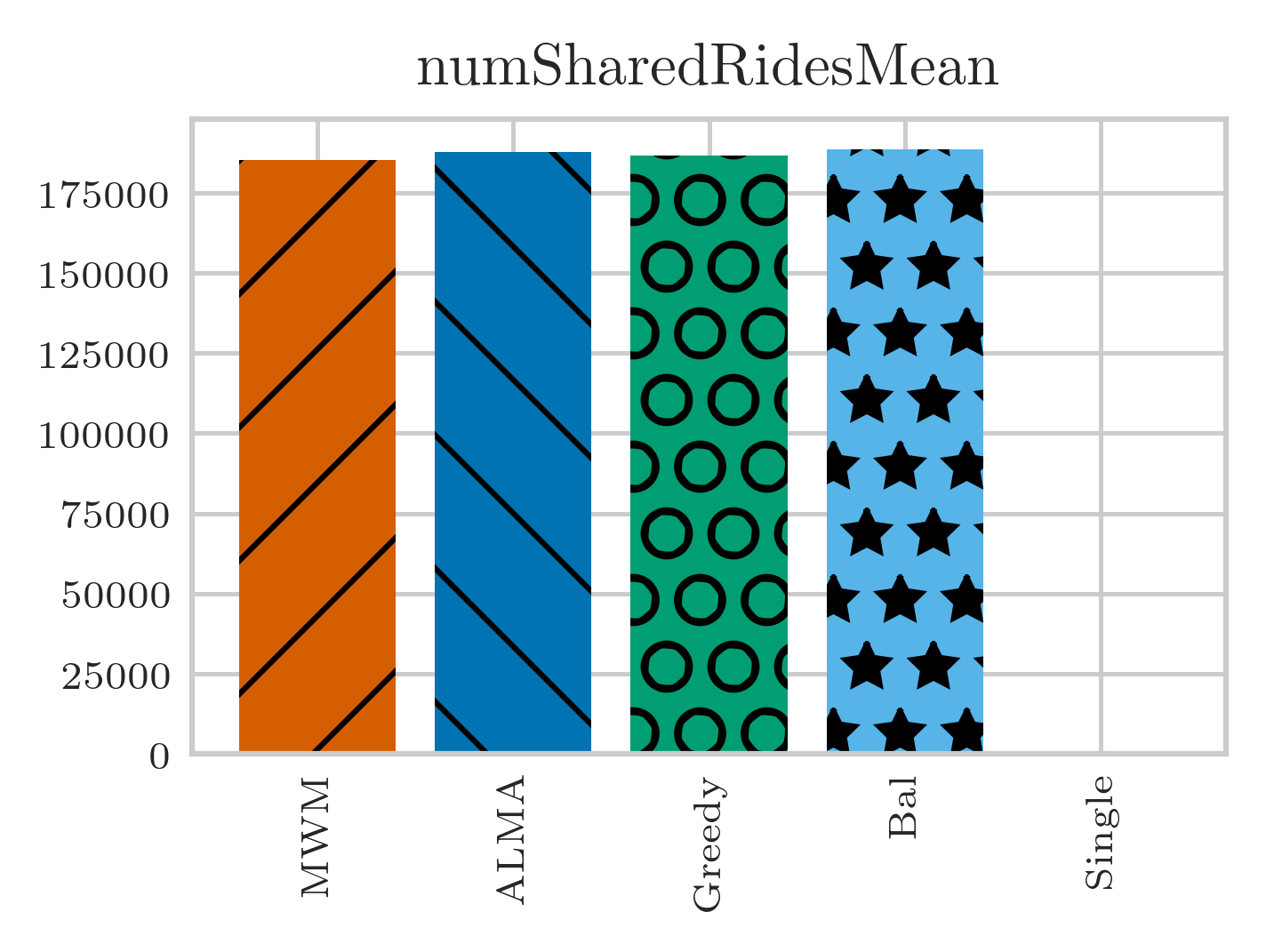}
		\caption{Number of Shared Rides}
		\label{fig_appendix: Jan15_numSharedRidesMean}
	\end{subfigure}
	~
	\begin{subfigure}[t]{0.23\textwidth}
		\centering
		\includegraphics[width = 1 \linewidth, trim={0.6em 0.6em 0.5em 1.8em}, clip]{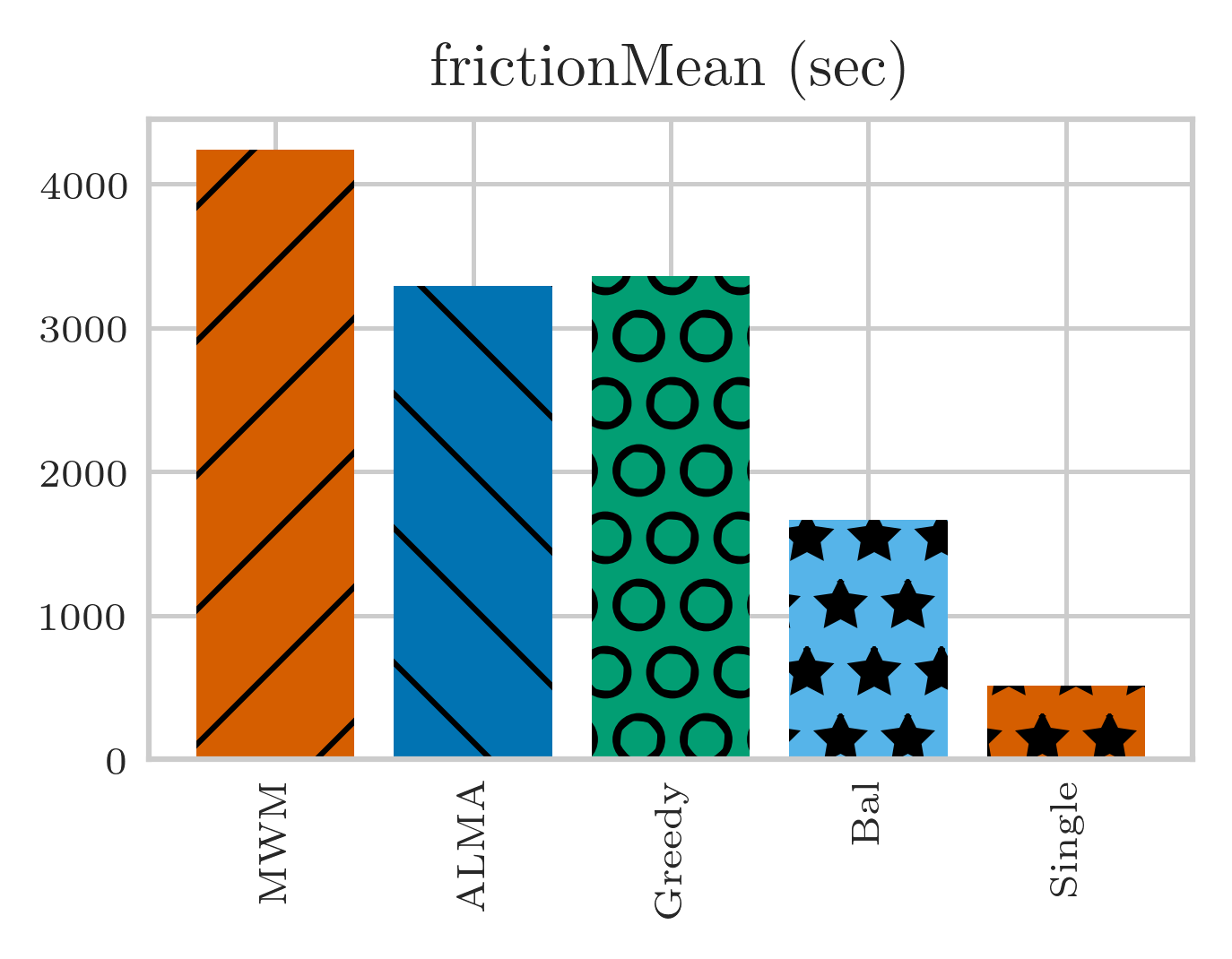}
		\caption{Frictions (s)}
		\label{fig_appendix: Jan15_frictionMean}
	\end{subfigure}%
	\caption{January 15, 2016 -- 00:00 - 23:59 (full day) -- Broader NYC Area -- \#Taxis = 6533 (base number).}
	\label{fig_appendix: Jan15}
\end{figure*}

\begin{table*}[b!]
\centering
\caption{January 15, 2016 -- 00:00 - 23:59 (full day) -- Broader NYC Area -- \#Taxis = 6533 (base number).}
\label{tb_appendix: Jan15}
\resizebox{\textwidth}{!}{%
\begin{tabular}{@{}lccccccccccccccccccc@{}}
\toprule
\textbf{} & \textbf{\begin{tabular}[c]{@{}c@{}}Distance\\ Driven (m)\end{tabular}} & \textbf{SD} & \textbf{\begin{tabular}[c]{@{}c@{}}Elapsed\\ Time (ns)\end{tabular}} & \textbf{SD} & \textbf{\begin{tabular}[c]{@{}c@{}}Time to\\ Pair (s)\end{tabular}} & \textbf{SD} & \textbf{\begin{tabular}[c]{@{}c@{}}Time to Pair\\ with Taxi (s)\end{tabular}} & \textbf{SD} & \textbf{\begin{tabular}[c]{@{}c@{}}Time to\\ Pick-up (s)\end{tabular}} & \textbf{SD} & \textbf{Delay (s)} & \textbf{SD} & \textbf{\begin{tabular}[c]{@{}c@{}}Cumulative\\ Delay (s)\end{tabular}} & \textbf{\begin{tabular}[c]{@{}c@{}}Driver\\ Profit (\$)\end{tabular}} & \textbf{SD} & \textbf{\begin{tabular}[c]{@{}c@{}}Number of \\ Shared Rides\end{tabular}} & \textbf{SD} & \textbf{Frictions (s)} & \textbf{SD} \\ \midrule
\textbf{MWM}             & 1.45E+09 & 0.00E+00   & 4.71E+13 & 0.00E+00     & 32.26 & 31.24                  & 0.00  & 0.00            & 244.90 & 433.03  & 40.01 & 131.68           & 317.17        & 1675.21 & 949.14                 & 1.85E+05 & 0.00      & 4238.58 & 10671.94 \\
\textbf{ALMA}            & 1.74E+09 & 5.28E+05   & 7.20E+12 & 1.37E+11     & 32.09 & 31.44                  & 0.00  & 0.00            & 394.30 & 701.06  & 32.17 & 133.83           & 458.57        & 1659.01 & 741.69                 & 1.88E+05 & 22.08     & 3286.78 & 8808.76  \\
\textbf{Greedy}          & 1.81E+09 & 2.36E+06   & 3.92E+12 & 3.31E+11     & 32.28 & 31.49                  & 0.00  & 0.00            & 427.74 & 750.89  & 37.92 & 137.07           & 497.93        & 1667.74 & 672.62                 & 1.87E+05 & 16.76     & 3357.00 & 9464.19  \\
\textbf{Bal}             & 1.57E+09 & 2.60E+05   & 2.42E+12 & 1.60E+11     & 4.97  & 18.17                  & 0.00  & 0.00            & 293.91 & 457.48  & 70.17 & 216.02           & 369.04        & 1736.02 & 153.93                 & 1.89E+05 & 0.00      & 1666.22 & 570.20   \\
\textbf{Single}          & 2.55E+09 & 0.00E+00   & 3.70E+13 & 0.00E+00     & 0.00  & 0.00                   & 62.85 & 753.27          & 298.52 & 1147.60 & 0.00  & 0.00             & 361.37        & 405.82  & 80.73                  & 0.00E+00 & 0.00      & 512.34  & 437.98   \\ \bottomrule
\end{tabular}%
}
\end{table*}

\begin{table*}[b!]
\centering
\caption{January 15, 2016 -- 00:00 - 23:59 (full day) -- Broader NYC Area -- \#Taxis = 6533 (base number). Each column presents the relative difference compared to the first line, i.e., the MWM (algorithm - MWM) / MWM, for each metric.}
\label{tb_appendix: Jan15Percentages}
\resizebox{\textwidth}{!}{%
\begin{tabular}{@{}lccccccccccccccccccc@{}}
\toprule
\textbf{} & \textbf{\begin{tabular}[c]{@{}c@{}}Distance\\ Driven (m)\end{tabular}} & \textbf{SD} & \textbf{\begin{tabular}[c]{@{}c@{}}Elapsed\\ Time (ns)\end{tabular}} & \textbf{SD} & \textbf{\begin{tabular}[c]{@{}c@{}}Time to\\ Pair (s)\end{tabular}} & \textbf{SD} & \textbf{\begin{tabular}[c]{@{}c@{}}Time to Pair\\ with Taxi (s)\end{tabular}} & \textbf{SD} & \textbf{\begin{tabular}[c]{@{}c@{}}Time to\\ Pick-up (s)\end{tabular}} & \textbf{SD} & \textbf{Delay (s)} & \textbf{SD} & \textbf{\begin{tabular}[c]{@{}c@{}}Cumulative\\ Delay (s)\end{tabular}} & \textbf{\begin{tabular}[c]{@{}c@{}}Driver\\ Profit (\$)\end{tabular}} & \textbf{SD} & \textbf{\begin{tabular}[c]{@{}c@{}}Number of \\ Shared Rides\end{tabular}} & \textbf{SD} & \textbf{Frictions (s)} & \textbf{SD} \\ \midrule
\textbf{MWM}             & 0.00\%  & --         & 0.00\%   & --           & 0.00\%    & 0.00\%                 & -- & --              & 0.00\%  & 0.00\%   & 0.00\%    & 0.00\%           & 0.00\%        & 0.00\%   & 0.00\%                 & 0.00\%    & --        & 0.00\%   & 0.00\%   \\
\textbf{ALMA}            & 20.13\% & --         & -84.69\% & --           & -0.53\%   & 0.64\%                 & -- & --              & 61.01\% & 61.90\%  & -19.59\%  & 1.63\%           & 44.58\%       & -0.97\%  & -21.86\%               & 1.34\%    & --        & -22.46\% & -17.46\% \\
\textbf{Greedy}          & 25.04\% & --         & -91.66\% & --           & 0.05\%    & 0.80\%                 & -- & --              & 74.66\% & 73.41\%  & -5.22\%   & 4.09\%           & 56.99\%       & -0.45\%  & -29.13\%               & 0.68\%    & --        & -20.80\% & -11.32\% \\
\textbf{Bal}             & 8.15\%  & --         & -94.85\% & --           & -84.61\%  & -41.85\%               & -- & --              & 20.01\% & 5.65\%   & 75.38\%   & 64.05\%          & 16.35\%       & 3.63\%   & -83.78\%               & 1.83\%    & --        & -60.69\% & -94.66\% \\
\textbf{Single}          & 75.86\% & --         & -21.44\% & --           & -100.00\% & -100.00\%              & -- & --              & 21.90\% & 165.02\% & -100.00\% & -100.00\%        & 13.94\%       & -75.78\% & -91.49\%               & -100.00\% & --        & -87.91\% & -95.90\% \\ \bottomrule
\end{tabular}%
}
\end{table*}

\clearpage


\begin{figure*}[t!]
\subsection{08:00 - 08:10 -- Manhattan} \label{Appendix Jan15Hour8to810Manhattan}
\end{figure*}

\begin{figure*}[t!]
	\centering
	\begin{subfigure}[t]{0.23\textwidth}
		\centering
		\includegraphics[width = 1 \linewidth, trim={0.6em 0.6em 0.5em 1.7em}, clip]{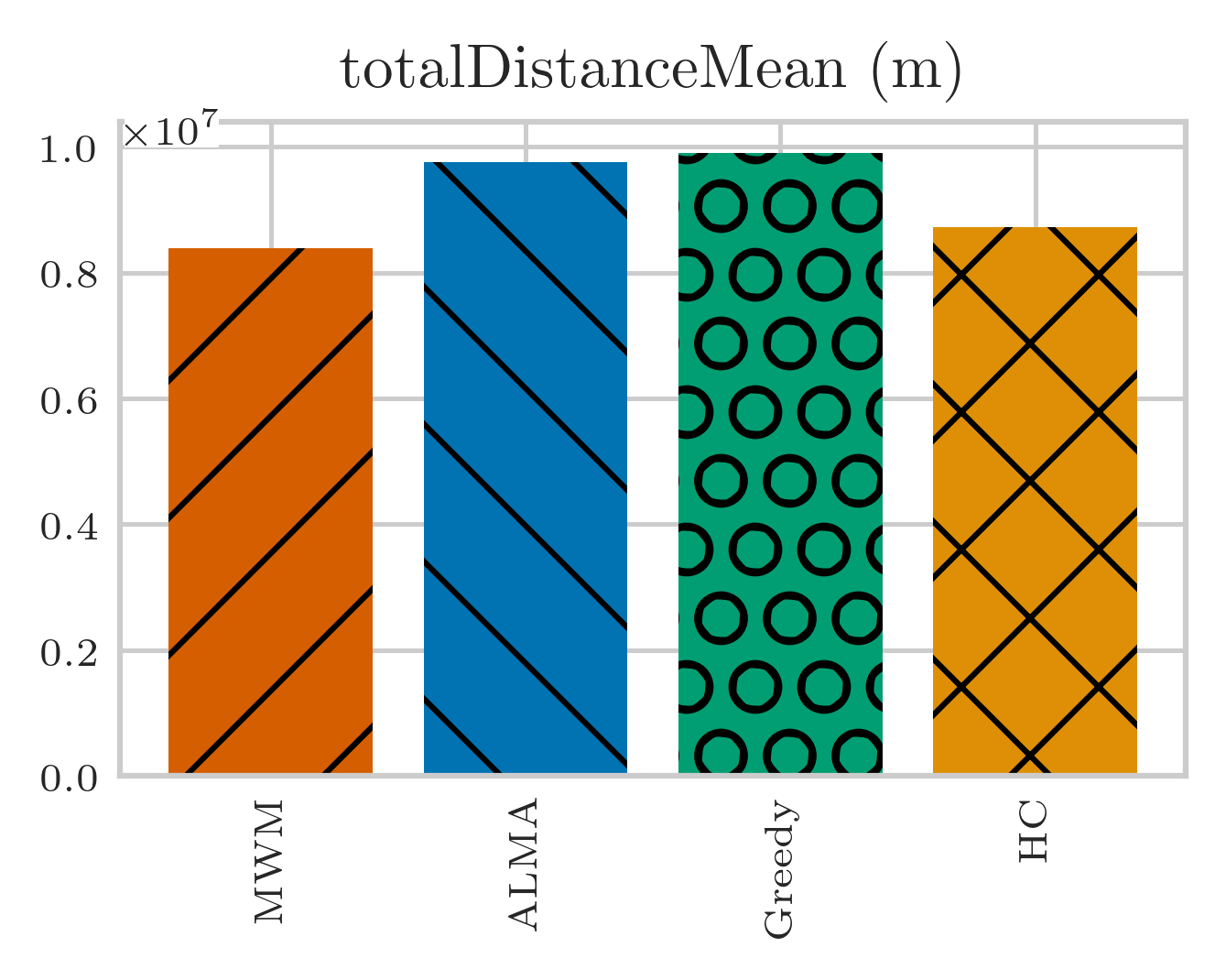}
		\caption{Total Distance Driven (m)}
		\label{fig_appendix: Jan15Hour8to810Manhattan_totalDistanceMean}
	\end{subfigure}
	~ 
	\begin{subfigure}[t]{0.23\textwidth}
		\centering
		\includegraphics[width = 1 \linewidth, trim={0.6em 0.6em 0.5em 1.8em}, clip]{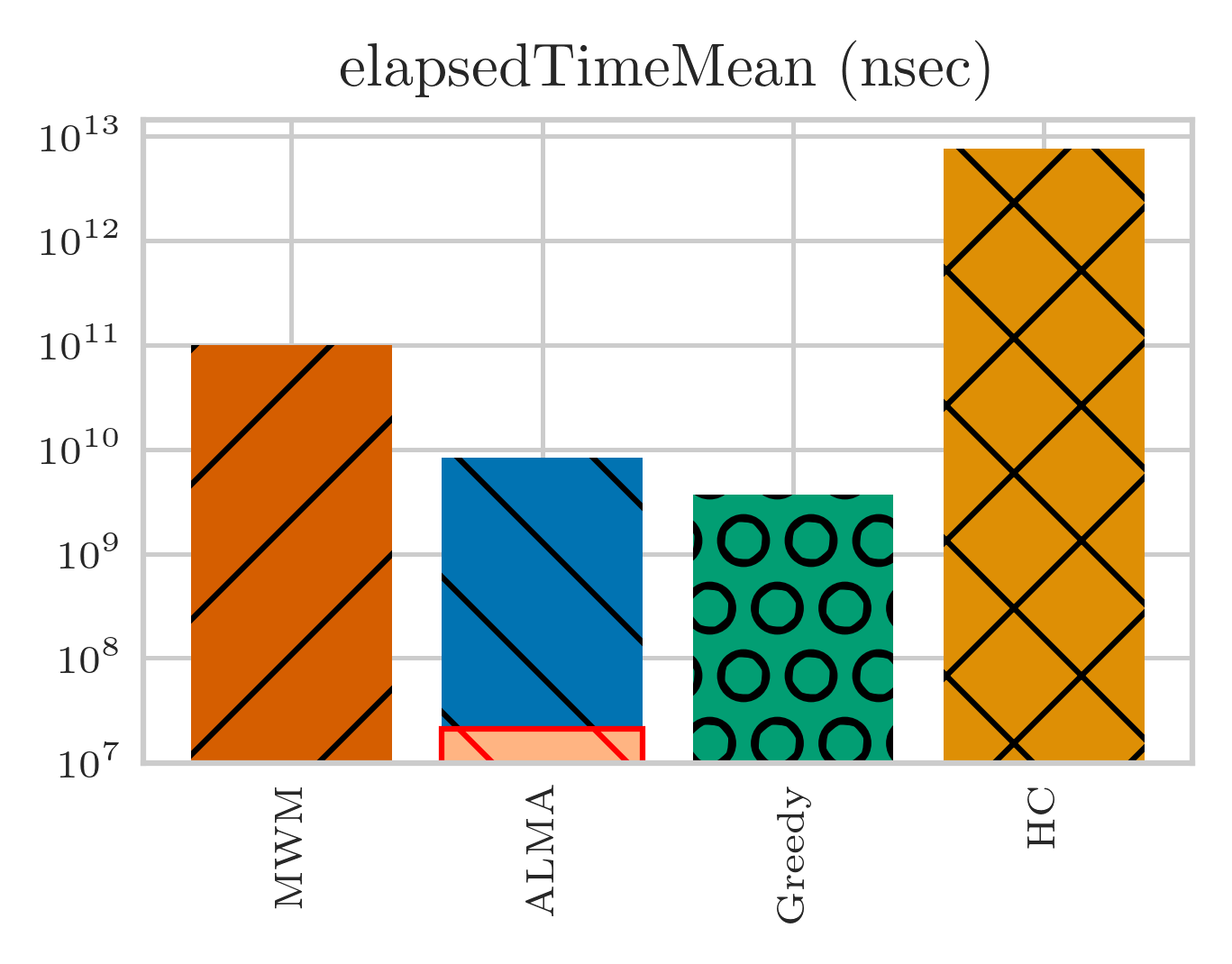}
		\caption{Elapsed Time (ns) [LOG]}
		\label{fig_appendix: Jan15Hour8to810Manhattan_elapsedTimeMean}
	\end{subfigure}
	~
	\begin{subfigure}[t]{0.23\textwidth}
		\centering
		\includegraphics[width = 1 \linewidth, trim={0.6em 0.6em 0.5em 1.8em}, clip]{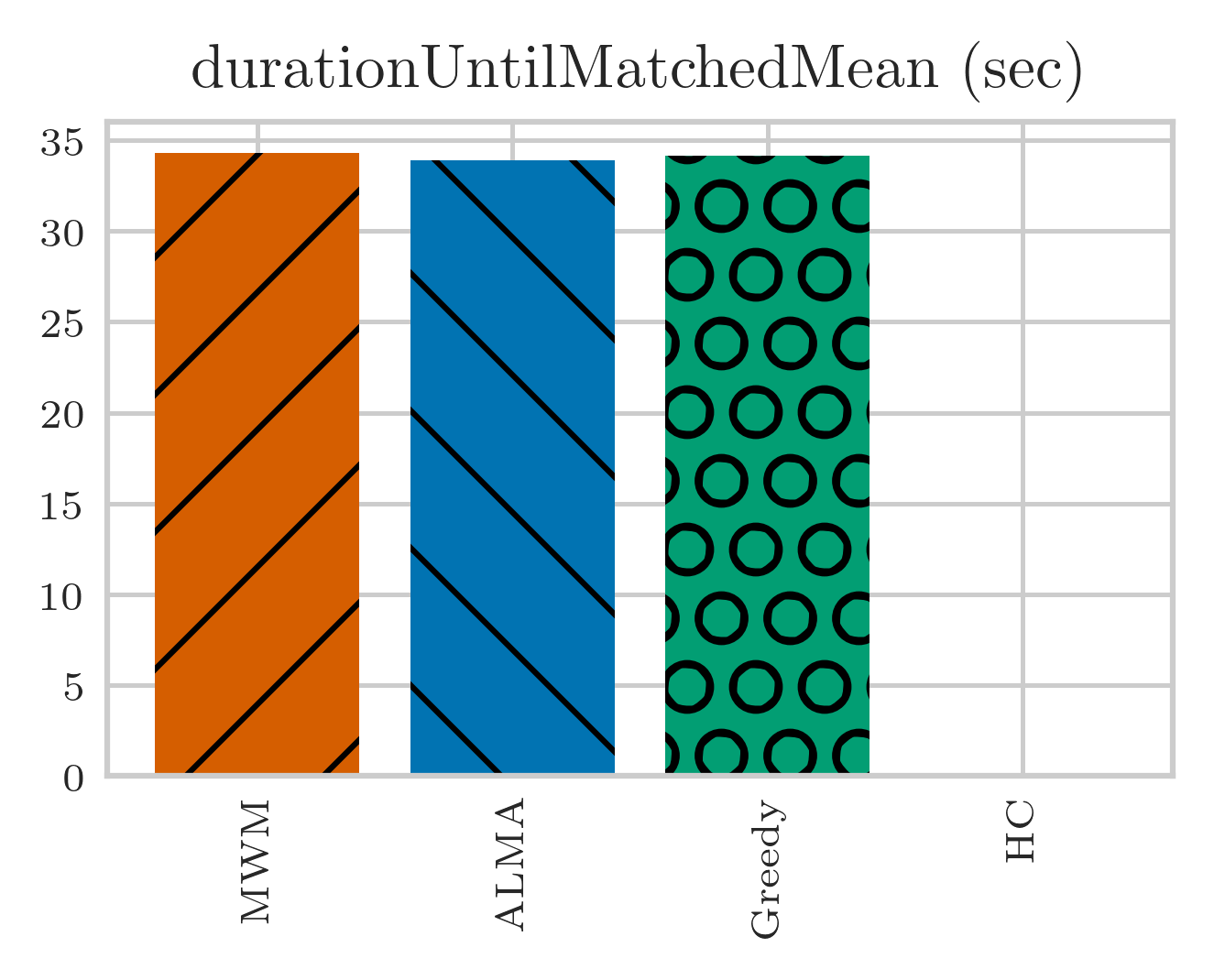}
		\caption{Time to Pair (s)}
		\label{fig_appendix: Jan15Hour8to810Manhattan_durationUntilMatchedMean}
	\end{subfigure}
	~
	\begin{subfigure}[t]{0.23\textwidth}
		\centering
		\includegraphics[width = 1 \linewidth, trim={0.6em 0.6em 0.5em 1.8em}, clip]{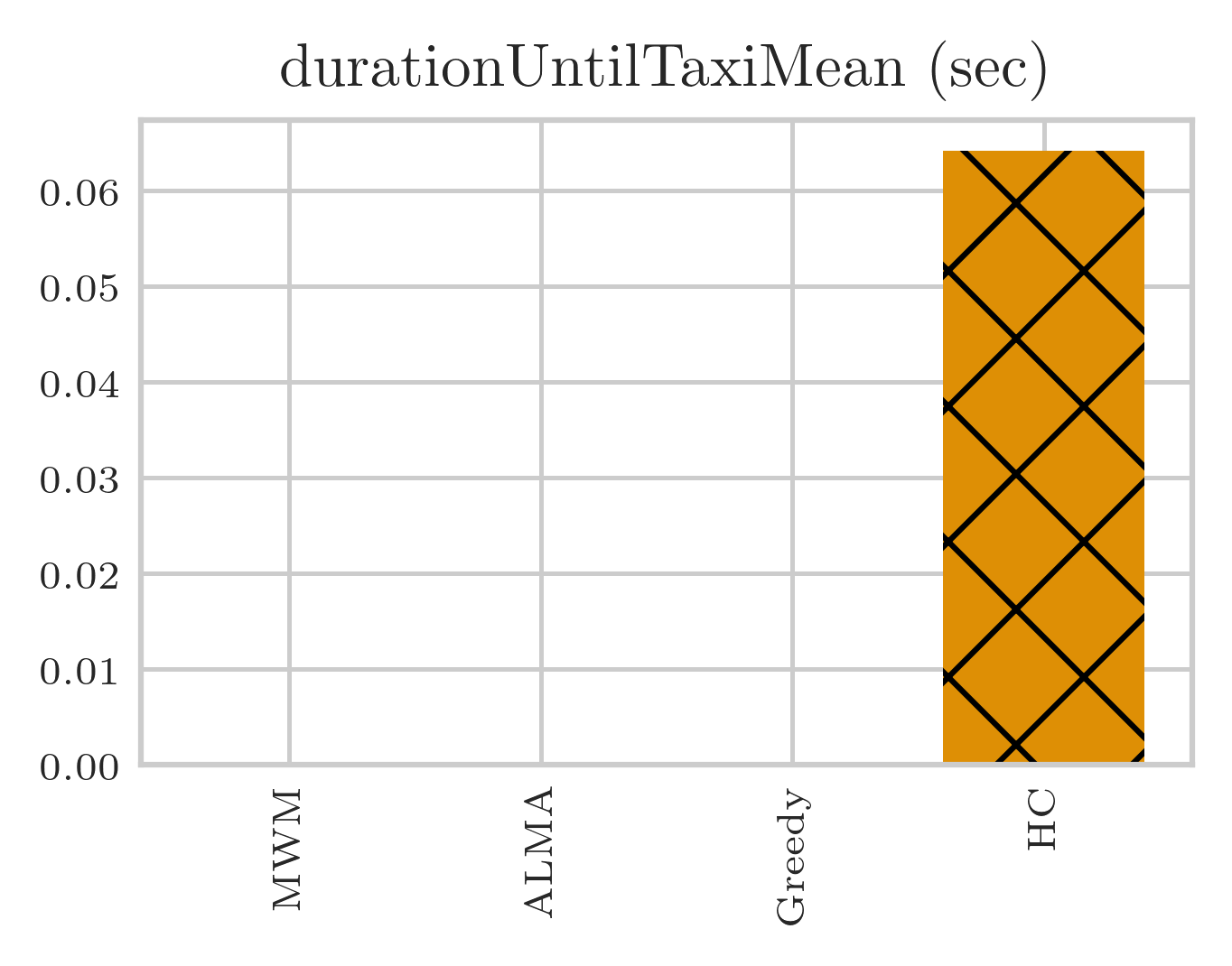}
		\caption{Time to Pair with Taxi (s)}
		\label{fig_appendix: Jan15Hour8to810Manhattan_durationUntilTaxiMean}
	\end{subfigure}

	\begin{subfigure}[t]{0.23\textwidth}
		\centering
		\includegraphics[width = 1 \linewidth, trim={0.6em 0.6em 0.5em 1.8em}, clip]{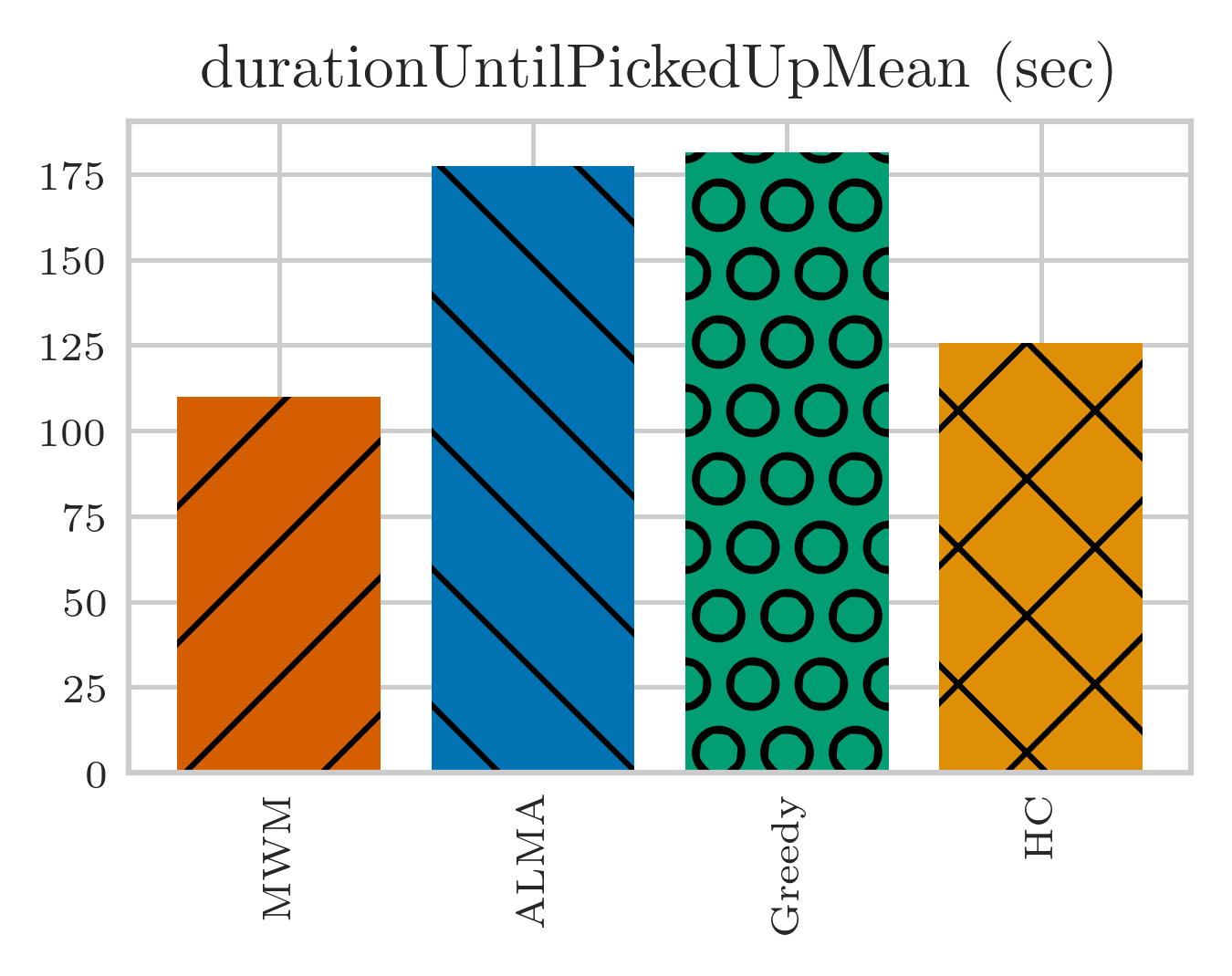}
		\caption{Time to Pick-up (s)}
		\label{fig_appendix: Jan15Hour8to810Manhattan_durationUntilPickedUpMean}
	\end{subfigure}
	~
	\begin{subfigure}[t]{0.23\textwidth}
		\centering
		\includegraphics[width = 1 \linewidth, trim={0.6em 0.6em 0.5em 1.8em}, clip]{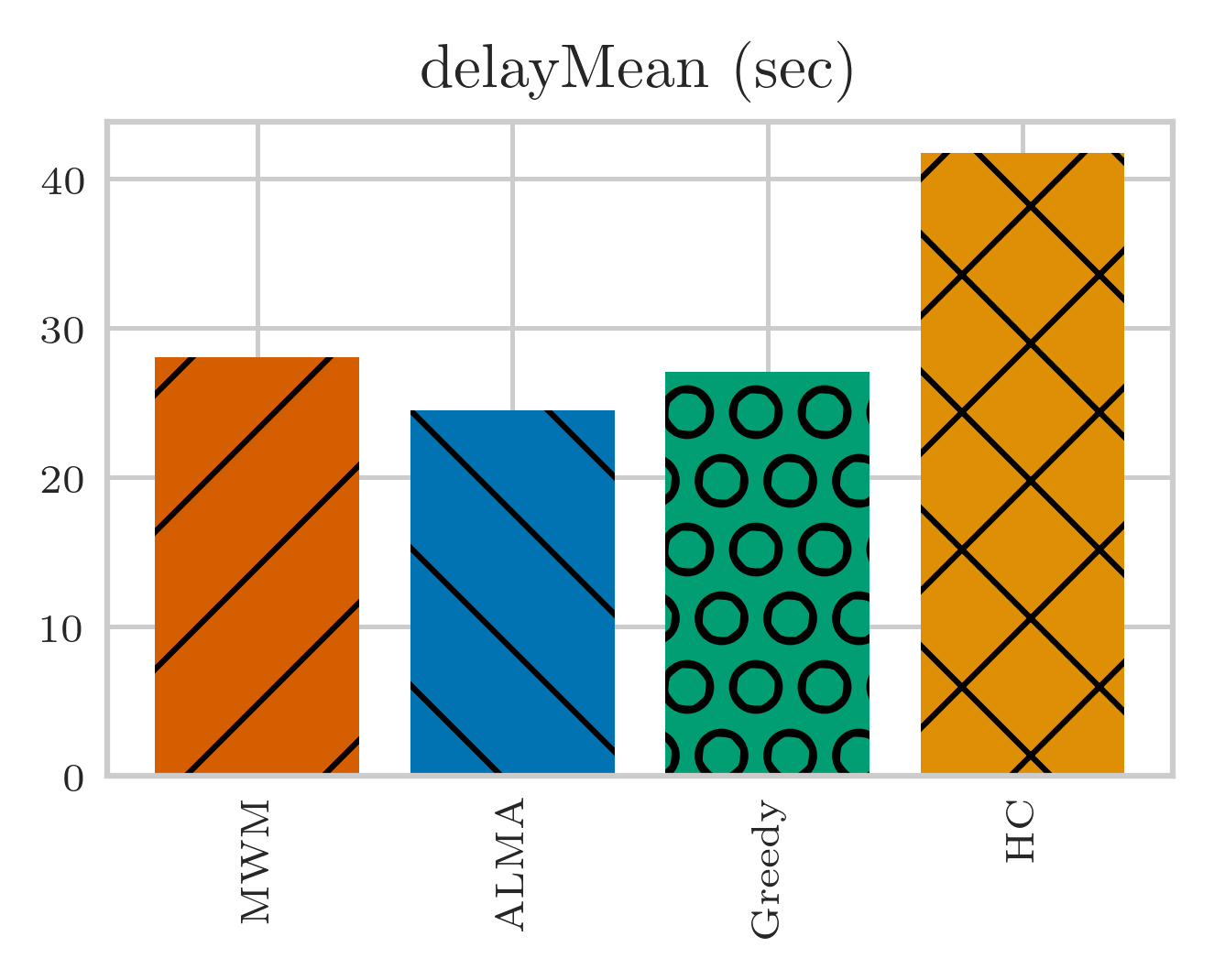}
		\caption{Delay (s)}
		\label{fig_appendix: Jan15Hour8to810Manhattan_delayMean}
	\end{subfigure}
	~
	\begin{subfigure}[t]{0.23\textwidth}
		\centering
		\includegraphics[width = 1 \linewidth, trim={0.6em 0.6em 0.5em 1.8em}, clip]{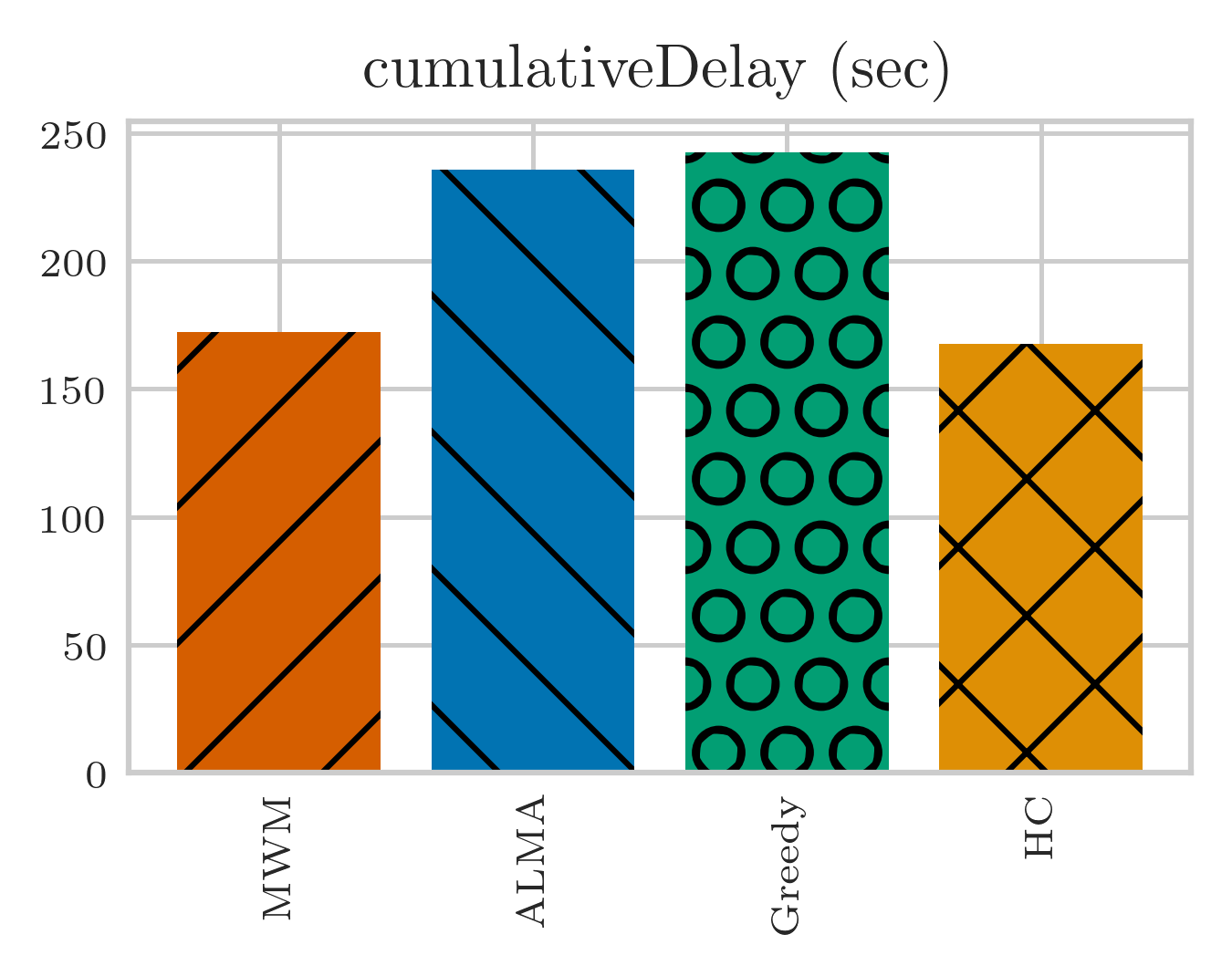}
		\caption{Cumulative Delay (s)}
		\label{fig_appendix: Jan15Hour8to810Manhattan_cumulativeDelay}
	\end{subfigure}
	~
	\begin{subfigure}[t]{0.23\textwidth}
		\centering
		\includegraphics[width = 1 \linewidth, trim={0.6em 0.6em 0.5em 1.8em}, clip]{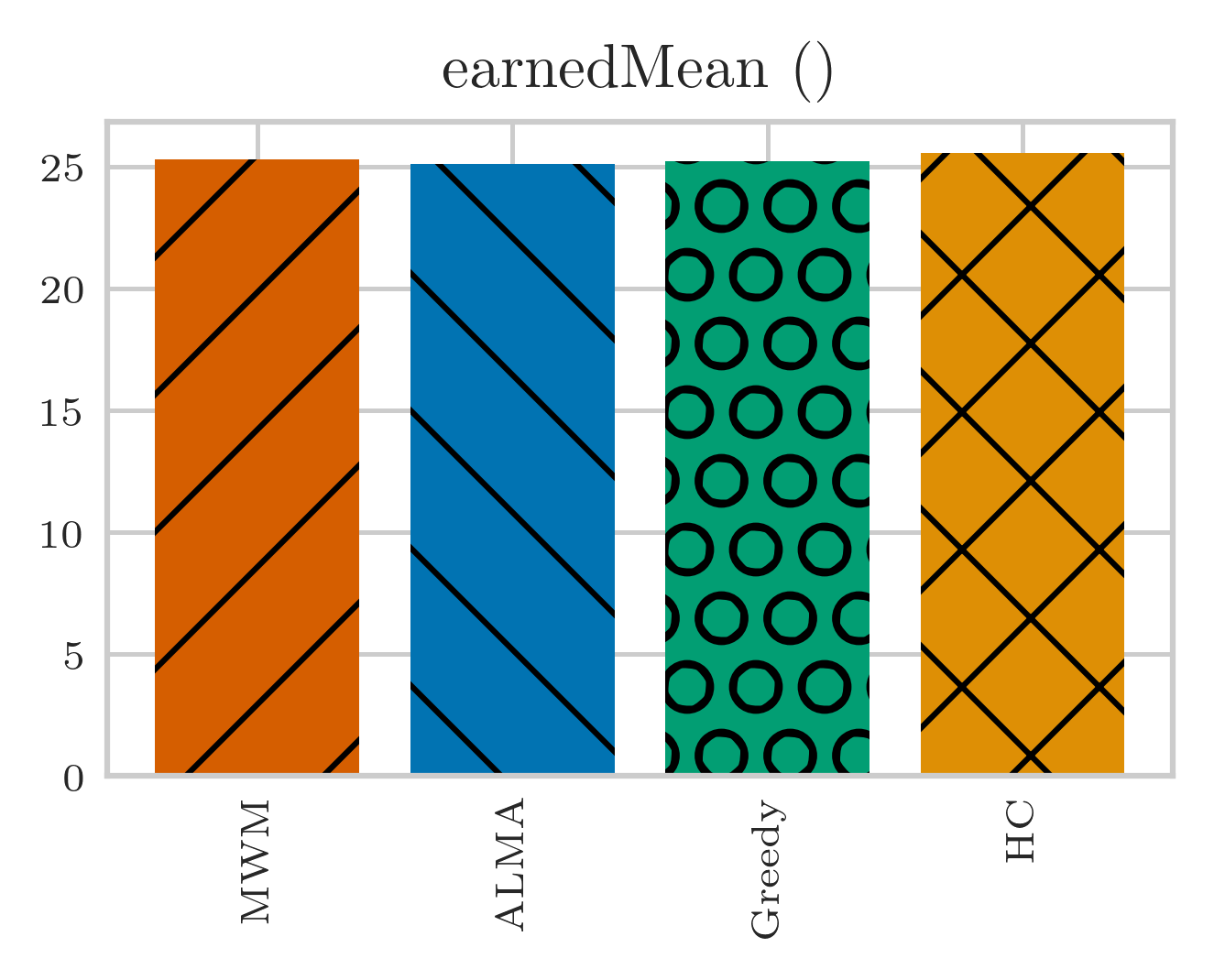}
		\caption{Driver Profit (\$)}
		\label{fig_appendix: Jan15Hour8to810Manhattan_earnedMean}
	\end{subfigure}

	\begin{subfigure}[t]{0.23\textwidth}
		\centering
		\includegraphics[width = 1 \linewidth, trim={0.6em 0.6em 0.5em 1.8em}, clip]{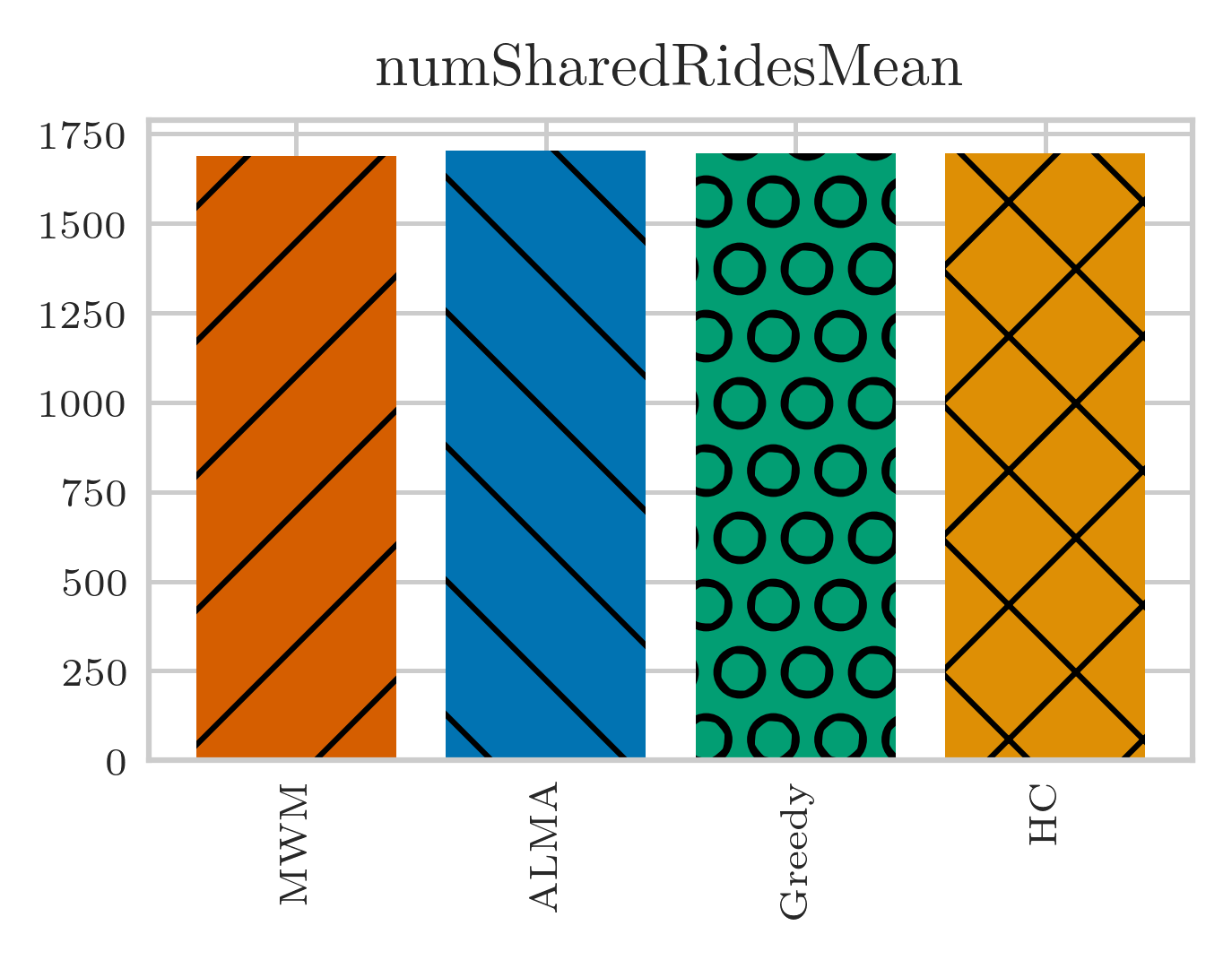}
		\caption{Number of Shared Rides}
		\label{fig_appendix: Jan15Hour8to810Manhattan_numSharedRidesMean}
	\end{subfigure}
	~
	\begin{subfigure}[t]{0.23\textwidth}
		\centering
		\includegraphics[width = 1 \linewidth, trim={0.6em 0.6em 0.5em 1.8em}, clip]{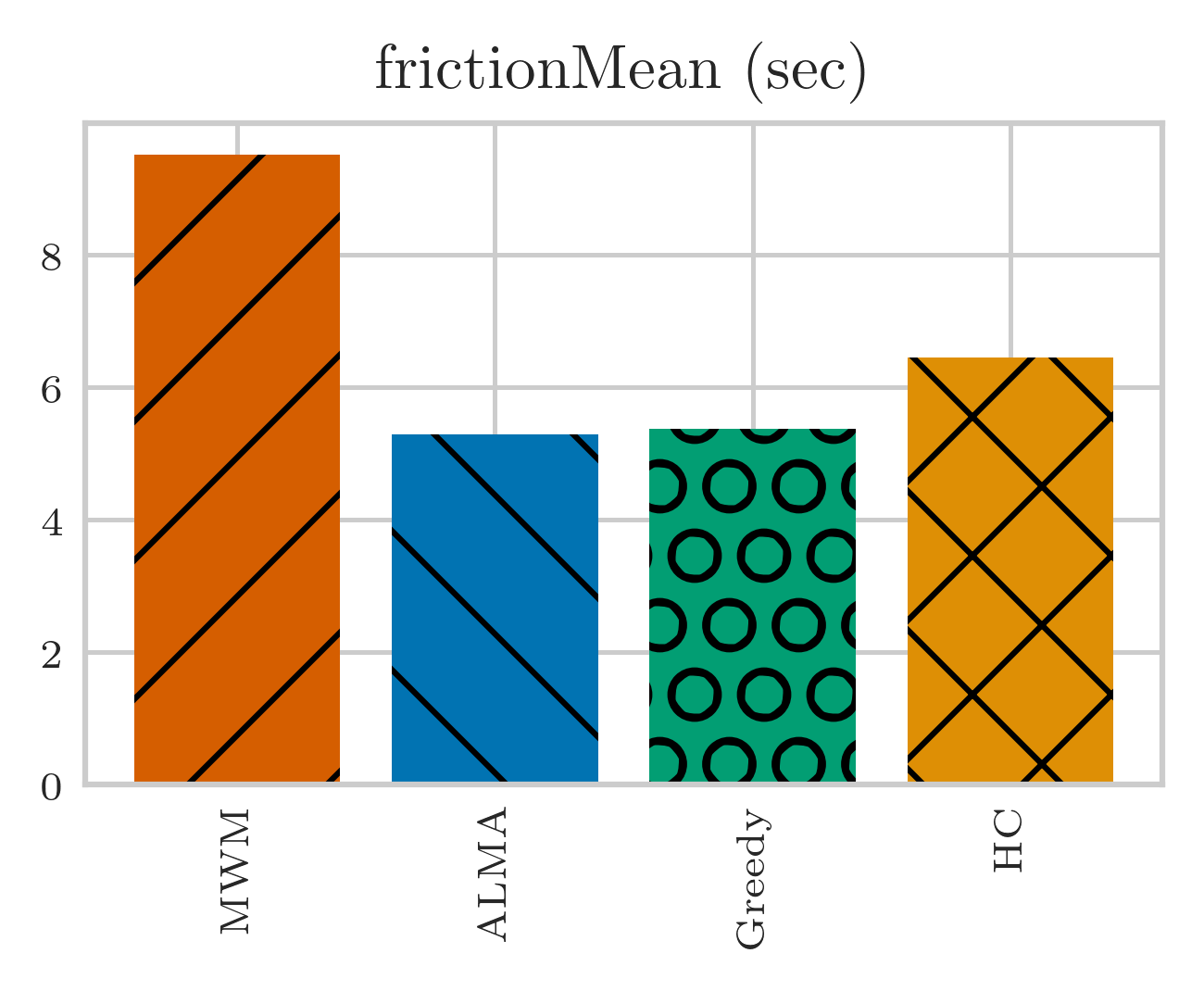}
		\caption{Frictions (s)}
		\label{fig_appendix: Jan15Hour8to810Manhattan_frictionMean}
	\end{subfigure}%
	\caption{January 15, 2016 -- 08:00 - 08:10 -- Manhattan -- \#Taxis = 2779 (base number).}
	\label{fig_appendix: Jan15Hour8to810Manhattan}
\end{figure*}

\begin{table*}[b!]
\centering
\caption{January 15, 2016 -- 08:00 - 09:00 -- Manhattan -- \#Taxis = 2779 (base number).}
\label{tb_appendix: Jan15Hour8to810Manhattan}
\resizebox{\textwidth}{!}{%
\begin{tabular}{@{}lccccccccccccccccccc@{}}
\toprule
\textbf{} & \textbf{\begin{tabular}[c]{@{}c@{}}Distance\\ Driven (m)\end{tabular}} & \textbf{SD} & \textbf{\begin{tabular}[c]{@{}c@{}}Elapsed\\ Time (ns)\end{tabular}} & \textbf{SD} & \textbf{\begin{tabular}[c]{@{}c@{}}Time to\\ Pair (s)\end{tabular}} & \textbf{SD} & \textbf{\begin{tabular}[c]{@{}c@{}}Time to Pair\\ with Taxi (s)\end{tabular}} & \textbf{SD} & \textbf{\begin{tabular}[c]{@{}c@{}}Time to\\ Pick-up (s)\end{tabular}} & \textbf{SD} & \textbf{Delay (s)} & \textbf{SD} & \textbf{\begin{tabular}[c]{@{}c@{}}Cumulative\\ Delay (s)\end{tabular}} & \textbf{\begin{tabular}[c]{@{}c@{}}Driver\\ Profit (\$)\end{tabular}} & \textbf{SD} & \textbf{\begin{tabular}[c]{@{}c@{}}Number of \\ Shared Rides\end{tabular}} & \textbf{SD} & \textbf{Frictions (s)} & \textbf{SD} \\ \midrule
\textbf{MWM}             & 8.38E+06 & 0.00E+00   & 9.92E+10 & 0.00E+00     & 34.32 & 30.93                  & 0.00 & 0.00            & 109.83 & 125.56 & 28.08 & 80.29            & 172.23        & 25.30 & 29.59                  & 1.69E+03 & 0.00      & 9.51 & 31.75 \\
\textbf{ALMA}            & 9.76E+06 & 6.10E+04   & 8.28E+09 & 1.93E+09     & 33.88 & 30.71                  & 0.00 & 0.00            & 177.21 & 216.99 & 24.49 & 75.40            & 235.58        & 25.09 & 26.38                  & 1.70E+03 & 9.76      & 5.29 & 22.26 \\
\textbf{Greedy}          & 9.91E+06 & 1.06E+04   & 3.66E+09 & 7.61E+08     & 34.16 & 30.88                  & 0.00 & 0.00            & 181.43 & 216.19 & 27.05 & 74.92            & 242.64        & 25.19 & 26.14                  & 1.70E+03 & 7.55      & 5.37 & 21.50 \\
\textbf{HC}              & 8.72E+06 & 0.00E+00   & 7.52E+12 & 0.00E+00     & 0.13  & 4.02                   & 0.06 & 2.78            & 125.69 & 155.09 & 41.77 & 106.45           & 167.65        & 25.56 & 29.61                  & 1.70E+03 & 0.00      & 6.45 & 27.19 \\ \bottomrule
\end{tabular}%
}
\end{table*}

\begin{table*}[b!]
\centering
\caption{January 15, 2016 -- 08:00 - 08:10 -- Manhattan -- \#Taxis = 2779 (base number). Each column presents the relative difference compared to the first line, i.e., the MWM (algorithm - MWM) / MWM, for each metric.}
\label{tb_appendix: Jan15Hour8to810ManhattanPercentages}
\resizebox{\textwidth}{!}{%
\begin{tabular}{@{}lccccccccccccccccccc@{}}
\toprule
\textbf{} & \textbf{\begin{tabular}[c]{@{}c@{}}Distance\\ Driven (m)\end{tabular}} & \textbf{SD} & \textbf{\begin{tabular}[c]{@{}c@{}}Elapsed\\ Time (ns)\end{tabular}} & \textbf{SD} & \textbf{\begin{tabular}[c]{@{}c@{}}Time to\\ Pair (s)\end{tabular}} & \textbf{SD} & \textbf{\begin{tabular}[c]{@{}c@{}}Time to Pair\\ with Taxi (s)\end{tabular}} & \textbf{SD} & \textbf{\begin{tabular}[c]{@{}c@{}}Time to\\ Pick-up (s)\end{tabular}} & \textbf{SD} & \textbf{Delay (s)} & \textbf{SD} & \textbf{\begin{tabular}[c]{@{}c@{}}Cumulative\\ Delay (s)\end{tabular}} & \textbf{\begin{tabular}[c]{@{}c@{}}Driver\\ Profit (\$)\end{tabular}} & \textbf{SD} & \textbf{\begin{tabular}[c]{@{}c@{}}Number of \\ Shared Rides\end{tabular}} & \textbf{SD} & \textbf{Frictions (s)} & \textbf{SD} \\ \midrule
\textbf{MWM}             & 0.00\%  & --         & 0.00\%    & --           & 0.00\%   & 0.00\%                 & -- & --              & 0.00\%  & 0.00\%  & 0.00\%   & 0.00\%           & 0.00\%        & 0.00\%  & 0.00\%                 & 0.00\% & --        & 0.00\%   & 0.00\%   \\
\textbf{ALMA}            & 16.40\% & --         & -91.65\%  & --           & -1.28\%  & -0.70\%                & -- & --              & 61.34\% & 72.81\% & -12.80\% & -6.09\%          & 36.78\%       & -0.83\% & -10.87\%               & 0.95\% & --        & -44.37\% & -29.90\% \\
\textbf{Greedy}          & 18.17\% & --         & -96.31\%  & --           & -0.45\%  & -0.15\%                & -- & --              & 65.18\% & 72.18\% & -3.67\%  & -6.68\%          & 40.88\%       & -0.43\% & -11.65\%               & 0.44\% & --        & -43.57\% & -32.29\% \\
\textbf{HC}              & 3.98\%  & --         & 7474.61\% & --           & -99.61\% & -87.01\%               & -- & --              & 14.44\% & 23.52\% & 48.74\%  & 32.59\%          & -2.66\%       & 1.02\%  & 0.05\%                 & 0.47\% & --        & -32.16\% & -14.37\% \\ \bottomrule
\end{tabular}%
}
\end{table*}

\clearpage


\begin{figure*}[t!]
\subsection{Dynamic Vehicle Relocation -- 00:00 - 23:59 (full day) -- Manhattan} \label{Appendix Jan15ManhattanRelocation}
\end{figure*}

\begin{figure*}[t!]
	\centering
	\begin{subfigure}[t]{0.23\textwidth}
		\centering
		\includegraphics[width = 1 \linewidth, trim={0.6em 0.6em 0.5em 1.7em}, clip]{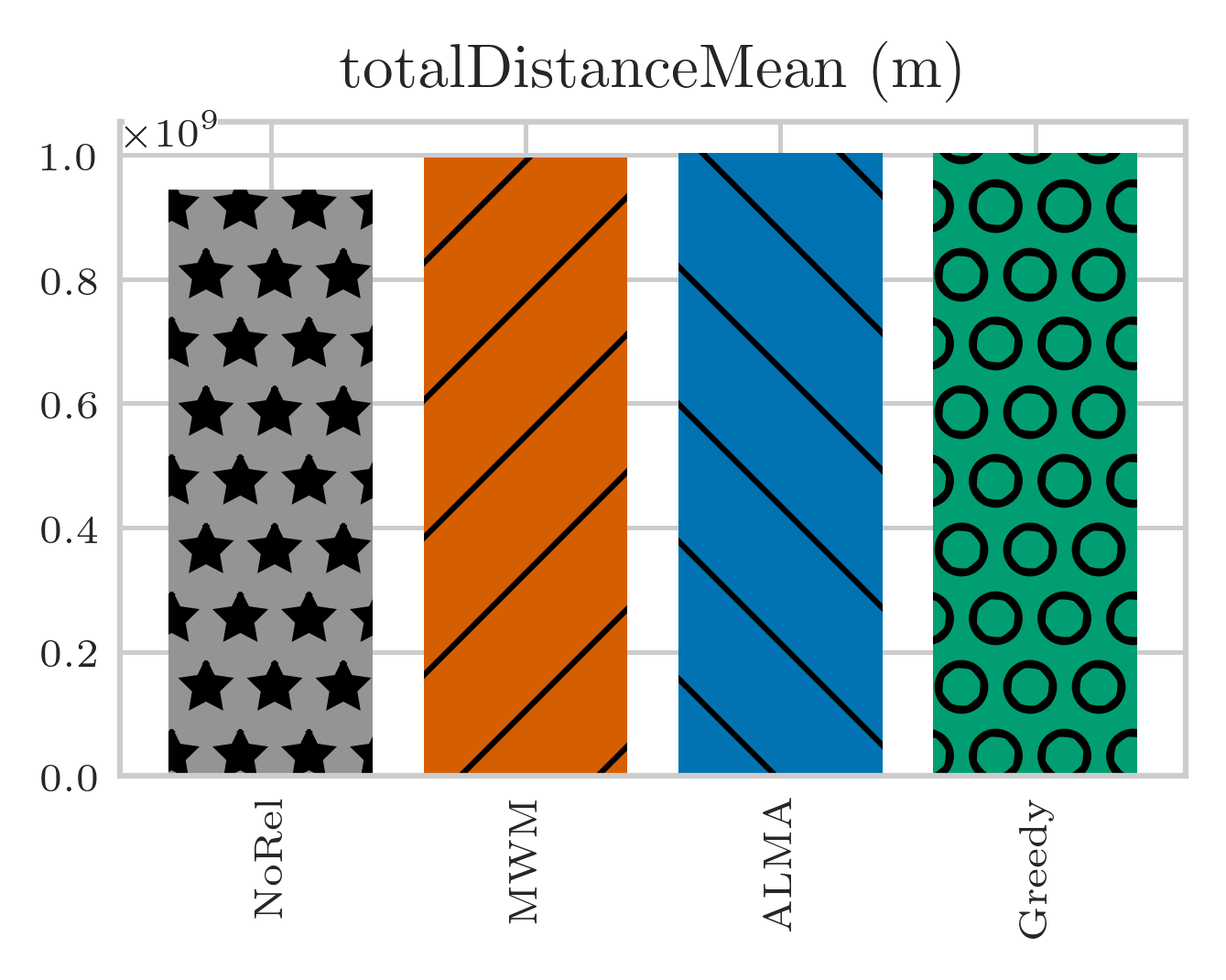}
		\caption{Total Distance Driven (m)}
		\label{fig_appendix: Jan15ManhattanRelocation/plots_totalDistanceMean}
	\end{subfigure}
	~ 
	\begin{subfigure}[t]{0.23\textwidth}
		\centering
		\includegraphics[width = 1 \linewidth, trim={0.6em 0.6em 0.5em 1.8em}, clip]{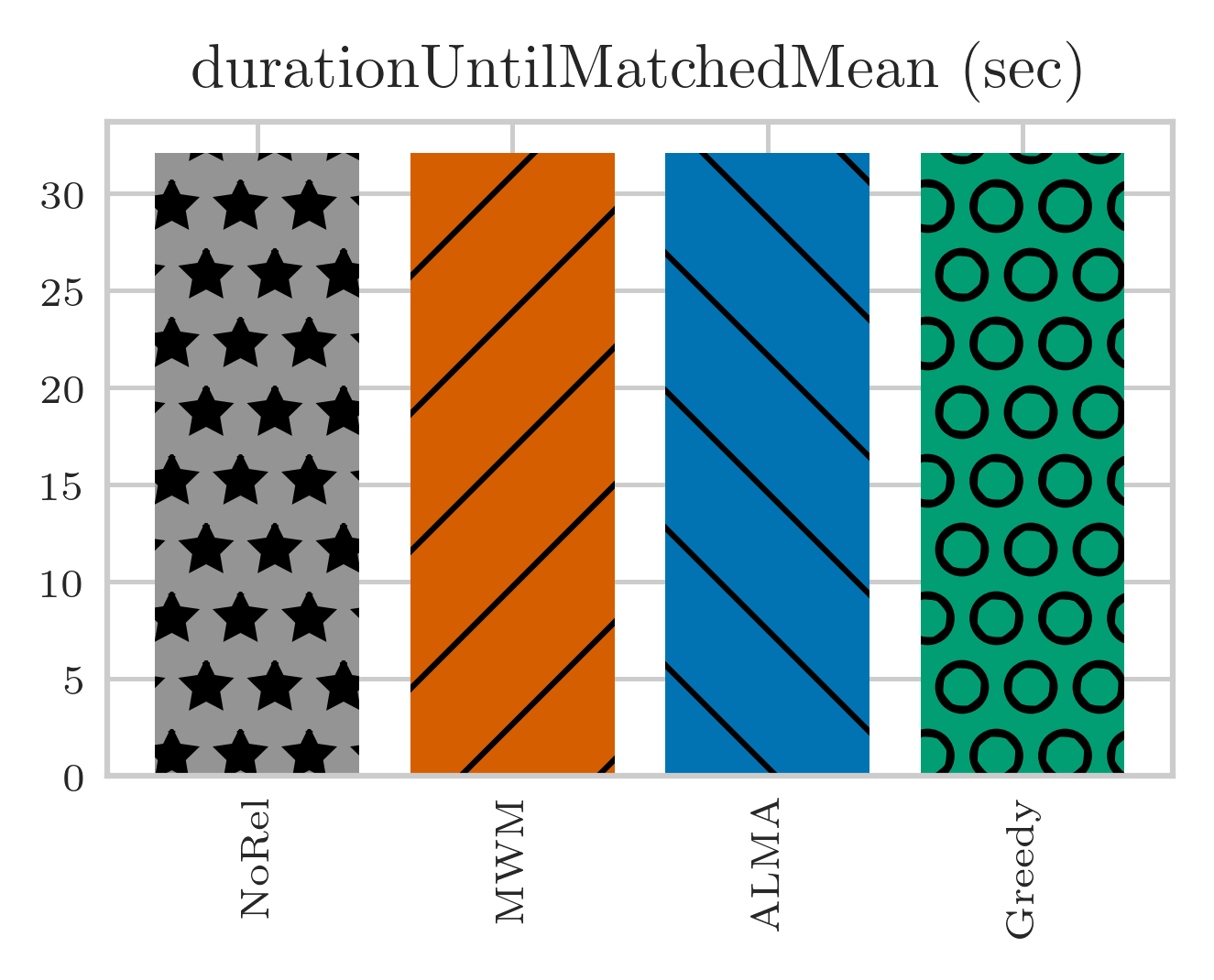}
		\caption{Time to Pair (s)}
		\label{fig_appendix: Jan15ManhattanRelocation/plots_durationUntilMatchedMean}
	\end{subfigure}
	~
	\begin{subfigure}[t]{0.23\textwidth}
		\centering
		\includegraphics[width = 1 \linewidth, trim={0.6em 0.6em 0.5em 1.8em}, clip]{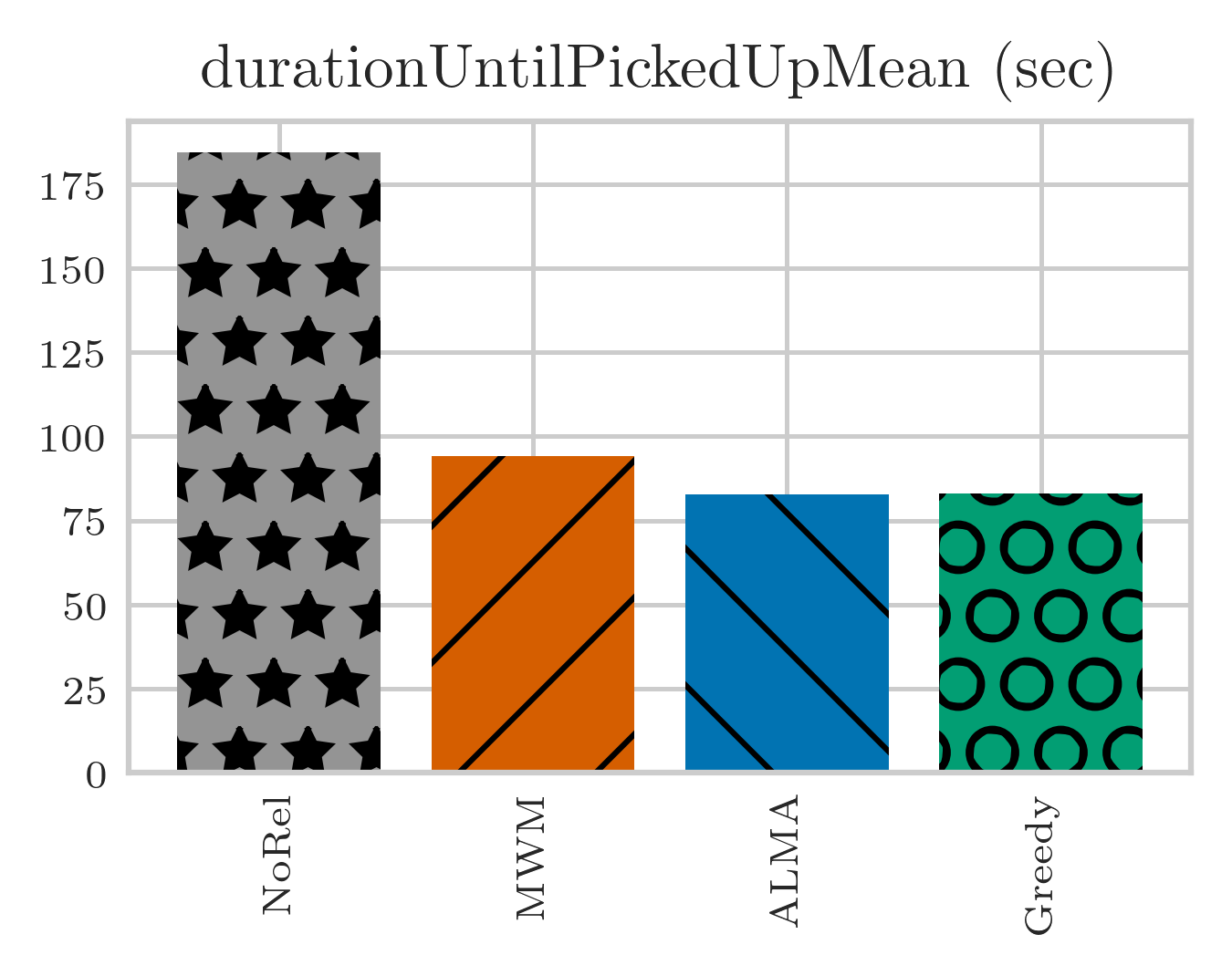}
		\caption{Time to Pick-up (s)}
		\label{fig_appendix: Jan15ManhattanRelocation/plots_durationUntilPickedUpMean}
	\end{subfigure}
	~
	\begin{subfigure}[t]{0.23\textwidth}
		\centering
		\includegraphics[width = 1 \linewidth, trim={0.6em 0.6em 0.5em 1.8em}, clip]{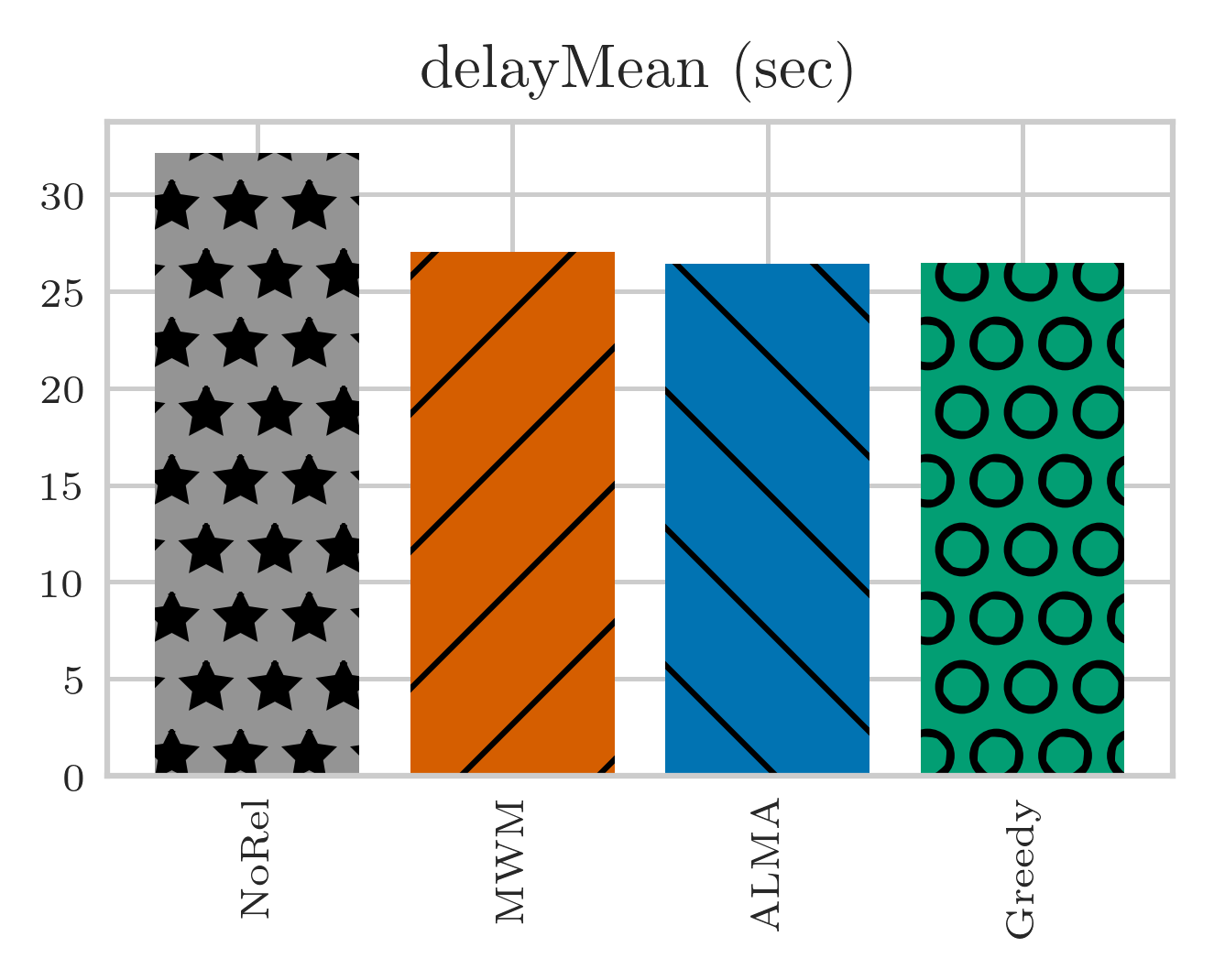}
		\caption{Delay (s)}
		\label{fig_appendix: Jan15ManhattanRelocation/plots_delayMean}
	\end{subfigure}

	\begin{subfigure}[t]{0.23\textwidth}
		\centering
		\includegraphics[width = 1 \linewidth, trim={0.6em 0.6em 0.5em 1.8em}, clip]{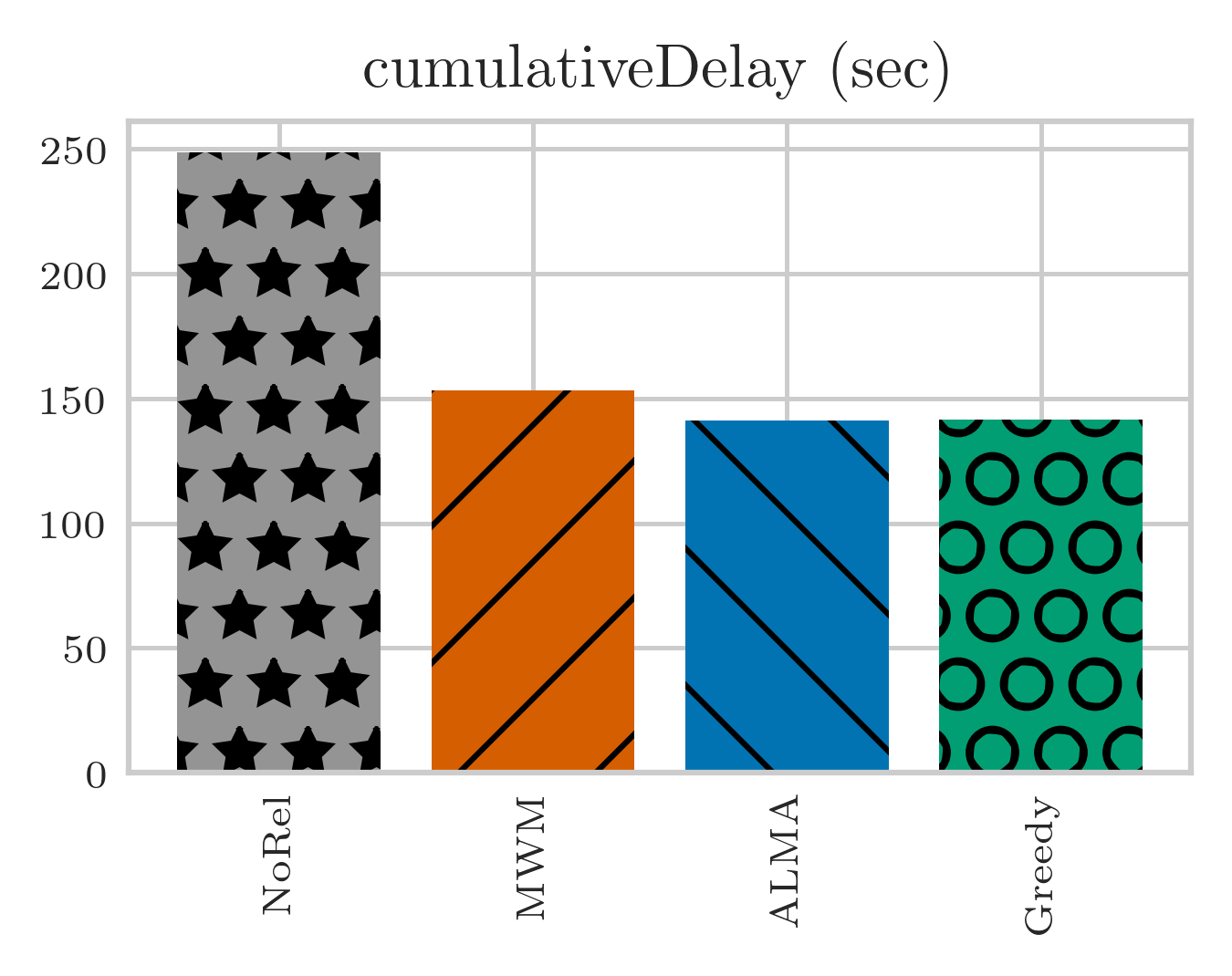}
		\caption{Cumulative Delay (s)}
		\label{fig_appendix: Jan15ManhattanRelocation/plots_cumulativeDelay}
	\end{subfigure}
	~
	\begin{subfigure}[t]{0.23\textwidth}
		\centering
		\includegraphics[width = 1 \linewidth, trim={0.6em 0.6em 0.5em 1.8em}, clip]{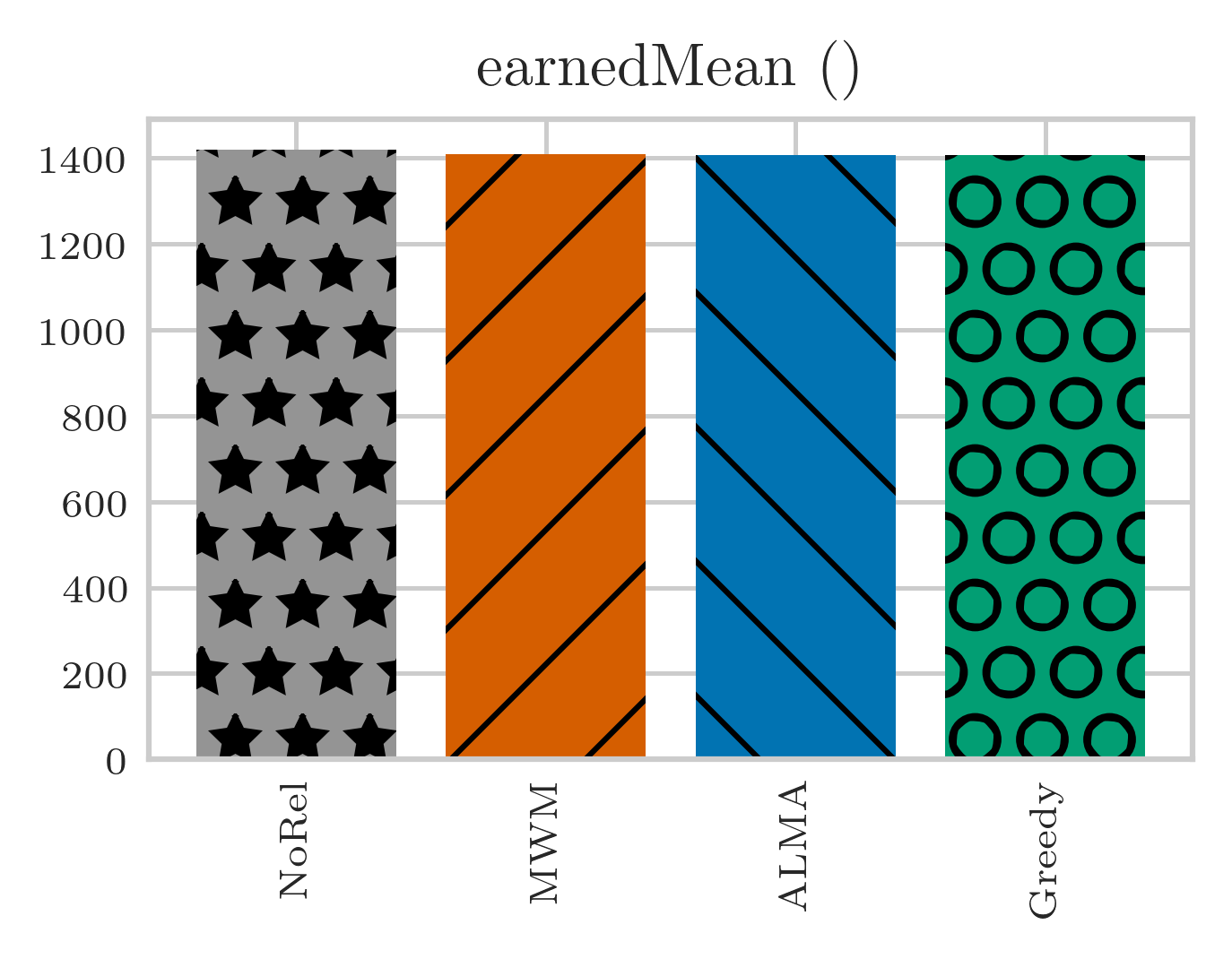}
		\caption{Driver Profit (\$)}
		\label{fig_appendix: Jan15ManhattanRelocation/plots_earnedMean}
	\end{subfigure}
	~
	\begin{subfigure}[t]{0.23\textwidth}
		\centering
		\includegraphics[width = 1 \linewidth, trim={0.6em 0.6em 0.5em 1.8em}, clip]{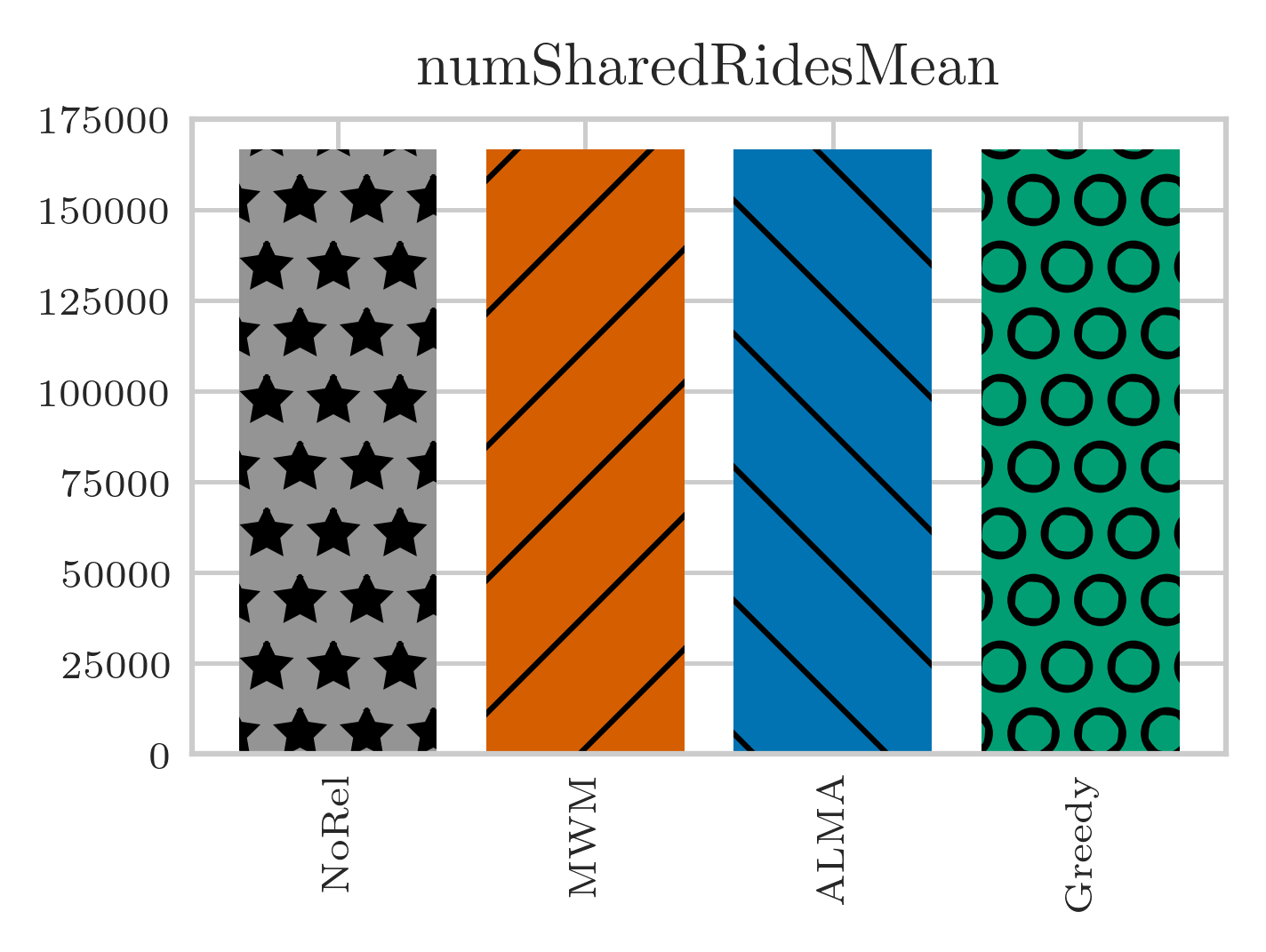}
		\caption{Number of Shared Rides}
		\label{fig_appendix: Jan15ManhattanRelocation/plots_numSharedRidesMean}
	\end{subfigure}
	~
	\begin{subfigure}[t]{0.23\textwidth}
		\centering
		\includegraphics[width = 1 \linewidth, trim={0.6em 0.6em 0.5em 1.8em}, clip]{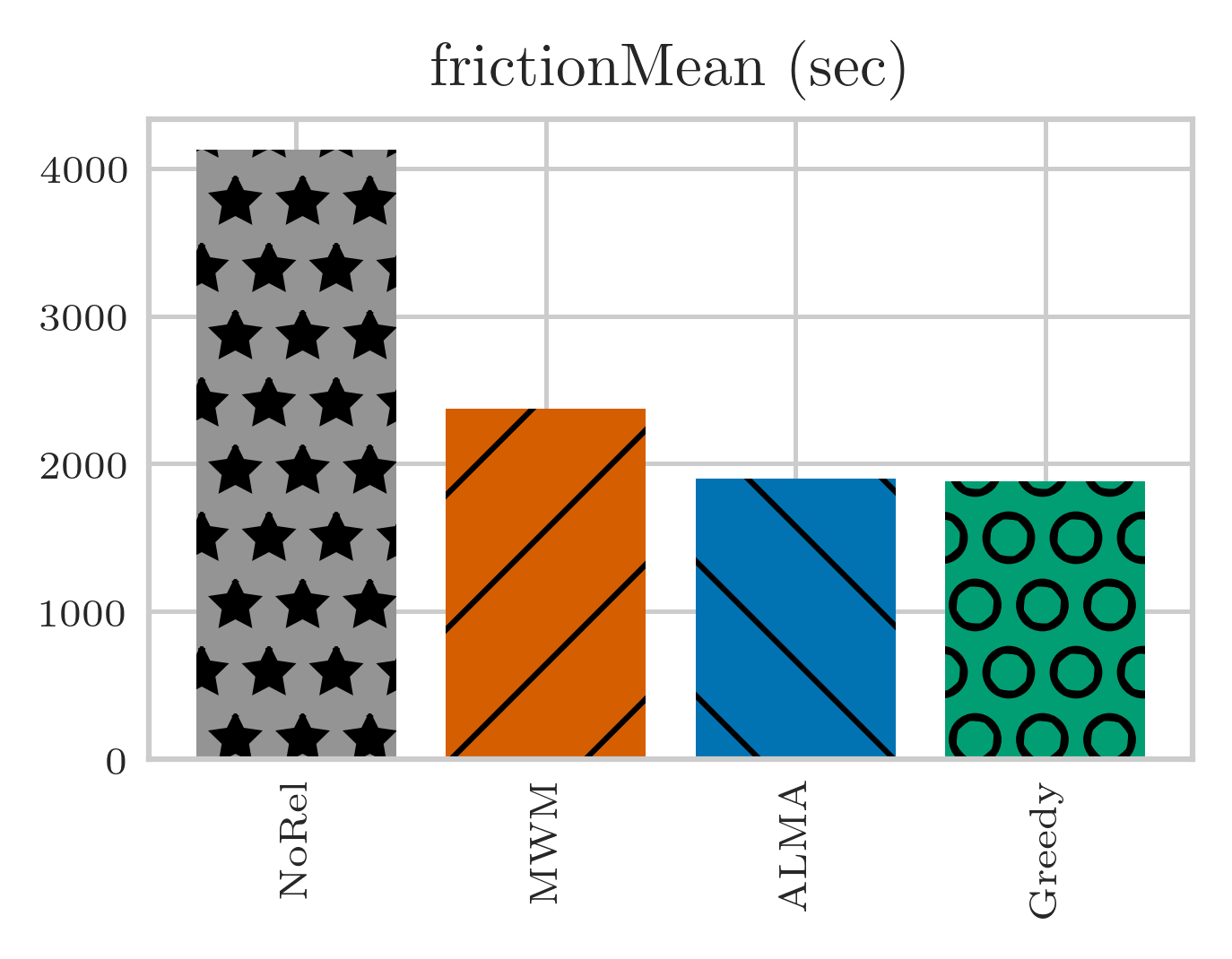}
		\caption{Frictions (s)}
		\label{fig_appendix: Jan15ManhattanRelocation/plots_frictionMean}
	\end{subfigure}%
	\caption{Dynamic Vehicle Relocation -- January 15, 2016 -- 00:00 - 23:59 (full day) -- Manhattan -- \#Taxis = 5081 (base number).}
	\label{fig_appendix: Jan15ManhattanRelocation}
\end{figure*}

\begin{table*}[b!]
\centering
\caption{Dynamic Vehicle Relocation -- January 15, 2016 -- 00:00 - 23:59 (full day) -- Manhattan -- \#Taxis = 5081 (base number).}
\label{tb_appendix: Jan15ManhattanRelocation}
\resizebox{\textwidth}{!}{%
\begin{tabular}{@{}lccccccccccccccccccc@{}}
\toprule
\textbf{} & \textbf{\begin{tabular}[c]{@{}c@{}}Distance\\ Driven (m)\end{tabular}} & \textbf{SD} & \textbf{\begin{tabular}[c]{@{}c@{}}Elapsed\\ Time (ns)\end{tabular}} & \textbf{SD} & \textbf{\begin{tabular}[c]{@{}c@{}}Time to\\ Pair (s)\end{tabular}} & \textbf{SD} & \textbf{\begin{tabular}[c]{@{}c@{}}Time to Pair\\ with Taxi (s)\end{tabular}} & \textbf{SD} & \textbf{\begin{tabular}[c]{@{}c@{}}Time to\\ Pick-up (s)\end{tabular}} & \textbf{SD} & \textbf{Delay (s)} & \textbf{SD} & \textbf{\begin{tabular}[c]{@{}c@{}}Cumulative\\ Delay (s)\end{tabular}} & \textbf{\begin{tabular}[c]{@{}c@{}}Driver\\ Profit (\$)\end{tabular}} & \textbf{SD} & \textbf{\begin{tabular}[c]{@{}c@{}}Number of \\ Shared Rides\end{tabular}} & \textbf{SD} & \textbf{Frictions (s)} & \textbf{SD} \\ \midrule
\textbf{NoRel}           & 9.45E+08 & --         & -- & --           & 32.10 & 30.84                  & 0.00 & 0.00            & 184.55 & 274.34 & 32.14 & 87.69            & 248.79        & 1420.37 & 895.19                 & 1.67E+05 & 0.00      & 4127.47 & 10597.84 \\
\textbf{MWM}             & 9.97E+08 & --         & -- & --           & 32.10 & 30.84                  & 0.00 & 0.00            & 94.21  & 129.01 & 27.01 & 70.81            & 153.33        & 1408.73 & 674.38                 & 1.67E+05 & 0.00      & 2372.57 & 5366.92  \\
\textbf{ALMA}            & 1.00E+09 & --         & -- & --           & 32.10 & 30.84                  & 0.00 & 0.00            & 82.72  & 114.61 & 26.42 & 69.31            & 141.24        & 1407.37 & 447.49                 & 1.67E+05 & 0.00      & 1898.47 & 2739.70  \\
\textbf{Greedy}          & 1.00E+09 & --         & -- & --           & 32.10 & 30.84                  & 0.00 & 0.00            & 82.99  & 114.65 & 26.44 & 69.29            & 141.53        & 1407.41 & 440.30                 & 1.67E+05 & 0.00      & 1880.72 & 2962.94  \\ \bottomrule
\end{tabular}%
}
\end{table*}

\begin{table*}[b!]
\centering
\caption{Dynamic Vehicle Relocation -- January 15, 2016 -- 00:00 - 23:59 (full day) -- Manhattan -- \#Taxis = 5081 (base number). Each column presents the relative difference compared to not using relocation (algorithm - NoRel) / NoRel, for each metric.}
\label{tb_appendix: Jan15ManhattanRelocationPercentages}
\resizebox{\textwidth}{!}{%
\begin{tabular}{@{}lccccccccccccccccccc@{}}
\toprule
\textbf{} & \textbf{\begin{tabular}[c]{@{}c@{}}Distance\\ Driven (m)\end{tabular}} & \textbf{SD} & \textbf{\begin{tabular}[c]{@{}c@{}}Elapsed\\ Time (ns)\end{tabular}} & \textbf{SD} & \textbf{\begin{tabular}[c]{@{}c@{}}Time to\\ Pair (s)\end{tabular}} & \textbf{SD} & \textbf{\begin{tabular}[c]{@{}c@{}}Time to Pair\\ with Taxi (s)\end{tabular}} & \textbf{SD} & \textbf{\begin{tabular}[c]{@{}c@{}}Time to\\ Pick-up (s)\end{tabular}} & \textbf{SD} & \textbf{Delay (s)} & \textbf{SD} & \textbf{\begin{tabular}[c]{@{}c@{}}Cumulative\\ Delay (s)\end{tabular}} & \textbf{\begin{tabular}[c]{@{}c@{}}Driver\\ Profit (\$)\end{tabular}} & \textbf{SD} & \textbf{\begin{tabular}[c]{@{}c@{}}Number of \\ Shared Rides\end{tabular}} & \textbf{SD} & \textbf{Frictions (s)} & \textbf{SD} \\ \midrule
\textbf{NoRel}           & 0.00\% & --         & -- & --           & 0.00\% & 0.00\%                 & -- & --              & 0.00\%   & 0.00\%   & 0.00\%   & 0.00\%           & 0.00\%        & 0.00\%  & 0.00\%                 & 0.00\% & --        & 0.00\%   & 0.00\%   \\
\textbf{MWM}             & 5.48\% & --         & -- & --           & 0.00\% & 0.00\%                 & -- & --              & -48.95\% & -52.97\% & -15.95\% & -19.25\%         & -38.37\%      & -0.82\% & -24.67\%               & 0.00\% & --        & -42.52\% & -49.36\% \\
\textbf{ALMA}            & 6.25\% & --         & -- & --           & 0.00\% & 0.00\%                 & -- & --              & -55.18\% & -58.22\% & -17.79\% & -20.96\%         & -43.23\%      & -0.92\% & -50.01\%               & 0.00\% & --        & -54.00\% & -74.15\% \\
\textbf{Greedy}          & 6.24\% & --         & -- & --           & 0.00\% & 0.00\%                 & -- & --              & -55.03\% & -58.21\% & -17.73\% & -20.98\%         & -43.11\%      & -0.91\% & -50.81\%               & 0.00\% & --        & -54.43\% & -72.04\% \\ \bottomrule
\end{tabular}%
}
\end{table*}

\clearpage


\begin{figure*}[t!]
\subsection{End-To-End Solution -- 00:00 - 23:59 (full day) -- Manhattan} \label{Appendix Jan15ManhattanRelocationEnd2End}
\end{figure*}

\begin{figure*}[t!]
	\centering
	\begin{subfigure}[t]{0.23\textwidth}
		\centering
		\includegraphics[width = 1 \linewidth, trim={0.6em 0.6em 0.5em 1.7em}, clip]{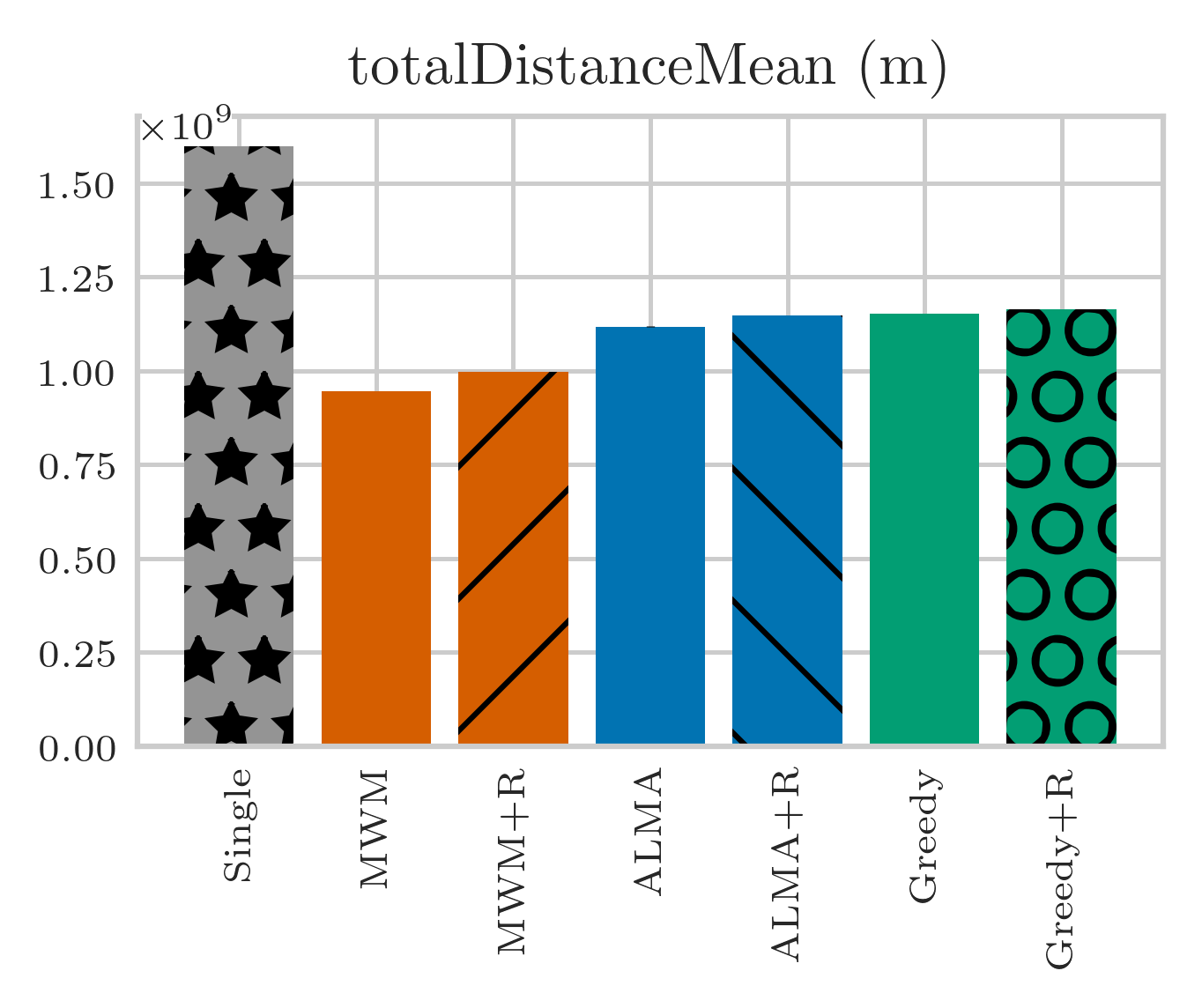}
		\caption{Total Distance Driven (m)}
		\label{fig_appendix: Jan15ManhattanRelocation/end2end_totalDistanceMean}
	\end{subfigure}
	~
	\begin{subfigure}[t]{0.23\textwidth}
		\centering
		\includegraphics[width = 1 \linewidth, trim={0.6em 0.6em 0.5em 1.8em}, clip]{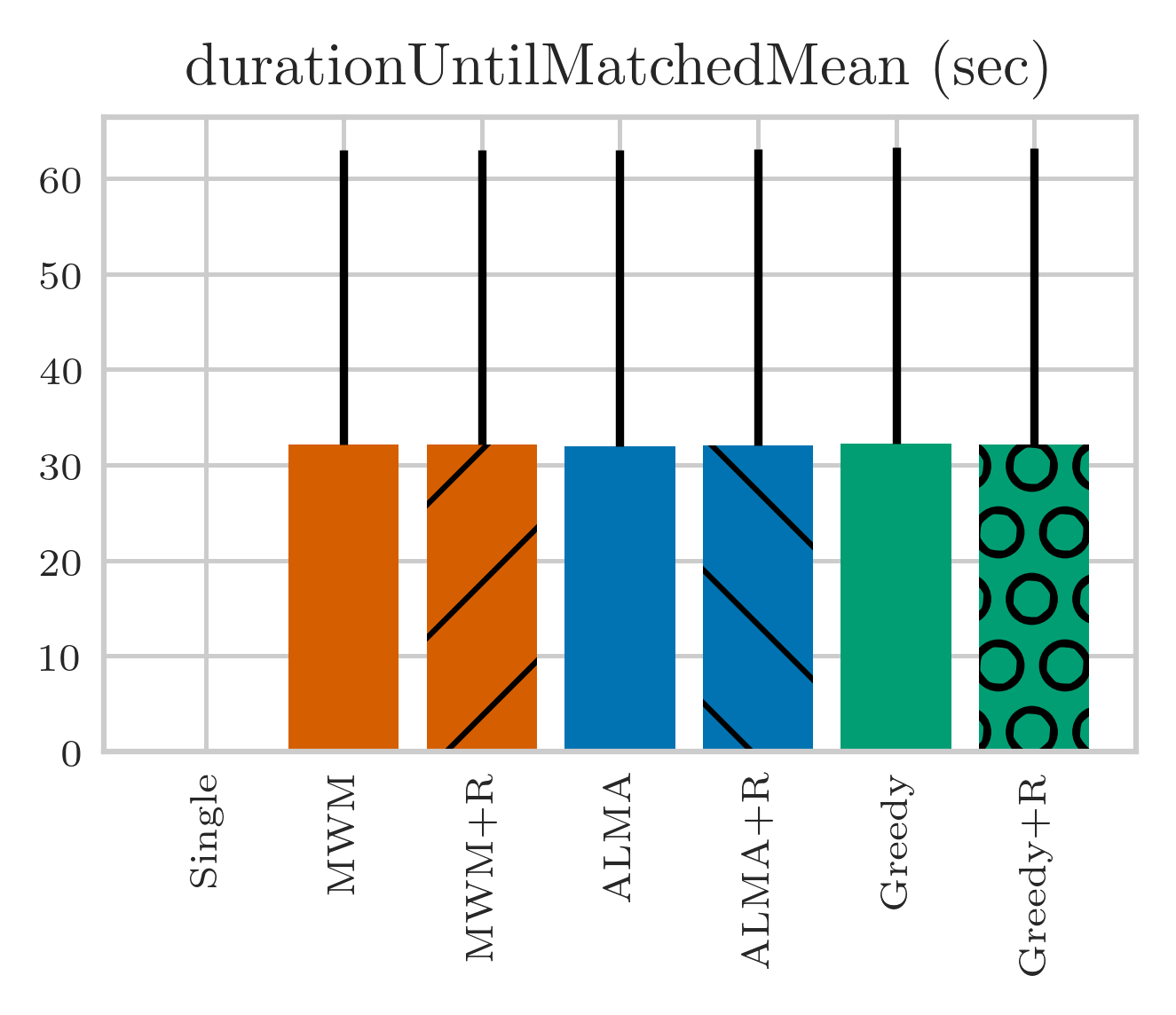}
		\caption{Time to Pair (s)}
		\label{fig_appendix: Jan15ManhattanRelocation/end2end_durationUntilMatchedMean}
	\end{subfigure}
	~
	\begin{subfigure}[t]{0.23\textwidth}
		\centering
		\includegraphics[width = 1 \linewidth, trim={0.6em 0.6em 0.5em 1.8em}, clip]{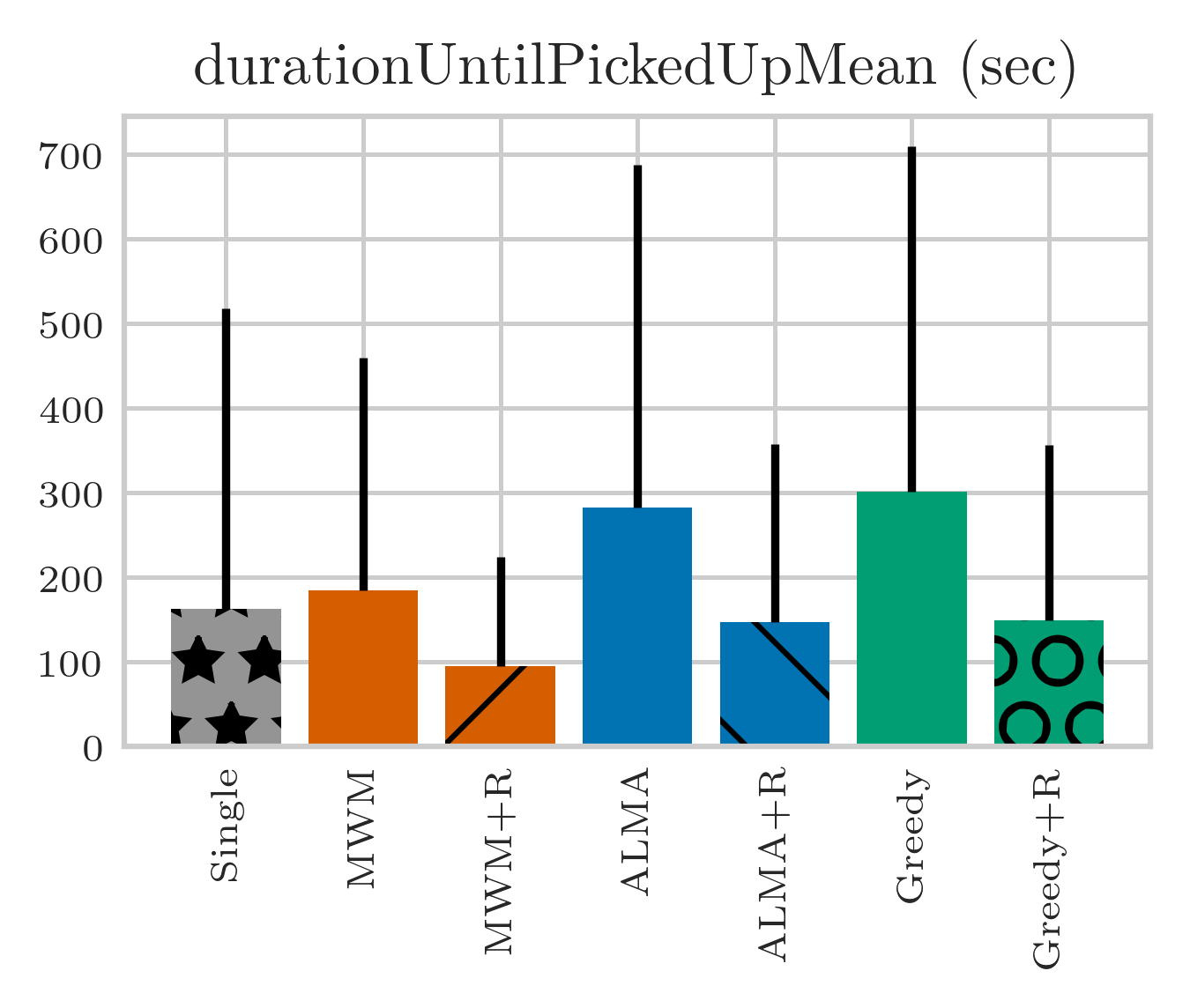}
		\caption{Time to Pick-up (s)}
		\label{fig_appendix: Jan15ManhattanRelocation/end2end_durationUntilPickedUpMean}
	\end{subfigure}
	~
	\begin{subfigure}[t]{0.23\textwidth}
		\centering
		\includegraphics[width = 1 \linewidth, trim={0.6em 0.6em 0.5em 1.8em}, clip]{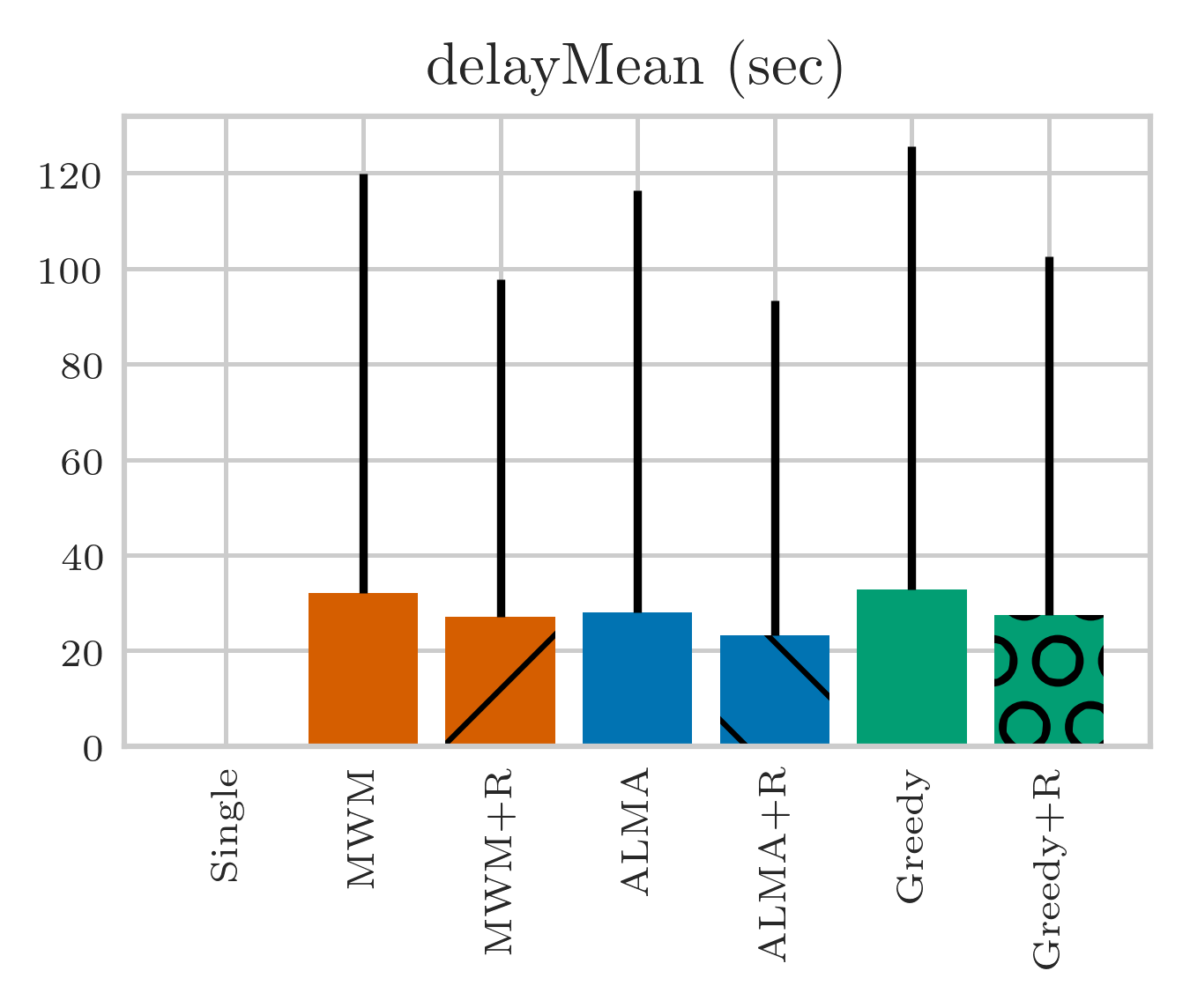}
		\caption{Delay (s)}
		\label{fig_appendix: Jan15ManhattanRelocation/end2end_delayMean}
	\end{subfigure}

	\begin{subfigure}[t]{0.23\textwidth}
		\centering
		\includegraphics[width = 1 \linewidth, trim={0.6em 0.6em 0.5em 1.8em}, clip]{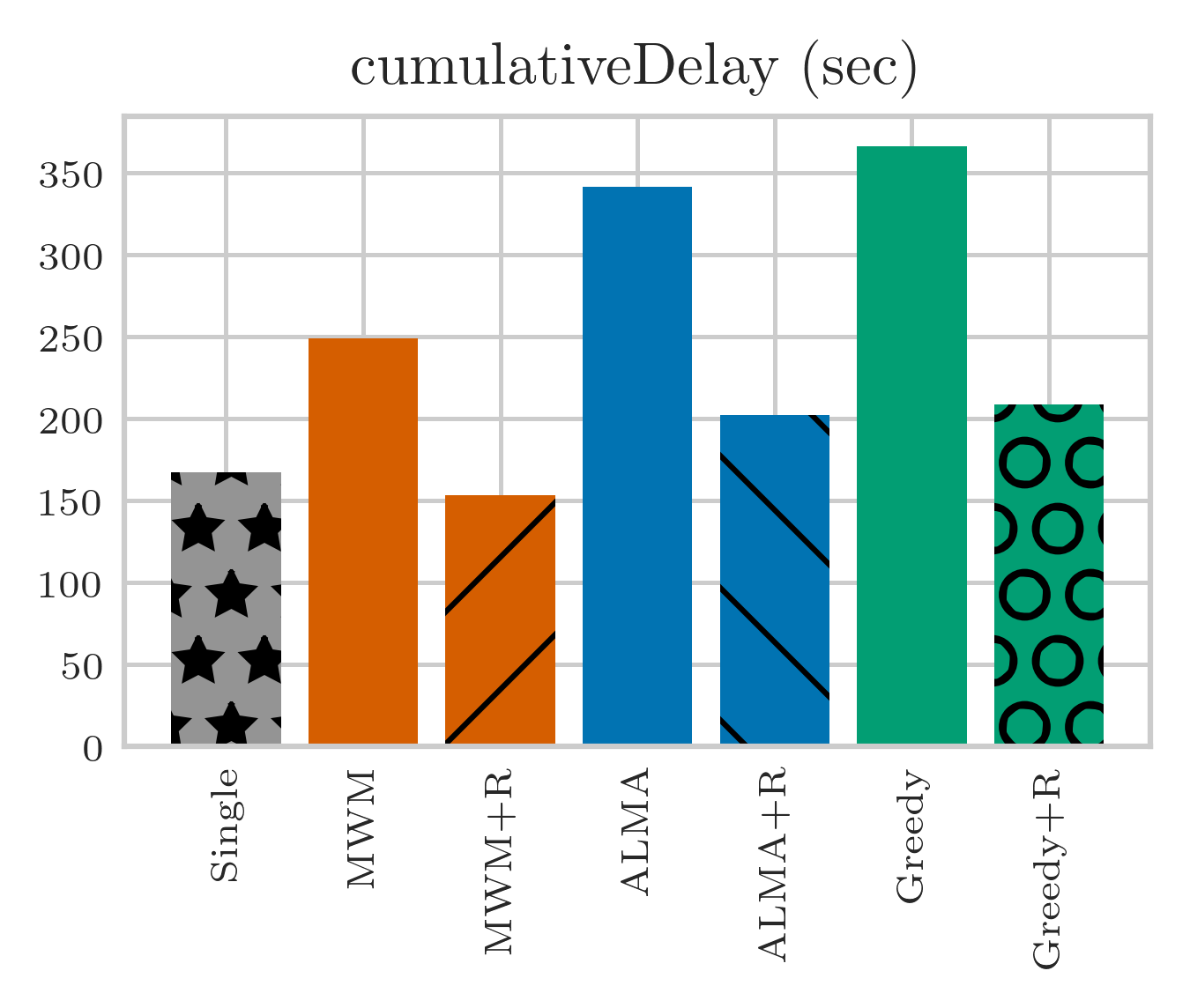}
		\caption{Cumulative Delay (s)}
		\label{fig_appendix: Jan15ManhattanRelocation/end2end_cumulativeDelay}
	\end{subfigure}
	~
	\begin{subfigure}[t]{0.23\textwidth}
		\centering
		\includegraphics[width = 1 \linewidth, trim={0.6em 0.6em 0.5em 1.8em}, clip]{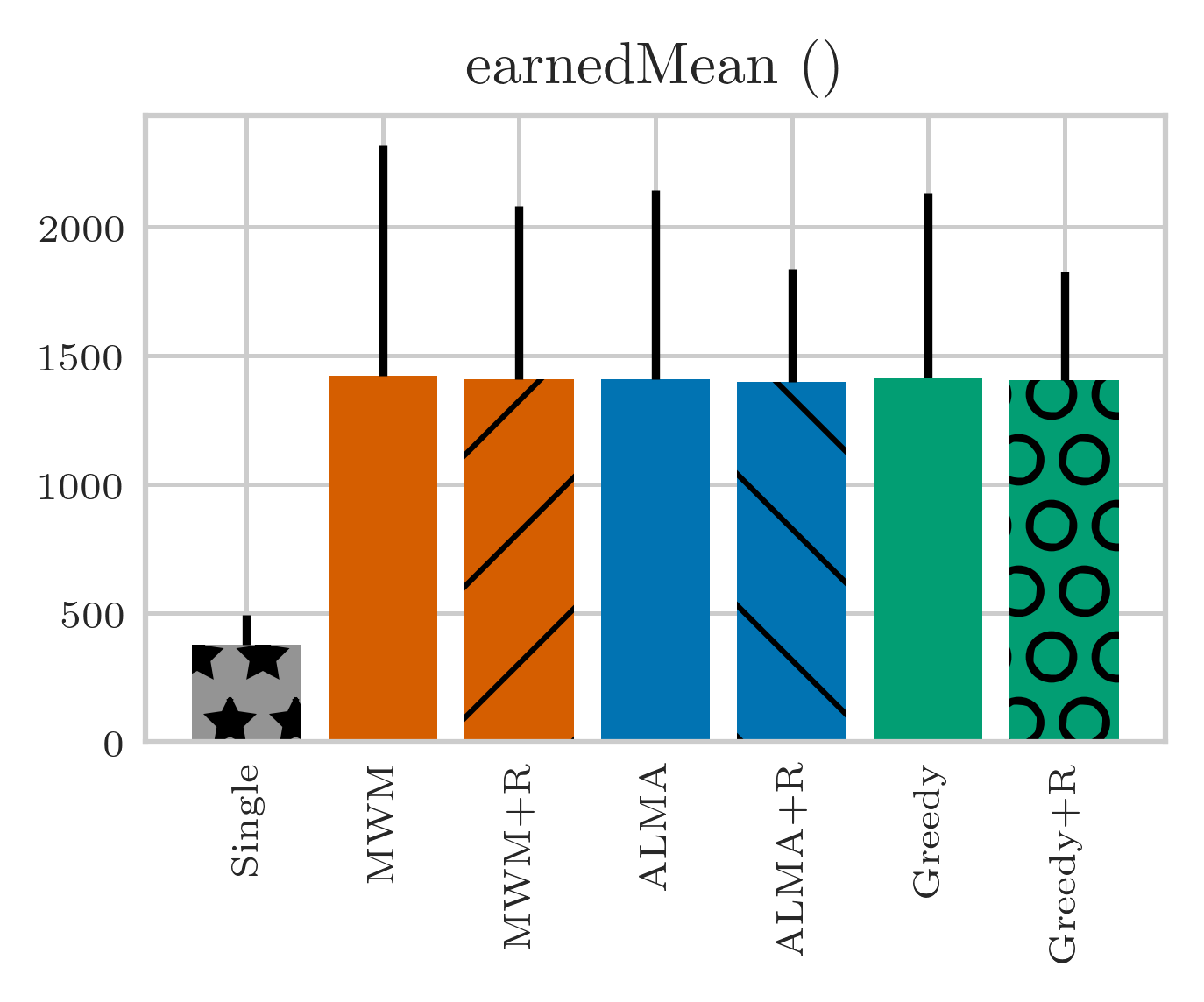}
		\caption{Driver Profit (\$)}
		\label{fig_appendix: Jan15ManhattanRelocation/end2end_earnedMean}
	\end{subfigure}
	~
	\begin{subfigure}[t]{0.23\textwidth}
		\centering
		\includegraphics[width = 1 \linewidth, trim={0.6em 0.6em 0.5em 1.8em}, clip]{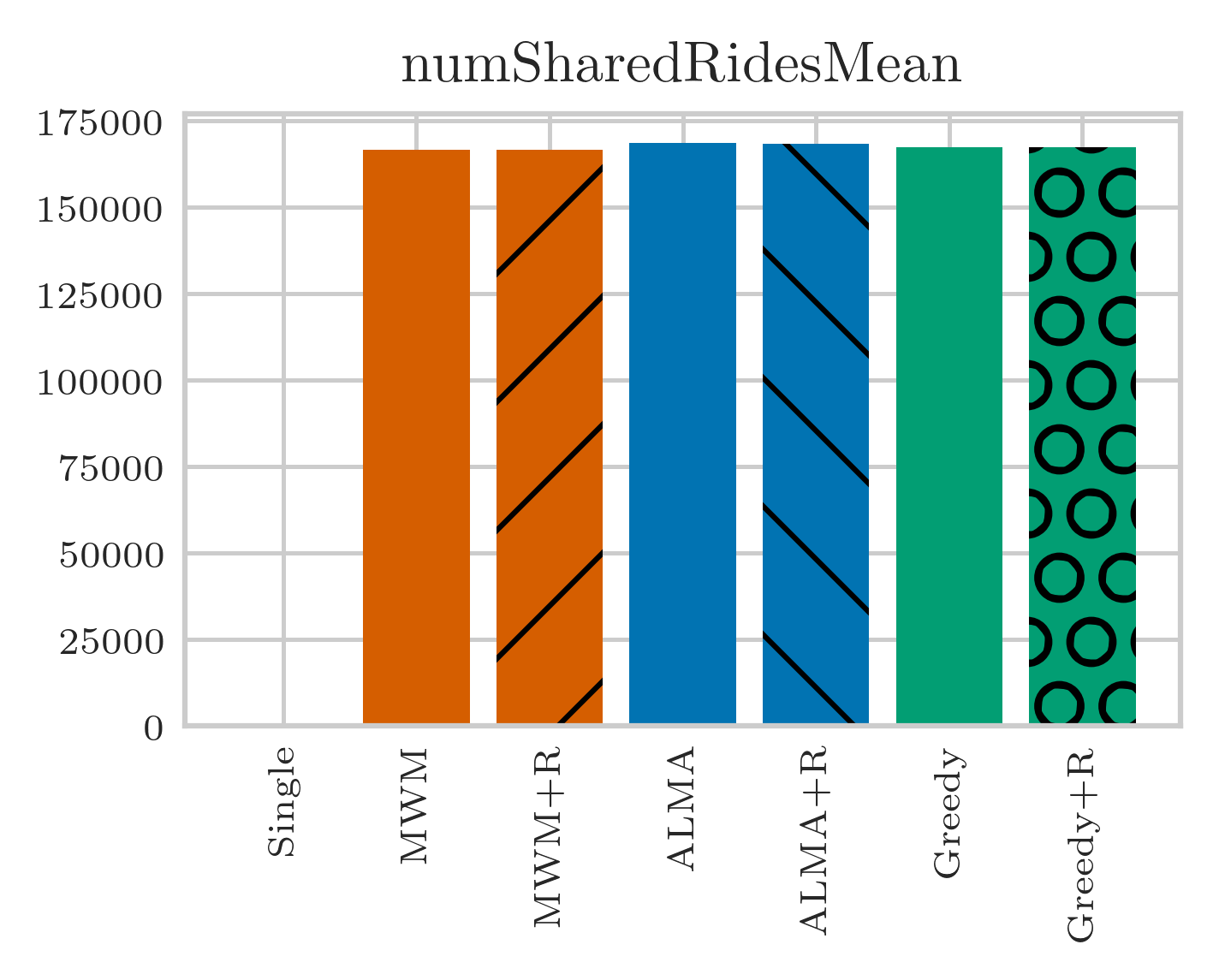}
		\caption{Number of Shared Rides}
		\label{fig_appendix: Jan15ManhattanRelocation/end2end_numSharedRidesMean}
	\end{subfigure}
	~
	\begin{subfigure}[t]{0.23\textwidth}
		\centering
		\includegraphics[width = 1 \linewidth, trim={0.6em 0.6em 0.5em 1.8em}, clip]{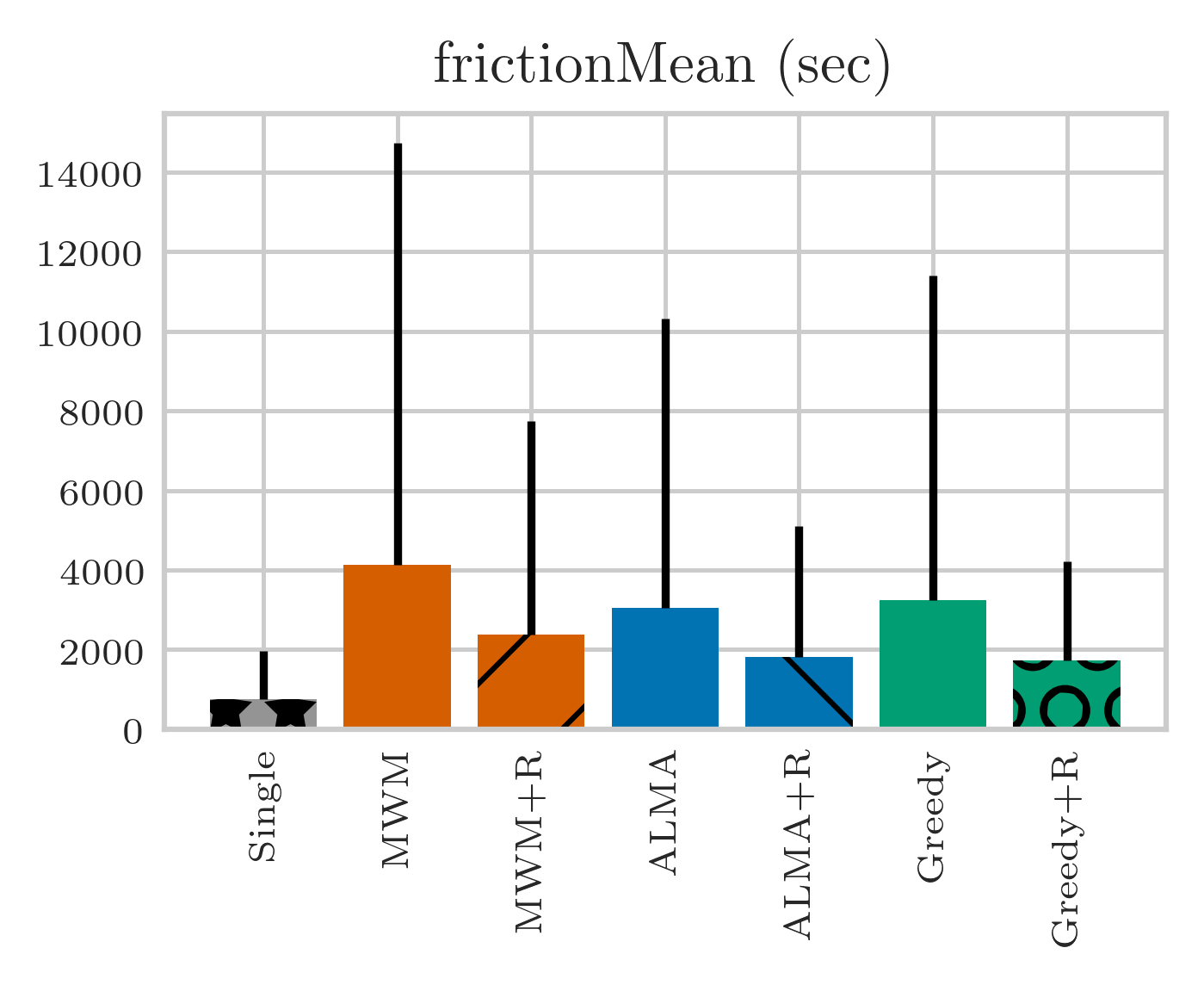}
		\caption{Frictions (s)}
		\label{fig_appendix: Jan15ManhattanRelocation/end2end_frictionMean}
	\end{subfigure}%
	\caption{End-To-End Solution -- January 15, 2016 -- 00:00 - 23:59 (full day) -- Manhattan -- \#Taxis = 5081 (base number)}
	\label{fig_appendix: Jan15ManhattanRelocation_end2end}
\end{figure*}

\begin{table*}[b!]
\centering
\caption{End-To-End Solution -- January 15, 2016 -- 00:00 - 23:59 (full day) -- Manhattan -- \#Taxis = 5081 (base number).}
\label{tb_appendix: Jan15ManhattanRelocation_end2end}
\resizebox{\textwidth}{!}{%
\begin{tabular}{@{}lccccccccccccccccccc@{}}
\toprule
\textbf{} & \textbf{\begin{tabular}[c]{@{}c@{}}Distance\\ Driven (m)\end{tabular}} & \textbf{SD} & \textbf{\begin{tabular}[c]{@{}c@{}}Elapsed\\ Time (ns)\end{tabular}} & \textbf{SD} & \textbf{\begin{tabular}[c]{@{}c@{}}Time to\\ Pair (s)\end{tabular}} & \textbf{SD} & \textbf{\begin{tabular}[c]{@{}c@{}}Time to Pair\\ with Taxi (s)\end{tabular}} & \textbf{SD} & \textbf{\begin{tabular}[c]{@{}c@{}}Time to\\ Pick-up (s)\end{tabular}} & \textbf{SD} & \textbf{Delay (s)} & \textbf{SD} & \textbf{\begin{tabular}[c]{@{}c@{}}Cumulative\\ Delay (s)\end{tabular}} & \textbf{\begin{tabular}[c]{@{}c@{}}Driver\\ Profit (\$)\end{tabular}} & \textbf{SD} & \textbf{\begin{tabular}[c]{@{}c@{}}Number of \\ Shared Rides\end{tabular}} & \textbf{SD} & \textbf{Frictions (s)} & \textbf{SD} \\ \midrule
\textbf{Single}          & 1.60E+09 & --         & -- & --           & 0.00  & 0.00                   & 5.59 & 99.05           & 161.77 & 355.37 & 0.00  & 0.00             & 167.36        & 376.90  & 116.72                 & 0.00E+00 & --        & 756.43  & 1216.88  \\
\textbf{MWM}             & 9.45E+08 & --         & -- & --           & 32.10 & 30.84                  & 0.00 & 0.00            & 184.55 & 274.34 & 32.14 & 87.69            & 248.79        & 1420.37 & 895.19                 & 1.67E+05 & --        & 4127.47 & 10597.84 \\
\textbf{MWM+R}           & 9.97E+08 & --         & -- & --           & 32.10 & 30.84                  & 0.00 & 0.00            & 94.21  & 129.01 & 27.01 & 70.81            & 153.33        & 1408.73 & 674.38                 & 1.67E+05 & --        & 2372.57 & 5366.92  \\
\textbf{ALMA}            & 1.12E+09 & --         & -- & --           & 31.98 & 31.01                  & 0.00 & 0.00            & 281.70 & 405.36 & 28.02 & 88.30            & 341.70        & 1406.70 & 736.46                 & 1.68E+05 & --        & 3047.45 & 7264.56  \\
\textbf{ALMA+R}          & 1.15E+09 & --         & -- & --           & 32.02 & 31.04                  & 0.00 & 0.00            & 146.68 & 210.60 & 23.22 & 70.06            & 201.92        & 1397.14 & 440.45                 & 1.68E+05 & --        & 1815.54 & 3295.10  \\
\textbf{Greedy}          & 1.15E+09 & --         & -- & --           & 32.18 & 31.05                  & 0.00 & 0.00            & 301.04 & 407.70 & 32.85 & 92.71            & 366.07        & 1414.66 & 719.44                 & 1.67E+05 & --        & 3242.70 & 8167.00  \\
\textbf{Greedy+R}        & 1.16E+09 & --         & -- & --           & 32.17 & 31.02                  & 0.00 & 0.00            & 148.88 & 207.00 & 27.46 & 75.18            & 208.52        & 1404.95 & 422.37                 & 1.67E+05 & --        & 1726.67 & 2487.20  \\ \bottomrule
\end{tabular}%
}
\end{table*}

\begin{table*}[b!]
\centering
\caption{End-To-End Solution -- January 15, 2016 -- 00:00 - 23:59 (full day) -- Manhattan -- \#Taxis = 5081 (base number). Each column presents the relative difference compared to the Singe Ride baseline (algorithm - Singe) / Singe, for each metric.}
\label{tb_appendix: Jan15ManhattanRelocation_end2endPercentages}
\resizebox{\textwidth}{!}{%
\begin{tabular}{@{}lccccccccccccccccccc@{}}
\toprule
\textbf{} & \textbf{\begin{tabular}[c]{@{}c@{}}Distance\\ Driven (m)\end{tabular}} & \textbf{SD} & \textbf{\begin{tabular}[c]{@{}c@{}}Elapsed\\ Time (ns)\end{tabular}} & \textbf{SD} & \textbf{\begin{tabular}[c]{@{}c@{}}Time to\\ Pair (s)\end{tabular}} & \textbf{SD} & \textbf{\begin{tabular}[c]{@{}c@{}}Time to Pair\\ with Taxi (s)\end{tabular}} & \textbf{SD} & \textbf{\begin{tabular}[c]{@{}c@{}}Time to\\ Pick-up (s)\end{tabular}} & \textbf{SD} & \textbf{Delay (s)} & \textbf{SD} & \textbf{\begin{tabular}[c]{@{}c@{}}Cumulative\\ Delay (s)\end{tabular}} & \textbf{\begin{tabular}[c]{@{}c@{}}Driver\\ Profit (\$)\end{tabular}} & \textbf{SD} & \textbf{\begin{tabular}[c]{@{}c@{}}Number of \\ Shared Rides\end{tabular}} & \textbf{SD} & \textbf{Frictions (s)} & \textbf{SD} \\ \midrule
\textbf{Single}          & 0.00\%   & --         & -- & --           & -- & --                     & 0.00\%    & 0.00\%          & 0.00\%   & 0.00\%   & -- & --               & 0.00\%        & 0.00\%   & 0.00\%                 & -- & --        & 0.00\%   & 0.00\%   \\
\textbf{MWM}             & -40.83\% & --         & -- & --           & -- & --                     & -100.00\% & -100.00\%       & 14.09\%  & -22.80\% & -- & --               & 48.66\%       & 276.85\% & 666.95\%               & -- & --        & 445.65\% & 770.90\% \\
\textbf{MWM+R}           & -37.59\% & --         & -- & --           & -- & --                     & -100.00\% & -100.00\%       & -41.76\% & -63.70\% & -- & --               & -8.39\%       & 273.76\% & 477.77\%               & -- & --        & 213.66\% & 341.04\% \\
\textbf{ALMA}            & -30.01\% & --         & -- & --           & -- & --                     & -100.00\% & -100.00\%       & 74.14\%  & 14.07\%  & -- & --               & 104.17\%      & 273.23\% & 530.96\%               & -- & --        & 302.88\% & 496.98\% \\
\textbf{ALMA+R}          & -28.17\% & --         & -- & --           & -- & --                     & -100.00\% & -100.00\%       & -9.33\%  & -40.74\% & -- & --               & 20.65\%       & 270.69\% & 277.36\%               & -- & --        & 140.02\% & 170.78\% \\
\textbf{Greedy}          & -27.86\% & --         & -- & --           & -- & --                     & -100.00\% & -100.00\%       & 86.10\%  & 14.73\%  & -- & --               & 118.73\%      & 275.34\% & 516.38\%               & -- & --        & 328.69\% & 571.14\% \\
\textbf{Greedy+R}        & -27.17\% & --         & -- & --           & -- & --                     & -100.00\% & -100.00\%       & -7.97\%  & -41.75\% & -- & --               & 24.59\%       & 272.76\% & 261.87\%               & -- & --        & 128.27\% & 104.39\% \\ \bottomrule
\end{tabular}%
}
\end{table*}
